\documentclass[60pt]{article}
\usepackage{amsmath,amssymb,amsthm,amscd,mathtools, graphicx, mathtext,color,wrapfig, color, verbatim, slashed,  float}

\usepackage{adjustbox}
\usepackage{dsfont}
\usepackage{tikz-cd}
\usepackage{bbm}
\usepackage{upgreek}
\usepackage{extarrows} 
\usepackage{diagbox}
\usepackage[mathscr]{eucal}
\usepackage{mathrsfs}
\usepackage{pifont}
\usepackage{graphicx}
\usepackage{cleveref}
\usepackage[font={small,sl}]{caption}
\usepackage[all]{xy}
\usepackage{wasysym}
\usepackage{tikz}
\usetikzlibrary{matrix,arrows}

\usepackage[numbers,sort&compress]{natbib}

\usepackage{tikzit}
\tikzstyle{w puncture}=[fill=none, draw=none, shape=circle, text={rgb,255: red,226; green,121; blue,46}, tikzit fill={rgb,255: red,226; green,121; blue,46}]
\tikzstyle{z puncture}=[fill=none, draw=none, shape=circle, text={rgb,255: red,84; green,174; blue,50}, tikzit fill={rgb,255: red,84; green,174; blue,50}]
\tikzstyle{intersection point}=[fill=black, draw=black, shape=circle, minimum size=3pt, inner sep=0pt]
\tikzstyle{sl3 w1 puncture}=[fill=none, draw=none, shape=circle, text={rgb,255: red,50; green,116; blue,181}, tikzit fill={rgb,255: red,50; green,116; blue,181}]
\tikzstyle{sl3 w2 puncture}=[fill=none, draw=none, shape=circle, text={rgb,255: red,226; green,121; blue,46}, tikzit fill={rgb,255: red,226; green,121; blue,46}]
\tikzstyle{sl3 w2 dot}=[fill={rgb,255: red,226; green,121; blue,46}, draw=none, shape=circle, inner sep=0pt, minimum size=2.5pt]
\tikzstyle{dot}=[fill={rgb,255: red,50; green,116; blue,181}, draw=none, shape=circle, inner sep=0pt, minimum size=2.5pt]
\tikzstyle{intersection point 2}=[fill={rgb,255: red,84; green,174; blue,50}, draw={rgb,255: red,84; green,174; blue,50}, shape=circle, minimum size=4pt, inner sep=0pt]
\tikzstyle{su2 puncture}=[fill=none, draw=none, shape=circle, tikzit fill={rgb,255: red,218; green,59; blue,38}, text={rgb,255: red,218; green,59; blue,38}]
\tikzstyle{intersection point 3}=[fill={rgb,255: red,218; green,59; blue,38}, draw={rgb,255: red,218; green,59; blue,38}, shape=circle, minimum size=4pt, inner sep=0pt]

\tikzstyle{brane}=[draw={rgb,255: red,50; green,116; blue,181}, -, thick]
\tikzstyle{map}=[dashed, ->, draw={rgb,255: red,50; green,116; blue,181}]
\tikzstyle{w puncture strand}=[-, draw={rgb,255: red,226; green,121; blue,46}, thick]
\tikzstyle{z puncture strand}=[-, draw={rgb,255: red,84; green,174; blue,50}, thick]
\tikzstyle{arrow}=[->]
\tikzstyle{I-brane}=[-]
\tikzstyle{squiggle arrow}=[->, draw={rgb,255: red,187; green,0; blue,187}]
\tikzstyle{correction map}=[draw={rgb,255: red,255; green,128; blue,0}, ->]
\tikzstyle{disk}=[-, draw=none, tikzit draw={rgb,255: red,255; green,128; blue,0}, fill={rgb,255: red,191; green,191; blue,191}]
\tikzstyle{disk outlined}=[-, draw=cyan, fill=none, thick]
\tikzstyle{sl3 e1 brane}=[-, draw={rgb,255: red,50; green,116; blue,181}, thick]
\tikzstyle{sl3 e2 brane}=[-, draw={rgb,255: red,226; green,121; blue,46}, thick]
\tikzstyle{sl3 e1 map}=[draw={rgb,255: red,50; green,116; blue,181}, dashed, ->]
\tikzstyle{sl3 e2 map}=[draw={rgb,255: red,226; green,121; blue,46}, dashed, ->]
\tikzstyle{sl3 w1 puncture strand}=[-, thick, draw={rgb,255: red,50; green,116; blue,181}]
\tikzstyle{sl3 w2 puncture strand}=[-, thick, draw={rgb,255: red,226; green,121; blue,46}]
\tikzstyle{disk2}=[-, fill={rgb,255: red,128; green,128; blue,128}, draw=none, tikzit draw=magenta]
\tikzstyle{brane with arrow}=[-{Stealth[length=1.8mm, width=2mm]}, very thick, draw={rgb,255: red,50; green,116; blue,181}]
\tikzstyle{red I-brane}=[-, draw={rgb,255: red,218; green,59; blue,38}, very thick]
\tikzstyle{red I-brane with arrow}=[draw={rgb,255: red,218; green,59; blue,38}, -{Stealth[length=1.8mm, width=2mm]}, very thick]
\tikzstyle{su2 puncture strand}=[-, draw={rgb,255: red,218; green,59; blue,38}, thick]
\tikzstyle{braid}=[-, very thick]
\tikzstyle{braid arrow}=[very thick, -{Stealth[length=1.8mm, width=2mm]}, draw=black]

\usetikzlibrary{decorations.markings}
\newcommand{\beq}{\begin{equation}}
\newcommand{\eeq}{\end{equation}}

\newcommand{\bC}{\ensuremath{\mathbb{C}}}

\newcommand{\nc}{\newcommand}
\nc{\mc}{\mathcal}
\newcommand{\fg}{\mathfrak{g}}

\newcommand{\fq}{\mathfrak{q}}

\newcommand{\Lfgh}{\widehat{^L\mathfrak{g}}}

\newcommand{\MDX}{\mathscr{D}_{\cal X}}
\newcommand{\MDx}{\mathscr{D}_{ X}}
\newcommand{\MDY}{\mathscr{D}_{\cal Y}}
\newcommand{\MDy}{\mathscr{D}_{ Y}}
\newcommand{\MDA}{\mathscr{D}_{\mathscr A}}
\newcommand{\MDa}{\mathscr{D}_{ A}}

\newtheorem{theorem}{Theorem}

\newtheorem{corrolary}{Corollary}
\newenvironment{theorem*}
 {\expandafter\def\expandafter\thetheorem\expandafter{\thetheorem*}\theorem}
 {\endtheorem}
\newenvironment{theorem^!}
 {\expandafter\def\expandafter\thetheorem\expandafter{\thetheorem^!}\theorem}
 {\endtheorem}
\newenvironment{corrolary*}
 {\expandafter\def\expandafter\thecorrolary\expandafter{\thecorrolary*}\corrolary}
 {\endcorrolary}

\usetikzlibrary{arrows,chains,matrix,positioning,scopes}
\makeatletter
\tikzset{join/.code=\tikzset{after node path={%
\ifx\tikzchainprevious\pgfutil@empty\else(\tikzchainprevious)%
edge[every join]#1(\tikzchaincurrent)\fi}}}
\makeatother
\tikzset{>=stealth',every on chain/.append style={join},
         every join/.style={->}}
\tikzstyle{labeled}=[execute at begin node=$\scriptstyle,
   execute at end node=$]
\DeclareMathAlphabet{\pazocal}{OMS}{zplm}{m}{n}
\newcommand{\B}{\pazocal{B}}

\begin{document}
\baselineskip=28pt  % a la harvmac
\baselineskip 0.7cm

\setcounter{tocdepth}{2}
\numberwithin{equation}{section}
\begin{titlepage}

%% Set the number of the title with 0

% change the footnote symbol
\renewcommand{\thefootnote}{\fnsymbol{footnote}}

\vskip 1.0cm

\begin{center}
{\LARGE \bf
 Homological Link Invariants 
\vskip 0.5 cm
from Floer Theory
\vskip 0.5 cm}

{\large
Mina Aganagic$^{1,2}$, Elise LePage$^{1}$, Miroslav Rapcak$^{3}$}

\medskip

\vskip 0.5cm
{\it
$^{1}$Physics Department, University of California, Berkeley, USA\\
$^{2}$Departments of Mathematics, University of California, Berkeley, USA\\
$^{3}$CERN, Theory Department,
 Geneva, Switzerland}

\end{center}

\vskip 0.5cm

%-----------------------------------------
\centerline{{\bf Abstract}}
\medskip

\noindent{}There is a generalization of Heegaard-Floer theory from ${\mathfrak{gl}}_{1|1}$ to other Lie (super)algebras $^L{\fg}$. 
The corresponding category of A-branes is solvable explicitly and categorifies quantum $U_q(^L\fg)$ link invariants. The theory was discovered in \cite{A1,A2}, using homological mirror symmetry. It has novel features, including equivariance and, if  $^L{\fg} \neq {\mathfrak{gl}}_{1|1}$, coefficients in categories.
In this paper, we give a detailed description of the theory and how it is solved in the two simplest cases: the ${\mathfrak{gl}}_{1|1}$ theory itself, categorifying the Alexander polynomial, and the ${\mathfrak{su}}_{2}$ theory, categorifying the Jones polynomial. Our approach to solving the theory is new, even in the familiar ${\mathfrak{gl}}_{1|1}$ case.

\noindent\end{titlepage}

\setcounter{page}{1} % don't number title page

\setcounter{section}{0}
%%%%%%%%%\\
\tableofcontents 
\newpage
\section{Introduction}\label{Intro}

In 1928, Alexander constructed a polynomial invariant of links in ${\mathbb R}^3$ \cite{Alexander}. In one of its formulations, it is defined by picking a projection of the link to a plane and the skein relation
\beq  
\includegraphics[scale=0.27]{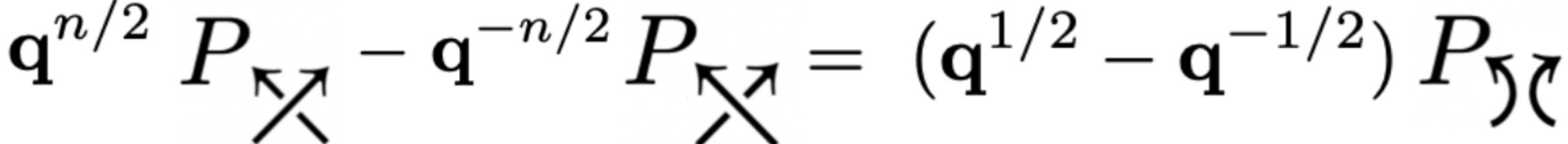}
\eeq
\noindent{}with $n=0$. 
The same skein relation with $n=2$ instead defines the polynomial link invariant discovered by Jones in 1988 \cite{JonesVFR}.

Witten showed that these invariants originate from Chern-Simons theory with gauge group based on a Lie algebra $^L{\fg}$ \cite{Jones}. In particular, the Jones polynomial comes from choosing $^L{\fg} = {\mathfrak{su}}_2$, with links colored by the defining, two-dimensional representation. The Alexander polynomial, comes from the same setting by taking 
$^L{\fg}$ to be the Lie superalgebra ${\mathfrak{gl}}_{1|1}$. The resulting link invariants are the $U_{\fq}(^L{\fg})$ quantum group invariants of Reshetikhin and Turaev \cite{RT}. 
\subsection{The categorification problem}
The quantum invariants of links are Laurant polynomials in ${\fq}^{1/2}$,
with coefficients  which are integers (up to an overall normalization). Crane and Frenkel \cite{CF} used this to suggest that link invariants may be shadows of a deeper, more fundamental theory, in which one can associate to a projection of the link to a plane
a bi-graded complex of abelian groups
$$C^{*, j}(K) = \ldots C^{i-1, j} (K)\xrightarrow{\partial^{i-1}} C^{i,j}(K) \xrightarrow{\partial^{i}} \ldots, $$ 
such that each homology group
${\cal H}^{i, j}(K) = {{\rm ker} \,{\partial^i}/ {\rm im}\, \partial^{i-1}}$ 
is an invariant of the link projection and whose equivariant Euler characteristic 
$$\chi(K) = \sum_{i,j} (-1)^i {\fq }^{j} {\rm dim} {\cal H}^{i,j}(K),
$$
is the quantum group link invariant. Independently, existence of such a theory was predicted by string theory. This was explained by Gukov, Schwarz and Vafa in \cite{GSV}, following Ooguri and Vafa in \cite{OV}.

\subsubsection{}
A toy model for categorification comes
from a Riemannian manifold $M$, whose Euler characteristic
\begin{equation*}
    {\chi (M)} = \sum _{k\in \mathbb{Z}} (-1)^{k} \mathrm{dim}\,{\mathcal H}^{k}(M)
\end{equation*}
is categorified by the cohomology
${\mathcal H}^{k}(M) = {\mathrm{ker} \;{d_{k}}/ \mathrm{im}\; d_{k-1}}$
of the de Rham complex
\begin{equation*}
    C^{*} = \cdots C^{k-1} \xrightarrow{d_{k-1}} C^{k} \xrightarrow{d_{k}} \cdots.
\end{equation*}
The Euler characteristic is, from the physics perspective, the partition
function of supersymmetric quantum mechanics with $M$ as a target space:
$\chi (M) = \mathrm{Tr} (-1)^{F} e^{-\beta H}$, with the Hamiltonian being
the Laplacian $H=d d^{*}+ d^{*} d$ and $d=\sum _{k} d_{k}$ the supersymmetry
operator. If $h$ is a Morse function on $M$, the complex can be replaced
by a Morse--Smale--Witten complex $C^{*}_{h}$ with the differential
$d_{h} = e^{h} d e^{-h}$. The complex $C^{*}_{h}$, as Witten showed
\cite{WittenM}, is the space of perturbative ground states of a
$\sigma $-model on $M$ with potential $h$. The action of the differential
$d_{h}$ is generated by instantons.

\subsubsection{}
Both the Alexander and the Jones polynomial have known categorifications; however, they have very different flavor.

A categorification of the Jones polynomial was discovered in 1999, by Khovanov \cite{Kh, KhICM}. Khovanov's construction is remarkable for being simple and completely explicit. However, unlike in our toy example of categorification of the
Euler characteristic of a Riemannian manifold, the construction is purely algebraic -- it does not originate from either geometry or physics. If we replace the Lie algebra ${\mathfrak{su}_2}$ by an arbitrary (ordinary) Lie algebra $^L{\fg}$, there is an abstract algebraic framework for categorification of quantum link invariants due to \cite{webster}, based on a derived category of modules of an algebra known as the KLRW algebra, constructed in \cite{KL1, R, webster}. Unlike Khovanov's construction, Webster's categorification exists only formally.

The Alexander polynomial is categorified by Heegaard-Floer theory, discovered in 2002 by Ozsvath and Szabo \cite{OS0a,OS0b,OS0c,OS1,OS}. It is based on Floer theory, or the A-model, with target $Y_{\mathfrak{gl}_{1|1}}$ which is a symmetric product of Riemann surfaces ${\cal A}$. 
Floer theory is modeled after the Morse theory
approach to supersymmetric quantum mechanics, applied to a theory in one higher dimension. The vector spaces underlying the chain complexes are spanned by intersection points of a pair of A-branes constructed from the link. The differential is defined by counting disk instantons. Unlike a generic A-model, Heegard-Floer theory turned out to be solvable explicitly \cite{grid, LOT}.

\subsubsection{}
The link categorification problem is to find a general framework for link homology theories that works uniformly for all Lie algebras and ideally originates from physics or geometry.  The theory should explain what link homology groups are and why they exist. In addition, it is not unreasonable to hope that it may be solvable.

Chern-Simons theory is a rare example of an explicitly solvable yet non-trivial quantum field theory. If one is fortunate enough, the theory whose Euler characteristic Chern-Simons theory computes should be solvable as well. Previous attempts \cite{KR1, KR2, SS,CK1, CK2, WF} gave theories that were either not general, not solvable, or both.

\subsection{The solution}
There is a generalization of Heegaard-Floer theory from ${\mathfrak{gl}}
_{1|1}$ to other Lie super-algebras $^L{\fg}$, discovered in \cite{A1, A2}. The theory manifestly categorifies quantum group link invariants. It provides a rare example of a category of branes which is solvable explicitly, yet highly non-trivial.

In this paper, we give an explicit description of this new Floer theory and how it is solved. We give an algorithm for computing link homologies that applies for any $^L{\fg}$. We describe the theory in detail in the two simplest cases, corresponding to ${\mathfrak{gl}_{1|1}}$ and to ${\mathfrak{su}_{2}}$. We prove explicitly Thm.~\ref{tinv}, that the theory gives homological link invariants and Thm.~\ref{tJ}, that it its Euler characteristics satisfy the skein relations (1.1). The approach to solving the theory is new, even in the familiar context of the ${\mathfrak{gl}_{1|1}}$ theory.

In the course of solving the theory, one also discovers that KLRW algebras arise naturally as algebras of opens strings in the Floer theory associated to simply laced Lie algebras $^L{\fg}$, as proven in \cite{ADZ}. We will describe the extension to other $^L{\fg}$, including to non-simply laced Lie algebras, and Lie superalgebras ${\mathfrak{gl}}(m|n)$ and ${\mathfrak{sp}}(m|2n)$  elsewhere.

\subsection{Acknowledgments}
M.A. is grateful to Vivek Shende and Peng Zhou for many discussions about Floer theory, and for collaboration on closely related work \cite{ADZ}. 
We thank the organizers and participants of the AIM workshop on ``Floer theory of symmetric products and Hilbert schemes", in December 2022, for a stimulating workshop which centered, in part, on this work. M.A. especially benefited from discussions with Mohammed Abouzaid, Robert Lipshitz and Ivan Smith during the workshop.
This research is supported by the NSF foundation grant PHY2112880 and by the Simons Investigator Award.

\section{Overview of the theory}
The section gives a brief overview of the theory for any $^L{\fg}$. In later sections, we will describe in detail the theories in the two special cases, corresponding to $^L{\fg} = \mathfrak{gl}_{1|1}$ and $\mathfrak{su}_{2}$.

We will study framed links in 
${\cal A} \times {\mathbb R}$
where ${\cal A} = {\mathbb C}^{\times}$ is an infinite complex cylinder. For this choice of ${\cal A}$, the theory will have two descriptions, related by mirror symmetry. Mirror symmetry will make complementary aspects of the theory easy to understand.

\subsection{Homological link invariants from A-branes}\label{how}
Pick $^L{\fg}$, which is either a simply laced Lie algebra, or the Lie superalgebra ${\mathfrak{gl}}(m|n)$. Pick also a collection of marked points on ${\cal A}$, or rather pairs of them, labeled by a minuscule representation $V_i$ of  $^L{\fg}$, and its conjugate $V_i^*$. The geometric structures below are associated to the zero weight subspace of $\bigotimes_i V_i \otimes V_i^*$, where $i$ runs over all pairs of punctures. The theory extends to non-simply laced Lie algebras $^L{\mathfrak{g}}$, and to ${\mathfrak{sp}(m|2n)}$, with some minor modifications. It is possible it exists more generally yet.

\subsubsection{}\label{choice}
A ``matching" is a collection of curves on ${\cal A}$ which have endpoints on a set of marked points, colored by pairs of conjugate representations. For every two such matchings, colored red and blue, we get a link in ${\cal A}\times {\mathbb R}$ by taking the red matching to overpass the blue.
Arbitrary links arise in this way. 

To assign a pair of matchings to a link, recall that any link has a representation as a ``plat" closure of a braid,  with a sequence of $d$ caps at the top and $d$ cups at the bottom. The result is due to Birman \cite{Birman}.
\begin{figure}[H]
\begin{center}
    \includegraphics[scale=0.23]{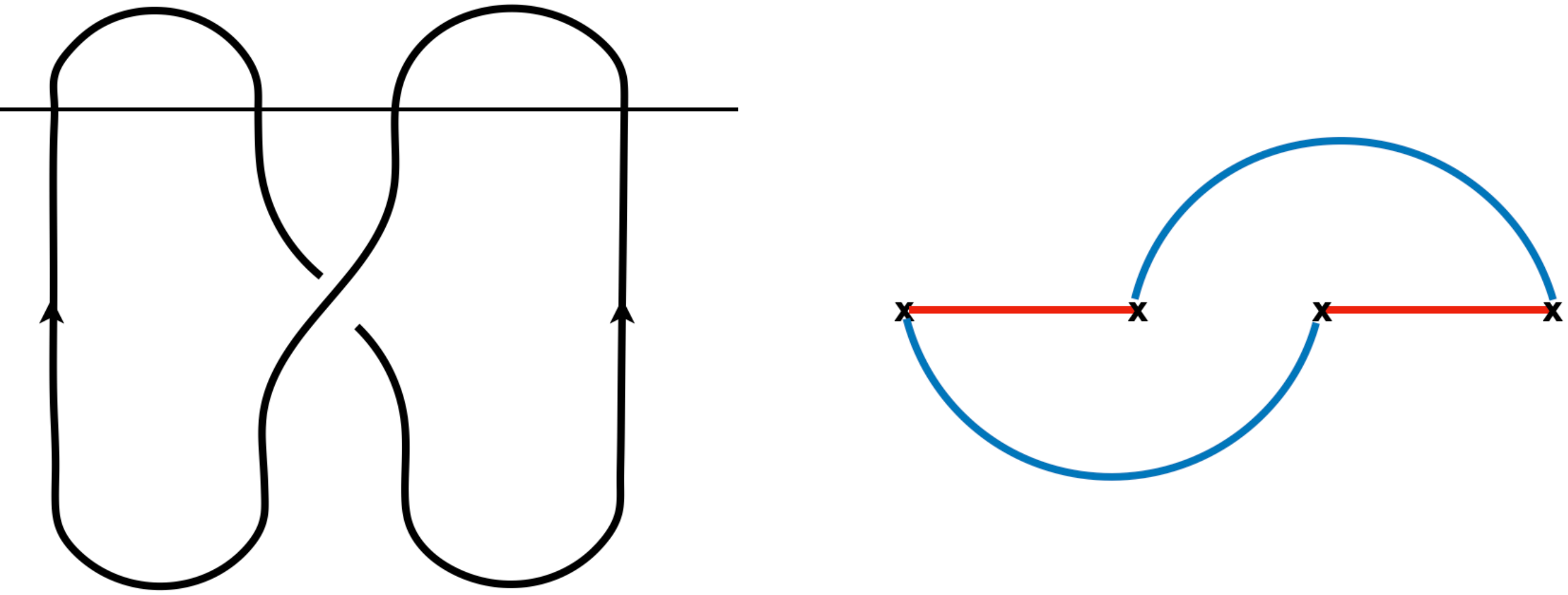}
    \caption{A pair of matchings derived from a plat closure representation of the unknot, for $d=2$.}\label{fT-1}
\end{center}
\end{figure}
To the top closure, we will assign a ``standard" red matching. It will be useful to isotope its endpoints to lie on a fixed circle in ${\cal A}$. To the bottom closure, we will assign the same standard matching but isotoped by the braid and colored blue. The braid moves are read from bottom to top.

\subsubsection{}
To the Riemann surface ${\cal A}$ with marked points colored by pairs of conjugate, minuscule representations of $^L{\fg}$, we will associate a category. The category 
$$
\MDy = D({\cal FS}(Y, W))
$$
is a category of equivariant A-branes on an exact symplectic manifold $Y=Y_{^L{\fg}}$ of $2c_1(K_Y)=0$, equipped with a choice of top holomorphic form $\Omega=\Omega_{^L{\fg}}$ and with a superpotential $W=W_{^L{\fg}}$. The symplectic form must be compatible with $\Omega$ in the sense that both give rise to the same real volume form $\Omega \wedge {\overline \Omega}$. A comprehensive introduction to derived Fukaya categories is \cite{Seidel}; a brief review is in \cite{Auroux}.

\subsubsection{}
The target $Y$ is a product of symmetric products, one for each node of the Dynkin diagram 
\beq\label{whY}
Y = \prod_{a=1}^{\rm{rk}{^L\fg}} Sym^{d_a}({\cal A}),
\eeq
with potential $W$ that couples them written in \cite{A2}. An important special feature of the theory is that $W$ is not single-valued.

\subsubsection{}
Objects of $\MDy$ are Lagrangians $L$ in $Y$, together with extra structures. Their grading is a choice of a lift of the phase of 
\beq\label{gradingG}\Omega e^{-W}|_L,
\eeq
to a single-valued function on the brane.  The choice of a lift of the phase of $\Omega|_L$ is the cohomological, or Maslov grading of the brane.  Since $W$ is not single-valued, the choice of a lift of the phase of $e^{-W}|_L$ leads to additional, equivariant, gradings in the theory. As always, the branes may be equipped with a local system and a choice of spin structure. We will see that both will play a role.

\subsubsection{}
To the blue and red segments we will assign a pair of A-branes
 $$I_{\mathcal U}, \;\; {\mathscr B} E_{\mathcal U}\; \in \MDy.$$ 
The branes depend on $^L{\fg}$ and on representations coloring the link components.
The choice of branes is explained by equivariant mirror symmetry when $^L{\fg}$ is simply laced.

\begin{figure}[H]
\begin{center}
    \includegraphics[scale=0.40]{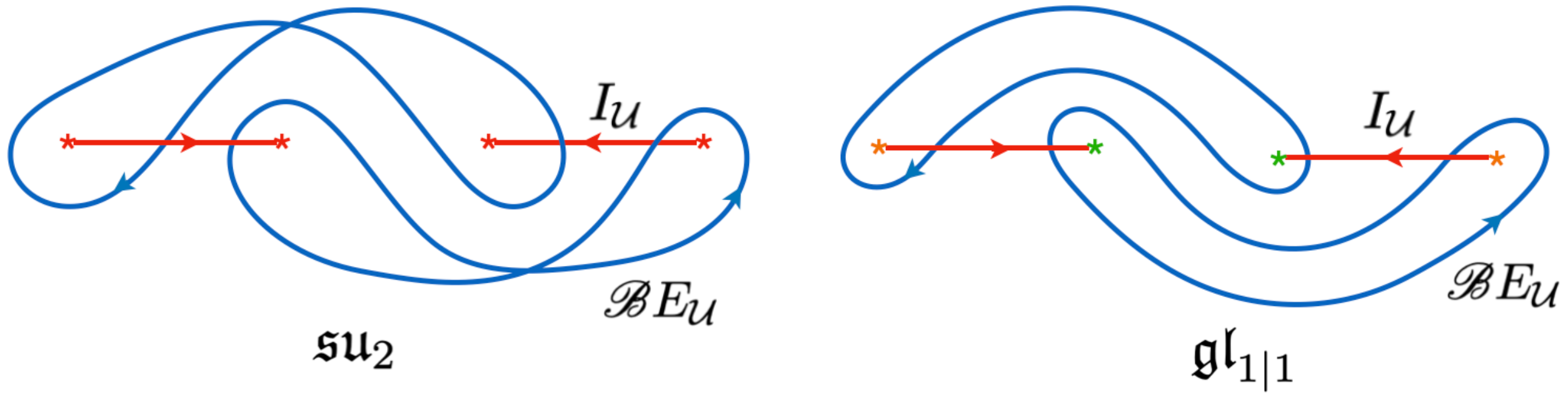}
    \caption{The branes corresponding to the unknot projection in Fig.~\ref{fT-1} in the $^L{\fg} =  {\mathfrak{su}}_2$ and ${\mathfrak{gl}}_{1|1}$ theories.}\label{B2as}
\end{center}
\end{figure}

Homological $U_{\fq}(^L{\fg})$ link invariants are the spaces of morphisms between the branes, viewed as objects of $\MDy$:
\beq\label{MP}
Hom^{*,*}_{\MDy}({\mathscr B} E_{\mathcal U},I_{\mathcal U}) = \bigoplus_{k,
 {\vec J} }
 Hom_{\MDy}({\mathscr B} E_{\mathcal U},I_{\mathcal U}[k]\{{\vec J}\}).
\eeq
Here, $[k]$ is ${\mathbb Z}$-valued cohomological, or Maslov degree.  The vector ${\vec J}$ is the equivariant degree, whose $J_0$ component is half integral, and related to ${\fq}$.    
The ${\rm rk}^L{\fg}$ additional equivariant ${\mathbb Z}$-gradings couple to holonomies of Chern-Simons gauge fields on the $S^1$.

\subsubsection{}
Spaces of morphisms between the branes in $\MDy$ are Floer cohomology groups \cite{Seidel}:
\beq\label{Fd}
Hom^{*,*}_{\MDy}({\mathscr B} E_{\mathcal U},I_{\mathcal U}) ={{\rm Ker}\, \delta_F}/{{\rm Im} \,\delta_F}.
\eeq
Floer theory, introduced in \cite{Floer}, generalizes the Morse theory approach to supersymmetric quantum mechanics, to a theory in one dimension up. The Floer complex is a vector space 
\beq\label{CF}
    hom_Y^{*,*}({\mathscr B} E_{\mathcal U},I_{\mathcal U})\equiv CF^{*,*}({\mathscr B} E_{\mathcal U},I_{\mathcal U}) = \bigoplus_{{\cal P} \in {\mathscr B} E_{\mathcal U} \cap I_{\mathcal U}} {\mathbb C} {\cal P},
\eeq
spanned by the intersection points of the branes,
keeping track of gradings, equipped with a differential $\delta_F$ 
\beq\label{Floer}
\delta_F: hom_Y^{*,*}({\mathscr B} E_{\mathcal U},I_{\mathcal U}) \rightarrow hom_Y^{*+1,*}({\mathscr B} E_{\mathcal U},I_{\mathcal U}).
\eeq
The space $hom^{*,*}_Y({\mathscr B} E_{\mathcal U},I_{\mathcal U})$ of morphisms between the branes in the ordinary Fukaya category is the space of perturbative supersymmetric ground states of the ${\cal N}=2$ Landau-Ginsburg model on an interval, with the pair of A-branes as boundary conditions.
The space $Hom^{*,*}_{\MDy}({\mathscr B} E_{\mathcal U},I_{\mathcal U})$ of morphisms in the derived Fukaya category  is spanned by the exact supersymmetric ground states.

\subsubsection{}\label{FTD}
The action of the Floer differential in \eqref{Floer} is defined
by counting holomorphic maps $y:{\rm D} \rightarrow Y$ from the unit disk ${\rm D}$ with two marked points to $Y$. The unit disk with two marked points is biholomorphic to an infinite strip of unit width.
The coefficient of ${\cal P}'$ in
$\delta_{F}{\cal P}$ is the count of maps
for which the two boundary components of ${\rm D}$ map to the branes ${\mathscr B} E_{\mathcal U}$ and $I_{\mathcal U}$ and the two marked points to ${\cal P}$ and ${\cal P}'$. Our orientation conventions are spelled out in \cite{A2} and are the same as in \cite{Auroux}. 

Holomorphic maps that contribute to the Floer differential, or the higher $A_{\infty}$ maps, have finite symplectic area. In a physically sensible A-model, the potential $W$ must pull back to a regular function on the disk ${\rm D}$ \cite{A2}. For theories with potentials which are regular this condition is vacuous, but not for us. It implies the equivariant degree of all Floer theory maps is $0$. Maps that contribute to the differential $\delta_F$ have Maslov index $1$. This is the index of the linearized Cauchy-Riemann operator for a fixed conformal structure on ${\rm D}$. After dividing by reparameterizations of the unit disk keeping the two boundary marked points fixed, the moduli space of maps that contribute becomes zero dimensional.

\subsubsection{}
There are no general means for computing the action of the Floer differential $\delta_F$ from its definition, rather, one has to approach the problem instanton by instanton. In Heegaard-Floer theory, for some very special link presentations \cite{grid}, one can compute the differential by arranging that only a few different kinds of disks appear; however, there is nothing canonical about it.

In sufficiently nice derived Fukaya categories, there is an alternate approach. In our setting, it will enable us to compute the action of differential, and hence the Floer cohomology groups, algebraically. 

In what follows, we will describe the category $\MDy$ explicitly, and how to solve it. We will give an algorithm for computing $Hom_{\MDy}^{*,*}({\mathscr B} E_{\mathcal U},I_{\mathcal U})$, and prove the resulting vector spaces are link invariants.

\subsubsection{}
To keep the length of the paper contained, will specialize to the two simplest cases, when the Lie algebra is either ${\mathfrak{su}}_2$ or ${\mathfrak{gl}}_{1|1}$, 
with links colored by their two-dimensional defining representations.  In both cases, 
$Y$ is the symmetric product of $d$ copies of ${\cal A}$
\beq\label{YwoD}
Y=Sym^{d}{\mathcal (A)},
\eeq
but with different $\Omega$ and $W$. A point in the symmetric product is $d$ unordered points on ${\cal A}$. Correspondingly, an object in $\MDy$, the category of $A$-branes on $Y$, is a product of $d$ one-dimensional Lagrangians on ${\cal A}$, which we take to be disjoint. 

In both cases, the $I_{\cal U}$ branes come from Lagrangians on $Y=Sym^d({\cal A})$ that are simply the product of the corresponding $d$ red segments on ${\cal A}$. The ${\mathscr B} E_{\cal U}$ branes are obtained by replacing the blue segments with products of $d$ figure eights in the ${\mathfrak{su}}_2$ case, or ovals for ${\mathfrak{gl}}_{1|1}$, as in Fig.~\ref{B2as}.

\subsection{Algebra of Thimbles}
The category of A-branes on $Y$ is generated by a finite set of branes. As the generating set, we will take the branes
$$
T_{\cal C} = T_{i_1} \times \ldots \times T_{i_d},
$$
which are products of $d$ real line Lagrangians on ${\cal A}$, where $d$ is the dimension of $Y$, colored by simple roots of the Lie algebra.

\begin{figure}[h!]
\begin{center}
     \includegraphics[scale=0.4]{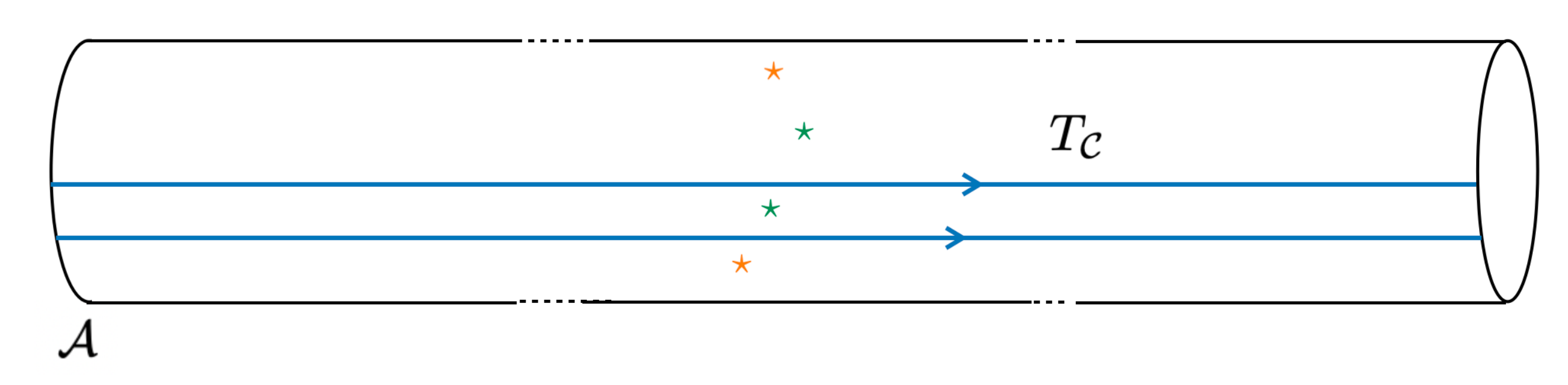}
 \caption{A $T_{\cal C}$-brane for $^L{\fg} =  {\mathfrak{gl}}_{1|1}$ with $d=2$.  }\label{T}
\end{center}
\end{figure}
There is a $T_{\cal C}$-brane for every inequivalent ordering of real line Lagrangians on ${\cal A}$. For $Y$ as in \eqref{whY}, every $T_{\cal C}$ brane has $d_a$ of its real line Lagrangians colored by a simple root $^Le_a$. 
Without changing the $T_{\cal C}$-brane, a one-dimensional component of it colored by a simple root $^Le_a$ can be isotoped through a marked point provided $\langle ^Le_a, \mu_j\rangle =0$, where $\mu_j$ is the highest weight labeling the puncture; it can pass through another real line Lagrangian colored by $^Le_b$ if $\langle ^Le_a, ^Le_b\rangle =0$.

\subsubsection{}
The $T$-branes should be thought of as thimbles \cite{HIV, WA} of the potential $W$, which $Y$ is equipped with, in a chamber of the parameter space \cite{A2}. 
The thimbles are the set of all initial conditions for upward gradient flows of ${\rm Re} (W)$, on which ${\rm Im} (W)$ is constant, and which asymptote to a critical point of the potential. 

One subtlety is that, to generate the category of A-branes with the $T$-branes, one has include to additional branes which one can think of as thimbles associated to ``critical points at infinity". Another subtlety, explained in section \ref{ssu2}, is that when $^{L}{\fg}\neq {\mathfrak{gl}_{1|1}}$, the $T_{\cal C}$-branes used to generate $\MDy$ get equipped with an additional structure, which is a local system of modules for a certain finite rank algebra.

\subsubsection{}
The category $\MDy$ of A-branes on ${Y}$ being generated by the $T$-brane 
\beq\label{Tgen}
T = {\bigoplus}_{\cal C} T_{\cal C},
\eeq
implies the equivalence
\beq\label{DA}
\MDy \cong \MDa
\eeq
where $\MDa$ is the derived category of modules of an algebra $A$,
\beq\label{Aa}
A = hom_Y^{*,*}(T,T), 
\eeq
which is the endomorphism algebra of the generator $T$.
The sum in \eqref{Tgen} is over all the thimbles, including those associated to critical points at infinity. $A$ is the algebra of open strings that begin and end on the branes. 

One of the virtues of this generating set of branes is that the algebra $A$ is computable explicitly and in a sense, as simple as possible. 
When $^L{\fg}$ is a Lie superalgebra, the algebra $A$ is an associative graded algebra with a differential. For ${\mathfrak{ gl}}_{1|1}$, we will describe the algebra in section \ref{gl11}. When $^L{\fg}$ is an ordinary simply laced Lie algebra, the algebra $A$ is even simpler, as it is an ordinary associative algebra in cohomological degree zero. For ${\mathfrak{ su}}_{2}$, it is described in section \ref{ssu2}. 

\subsubsection{}
One should think of the equivalence in \eqref{DA} as one half of homological mirror symmetry,
\beq\label{DAX}
\MDx \cong \MDa \cong \MDy.
\eeq
Here $X$ is the mirror of $Y$, and $\MDx$ is the category of B-type branes on $X$, working equivariantly with respect to a torus ${\mathbb T}$. When $^L{\fg}$ is an ordinary Lie algebra, $\MDx$ is the derived category of ${\mathbb T}$-equivariant coherent sheaves on $X$; in Lie superalgebra cases, it is a category of matrix factorizations.  The $T$-branes on $Y$ map to vector bundles on $X$ with the same label. Mirror to adding to $Y$ its multi-valued potential $W$ is working equivariantly with respect to the torus ${\mathbb T}$. Conversely, turning off the ${\mathbb T}$-action sets $W$ to zero.

\subsubsection{}\label{nocritical}
Every $T_{\cal C}$-brane associated to an honest critical point of $W$ has a dual $I_{\cal C}$-brane, which satisfies that the only non-vanishing Homs from the $T$-branes to the $I$-branes are 
\beq\label{dual}
hom_Y(T_{\cal C}, I_{{\cal C}'}) ={\delta}_{{\cal C}, {\cal C}'} {\mathbb C},
\eeq
in Maslov and equivariant degree zero. 
The $I_{\cal C}$-brane is its dual thimble, corresponding to the set of all initial conditions of downward gradient flows of ${\rm Re} \;W$. Correspondingly, a $T_{\cal C}$-brane intersects only the dual $I_{{\cal C}'}$-brane, and only once, at the critical point of the potential, leading to \eqref{dual}.
Among the $I$-branes, one finds included the $I_{\cal U}$-branes that serve as caps. There is a Koszul dual way to generate $\MDy$, in terms of $I$-branes instead of $T$-branes \cite{A2}.

\subsubsection{}
The equivalence of categories in \eqref{DA} comes from the Yoneda functor 
\beq\label{Yoneda}
    hom_Y(T, -):{\MDy} \rightarrow \MDa,
\eeq
which maps A-branes on $Y$ to complexes of modules of the algebra $A$. It maps $T_{\cal C}$-branes to projective modules of the algebra $A$, and the dual $I_{\cal C}$-branes to simple modules. A projective module is a direct summand of a free module. Since an algebra is always a free module for itself, the free module whose summand is $T_{\cal C}$ corresponds to the $T=\bigoplus_{\cal C} T_{\cal C}$ brane. 

\subsection{Floer complexes from $T$-brane algebra}\label{tores}
Like every brane in $\MDy$, the braided cup branes have a description as a complex,  every term of which is a direct sum of $T$-branes. The complex $({\mathscr B}E(T), \delta)$ is a direct sum brane ${\mathscr B}E(T)$, together with a differential $\delta$. The complex describes how to get the ${\mathscr B} { E}_{\mathcal U} \in \MDy$ brane
by starting with the direct sum of the generators
\beq\label{aproB}
 {\mathscr B} { E}(T)=\bigoplus_k  {\mathscr B} { E}_k(T) [k],
\eeq
and deforming differential $\delta \in A = hom_Y^{*,*}(T,T)$
away from zero. The differential $\delta$ acts on the brane as a cohomological degree one and equivariant degree zero operator 
$$\delta:  {\mathscr B} { E}(T)\xrightarrow{\delta} {\mathscr B} { E}(T)[1]$$
that
squares to zero in the algebra $A$ (in an appropriate sense).

\subsubsection{}
Per definition, the direct sum brane ${\mathscr B} { E}(T)$ together with the differential $\delta$, is equivalent to the brane we started with, as object of ${\MDy}$
\beq\label{PR}
{\mathscr B} E_{\cal U} \cong ( {\mathscr B} { E}(T), \delta).
\eeq
 Since the $T$-branes map to projective modules of the algebra $A$ by the Yoneda functor \eqref{Yoneda}, the complex $( {\mathscr B} { E}(T), \delta)$ is a projective resolution of the ${\mathscr B} E_{\cal U}$ brane. In particular, this means that all Homs of the ${\mathscr B} E_{\cal U}$-brane,  viewed as a Lagrangian on $Y$ are equivalent in  $\MDy$ to the Homs of the  complex  $( {\mathscr B} { E}(T), \delta)$ resolving the brane.

\subsubsection{}
Because the $I_{\cal U}$-brane is one of the simples of the algebra $A$,
from the projective resolution of the ${\mathscr B}E_{\cal U}$-brane we get for free a complex of vector spaces
\begin{equation}\label{resolutionhom}
    (\hom^*_Y\bigl( {\mathscr B} { E}(T), I_{\mathcal U}\{\vec J\}\bigr), \delta_F),
\end{equation}
one for each $\{\vec J\}$, whose cohomology is the link homology group we are after:
\begin{equation}\label{final}
\mathrm{Hom}_{\mathscr{D}_{Y}}\bigl({\mathscr B} E_{\mathcal U},I_{\mathcal U}[k]
\{ {\vec J}\}\bigr) = H^{k}\bigl(\hom^*_Y({\mathscr B}E_{\mathcal U}(T), I_{\mathcal U}\{ {\vec J}\})\bigr).
\end{equation}
The complex of vector spaces in \eqref{resolutionhom} is obtained from the (twisted) complex of branes $({\mathscr B}E(T), \delta)$  by applying
the functor  
\beq\label{hI} hom^{*}_Y( -, I_{\cal U}\{\vec J\}) =\bigoplus_{k\in {\mathbb Z}}\hom_Y( -, I_{\cal U}[k]\{\vec J\}),
\eeq
and making use of the duality between $T$-branes and the
$I$-branes in \eqref{dual}.  The fact that the $k$-th cohomology group of the complex on the right hand side in \eqref{final} agrees with the Hom space in $\MDy$ on the left is a standard piece of homological algebra. We will explain this in section \ref{FR} in more detail.

\subsubsection{}\label{tojust}
The complex we get in~\eqref{resolutionhom} is the Floer complex. By construction of the direct sum brane ${\mathscr B}E(T)$ from section~\ref{Algorithm}, the vector space we get at the $k$-th term of the complex, in equivariant degree
$\vec{J}$, can be identified with the space spanned by the intersection points of the $\mathscr{B} E_{\mathcal U}$-brane and the
$I_{\mathcal U}$-brane, in the same degree
\begin{equation}\label{atop}
hom_Y( {\mathscr B} { E}_{\mathcal U}, I_{\mathcal U}[k]\{\vec{J}\}) =\hom_Y\bigl( {\mathscr B} { E}_{k}(T), I_{\mathcal U}\{\vec{J}\}\bigr).
\end{equation}
Since, in addition, the cohomology of the complex coincides with Floer cohomology, it follows that the functor in \eqref{hI} maps the brane differential $\delta$, to the Floer differential $\delta_F$ in \eqref{Floer}. Thus, knowing $\delta$, we obtain the Floer differential $\delta_F$ without counting holomorphic curves.

\subsubsection{}
It follows that the link homology $\mathrm{Hom}^{*,*}_{\mathscr{D}_{Y}}\bigl({\mathscr B} E_{\mathcal U},I_{\mathcal U} \bigr)$ captures only a small part of the geometry
of ${\mathscr B} E_{\mathcal U}$, the braided cup brane. Because the
$T$-branes are dual to the $I$-branes by~\eqref{dual}, almost all terms in
the complex~\eqref{resolutionhom} vanish. 
The cohomology~\eqref{final} of the complex that remains is the $U_{\mathfrak q}(^L{\mathfrak g})$ link homology.

\subsection{Decategorification}\label{decat}
The category of A-branes $\MDy$ manifestly categorifies  $U_{\mathfrak{q}}(^{L}{\mathfrak{g}})$ link invariants \cite{A1,A2}.

\subsubsection{}\label{sCFT}
A conformal block of $\widehat{^L\mathfrak{g}}$ on ${\cal A}$ corresponding to a brane $L\in \MDy$ has an integral representation 
\beq\label{cb}
{\cal V}_{\alpha}[{ L}] = \int_L \;\Phi_{\alpha}\; \Omega \; e^{-W},
\eeq
developed by Feigin and Frenkel \cite{FF} and Schechtman and Varchenko \cite{SV1, SV2} (see  \cite{EFK} for review) for simple Lie algebras $^L{\fg}$. ${\cal V}_{\alpha}[{ L}]$ are solutions of the trigonometric version of the Knizhnik-Zamolodchikov \cite{KZ} equation of $\Lfgh$ because ${\cal  A}$ is an infinite cylinder. 
It depends on the brane only through its $K$-theory class $[L]$.  

The integral differs from the equivariant central charge function ${\cal Z}:{\MDy}\rightarrow {\mathbb C}$ of the A-brane category
\beq\label{zb}
{\cal Z}[{ L}] = \int_L \Omega \; e^{-W}
\eeq
by insertions of chiral ring operators ${\Phi_{\alpha}}$ \cite{A2}. These are rational functions on $Y$, so the actions of monodromies on \eqref{cb} and \eqref{zb} coincide. The equivariant central charge functions are also conformal blocks, this time of a $W$-algebra associated to the Lie algebra $\fg$, which is Langlands dual to $^L{\fg}$ \cite{AFO}. 

Conformal blocks corresponding to collections of caps and cups come from $K$-theory classes of branes  $E_{\mathcal U}$ and $I_{\mathcal U}$ in $\MDy$, which are proportional, with constant of proportionality that is a ${\fq}$-number.

\subsubsection{}\label{sCFT2}
If the braiding functor takes $L$ to ${\mathscr B}{ L} \in \MDy$, its action on K-theory classes of branes $[{\mathscr B}{ L}] ={\mathfrak B}[{ L}] $ is obtained from Picard-Lefshetz theory. In our case, ${\mathfrak B}$ represents the action of a braid group element on the space of conformal blocks,
\beq\label{BV}
    {\cal V}_{\alpha}[{\mathscr B}{ L}] = {\mathfrak B}{\cal V}_{\alpha}[{ L}],
\eeq
so by the
Kohno-Drinfeld theorem \cite{Drinfeld}, ${\mathfrak B}$ is an element of the $U_{\fq}(^L\fg)$ quantum group.
It is a product of R-matrices, which describe exchanges of neighboring pairs of vertex operators associated with elementary braid moves.

\subsubsection{}\label{sCFT3}
The $U_{\fq}(^L\fg)$ invariant of a link described as plat closure of a braid is the matrix element of ${\mathfrak B}$ between conformal blocks $[E_{\cal U}]$ and $[I_{\cal U}]$ corresponding to caps and cups. This matrix element is, per construction, the equivariant intersection number of the branes: \begin{equation}\label{EG}
    \chi ({\mathscr B} E_{\mathcal U},I_{\mathcal U}) = \bigoplus _{{\mathcal P} \in {\mathscr B} E_{\mathcal U} \cap I_{\mathcal U}} (-1)^{M({\mathcal P})}{{\mathfrak{\vec q}}}^{{\vec J}({\mathcal P})}.
\end{equation}
In addition to the Maslov, or cohomological grading, we keep track of the equivariant gradings coming from the non-single-valued potential. 
Above,  $M({\mathcal P})$ and ${\vec J}({\mathcal P})$ are the Maslov and equivariant gradings of the point ${\mathcal P}$.
The equivariant intersection number is also, per definition, the graded Euler characteristic of the homology $Hom^{*,*}_{\MDy}({\mathscr B} E_{\mathcal U},I_{\mathcal U})$ in \eqref{MP},
\beq\label{ECh}
    \chi ({\mathscr B} E_{\mathcal U},I_{\mathcal U}) = \bigoplus_{M, {\vec J}} (-1)^{M} {{\mathfrak{\vec q}}}^{{\vec J}}  {\rm dim}_{\mathbb C}( Hom_{\MDy}({\mathscr B} E_{\mathcal U}, I_{\mathcal U}[M]\{{\vec J}\}))
\eeq
where the dimension is that of homology group over ${\mathbb C}$, or over ${\mathbb Q}$, ignoring torsion.

\subsubsection{}\label{sreduced}
The quantum link invariants depend on choice of normalization, which one can take to be the value of the invariant for the unknot. One common choice assigns to the unknot the quantum dimension of the representation coloring it. Another assigns to it $1$. The link invariant that assigns $1$ to the unknot is known as ``reduced", as opposed to the ``unreduced" invariant that assigns it the quantum dimension.

\begin{figure}[H]
	\centering
	\includegraphics[scale=0.45]{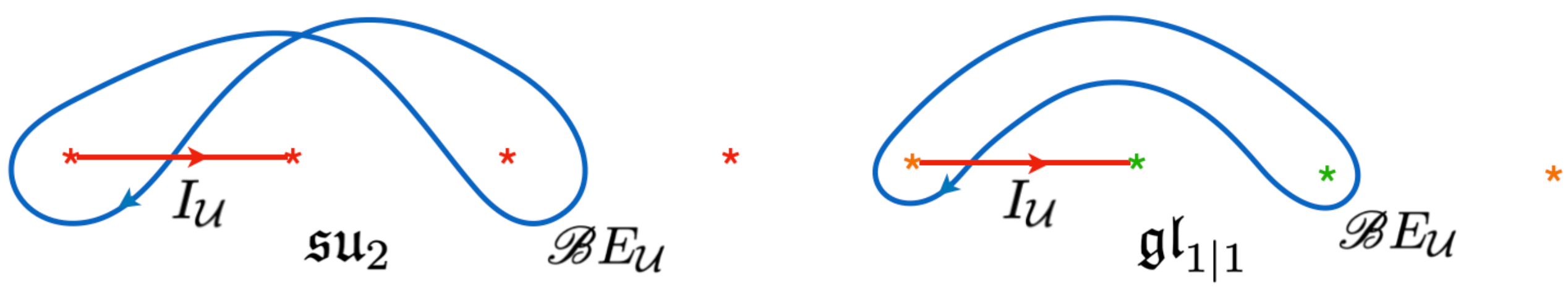}
	\caption{The branes for reduced unknot for ${\mathfrak{su}_2}$ and ${\mathfrak{gl}_{1|1}}$, respectively. }\label{freduced}
\end{figure}

One way to get the reduced $U_{\fq}(^L{\fg})$ invariant of a link $K$ is to start with a presentation of the link as a closure of a $(V, V^*)$ tangle, which has one strand open at the top and one at the bottom, colored by representations $V$ and $V^*$. The reduced invariant of the link is the invariant of the tangle. The ratio of the original and reduced invariant is the quantum dimension of the representation  $V$. In our framework, cutting a strand translates to removing a pair of blue and red segments that meet at a puncture. By construction, this leaves a puncture which one can take to infinity without crossing any of the remaining matchings \cite{M}. For example, for the unknot and $^L{\fg}={\mathfrak{su}_2}$ or ${\mathfrak{gl}_{1|1}}$, this changes the brane configurations from Fig. \ref{B2as} to Fig. \ref{freduced}.

\subsubsection{}
Representation theory sourced by KZ equations \cite{KZ} and their quantum qKZ counterparts \cite{qKZ}, based on bosonic Lie algebras, has played a central role in mathematical physics since their discoveries. The extension of the theory to Lie superalgebras is not as well developed, with the exception of the extension of Drinfeld's theorem to superalgebras from \cite{EG}. It should be only a matter of time before this is remedied. For bosonic simply laced Lie algebras, the contemporary understanding of the theory is due to Maulik and Okounkov \cite{MO}, followed by \cite{Or, OkS, Or2, ESE, AO, Or3} and building on work of Nekrasov and Shatashvili \cite{NS}. It is based on
quantum geometry of certain holomorphic symplectic varieties that arise as moduli spaces of three-dimensional quiver gauge theories with ${\cal N}=4$ supersymmetry. The relation between representation theory of quantum groups and quantum geometry naturally extends to a larger set, coming from moduli spaces of certain very special three dimensional quiver gauge theories with only ${\cal N}=2$ supersymmetry. In one direction, this leads to extension of the theory to ${\mathfrak{gl}(m|n)}$, development of which commenced in \cite{RR}. These gauge theories can be shown to lead to equivariant central charge functions \eqref{zb} coming from conformal blocks of the corresponding super $W$-algebra from \cite{Lit}. The other direction of extension is to Verma modules \cite{AT}. 

While the conceptual explanation for why the theory decategorifies to $U_{{\fq}}(^L{\fg})$ link invariants will ultimately be along the lines 
presented in this subsection for any $^L{\fg}$, there is an alternative, direct way to understand it, once one picks a specific Lie algebra. 
%For ${\mathfrak{gl}_{1|1}}$, we describe this in section \ref{gl11 skein}.

\subsection{Isotopy invariance and skein}\label{Inv}
It is manifest at the outset that the theory gives rise to homological invariants of braids. It is true, though not manifestly so, that it also gives rise to homological invariants of links, invariant under ambient isotopy:
\begin{theorem}\label{tinv}
    The homology groups $Hom_{\MDy}^{*,*}({\mathscr B}E_{\cal U}, I_{{\cal U}})$ are invariants of links.
\end{theorem}
To prove the theorem, we need to verify homology groups $Hom^{*, *}_{\MDy}({\mathscr B}E_{\cal U}, I_{\cal U})$  satisfy Markov-type moves for plat closures due to Birman \cite{Birman} and Bigelow \cite{bigelow}. In \cite{Birman, bigelow}, it is proven that two oriented braids have isotopic plat closures if and only if they differ by the sequence of moves in Figs.~\ref{MI}, \ref{MII}, and \ref{stab}, and an analogous set with the tops and bottoms of Figs.~\ref{MI} and \ref{stab} exchanged.

A proof of Thm.~\ref{tinv} was given in \cite{A1,A2}, for all simply laced Lie algebras. Using the results of this paper, we will give a more direct proof of the theorem for ${\mathfrak{gl}_{1|1}}$ and ${\mathfrak{su}_2}$ theories.  We briefly sketch the strategy here. The detailed proofs are in section \ref{Mgl11} for ${\mathfrak{gl}_{1|1}}$ and in section  ${\mathfrak{su}_2}$ for \ref{Msu2}. They rely on a technical result from appendix \ref{SM}.

\subsubsection{}\label{thm1s}
\begin{figure}[H]
	\centering
	\includegraphics[scale=0.25]{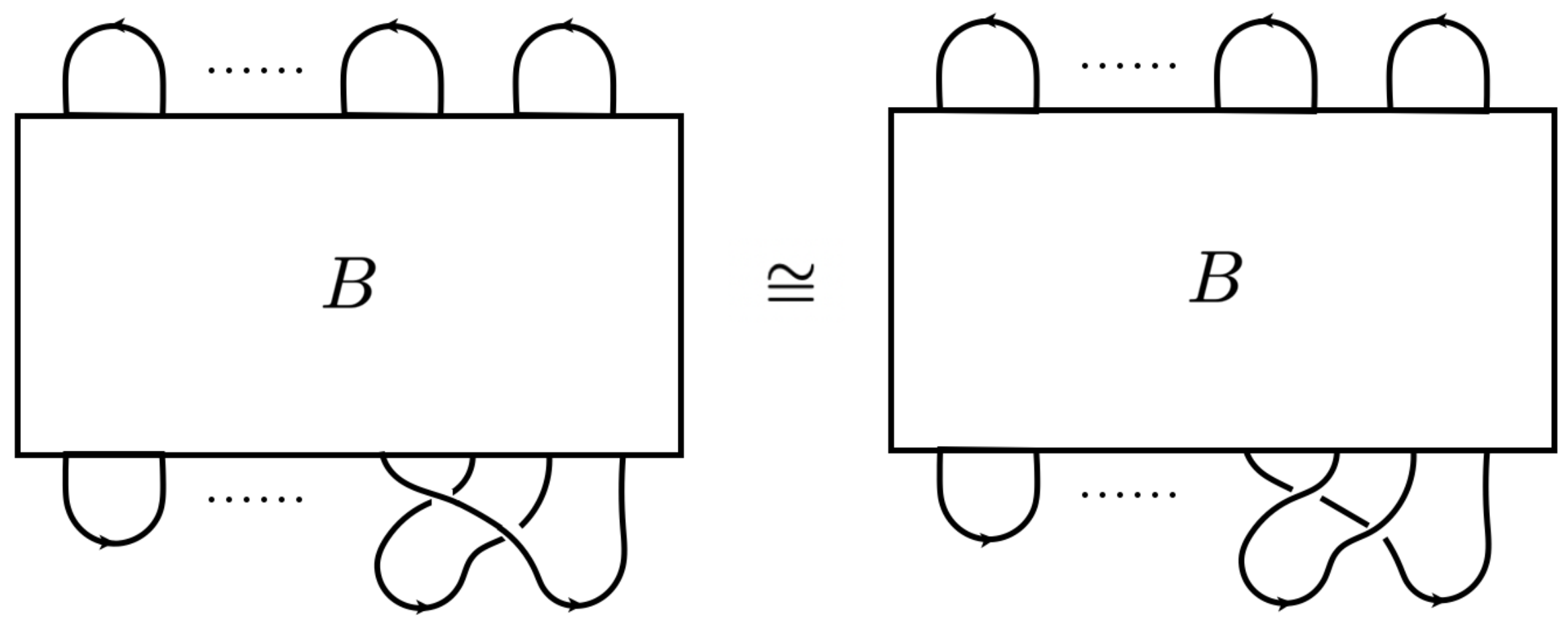}
	\caption{Markov-type move for plat closure of a braid $B$ }\label{MI}
\end{figure}

\begin{figure}[H]
	\centering
	\includegraphics[scale=0.27]{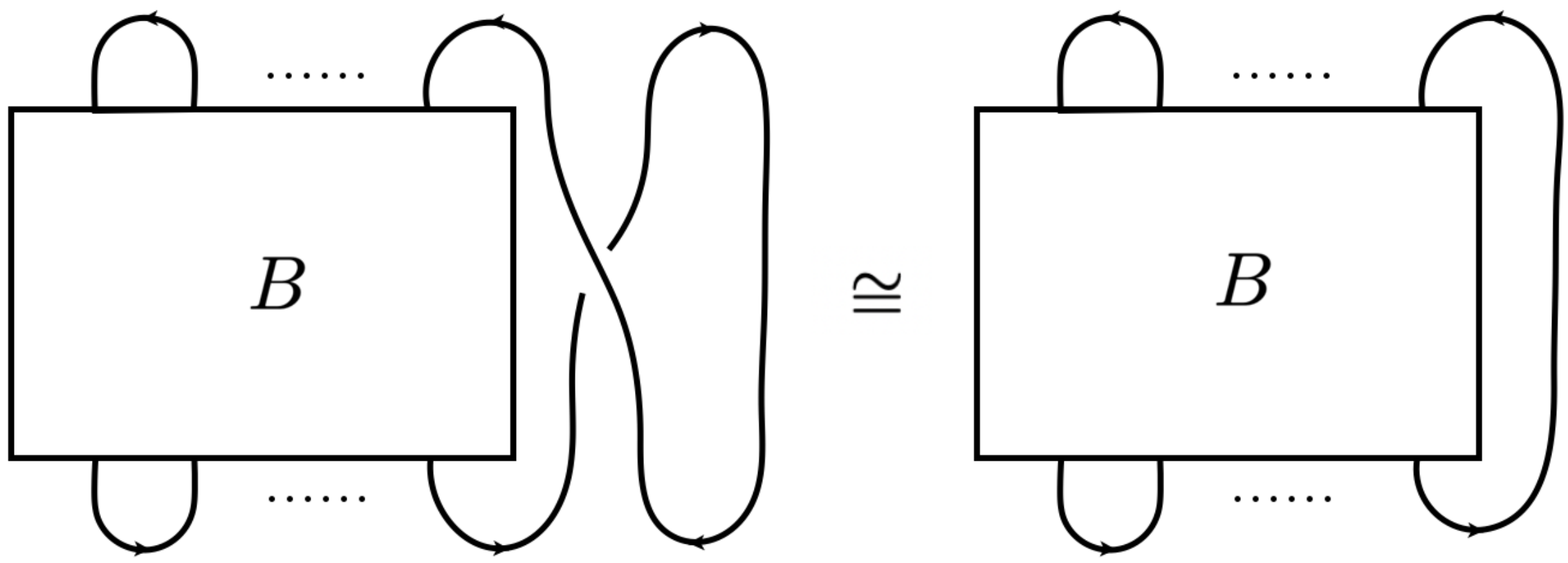}
	\caption{Stabilization, or Markov II-type move}\label{MII}
\end{figure}

\begin{figure}[H]
	\centering
	\includegraphics[scale=0.4]{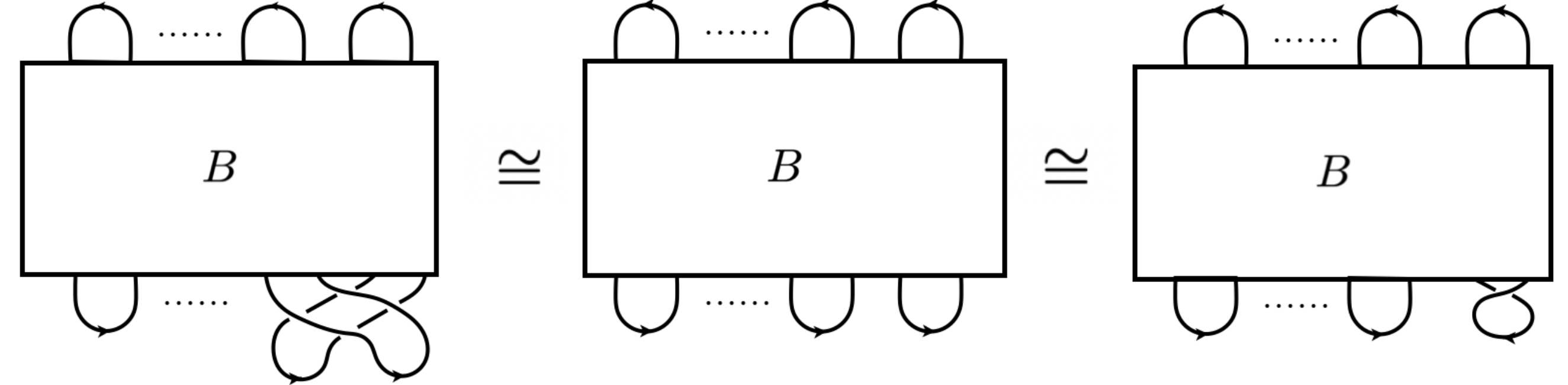}
	\caption{Further Markov-type moves.}\label{stab}
\end{figure}

The key fact used in the proof of Thm.~\ref{tinv} is that moves in Figs.~\ref{MI}, \ref{MII} and \ref{stab} hold as a consequence of equivalences satisfied by the $E_{\cal U}$-branes, viewed as objects of ${\MDy}$, in other words, even before taking Homs. Equivalences of $E_{\cal U}$-branes under those moves 
\noindent{}imply that all Homs involving the branes are  invariant, and $Hom^{*,*}_{\MDy}({\mathscr B} E_{\cal U}, I_{\cal U})$ in particular. We also learn that homology groups $Hom^{*,*}_{\MDy}({\mathscr B} E_{\cal U}, E_{\cal U})$ are invariant under the remaining Birman moves, obtained from moves in Figs.~\ref{MI} and \ref{stab} by exchanging their top and bottom. The groups $Hom^{*,*}_{\MDy}({\mathscr B} E_{\cal U}, E_{\cal U})$ are not, however, invariant under the stabilization move in Fig.~\ref{MII} -- only $Hom^{*,*}_{\MDy}({\mathscr B} E_{\cal U}, I_{\cal U})$ are. Finally, we prove that $Hom^{*,*}_{\MDy}({\mathscr B} E_{\cal U}, E_{\cal U})$ coincides with $Hom^{*,*}_{\MDy}({\mathscr B} E_{\cal U}, I_{\cal U})$ times a fixed graded vector space of rank $2^d$. It follows that $Hom^{*,*}_{\MDy}({\mathscr B} E_{\cal U}, I_{\cal U})$ are invariant under all Markov-type moves for plat closures, and hence that they are link invariants. This proves the theorem.

\subsubsection{}\label{NormI}
The Chern-Simons or $U_{\fq}(^L{\fg})$ invariants are invariants of framed links \cite{Jones, RT}. Framing is a choice of a normal vector field to the knot, which makes the knot into a ribbon. Our category of branes, and conformal field theory, give all links a definite framing. This is the vertical or blackboard framing \cite{Jones}, where the vector fields are everywhere normal to the plane of the projection. 
In vertical framing, the second move in Fig.~\ref{stab} does not hold as stated. Rather, the move adds a full twist to the ribbon, changing its framing by one unit.  In $\MDy$, change of framing is reflected in a shift of equivariant and homological grading of the ${\mathscr B}E_{\cal U}$ brane.
An exception is the ${\mathfrak{gl}_{1|1}}$ theory for which the framing acts trivially.

We will change the framing of links from vertical to zero framing, so that the second relation in Fig.~\ref{stab} holds on the nose. A link in zero framing differs from the vertical one by adding $w$ twists to the corresponding ribbon. The writhe $w$ is the number of positive \includegraphics[height=\fontcharht\font`\B]{positive} minus the number of negative \includegraphics[height=\fontcharht\font`\B]{negative} crossings in an oriented link. See section \ref{Msu2} for more detail.

\subsubsection{}
The Euler characteristic of the theory is the corresponding quantum link invariant on general grounds from section \ref{sCFT}, so in particular, we have:

\begin{theorem}\label{tJ}
    The equivariant Euler characteristics of $Hom_{\MDy}^{*,*}({\mathscr B}E_{\cal U}, I_{{\cal U}})$ satisfy the skein relations in eqn.~(1.1), with $n=0$ for ${\mathscr D}_{Y_{\mathfrak{gl}_{1|1}}}$, and $n=2$ for ${\mathscr D}_{Y_{{\mathfrak{su}_2}}}$.
\end{theorem}
\noindent{}Once we pick a specific Lie algebra, there is a simple, direct way to prove Thm.~\ref{tJ}. The proof is in section \ref{gl11 skein} for the $\mathfrak{gl}_{1|1}$ theory and in section \ref{su2 skein} for the $\mathfrak{su}_{2}$ theory. 
\begin{figure}[h]
	\centering
	     \includegraphics[scale=0.4]{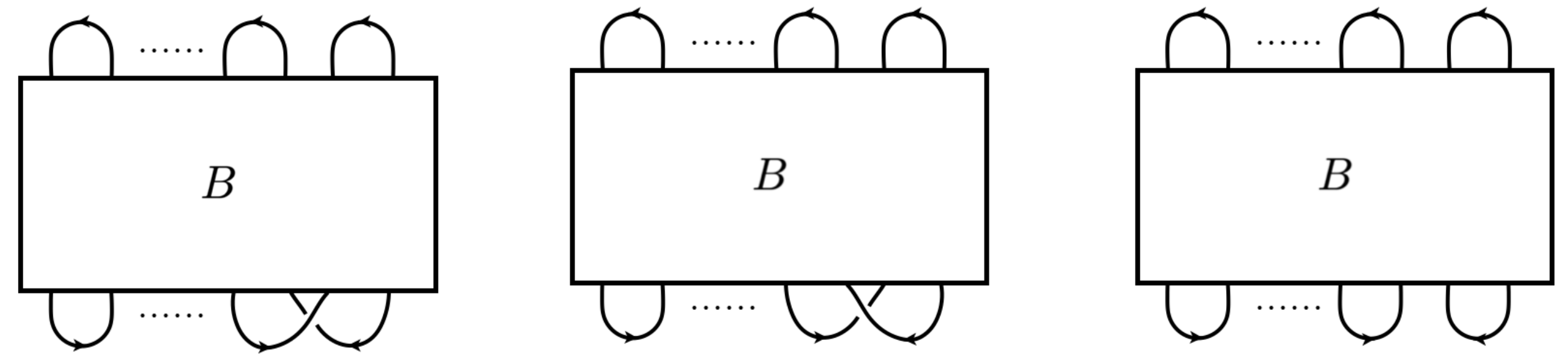}
	\caption{ Plat closures for the skein relation in eqn.~(1.1)}\label{Skeinb}
\end{figure}
\noindent{}The ${\fq}$ variable used throughout the text is related to the ${\bf q}$ in eqn.~(1.1) by a simple redefinitions spelled out in eqn.~\eqref{glred} for $\mathfrak{gl}_{1|1}$ and in eqn.~\eqref{sured} for $\mathfrak{su}_{2}$. Since the theory produces homological invariants of links, to prove the theorem, it suffices to consider only the very special link presentations in Fig.~\ref{Skeinb}. The theorem is proven by showing that K-theory classes of the braided $E_{\cal U}$-branes associated to the very bottom of the diagrams in Fig.~\ref{Skeinb} satisfy the skein relation.

\section{Resolutions of branes in $\mathscr{D}_{Y_{^L\mathfrak{g}}}$}\label{Algorithm}
A key result of the paper is an algorithm for computing projective resolutions, which describe branes $L\in \mathscr{D}_{Y_{^L\mathfrak{g}}}$ as complexes of $T$-branes in~\eqref{PR}.  Applied to the braided cap brane $L={\mathscr B}E_{\cal U}$, this lets us compute the $U_{\mathfrak{q}}(^{L}{\mathfrak{g}})$ link homology, as described in section \ref{tores}.

\subsection{Finding resolutions}
In general, finding a projective resolution of a brane $L\in \MDy$
requires solving two difficult problems. Each is in principle solvable, though typically not in practice.

\subsubsection{}
The first problem is to compute which right module of the algebra $A$ the brane $L\in \MDy$ gets mapped to by the Yoneda functor 
$$
hom_Y(T,-): {\MDy} \longrightarrow \MDa.
$$
Apriori, computing the module requires counting disk instantons, which is a problem with no general solution.
The second problem is to find the projective resolution of this module, as a complex of $T$ branes $(L(T), \delta)$, as in~\eqref{PR}. This problem is known to be
solvable, however, only formally so, by infinite bar resolutions.

\subsubsection{}
We will solve the two problems simultaneously for any $L\in \MDy$, using features specific to our theory. Among the special features of our theory is the fact that Lagrangians of interest on $Y$ are products $L=L_1\times \ldots \times L_d$ of one-dimensional Lagrangians $L_i$ on ${\cal A}$, where $d = \sum_{a=1}^{\rm{rk}^L{\fg}} d_a$, and that the thimble algebra $A=hom_Y^{*,*}(T,T)$ is explicitly known. 

The further key feature is that the theory on $Y$ has an even simpler cousin, corresponding to working on $Y$ with a special divisor $D_O$ deleted:
\beq\label{Y0def}
    Y_O =Y\backslash D_O.
\eeq
The divisor $D_O$ contains a union of diagonals, one for each copy of symmetric product in $Y=\prod_{a=1}^{\rm{rk}^L{\fg}}\, Sym^{d_a}({\cal A})$. Along a diagonal, a pair of points coincide on ${\cal A}$. By removing the diagonals one removes all aspects of the theory on $Y$ that are difficult to understand. We will solve the problem on $Y$ by first solving it on $Y_O$, and then describing the effect of filling $D_O$ back in.

\subsubsection{}
The category of A-branes on $Y_O$ is derived equivalent to the category of modules of the algebra 
\beq\label{A0def}
    A_O= hom_{Y_O}^{*,*}(T,T),
\eeq
which is the endomorphism algebra of $T$-branes on $Y_O$,
\beq\label{DA0}
    \mathscr{D}_{A_O} \cong \mathscr{D}_{Y_O}.
\eeq
The algebras $A$ and $A_O$ have the same elements, because $T$-branes are disjoint from $D_O$, but they have different $A_{\infty}$ products - in our case, what differs is the associative product or the differential. 
As we will see, the algebra $A_O$ is computable by elementary means, thanks to the very special features of $Y_O$.

\subsubsection{}
Resolutions in terms of $T$-branes of branes $L$ on $Y_O$, which are products of one-dimensional Lagrangians $L=L_1\times \ldots \times L_d$, turn out to be computable by elementary means. The result is $(L(T), \delta_O)$, the 
Both the direct sum brane $L(T)$ together which is a direct sum of $T$-branes, together with the differential $\delta_O$ that squares to zero in the algebra $A_O$.  Both $L(T)$ and $\delta_O$ are deduced simply from the $d$-tuple of one dimensional Lagrangians $L_1, \ldots, L_d$.

\subsubsection{}
In going from $Y_O$ to $Y$, we put $D_O$ back in.  This leaves the $T$-branes unchanged but replaces the algebra $A_O$ by $A$, and the differential $\delta_O$ by $
\delta$, which are their deformations.
Since algebra $A$ is known exactly, finding the differential $\delta$ is an algebraic problem, with a straight-forward solution.

Below, we will explain this in a general setting, applicable for any $^L{\fg}$. In later sections, we will describe in detail the two special cases of
$^L{\fg} = {\mathfrak{gl}}_{1|1}$ and  ${\mathfrak{su}}_{2}$.

\subsection{The geometric origin of the algorithm}\label{Aone}
Consider the theory in one complex dimension, for which $Y={\cal A}$. A key simplification of the one-dimensional world is that cones over all morphisms are geometric: they correspond to taking connected sums of Lagrangians over their intersection points.

\subsubsection{}\label{coneone}
Suppose two graded one-dimensional Lagrangians $L'$ and $L''$ on ${\cal A}$ intersect over a point $p \in L'\cap L''$. More precisely, let $p$ be an element of the Floer complex
\beq\label{pe}
    p \in hom_{Y}(L', L'').
\eeq 
We get a new one-dimensional Lagrangian $L$ on ${\cal A}$, by taking the connected sum of $L' [1] \oplus L''$ over $p$.
The resulting Lagrangian $L$ is equivalent, as an object of $\MDy$, to the complex
\beq\label{cright}
    L\;\; \cong\;\;  L' \xrightarrow{\;\;p\;\;} L''.
\eeq
The complex is the direct sum of $L' [1] \oplus L''$ with differential deformed by $p$ in \eqref{pe}. 
\begin{figure}[H]
	\centering
	     \includegraphics[scale=0.35]{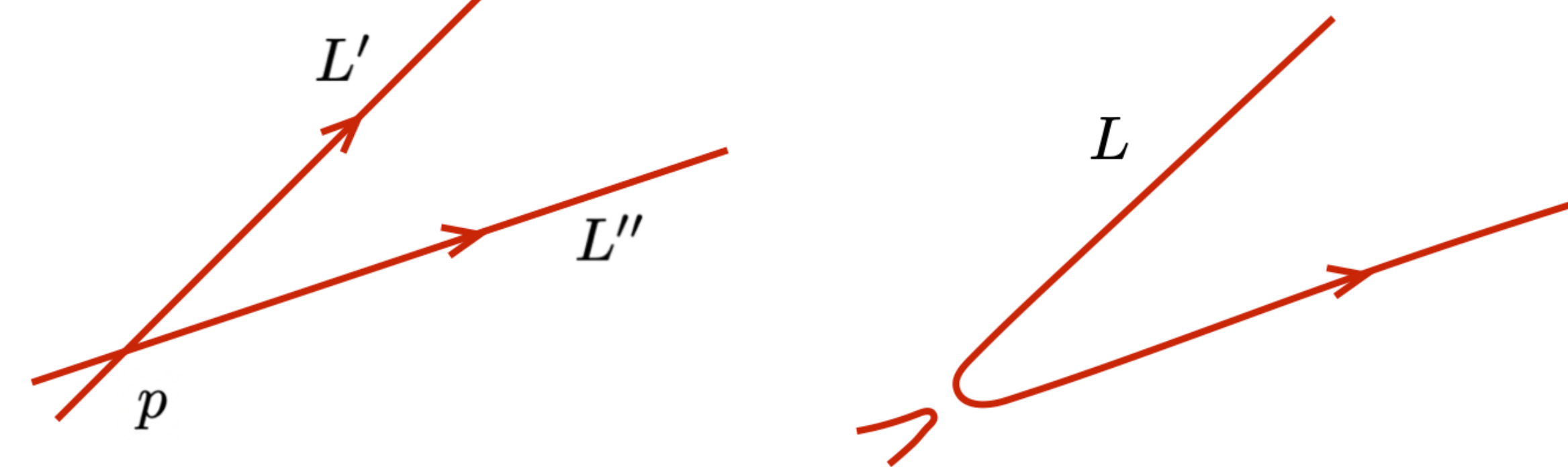}
	\caption{A cone $L' \xrightarrow{p} L''$ is a brane equivalent to $L$, the connected sum of $L'[1]$ and $L''$.}
	\label{cone1}
\end{figure}
\noindent{}Conversely, any brane $L$ which can be broken into  $L' [1] \oplus L''$ intersecting over $p \in hom_{Y}(L', L'')$ is given by a complex of the form  \eqref{cright}.

\subsubsection{}
In general, our target $Y$ is a (product of) symmetric products of ${\cal A}$, of dimension $d$.
Take a pair of branes $L', L''$ on $Y$, each of which is a product of $d$ one-dimensional Lagrangians on ${\cal A}$. Moreover, assume that $L'$ and $L''$ coincide up to one of their factors.
Take, for example,
\beq\label{exL}
    L' =  L_1' \times L_2 \times \ldots \times L_d, \qquad L'' =  L''_1\times L_2 \times \ldots \times L_d.
\eeq
Suppose the one-dimensional Lagrangians $L_1'$ and $L_1''$ intersect over a point $p.$
If 
\beq\label{pd}
    p \;\in\; hom_{Y}(L_1', L_1''),
\eeq 
we can get a new Lagrangian $L_1$ on ${\cal A}$, by starting from $L_1' [1] \oplus L_1''$ and
deforming the differential using $p$, as in \eqref{cright}. We also get a new Lagrangian $L$ on $Y$ given by
\beq\label{Lpp}
    L =L_1\times L_2 \times \ldots \times L_d, 
\eeq
obtained from $L' [1] \oplus L''$ by deforming the differential using 
\beq\label{pD}
    {\cal P} \;=\;(p, id_{L_2}, ..., id_{L_d}) \;\;\in \;\;hom_Y(L', L''),
\eeq
where $id_{L_j}$ is the identity morphism of the one-dimensional $L_j$ brane.  
The result is the complex is
\beq\label{LCr}
    L \;\;\cong\;\; L' \xrightarrow{\;{\cal P}\;} L'' ,
\eeq
with $L'$ placed in degree $[1]$, and $L''$ in degree $[0]$. 
\begin{figure}[H]
	\centering
	     \includegraphics[scale=0.35]{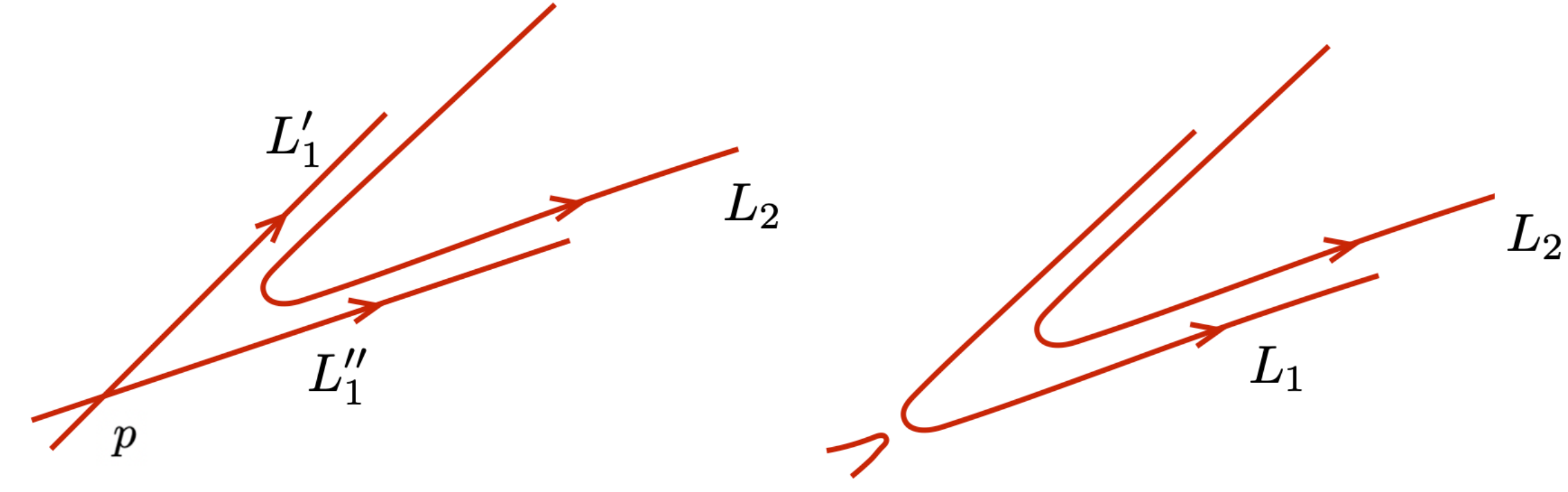}
	\caption{A cone over ${\cal P} = (p, id_{L_2})$, for $d=2$.}
	\label{cone2}
\end{figure}
Conversely, the $L \in \MDy$ brane is equivalent to the complex, obtained from 
 $L' [1] \oplus L''$ 
by deforming the differential, with the map coming from a one-dimensional intersection point in one of its factors as in \eqref{pD}.

\subsection{The algorithm, part I}
Let $L$ be a product of $d$ one-dimensional compact Lagrangians on $Y$, as in \eqref{Lpp}. By iterating the construction we just described, $L$ can be broken up into $L(T)$, a direct sum of $T$-branes, each of which is a product of $d$ real-line Lagrangians on ${\cal A}$.

\subsubsection{}
The $L(T)$-brane and the induced differential can be represented as a $d$-dimensional rectangular grid. The grid is a ``product" of $d$ one-dimensional grids, each corresponding to resolution of a factor brane in $L$. 
The $L(T)$-brane is a sum of $d$-dimensional $T$-branes, one for each grid site.  
We will denote by $\delta_O$ the sum of terms associated to one-dimensional edges of the grid.  
Each edge corresponds to an intersection point of $T$-branes, and is oriented in the direction of the corresponding $hom$. Like in equation \eqref{pD}, the intersection points are identity in all but the one slot corresponding to the edge direction. Every intersection point translates to a specific element of the algebra.

\subsubsection{}\label{LTgrade}
The relative gradings of the $T$-brane summands in $L(T)$ are determined by the gradings of algebra elements of the grid's edges. Start by picking a $T$-brane at some vertex of the grid. Given this, the maps on the edges of the grid determine the relative grading of all the other $T$-branes in the complex.  The difference in equivariant gradings of a $T$-brane an edge points to, and the $T$-brane the edge points from, is the equivariant degree of the element on the edge. The difference in their cohomological degrees is the cohomological degree of the edge element minus $1$.
The choice of absolute grading is equivalent to the choice of grading of $L$, viewed as an object of the category $L\in \MDy$.

\subsubsection{}\label{trial}
Starting from the grid, take the direct sum of $T$-branes on its sites to get
\beq\label{dsb}L(T) = \bigoplus_{k} L_k(T)[k],
\eeq 
where $L_k(T)[k]$ stands for the contribution of $T$-branes in cohomological degree $k$, and various equivariant degrees ${\vec J}$ which we suppress. 
Taking the sum of all the edge elements, we get a
cohomological degree one operator 
\beq\label{dsd}
\delta_O: L(T) \rightarrow L(T)[1].
\eeq
Per construction, cones over all morphisms in $\delta_O$ are geometric; they describe taking connected sums of $T$-branes over their intersection points.

\subsubsection{}
For $(L(T),\delta)$ to define a brane, an object of $\MDy$, $\delta$ needs to be a differential. A differential is a degree one map $\delta: L(T) \rightarrow L(T)[1]$ which squares to zero in an appropriate sense. 
If $A$ is an ordinary associative algebra, which is the case if $^L{\fg}$ is an ordinary Lie algebra like ${\mathfrak{su}_2}$, $\delta$ is a differential if 
\beq\label{sqo}
    \delta\cdot \delta=0,
\eeq
so $\delta$ squares to zero in the algebra in the usual sense. If $A$ is a differential graded algebra, as is the case when $^L{\fg}$ is a Lie superalgebra like ${\mathfrak{gl}_{1|1}}$, $\delta$ is a differential if it squares to zero in the sense of ``twisted complexes":
\beq\label{sqt}
    \delta\cdot_{{\rm Tw}} \delta=0,
\eeq
the precise statement of which we will give in a later section.

\subsubsection{}
In general, the operator $\delta_O$ we just found is not the full differential describing the brane $L\in \MDy$ as a complex of $T$-branes.
This is because $\delta_O$ may fail to square to zero in the sense of \eqref{sqo} or \eqref{sqt}, as appropriate. The fact that $\delta_O$ in general fails to define a differential on $Y$ is unsurprising since it is essentially a product of differentials on $d$ copies of ${\cal A}$, but 
there are maps to $Y$ that are not inherited from maps to the product. Such maps have non-zero intersection the divisor of the diagonal $\Delta $ in $Y$. (In the language of Coulomb branches, the diagonal $\Delta$ is not trivial if the original gauge theory is not abelian.)

\subsubsection{}
While the operator $\delta_O$ may not correspond to a differential of a brane on $Y$, it defines a differential of a brane on $Y_O = Y\backslash D_O$. On $Y_O$, all maps that contribute to $A_\infty$ relations are products of one dimensional maps, as a result of deleting $\Delta$. 

Thus, $L(T)$ together with the differential $\delta_O$ describes the resolution of the brane $L$ viewed as an object of ${\mathscr D}_{Y_O},$ the category of A-branes on $Y_O$. As we will see, even though $\delta_O$ may not coincide with the true differential $\delta$ on $Y$, it determines it uniquely, up to gauge equivalence. Gauge equivalences are maps $e^b\in End(L(T))$ which take $\delta$ to $e^b\delta e^{-b
}$. These preserve the cohomology of the differential, and take the brane $(L(T), \delta)$ to an equivalent one.

\subsection{The algorithm, part II}\label{simpleA}
The category of branes on $Y$ is a deformation of the category of branes on $Y_O$ corresponding to filling in $D_O$. Having found the resolution of the brane $L$ on $Y_O$, we can determine a resolution on $Y$, as a solution to a deformation problem. This approach to understanding the categories of A-branes, namely solving the theory in the complement of a divisor and then filling it back in, was pioneered in \cite{Seideld} and is central to the subject. 

The deformation that takes the theory on $Y_O$ to the theory on $Y$ is simple to describe since both the $T$-branes, which we use to generate the category, and the product Lagrangians $L$ are disjoint from the divisor $D_O$ we deleted. In particular, this means that the generators of the corresponding Floer complexes are the same both on $Y$ and on $Y_O$, and only the $A_{\infty}$ maps involving them may change.

\subsubsection{} 
For simplicity, assume we have deleted only the diagonal $\Delta\subset D_O$. Let $Y_0 = Y\backslash \Delta$, let $\delta_0$ be the differential of the brane on $Y_0$, and denote by $\hbar$ be the parameter that keeps track of the intersection of the map to $Y$ with the diagonal $\Delta$.
Every holomorphic map to $Y$ intersects the diagonal some number $i\geq 0
$ times, and gets a weight $\hbar^{i}$. 
The theory on $Y_0$ is obtained from the theory on $Y$ by setting $
\hbar$ to zero,
since this is the same as restricting to maps that do not intersect the diagonal $\Delta$. Conversely, the theory on $Y$ is a deformation of the theory on $Y_0$, corresponding to setting $\hbar\neq 0.$

\subsubsection{}
The algebra $A_{\hbar}$ on $Y$ being a deformation of the algebra $A_{\hbar=0}$ on $Y_0$ means they coincide as vector spaces, and that, given any two of their elements $a,b$, the product $\cdot$ and differential $\partial$ are each polynomials in $\hbar$: 
\beq\label{defp1}
    (a\cdot b)_{\hbar} = (a\cdot b)_0+(a\cdot b)_1 \hbar +\ldots = \sum_{k\geq 0} (a\cdot b)_k \, \hbar^k,
\eeq
where $(a\cdot b)_0$ is the product of $a$ and $b$ viewed as elements of $A_{0}$, and 
\beq\label{defd1}
    (\partial a)_{\hbar} = (\partial a)_1 \hbar+\ldots = \sum_{k\geq 1} (\partial a)_k \, \hbar^k.
\eeq
The algebra $A_0$ is always an ordinary associative algebra, with vanishing differential $(\partial a)_0=0$. The algebra $A_{\hbar}$ is known explicitly, as are all the terms in equations \eqref{defp1} and \eqref{defd1}. If $^L{\fg}$ is the superalgebra ${\mathfrak{gl}_{1|1}}$, only the differential deforms. If $^L{\fg}$ is ${\mathfrak{su}_{2}}$, or more generally an ordinary Lie algebra, only the product deforms, since all the elements of the algebra have cohomological degree zero. For a general Lie superalgebra, both the product and the differential $\partial$ deform.

\subsubsection{}
The fact that the series in \eqref{defp1} and \eqref{defd1} always truncate follows from the algebra $A_{\hbar}$, but it can be understood apriori. 
At $\hbar=0$, the theory turns out to have an extra symmetry under which all algebra elements have non-negative degrees, while $\hbar$ has degree $1$. Setting $\hbar\neq 0$, the symmetry is broken, as $\hbar$ transforms under it. The terms that can appear on the right hand sides of \eqref{defp1} and \eqref{defd1} are constrained by the symmetry, even a broken one, and finiteness follows.

\subsubsection{}\label{dfinal}
Since the brane $L$ does not intersect the diagonal, the set $L(T)$ of $T$-branes that resolve it are the same in $Y$ and $Y_0= Y\backslash \Delta$. In going from $Y_0$ to $Y$, only the brane differential may change. 

The differential $\delta$ corresponding to the brane on $Y$ is an $\hbar$ deformation of the differential $\delta_0$ on $Y_0$. It is obtained from $\delta_0$ by setting
\beq\label{nf}
    \delta_{\hbar} = \delta_0+\hbar \delta_1+\ldots =\sum_{k=0} \hbar^k \delta_k, 
\eeq
and solving a set of equations which come from imposing either \eqref{sqo} or \eqref{sqt}, depending on whether $A_{\hbar}$ is an ordinary associative algebra, or a differential graded one. At order $\hbar^k$, the equation solved by $\delta_k$, given $\delta_0, \ldots , \delta_{k-1}$ is a linear inhomogenous equation, of the form 
\beq\label{klin}
    \{\delta_0, \delta_k\}_0 = D_k(\delta_0, \ldots, \delta_{k-1}),
\eeq
where the left hand side is a product of $\delta_0$ and $\delta_k$ in $A_0$, and $D_k(-)$ is an operator depending only on lower order terms $\delta_0,\ldots, \delta_{k-1}.$

\subsubsection{}\label{dunique}
In general, the deformation problem may not have any solutions, and if it does, the solution may not be unique, see e.g. \cite{Seideld}. 
 
Existence of solutions in our case follows from the fact that complexes $(L(T), \delta)$ and $(L(T), \delta_0)$ describe a resolution of the same product Lagrangian $L$ on $Y$ and on $Y_0 = Y\backslash \Delta$, which avoids the diagonal.

Uniqueness of the solution is a consequence of equivariance. 
On $Y_0$, equivariance forbids any deformation of $\delta_0$ that does not simply scale its elements. This is because the graded algebra $A_0 = \bigoplus_{{\cal C}, {\cal C}'}hom^{*,*}_{Y_0}(T_{\cal C}, T_{{\cal C}'})$ is non-degenerate: in every fixed degree, the $hom^{*,*}_{Y_0}(T_{\cal C}, T_{{\cal C}'})$ is at most one dimensional. 

Since $\delta_0$ is a differential of a product complex, the meaning of the remaining freedom to scale components of $\delta_0$ is easily understood. All scaling actions of the form $\delta_0 \rightarrow e^b\delta_0 e^{-b}$, for some $b\in End(L(T))$, are gauge equivalences which do not change the brane. As we will see, any remaining freedom to scale components of $\delta_0$, turns out to correspond to changing the complexified flat connection on a brane of topology of $(S^1)^d$.
 There is always a unique deformation of $\delta_0$, the differential of the brane on $Y_0$, to the differential $\delta$ of the brane on $Y$ that keeps the flat connection fixed.

\subsubsection{}
The deformation idea is powerful and we will use it more than once if $D_O$ has other components besides $\Delta$. For example, while in the ${\mathfrak{gl}_{1|1}}$ theory $D_O=\Delta$, in the ${\mathfrak{su}_{2}}$ theory $D_O$ is the union of $\Delta$ and the divisors $D_{a_i}$ of punctures on ${\cal A}$. The deformation theory works in the same way for all components of $D_O$, although practically computing the effect of the filling divisors of punctures is simpler than computing the effect of filling in the diagonal $\Delta$, as the maps one ends up counting are simpler.
\subsection{Floer differential from resolutions}\label{FR}
The projective resolution of $L\in \MDy$ 
\beq\label{cong}
    L \cong (L(T), \delta)
\eeq
encodes, with no further work, the Floer complex of $L$ with any $I_{\cal C}$ brane, together with the Floer differential acting on it.

\subsubsection{}
By isotoping the Lagrangian brane $L$ to one obtained from $L(T)$ by taking connected sums over intersection points of $T$-branes at infinity on ${\cal A}$, we get a one to one correspondence between the set of graded intersection points of 
$L$ with any $I_{\cal C}$-brane, and the set of graded intersection points of $L(T)$ with the same brane: 
\begin{equation}\label{atopa}
    hom_Y( L, I_{\mathcal C}[k]\{\vec{J}\}) =\hom_Y\bigl( L(T), I_{\mathcal C}[k]\{\vec{J}\}\bigr). 
\end{equation}
This follows because all the intersection points are in the interior of ${\cal A}$, and $L$ differs from $L(T)$ only at its infinity. Moreover, the correspondence in \eqref{atopa} preserves the gradings by construction of $L(T)$: the relative degrees of summands in $L(T)$ are determined from the intersection points of $T$ branes at infinity whose cones we take to get $L$.

\subsubsection{}
Because complex $(L(T), \delta)$ is a projective resolution of the brane $L$ in $\MDy$, the $k$-th cohomology group of the complex obtained by applying the functor $hom^*_Y(-, I_{\cal C}\{{\vec J}\})$ to it is, per definition, the space $Hom_{\MDy}(L,I_{\cal C}[k]\{\vec J\})$. The same functor sends the brane $L(T)$ to the set of intersection points of degree $[k]\{\vec J\}$, so it must send the brane differential $\delta$ to the Floer differential. 

The functor whose action on $L\in \MDy$ the algorithm computes, defined by finding a projective resolution of $L$, applying the $hom_Y(-,I_{\cal C}\{\vec J\})$ to that, is the derived functor usually denoted by the ${\bf L}hom(-, I_{\cal C}\{\vec J\})$. From this perspective, the identification of the resulting complex with the Floer complex is a tautological consequence of thinking of $L$ as an object of the derived Fukaya category $\MDy$, generated by the $T=\bigoplus_{\cal C} T_{\cal C}$ brane.
 
The problem we solved in $\MDy$ arizes the B-side of mirror symmetry too.  The ${\bf L}hom$ functor defines the $Hom$'s in the derived category of coherent sheaves, see e.g. \cite{caldararu}. One needs to understand its action to compute the $Hom$'s.

\subsubsection{}\label{dfd}
Computing $\delta_F$ given $\delta$ requires no work. Since the $T$-branes and the $I$-branes are dual in the sense of \eqref{dual}, the functor $hom^{*,*}_Y(-, I_{\cal C})$ sends all terms in the differential $\delta$ to zero, apart from those coming from the identity maps of its dual $T_{\cal C}$-brane.  
 
The terms in the complex $(L(T), \delta)$ coming from the identity map of the $T_{\cal C}$-brane are of the form
\beq\label{tc}
    T_{\cal C}\{\vec J\} \in L_k(T)\;\;  \xrightarrow{ c \cdot id_{T_{\cal C}} } \;\; T_{\cal C} \{\vec J\} \in L_{{k-1}}(T),  
\eeq
where $L(T) = \bigoplus_k L_k(T)[k]$ as in \eqref{dsb}. The functor $hom^*_Y(-, I_{\cal C}\{ \vec J\})$ sends \eqref{tc} to a pair of intersection points in degrees $[k-1]\{\vec J\}$ and $[k]\{\vec J\}$, 
\beq\label{tcr}
    {\cal P} \in hom_Y(L, I_{\cal C}[k-1]\{{\vec J}\})  \xrightarrow{\;\;\; c \;\;\;} {\cal P}' \in hom_Y(L, I_{\cal C}[k]\{{\vec J}\}),
\eeq
and evaluates the differential between them. The direction of the differential changes from \eqref{tc} to \eqref{tcr}, as the functor $hom_Y^{*,*}(-,I_{\cal C})$ is contravariant. The coefficient $c$ is count of holomorphic maps interpolating from ${\cal P}$ to ${\cal P}'$ and contributing to the Floer differential.

\subsubsection{}
While the differential $\delta_F$ obtained algebraically should agree with the Floer differential on general grounds we explained here and in section \ref{tores}, providing a direct proof of this is worthwhile, as it amounts to verifying the definition of Floer theory on $Y$ in terms the maps that contribute to its amplitudes.

\section{ The ${\mathfrak{gl}_{1|1}}$  theory}\label{gl11}
We will now illustrate the general theory in the context of $^L{\fg} = {\mathfrak{gl}_{1|1}}$ and links colored by its two-dimensional representation. The result is a theory categorifying the Alexander polynomial. The theory should coincide with the familiar Heegaard-Floer theory, the flavor of it known as ${\widehat{HFK}}$. Details of our formulation of the theory are new however, as is our approach to solving it. It is distinct from  previous approaches in \cite{grid, LOT}, which are based on genus $\geq 1$ Heegaard surfaces, while ours is for ${\mathcal A} = {\mathbb C}^\times$, which is a genus zero surface  with punctures.

\subsection{$Y_{\mathfrak{gl}_{1|1}}$ and its potential}
Take an oriented link $K$, colored by the defining representation $V$ of $\mathfrak{gl}_{1|1}$. As in section 2, from a plat closure representation of the link with $d+1$ caps and cups at the top and the bottom, we get a configuration of $2(d+1)$ punctures on ${\cal A}$. We will call half of the punctures, those colored by the highest weight $\mu_+$ of $V$, the even punctures and place them at $y=a_{2i}$, where $i$ runs from $1$ to $d$. The odd punctures are at $y=a_{2i-1}$ and are colored by the highest weight $\mu_-$ of the conjugate representation $V^*$. To get unreduced link homology, we would work with the target space which is the symmetric product of $d+1$ copies of ${\cal A}$, since the weight zero subspace of $(V \otimes V^{*})^{\otimes (d+1)}$ is $d+1$ levels below the highest weight. To get reduced link homology we will delete a pair of matchings and the branes associated to them but not the corresponding punctures. 

The target is, as in \eqref{YwoD}, the symmetric product of $d$ copies 
$$Y = Sym^d({\cal A}).$$
of the Riemann surface ${\cal A}$ we started with. A point in $Y$ is $d$ unordered points on ${\cal A}$. 

While one commonly introduces gradings in Heegaard-Floer theory ``by hand", it is better to think of them as coming from a Landau-Ginsburg potential $W_{{\mathfrak{gl}_{1|1}}}: Y \rightarrow {\mathbb C}$ and the top holomorphic form $\Omega_{{\mathfrak{gl}_{1|1}}}$, described below. The result is a version of Heegaard-Floer theory known as ``hat" or $\widehat{HF}$ for the way the punctures on ${\cal A}$ will end up being treated.

\subsubsection{}\label{Ogl11}
The top holomorphic form $\Omega = \Omega_{{\mathfrak{gl}_{1|1}}}$, which gives rise to the Maslov ${\mathbb Z}$-grading in the theory, coincides with the standard top holomorphic form on $Sym^d({\mathbb C}^{\times})$, 
\beq\label{OG11}
    \Omega = \bigwedge_{\alpha =1}^{d} \frac{dy_{\alpha}}{y_{\alpha}} \cdot   g(y_{\alpha} )\cdot \prod_{\alpha<  \beta}(y_{\alpha}- y_{\beta}),
\eeq
apart from the first order poles at all the odd, or ``-" punctures:
\beq\label{gy}
g(y) = \prod_{i=1}^{d+1} \frac{1}{(1-y/a_{2i-1})} \cdot \prod_{ i<j} {\displaystyle \;(1\,- \, a_{2i-1}/a_{2j-1})}.
\eeq
Since one should regard $\Omega\wedge {\overline \Omega}$ as the volume form on $Y$, the odd punctures should be treated as coming from infinitely long cylindrical ends.

\subsubsection{}\label{Wgl11}
The potential 
\beq\label{sup}
    W = W_{{\mathfrak{gl}_{1|1}}} = \lambda_{0} W^{0} + \lambda_{1} W^{1}, 
\eeq
is a sum of two multi-valued holomorphic functions on $Y$ and gives rise to a pair of equivariant gradings in the theory. We will take $\lambda_0, \lambda_1$ to be complex parameters, and choose normalizations so that 
\beq\label{cform}
c^0=dW^0/2\pi i,\qquad c^1=dW^1/2\pi i \;\;\in\;\; H^1(Y, {\mathbb Z}),
\eeq
are one forms with integer periods. $W^0$ is associated to the ${\fq}$-grading, and $W^1$ to the equivariant grading coming from ${\cal A}$ being an infinite cylinder. The parameter  $\lambda_0$ is related to ${\fq}$ by
\beq\label{qdef}
{\fq} = e^{2\pi i \lambda_0},
\eeq 
and $\lambda_1$ is associated to the holonomy around the $S^1$.
Explicitly:
$$
W^{0} = \ln f, \qquad \qquad W^1 =\ln  \prod_{\alpha =1}^{d}y_{ \alpha}.
$$
where
\beq\label{dressed}
f (y)=   \prod_{ i =1}^{d+1} \prod_{\alpha=1}^d   \frac{(1\,- \, a_{2i}/y_{\alpha})}{(1- a_{2i-1}/y_{\alpha})} \cdot \prod_{ i<j} \frac{(1\,- \, a_{2i-1}/a_{2j-1})^{\frac{1}{2}}}{(1- a_{2i}/a_{2j})^{\frac{1}{2}}} .
\eeq
The different contributions of the odd and the even punctures reflect the fact the even ones are associated to $V$ and the odd ones to $V^*$. 
(An invariant way to write $f (y)$, in terms of highest weights and simple roots of Lie algebra ${^L\fg}$ is given in \cite{A2}.)

Terms in $W$ and $\Omega$ that depend only on the positions of punctures become relevant once we consider the action of braiding on $\MDy$, where the $a$'s start to vary.

\subsection{The category of A-branes on $Y_{\mathfrak{gl}_{1|1}}$}
The category of A-branes on $Y=Y_{\mathfrak{gl}_{1|1}}$ is an example of a Fukaya category with potential $W$, with a somewhat novel feature of equivariance, due to $W$ being not single valued. Here, we briefly review the general features of the theory.

\subsubsection{}
The objects of $\MDy$ are based on graded Lagrangian branes in $Y$. 
Every Lagrangian $L$ on $Y$ is an unordered product of $d$ one dimensional Lagrangians $L_{\alpha}$ on ${\cal A}$:
\beq\label{Lo}
    L= L_1 \times \ldots \times L_d,
\eeq
since a point on $Y$ is a $d$-tuple of unordered points on ${\cal A}.$
The Maslov grading of $L$ is the lift of the phase of $\Omega^{\otimes 2}$ to a real valued function $2{\varphi}:L\rightarrow {\mathbb R}$.  The equivariant grading is the choice of a lift of the phase of $e^{-W}$ to a function on $L$. For such a lift to exist, the one forms $c^0$ and $c^1$ in 
\eqref{cform} must have vanishing periods restricted to $L$.

\subsubsection{}
An example of an object in $\MDy$ is the braided cup brane:
\beq\label{BE11}
    {\mathscr B} E_{\cal U} =  {\mathscr B}E_1 \times \ldots \times {\mathscr B}E_d, 
\eeq
obtained by replacing every blue segment of a link, ending on a pair of $+$ and $-$ punctures, by an oval containing it.  Another example of an object of $\MDy$ is the cap brane
\beq\label{I11}I_{\cal U} = I_1 \times \ldots \times I_d
\eeq
which is a product of $d$ non-intersecting intervals,  where $I_j$  ends on pair of $+$ and $-$ punctures at $y=a_{2j-1}$ and $y= a_{2j}$.  Such cup and cap branes arise from a link $K$ as the output of the construction from section \ref{how}.

\subsubsection{} 
The ${\mathscr B}E_{\cal U}$ brane is an object of ${\MDy}$, even though it bounds disks on ${\cal A}$. The phase of $\Omega$ is well defined on the brane as winding of the phase of $dy_i$ around the $i$-th factor brane ${\mathscr B}E_i$ gets canceled by the winding of the phase of $g(y_i)$ around the negative puncture inside it. By contrast, a Lagrangian which bounds a generic disk on ${\cal A}$ does not give rise to a brane in $\MDy$, as it does not admit Maslov grading.

\subsubsection{}\label{flow}
The $I_{\cal U}$ branes are allowed to end at the punctures on ${\cal A}$ due to the fact $W$ has logarithmic singularities there. In general, Lagrangians $L$ with boundaries on $Y$ do not give rise to objects of $\MDy$, since in particular, Floer complexes involving them would not be well defined. Lagrangians ending on the punctures of ${\cal A}$ avoid this as follows.

In Floer theory on $Y$ with potential $W$, elements of the Floer complex $hom_Y^{*,*}(L, L')$ are really Reeb flows. They are solutions to equations for flows
$$
\partial_s y =X_W,
$$
that start at $s=0$ on $L$ and end at $s=1$ on $L'$. Here, $X_W$ is the vector field generated by the Hamiltonian $H_W = \frac{1}{2} {\rm Re} W$. Explicitly, $X_W^i =\frac{i}{2} g^{i {\overline j}} \partial_{{\overline y^j}} W$ where $g_{i {\overline j}}$ is the Kahler metric on $Y$ \cite{GMW}.  One can set $X_W$ to zero in the flow equation, thereby trading solving flow equations for finding intersection points of Lagrangians, at the expense of replacing $L$ by a Lagrangian $L^{W}$ obtained from $L$ by time one flow of $X_W$. When $W$ has logarithmic singularities, this introduces an infinite amount of wrapping of $L^{W}$ at the punctures it ends on.  As a result, Floer complexes of $I_{\cal C}$-branes are well defined, as any brane that intersects the $I_{\cal C}$-brane at the puncture, intersects it infinitely many times.

\subsubsection{}\label{lag} 
Given a pair of Lagrangians $L= L_1\times \ldots \times L_d$ and $L'= L'_1\times \ldots \times L'_d$, their Floer complex \eqref{CF} is generated by intersection points
$${\cal P} \subset L \cap L'.$$ 
The intersection points ${\cal P}$ are $d$-tuples of points on ${\cal A}$ that lie on intersection of the one dimensional factor branes, up to permutation: 
\beq\label{intp}
    {\cal P} =({ p}_1, \ldots , p_d) , \qquad p_{\alpha} \in L_{\alpha} \cap L'_{\sigma(\alpha)},
\eeq
where $\sigma$ is any element of $S_d$, the symmetric group of $d$ elements. One needs to allow for an arbitrary permutation $\sigma$ in \eqref{intp} as, in going around any closed loop in $Y$, the corresponding $d$ points on ${\cal A}$ come back to themselves only up to the action of the symmetric group.  The action of the Floer differential $\delta_F$ \eqref{Floer} is best described using a formulation of Floer theory specialized to targets $Y$ which are symmetric products or Hilbert schemes \cite{L}.

\subsubsection{}\label{cylindrical}
Floer theory with target $Y$ that is symmetric product space (or a Hilbert scheme) has an alternative ``cylindrical" formulation \cite{L}. It relies on the fact that every holomorphic map to from a disk ${\rm D}$ to $Y=Sym^d({\cal A})$ corresponds to Riemann surface that is embedded in the product of ${\rm D}$ and ${\cal A}$, 
$$S \;\; \subset \;\; \rm{D}\times {\cal A},$$
and that has a projection to ${\rm D}$ as a $d$-fold cover. For any point $z\in {\rm D}$ on $S$, its preimage on $S$ determines a $d$-tuple of points $y_1(z), \ldots, y_d(z)$ on ${\cal A}$, and hence a map from $z\in {\rm D}$ to a point $y(z)= (y_1(z), \ldots, y_d(z))\in Y$, and vice versa. (The cylindrical formulation of Floer theory should be thought of as an open Gopakumar-Vafa theory \cite{GV}. Its existence follows from string theory realization of the theory \cite{A3}.)

\subsubsection{}
Projection of $S$ to ${\cal A}$ defines a domain $A=A(y)$ on the Riemann surface with boundaries on one-dimensional Lagrangians. Many aspects of the map $y: {\rm D} \rightarrow Y$ can be reconstructed from a domain $A$, with little work.  The Maslov index and the equivariant degree of the map $y: {\rm{D}}\rightarrow Y$ can be recovered from the 2-chain $A$, using \eqref{indcompare},  \eqref{Jgrade} and \eqref{indgen}. Furthermore, any domain $A$ that has both positive and negative multiplicities in some connected components cannot come from a holomorphic map, since holomorphic maps are orientation preserving.

\subsubsection{}\label{Wreg}
In the ${\mathfrak{gl}_{1|1}}$ theory, the Floer differential $\delta_F$ and all other $A_{\infty}$ products, count only maps that project to domains on ${\cal A}$ containing no punctures.

Apriori, as described in section \ref{FTD}, the A-model counts maps $y: {\rm D} \rightarrow Y$ which have finite symplectic area and pull the Landau-Ginzburg potential $W_{{\mathfrak{gl}}_{1|1}}$ back to a regular function on the disk ${\rm D}$.  In the current context, the first of these conditions implies that maps are not allowed to pass through the negative punctures on ${\cal A}$, as near every such puncture $\Omega$ has a simple pole per eqn.~\eqref{OG11}. The simple pole translates into the negative punctures being at infinity of $Y$, since $\Omega\wedge{\overline \Omega}$ is proportional to the real volume form on $Y$. Correspondingly, any map to $Y$ passing through a negative puncture has an infinite area, and the $e^{-Area}$ factor sets its contribution to any of the $A_{\infty}$ amplitudes to zero. Every holomorphic map to $Y$ which pulls the Landau-Ginzburg potential $W$ back to a regular function on the disk ${\rm D}$ has equivariant degree zero. For this to be the case, the map to $Y$ must project to a domain on ${\cal A}$ that contains equal numbers of positive and negative punctures. The two conditions together imply the domains the map projects to contain no punctures at all.

The version of Heegaard-Floer theory in which the only maps to $Y$ that contribute to the Floer differential or to any other $A_{\infty}$ maps are those that do not pass through any of the punctures is known as $\widehat{HF}$, of ${\widehat {HFK}}$.

\subsubsection{}\label{local} 
Maslov index of any map $y: {\rm D}\rightarrow Y$ from a $2$-pointed disk ${\rm D}$ to $Y$, which interpolates from ${\cal P}$ to ${\cal P'}$ viewed as elements of $hom_Y^{*,*}(L, L')$,  is
given by
\beq\label{FI}{ ind}(y) = M({\cal P}')- M({\cal P})= \int_{\partial {\rm D}} y^*d\varphi/{\pi},
\eeq
where ${\varphi}$ the phase of the holomorphic form $\Omega$, and our orientation conventions are as in \cite{A2}.  See section 4 of \cite{A2}, for definition of the absolute grading $M({\cal P})$. 

The index $ind(y)$ be computed \cite{R, L} from the domain $A$ on the Riemann surface $ {\cal A}$ that the map to $Y$
projects to:
\beq\label{indcompare}
{ ind}(y) =  2 e(A) +i(A) - 2n_-(A).
\eeq
The three terms in the formula correspond to the three factors which contribute to $\Omega$ in \eqref{OG11}. The Euler measure $e(A)$ in \eqref{indcompare} is 
the contribution to the index coming from the $\prod_\alpha dy_{\alpha}/y_{\alpha}$ factor in $\Omega$. 
It equals 
\beq\label{Euler}
e(A) = \chi(A) - \#\textup{acute}/4 +  \#\textup{obtuse}/4,
\eeq
where $\#\textup{acute}$ is the number of acute and $\#\textup{obtuse}$ the number of obtuse angles at vertices of $ A$. The vertices are the intersection points of the one dimensional Lagrangians in $L$ and $L'$ which enter ${\cal P}$ and ${\cal P}'$. The second term in \eqref{indcompare}  is the contribution to the phase of $\Omega$ coming from the factor $\prod_{\alpha<\beta} (1-y_{\alpha}/y_{\beta})$ in \eqref{OG11}. It equals the intersection $i(A)$ of the map to $Y$ with the diagonal $\Delta$
\beq\label{iD}
 i(A) = n_{{\cal P}}(A) + n_{{\cal P}'}(A) - e(A),
\eeq
where 
$n_{\cal  P}(A) = \sum_{a=1}^d n_{p_{\alpha}}(A)$ and where $n_{p_{\alpha}}(A)$ is $1/4$ of the sum of multiplicities of $A$ in the four corners at $p_{\alpha}$.
The last contribution to \eqref{indcompare} is the multiplicity of  the domain $A$ at the ``$-$" punctures:
$$n_-(A) = \sum_{j=1}^{d+1} n_{a_{2j-1}}(A),$$
where $n_{a}(A)$ is the multiplicity of $A$ at a point $y=a$ on ${\cal A}$. It comes from the factor $g(y)$ in $\Omega$.

For a map to contribute to the Floer differential it has to have Maslov index $ind(y)=1$, and $n_-(A)$ which vanishes per section \ref{Wreg}. By contrast, for the purpose of computing relative degrees of a pair of intersection points, any topological disk interpolating between them will do.

\subsubsection{}\label{Jdeg}
The equivariant degree ${\vec J} = (J_0, J_1)$ of a map that interpolates from
${\cal P}$ to ${\cal P}'$ is given by
\beq\label{releq}
    J_i(y) = J_i({\cal P}')- J_i({\cal P}) = -\int_{\partial {\rm D}} y^* c^i.
\eeq 
where $c^i$ are the one forms in \eqref{cform}. While the absolute equivariant degree $J_i({\cal P})$ of any one intersection point ${\cal P}$ depends on equivariant gradings of the Lagrangians $L, L' \in \MDy$ \cite{A2}, their relative degree \eqref{releq} depends only on $A$. 
In terms of domain the $A$ corresponding to $y$, the $c^0$ equivariant degree is given by
\beq\label{Jgrade}
    J_0(y) =   n_-(A)  - n_+(A)  
\eeq
Above, $n_+(A)$ is the multiplicity of $A$ at the $+$ punctures, $n_+(A) = \sum_{j=1}^{d+1} n_{a_{2j}}(A)$. 
The $c^1$ equivariant degree equals
\beq\label{J1}
    J_1(y)= n_{\infty}(A)-n_0(A),
\eeq
and automatically vanishes for all domains $A$ with compact support. 
A map to $Y$ that contributes to any A-model amplitude necessarily has all equivariant degrees zero, as it avoids all the punctures.

\subsubsection{}
The Floer differential $\delta_F=m_1$ is the first in the sequence of $A_{\infty}$-operations, obtained by counting holomorphic maps from a disk ${\rm D}$ to $Y$ which take as an input $\ell$ incoming strings/intersection points, and output one. The corresponding index (the difference between the Maslov indices of the outgoing and incoming intersection points) equals
\beq\label{indgen}
ind(y) = 2e(A) +  i(A) - 2n_-(A)- (\ell-1)d/2.
\eeq
Maps that can contribute to $m_{\ell}$ are rigid holomorphic maps, for which $ind(y)=2-\ell$, and which have $n_{a_{j}}(A)=0$, for all $j$. All such maps have equivariant degrees zero.

\subsection{The $\mathfrak{gl}_{1|1}$ $T$-branes and their algebra}\label{TAg}
The category of A-branes on $Y_{\mathfrak{gl}_{1|1}}$  with potential $W_{\mathfrak{gl}_{1|1}}$ is generated by a finite set of branes. For us, the relevant ones are the $T$-branes 
\begin{equation*}
    T_{\mathcal C} = T_{i_{1}} \times \dots \times T_{i_{d}} \in \mathscr{D}_{Y},
\end{equation*}
which are products of $d$ real lines on a cylinder (which one should regard as colored by the one simple root $^L{e}$ of $\mathfrak{gl}_{1|1}$). Passing a component of $T_{\cal C}$ across a puncture changes the brane, as its equivariant grading changes.
The direct sum of $T$-branes
\begin{equation*}
    T=\bigoplus _{\mathcal C} T_{\mathcal C} \in \mathscr{D}_{Y},
\end{equation*}
over all possible orderings, is the generator of $\mathscr{D}_{Y}$.  We are including in the sum, some branes which are actually zero objects of $\MDy$. This apparent redundancy is a crucial feature, resulting in a simple and explicit algorithm to obtain projective resolutions of branes in $\MDy$. (The zero objects of $\MDy$ correspond to $T_{\cal C}$ branes with an even number of components between a pair of punctures.)

\subsubsection{}
The endomorphism algebra of the $T$-branes,
$$A
= hom_Y^{*,*}(T,T),
$$
which stands for:
$$
A= \bigoplus_{{\cal C}, {\cal C}'} \bigoplus_{M , {\vec J}} hom_Y(T_{\cal C},T_{{\cal C}'} [M]\{\vec J\}),
$$
is as a vector space, spanned by the generators of the Floer complex, which are the intersection points of $T$-branes. Explicitly
\beq\label{CFT}
    hom_Y(T_{\cal C},T_{{\cal C}'} [M]\{\vec J\}) 
\eeq
is the span of intersection points of Maslov index $M$ and equivariant degree ${\vec J}$.

\subsubsection{}\label{wrapped}
\begin{figure}[H]
\begin{center}
    \includegraphics[scale=0.37]{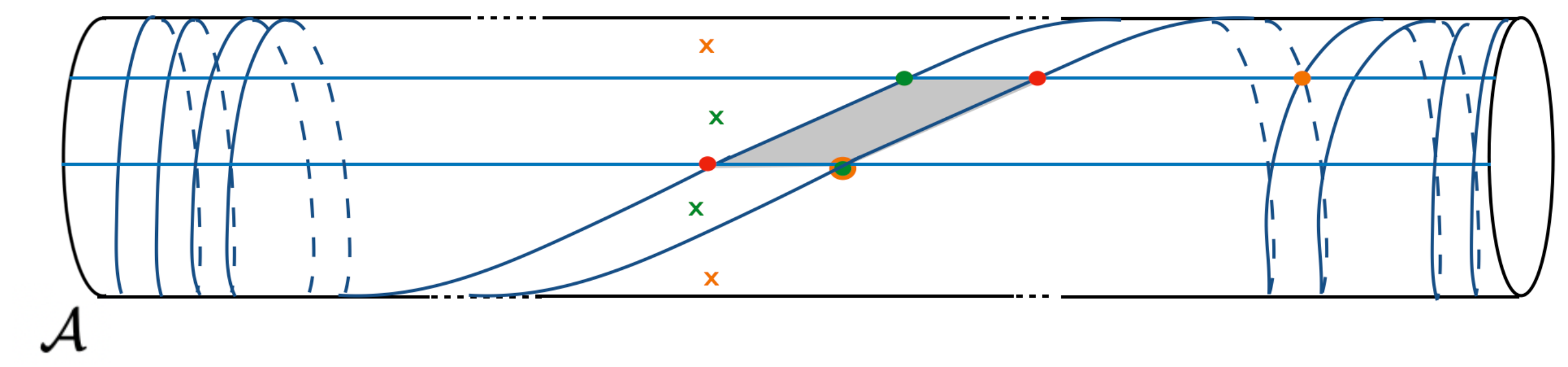}
    \caption{The Riemann surface ${\cal A}$ for $d=2$. Three intersection points of $T^{\zeta}_{\cal C}$ and $T_{{\cal C}'}$ branes are shown, indicated as the pair of green, red and orange dots.}
    \label{gl11A}
\end{center}
\end{figure}

The $T$-branes are not compact, so defining Floer complexes between them requires an additional step.
The Floer complex 
\beq\label{ACP}
    hom_Y(T_{\cal C},T_{{\cal C}'} [M]\{\vec J\}) = CF^{M,\vec J}(T^{\zeta}_{\cal C},T_{{\cal C}'}).
\eeq
is obtained by replacing the $T_{\cal C}$ brane with ${ T}_{\cal C}^{\zeta}$, 
in which each of the $d$ real line Lagrangians in $T_{\cal C} = T_{i_1}\times \ldots \times T_{i_d}$ is taken to make an infinitesimal angle $+\zeta$ with the axis of the cylinder \cite{HIV}, as in Fig.~\ref{gl11A}. The ${ T}_{\cal C}^{\zeta}$ brane winds infinitely may times around both asymptotic ends, and since $\zeta$ is infinitesimally small, all the intersections points in ${ T}^{\zeta}_{\cal C}\cap { T}_{{\cal C}'}$ occur near the two infinities on ${\cal A}$.
The deformation that replaces $T$ by $T^{\zeta}$ has the same effect as turning on a Hamiltonian $H_{\zeta}$ which one takes to vanish in the interior and behave near the infinities as $H_{\zeta} \sim \zeta \sum_{\alpha} |Y_{\alpha}|^2 $. Here, $Y$ is the cylindrical coordinate on ${\cal A}$,  $Y\sim Y+ 2\pi i$, related to the coordinate $y$, $y\in {\mathbb C}^{\times}$, which we have been using so far by $y=e^Y$. On the cover where we open up the cylinder to a plane, the time $1$ flow of $H_{\zeta}$ generates a rotation by an angle $+\zeta$. This leads to the definition of ${\MDy}$ which is known as the (derived) wrapped Fukaya category of $Y$, described in \cite{Auroux} for example. 

In general, one expects that \cite{GMW}, provided $|dW|^2\gg |W|$ near an infinity of $Y$, the superpotential effectively compactifies $Y$, and the deformation we introduced here is not necessary.  As is, to ensure the invariance of the theory under Hamiltonian symplectomorphisms, one needs to introduce the additional  $\zeta$-deformation, as we did above. In principle, one should further replace $T_{\cal C}^{\zeta}$ by a Lagrangian obtained from it by time one flow of $H_W={\frac{1}{2}}{\rm Re} W$, as in section \ref{flow}. This has no further effect however, since the definition of the $\zeta$-deformation is designed to ensure invariance under Hamiltonian symplectomorphisms. 
(The need to perturb by $H_{\zeta}$ at infinity of ${\cal A}$, and at the odd punctures, but not near the even punctures, is due to different behavior of the Kahler form, which enters into $|dW|^2$. We have been assuming the Kahler form is the one compatible with the volume form coming from
$\Omega \wedge {\overline \Omega}$.)

\subsubsection{}
The A-model being a theory of open strings, a 
generator of the Floer complex corresponding to the intersection point
\beq\label{oP}
    {\cal P}\in hom_Y(T_{\cal C},T_{{\cal C}'} [M]\{\vec J\}),
\eeq 
is fundamentally a flow $\partial_s y = X^{\zeta}$ that begins on $T_{\cal C}$ at $s=0$ and ends on $T_{{\cal C}'}$ at $s=1$, generated by the vector field coming from the Hamiltonian $H_{\zeta}$. Using the invariance of the theory under Hamiltonian symplectomorphisms of individual branes, we have been trading the flow
for the intersection point 
$${\cal P} \in T^{\zeta}_{\cal C} \cap T_{\cal C},$$
where $T^{\zeta}_{\cal C}$ is the image of $T_{\cal C}$ under the time one flow of the Hamiltonian. 

\noindent{}Now, we will reverse the above construction, to get an economical, graphical representation of the generators of $A$, as follows.

\subsubsection{}
The intersection point ${\cal P}$, or rather a time-one flow corresponding to it, maps to a configuration of $d$ blue strings on a cylinder whose vertical direction parameterizes the flow, and whose base corresponds to the $S^1$ in ${\cal A}$. 
 The strings start at $s=0$, from the positions of $d$ one-dimensional components of $T_{\cal C}$ on the $S^1$ in ${\cal A}$,
and end at $s=1$, where the corresponding components of $T_{{\cal C}'}$ are.  
The path the strings take from $s=0$ to $s=1$ is determined uniquely by ${\cal P}$. The $+$ and $-$ punctures on the cylinder translate, respectively, to additional orange and green strings, whose positions on the $S^1$ are those of the corresponding punctures and independent of $s$.
 
\begin{figure}[h!]
\begin{center}
  \includegraphics[scale=0.33]{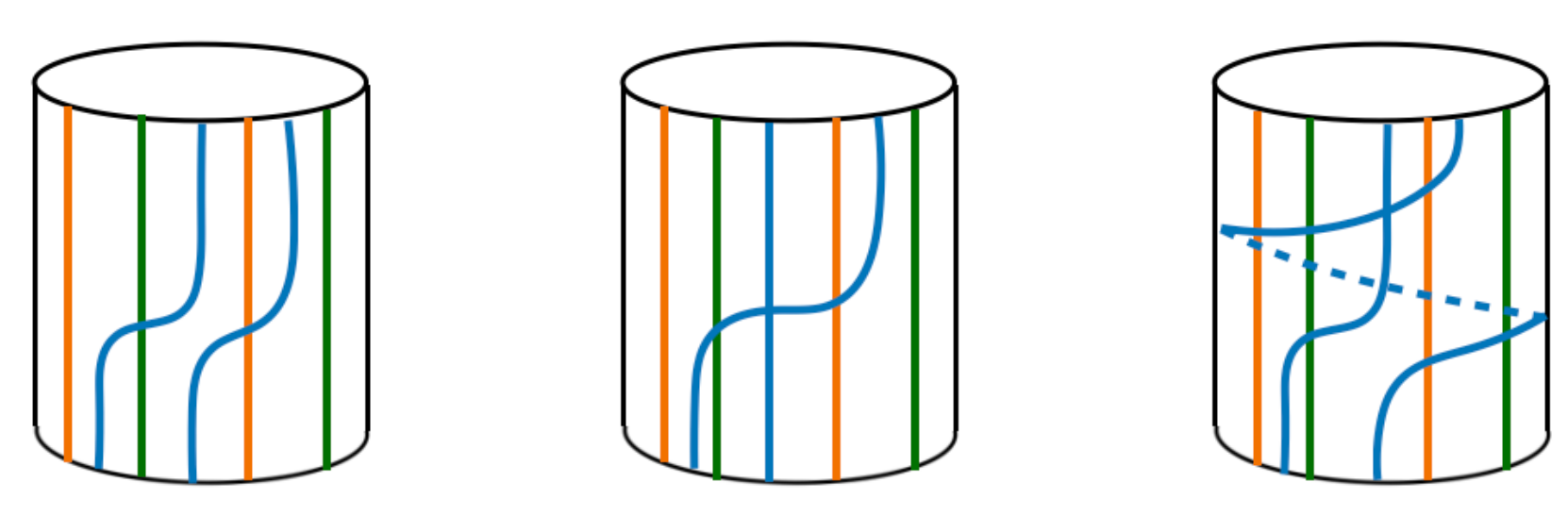}
 \caption{String diagrams corresponding to, respectively the green, red and orange intersection points, from Fig.~\ref{gl11A}.}\label{gl11A2}
\end{center}
\end{figure}

\subsubsection{}
The relative gradings of intersection points ${\cal P}$ and ${\cal P}'$ in  $T^{\zeta}_{\mathcal C}\cap T_{{\mathcal C}'} $ can be computed from a domain $A$ on the Riemann surface interpolates between them, using expressions \eqref{indcompare} and \eqref{Jgrade}. The absolute grading depend on the choice of relative grading of the $T_{\cal C}$ and $T_{{\mathcal C}'}$. We will use the same conventions as in \cite{A2}.

\begin{figure}[H]
\begin{center}
 \hbox{  \includegraphics[scale=0.45]{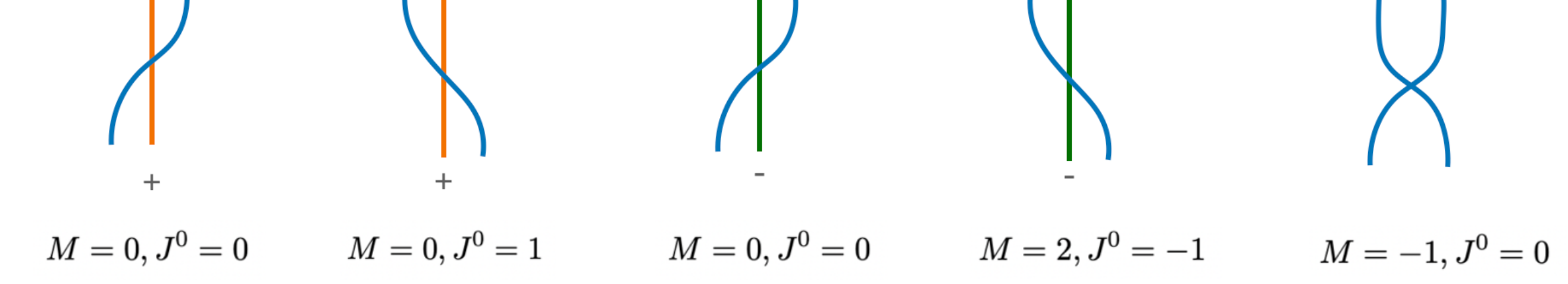}}
 \caption{The string bits out of which any string diagram is composed. $M$ and $J^0$ are their Maslov and ${\fq}$ equivariant degrees. The $J^1$ equivariant degree is given by the net winding number of the strands around the cylinder.}\label{gl11B}
\end{center}
\end{figure}

The result is that the Maslov ${M}({\cal P})$ and ${J}^0({\cal P})$ equivariant degrees of any ${\cal P} \in T^{\zeta}_{\mathcal C}\cap T_{{\mathcal C}'} $ can be read of from the corresponding string diagram, as the sum of degrees of all the string bits, per Fig.~\ref{gl11B}.
The $J^{1}({\cal P})$ equivariant degree is the net winding number of the string diagram around the cylinder. 

\subsubsection{}
The Floer product, which reads
\beq\label{product}
m_2: hom_Y^{*,*}(T_{{\cal C}'}, T_{{\cal C}''})
\cdot hom_Y^{*,*}(T_{\cal C}, T_{{\cal C}'})   \rightarrow hom_Y^{*,*}(T_{\cal C}, T_{{\cal C}''}), 
\eeq
translates into stacking cylinders and rescaling, per construction. The composition of cylinders or $m_2(\cal P',\cal P)$ leads to an algebra $A$ with product $\cdot$ defined by \cite{Seidel}:
\beq\label{defp}
 {\cal P}'\cdot {\cal P}  = (-1)^{M(\cal P)} m_2(\cal P',\cal P).
\eeq
The product is defined by counting  maps $y:{\rm D} \rightarrow Y$, from a three pointed disk to $Y$, with three boundary components mapping  to $T_{\cal C},$ $T_{{\cal C}'}$ and $T_{{\cal C}''}$.

\subsubsection{}
The algebra product can be computed explicitly, as follows \cite{Auroux1, Auroux2}.  For a map to contribute to $m_2$, its Maslov index
(obtained from \eqref{indgen} with $\ell=2$):
\beq\label{ind2}
ind(y) = 2e(A) + i(A) - 2n_-(A)-d/2,
\eeq
must vanish, $ind(y)=0$. Maps from a $3$-pointed disk to $Y$ with boundaries on the $T$-branes are shown in \cite{Auroux1, Auroux2} to have $e(A)=d/4$. They also have $n_-(A)=0$, as explained in section \ref{Wreg}. It follows that maps that contribute to $m_2$ in \eqref{product} have
$$
i(A)=0.
$$
The quantity $i(A)$ is the branching index of the projection to ${\rm D}$ \cite{R,L}, which vanishes for maps $y:{\rm D} \rightarrow Y$ which are disconnected $d$-tuples of maps from ${\rm D}$ to ${\cal A}$. 
It follows that the algebra product is inherited from $d$ copies of a one dimensional theory. 
\subsubsection{}\label{tcount}
The algebra product $A$ introduces relations between its elements. The relations say that the product of two string diagrams is either a taut diagram or it vanishes. A taut diagram has no ``excess" intersection, one which can be removed by isotoping the diagrams keeping the endpoints fixed. Explicitly: 
\begin{figure}[H]
\begin{center}
    \includegraphics[scale=0.3]{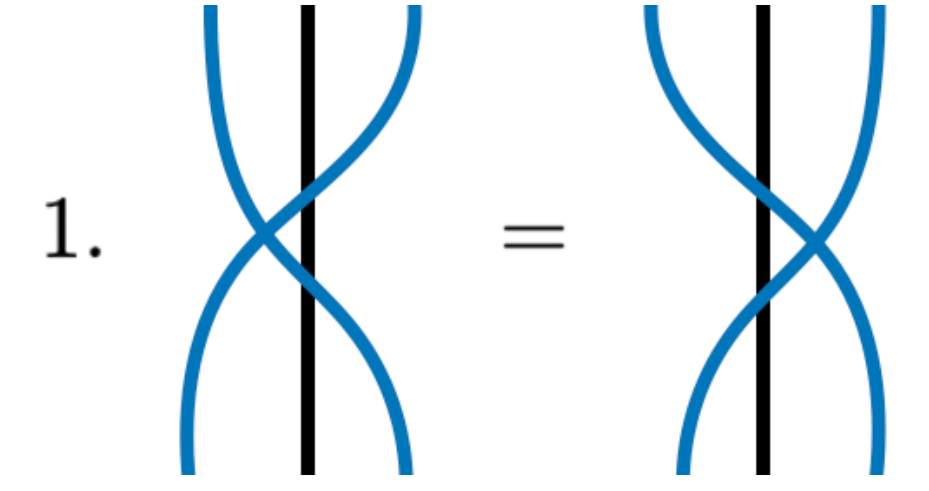}
\end{center}\label{gl11rel1}
\end{figure}
\begin{figure}[H]
\begin{center}
\vskip -1cm
    \includegraphics[scale=0.33]{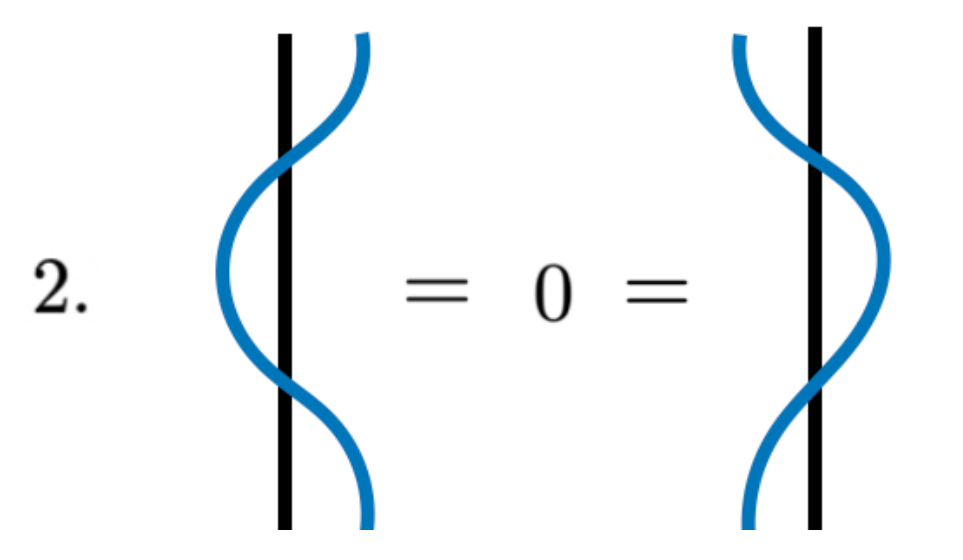}
\end{center}\label{gl11rel2}
\vskip -0.7cm
\end{figure}
\noindent{}The diagrams above compute $m_2$, and hence the algebra product. The black string can stand for either a blue string, colored by the simple root, or the orange and green strings colored by the highest weights of $V$ and $V^*$. 

The products are based on counts of embedded triangles in Floer theory with target which is effectively $d$ copies of ${\cal A}$, as described in \cite{Auroux1, Auroux2} and reviewed in \cite{A2}. Every such triangle contributes a factor of $\pm 1$ to the amplitude, with the sign that depends on the orientations of the branes at the boundaries \cite{Seidel} (see also \cite{Douglas}). The orientations of $T$-branes are such that the triangles which contribute to $m_2({\cal P}, {\cal P}')$ always contribute $+1$.

\subsubsection{}
The Floer differential  
\beq
    m_1:\;\; hom_Y^{*,*}(T_{{\cal C}}, T_{{\cal C}'}) \;\; \rightarrow\;\; hom_Y^{*+1,*}(T_{\cal C}, T_{{\cal C}'}), 
\eeq
turns the algebra $A$ into a differential graded algebra. The algebra differential $\partial$ is defined by
\beq\label{defd}
    \partial {\cal P} = (-1)^{M({\cal P})} m_1({\cal P}).
\eeq
Maps that contribute to $m_1$ have Maslov index 
$$
ind(y) = 2e(A) +  i(A) - 2n_-(A),
$$
from \eqref{indcompare}, equal to $ind(y)=1$. Every two-pointed map from ${\rm D}$ to $Y$ with boundaries on $T$-branes necessarily projects to domains $A$ that are (products of) rectangles and isolated points. The latter are images of constant maps from a disk ${\rm d}$ to ${\cal A}$. For such domains, the Euler measure from \eqref{Euler} vanishes $e(A) =0$ (every rectangle has Euler characteristic one, and $4$ acute angles). Maps also have $n_-(A)=0$ as they are not allowed to pass through the punctures. 

It follows that maps with Maslov index one all have
$$
i(A) =1,
$$
so they intersects the diagonal in the symmetric product exactly once. Such maps have an image on ${\cal A}$ which consists of exactly one empty rectangle and the corresponding number of maps to points. Fortunately, the count of this kind of a map is known. See e.g. \cite{grid}. This translates into the differential $\partial$ which 
acts on any diagram by removing a crossing locally:
\begin{figure}[H]
\begin{center}
    \includegraphics[scale=0.33]{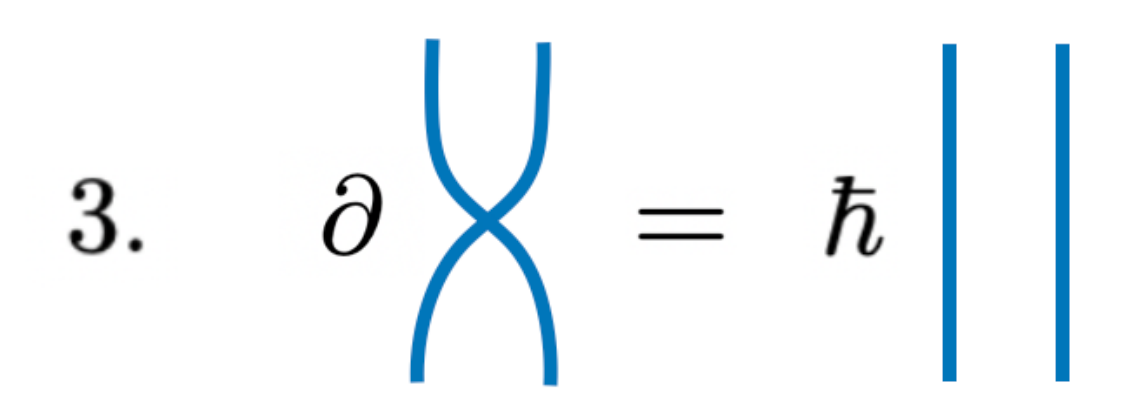}
\end{center}
\vskip -1cm
\end{figure}

\subsubsection{}
As in section \ref{Algorithm}, it is useful to introduce a parameter $\hbar$, which keeps count of the intersection of a map to $Y$ with the diagonal $\Delta$. The value of $\hbar$ is immaterial, as long as it is not zero, as it can be absorbed into the definition of the algebra element described by the crossing of two blue strands. The algebra 
$$A_{0} = hom_{Y_0}^{*,*}(T,T),$$
obtained from $A_{\hbar}$ by setting $\hbar$ to zero, is the endomorphism algebra of $T$-branes on $Y_0 = Y\backslash \Delta$. The algebra $A_0$ differs from $A_{\hbar}$ only in that its differential vanishes identically. The algebra $A_0$ plays a central role in the algorithm from section \ref{Algorithm}, applied to ${\mathfrak{gl}_{{1|1}}}$, which we will revisit below. The divisor $D_O$ from \eqref{Y0def} coincides with $\Delta$ in the present setting, so in particular $A_O=A_0$ and $Y_O=Y_0$.

\subsubsection{}
All the higher $A_{\infty}$ products $m_{\ell}$ with $\ell>2$ incoming strings and one outgoing one string vanish. Maps that contribute to $m_{\ell}$ are rigid holomorphic maps for which the Maslov index equals $ind(y)=2-\ell$. From the formula for the index $ind(y)$ given in \eqref{indgen}, together with the fact that for $\ell+1$ pointed maps bounded by $T$-branes $e(A)= (\ell-1)d/4$ \cite{Auroux1, Auroux2} and $i(A)$ is non-negative, it follows that to contribute to $m_{\ell>2}$, a map must have $n_{-}(A)>0$. Such a map intersects the divisor of the negative punctures at least once. The divisor is deleted from $Y$, so the map vanishes.

\subsubsection{}
 The algebra $A_{\hbar}$ is an associative graded algebra with a differential. It is a version of the algebra discovered by Lipshitz, Osvath, and Thurston \cite{LOT}, adapted for the Heegaard surface that is a cylinder with punctures. 

The signs in the relations between Floer operations and the algebra product \eqref{defp} and the differential \eqref{defd} are important. They are chosen so that the $A_{\infty}$ relations of the Fukaya category translate into the standard relations of the differential graded algebra with a product that is associative.  See appendix \ref{A} for details. In particular, some products which are equal when computed by $m_2$ as described in section \ref{tcount}, are equal only up to a sign computed by the $\cdot$ product.

\subsection{Projective resolutions}\label{PRg}
Because the algebra $A$ has elements of non-zero Maslov degree, the complexes
resolving the ${\mathscr B} { E}_{\mathcal U}$ branes (or any other brane) are really 
twisted complexes.

\subsubsection{}
Resolution of the ${\mathscr B} { E}_{\mathcal U} \in \MDy$ brane in terms of ${T}$-branes, 
$${\mathscr B} { E}_{\mathcal U} \cong ( {\mathscr B} { E}(T), \delta)
$$is a direct sum brane
\beq\label{aproBgl}
    {\mathscr B} { E}(T)=\bigoplus_k  {\mathscr B} { E}_k(T) [k]
\eeq
together with a differential $\delta$, which is a cohomological degree $1$ and equivariant degree zero operator 
\beq\label{tdb}
    \delta:  {\mathscr B} { E}(T)\rightarrow {\mathscr B} { E}(T)[1]
\eeq
that squares to zero. Informally, we get the direct sum brane by stretching the ${\mathscr B} { E}_{\mathcal U} $ brane straight along the cylinder ${\cal A}$ and breaking it at both ends at infinity to a direct sum of $T$-branes. The original brane ${\mathscr B} E_{\cal U}$ is recovered by turning on the differential $\delta$. Writing this out explicitly, $\delta$ is a sum of algebra $A$ elements
\beq\label{deltae}
    \delta =\sum_{k, \ell} d_{k, \ell} \in A
\eeq
corresponding to Homs between $T$-branes in the direct sum brane:
\beq\label{degdeltae}
    d_{k, \ell} \in hom_Y( {\mathscr B} { E}_k(T)[k],  {\mathscr B} { E}_{\ell}(T)[\ell+1]).
\eeq

\subsubsection{}\label{close}
The twisted differential $\delta$, viewed as acting on the brane \eqref{tdb} in $\MDy$, squaring to zero means that it satisfies the Maurer-Cartan equation
\beq\label{asquare}
    \partial \delta + \delta \cdot \delta = 0.
\eeq
This is true only up to signs, whose details are quite technical but crucial, unless one is satisfied with working over ${\mathbb Z}_2$ coefficients. The sign assignments in the category $\MDy$ of twisted complexes are explained in appendix \ref{ssigns}. The upshot is that $\delta$
satisfies
\beq\label{square}
    m^p_1(\delta) + m^p_2(\delta, \delta) =0, 
\eeq
where $m^p_{\ell}$ is a twist of the Floer theory $m_{\ell}$ from \eqref{defp} and \eqref{defd} by signs. The condition \eqref{square} is as simple as it is only because all $m_{\ell>2}$ maps vanish in our algebra. For a general $A_{\infty}$ algebra, all higher morphisms would contribute and the condition would be $\sum_{\ell=1}^{\infty} m^p_\ell(\delta, \ldots, \delta) =0$ instead.
For simplicity, we will write \eqref{square} simply as \eqref{asquare}.

\subsubsection{}
The algorithm from section \ref{Algorithm} starts by giving the projective resolution of any one component of the ${\mathscr B} E_{\cal U}$ brane in the $d=1$ theory, as described in section \ref{Aone}. 
It recursively constructs the resolution of the brane in ${\mathscr D}_{Y_0}$, by taking a product with one factor brane at a time. The result of that section is the direct sum brane ${\mathscr B} { E}(T)$
and the differential 
$$\delta_0:  {\mathscr B} { E}(T)\rightarrow {\mathscr B} { E}(T)[1],$$
which is a cohomological degree $1$ and equivariant degree zero operator that squares to zero in algebra $A_0$. The differential $\delta_0$, per construction, knows everything there is to know about the brane ${\mathscr B} E_{\cal U}$, viewed as a product of $d$ Lagrangians on ${\cal A}$. Since $A_0$ is an ordinary associative algebra, the differential squaring to zero means that
\beq\label{dap} 
\delta_0\cdot 
\delta_0 =0,
\eeq
 up to sign twists. The relation \eqref{dap} holds both in $A_0$ and in $A_{\hbar}$, since $A_0$ and $A_{\hbar}$ differ only in the differential.

\subsubsection{}
The complex ${\mathscr B} { E}(T)$ with differential $\delta_0$  describes the resolution of the brane ${\mathscr B} E_{\cal U}$ in $\mathscr{D}_{Y_0}$. The full differential $\delta$ of the brane in $\MDy$ is obtained by deforming $\delta_0$ per section \ref{dfinal}:
$$
\delta = \delta_0+ \hbar \delta_1+ \ldots
$$
and asking for the full differential to satisfy \eqref{square}. For example, $\delta_1$ is a solution to the linear equation
$$
\partial \delta_0 + \delta_0 \cdot \hbar \delta_1 + \hbar \delta_1\cdot \delta_0 =0.
$$

\subsubsection{}\label{sec:signs1}
To get a Floer homology theory with ${\mathbb Z}$-coefficients we need to specify the spin structure on the ${\mathscr B}E_{\cal U}$ brane we started with \cite{FOOO, OF}.
The choice of spin structure is necessary to orient the moduli spaces of holomorphic maps Floer theory is based on. For us, this translates into a ${\mathbb Z}_2$ worth of choices of sign for every component of the brane of $S^1$ topology. From perspective of its resolution, the $({\mathscr B}E(T), \delta)$ brane, to get a ${\mathbb Z}$ graded theory, we need to specify signs of maps in the differential $\delta$. The two ambiguities must match, so the inequivalent choices of signs in $\delta$ should correspond to the inequivalent choices of spin structure in the original ${\mathscr B}E_{\cal U}$ brane.

Since $\delta$ arises as a deformation of $\delta_0$, to specify signs in $\delta$ it suffices to specify the signs in $\delta_0$. Because $\delta_0$ is a differential of a product complex, signs in it are determined from signs in its factor complexes, by a slight variant of the standard procedure of forming a product complex. There is, as expected, a ${\mathbb Z}_2$ worth of choices signs in a complex describing a one dimensional factor brane, as only the product of all signs turns out to be gauge invariant. The departure from the standard sign assignment on a product complex is needed because $\delta_0$ is a differential of a twisted, not ordinary complex. The sign assignments in $\MDy$ are explained thoroughly in appendix \ref{ssigns}. They follow verbatim the general prescription from \cite{Seidel}.

For definiteness, out of $2^d$ possible sign choices, we choose the one for which the product of all the maps in the differentials of one dimensional factor branes is even. 

\subsubsection{}
The different choices of spin structure on the ${\mathscr B} E_{{\cal U}}$ brane of $(S^1)^d$ topology may be related by changing holonomies of the flat $U(1)$ connection around the $S^1$'s by $e^{i\pi}$, as in \cite{Cho, Cho2}. 
The moduli of flat connections on the ${\mathscr B}E_{\cal U}$ brane is complexified by the symplectic area of $d$ annuli swept out by deforming its one-dimensional factor branes. 

In terms of its resolution, 
changing the holonomy of the complexified flat $U(1)$ connection on the ${\mathscr B}E_{\cal U}$ brane should correspond to scaling the maps in $\delta_0$ by ${\mathbb C}^{\times}$-valued parameters, modulo gauge transformations of ${\mathscr B}E(T)$, the direct sum brane, as in section \ref{dunique}. Gauge transformations send the differential  $\delta_0$ to $e^b \delta_0 e^{-b}$ where  $e^b \in End({\mathscr B}E(T))$. It is elementary to show that for every one dimensional factor brane, we get exactly one ${\mathbb C}^\times$-valued gauge invariant scaling parameter, which should be identified with the choice of holonomy.

The differential $
\delta$ of the brane on $Y$ is, per definition, the deformation of $\delta_0$ keeping the flat connection fixed. Homology groups $Hom_{\MDy}^{*,*}({\mathscr B}E_{\cal U}, I_{\cal U})$ being invariant under the swiping move from section \ref{Mgl11} and appendix \ref{Markovgl}, should be independent of the choice of complexified flat connection one starts with, as long as one stays away from infinities in the moduli space, at which topology of the brane may change.

\subsection{Simple examples}
We will now illustrate the algorithm, starting with the simplest one-dimensional examples. 

\subsubsection{}\label{simplest}
\begin{figure}[H]
\begin{center}
  \includegraphics[scale=0.42]{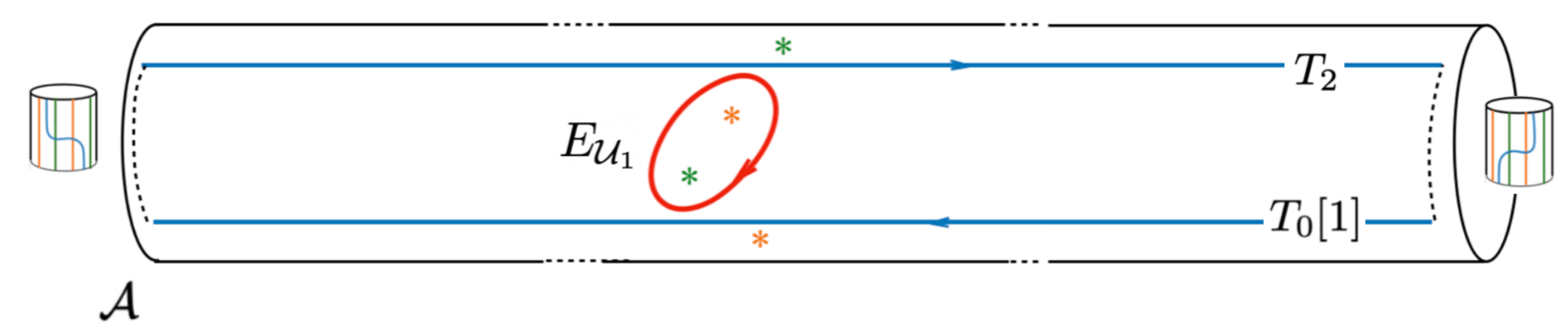}
 \caption{ The $E_{{\cal U}_1}$ brane is the connected sum of the two $T$-branes over the intersection points corresponding to the open strings/Reeb chords shown. }\label{gl11ex1}
 \vskip -1cm
\end{center}
\end{figure}

Consider a single cup brane in a theory with $d=1$. The $E_{{\cal U}_1}$ brane in Fig.~\ref{gl11ex1} is one of the four cup branes in the theory on $Y={\cal A}$ with two even and two odd punctures. By isotoping the $E_{{\cal U}_1}$ brane, and allowing it to break at $y=0$ and $y=\infty$ we get the direct sum brane:
\beq\label{gl1cup1}{E}_{{\cal U}_1}(T) = T_0[1] \oplus T_2.
\eeq
The cohomological degree shift comes from the flip of orientation, per Fig.~\ref{gl11ex1}. 
{\noindent }In fact, the construction we are about to describe fixes all the degrees for us from the geometry of the brane.
To recover the $E_{{\cal U}_1}$ brane, we take the connected sum of of $T_0[1]$ and $T_2$, over a specific pair of their intersection points at $y=\infty$ and $y=0$. 
The intersection point at $y=\infty$ corresponds to the following element of the algebra $A$:
\vspace*{-5pt}
\begin{figure}[H]
\begin{center}
  \includegraphics[scale=0.5]{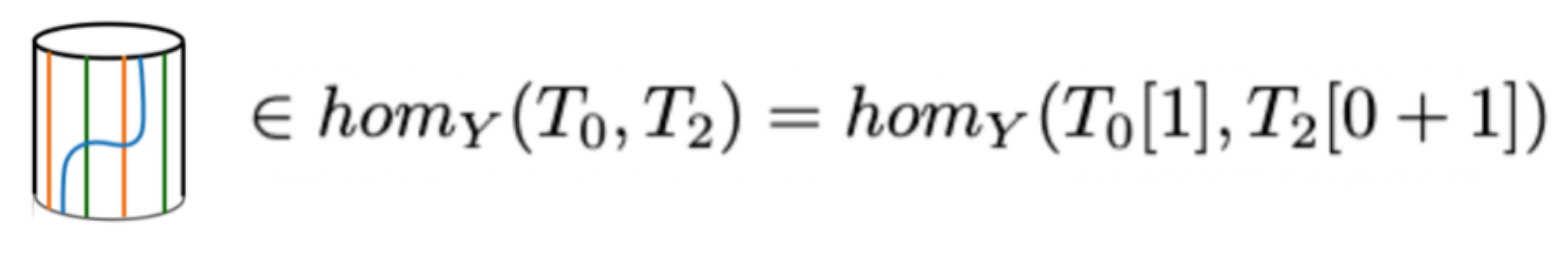}
\end{center}
\vspace*{-20pt}
\end{figure}
{\noindent }which encodes the Reeb flow from $T_0$ to $T_1$ that we used in Fig.~\ref{gl11ex1} to take the connected sum.
The intersection point at  $y=0$,  corresponds to a blue string going the other way:
\begin{figure}[H]
\begin{center}
  \includegraphics[scale=0.5]{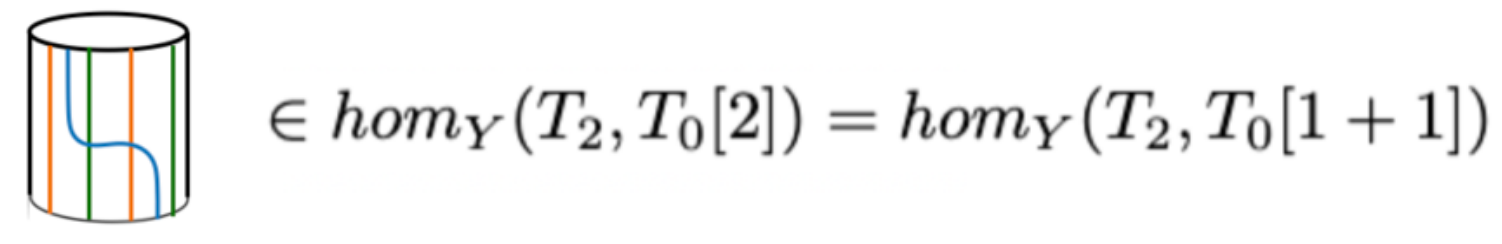}
\end{center}
\vspace*{-20pt}
\end{figure}
{\noindent }The orientations of the maps in the complex follow from Fig.~\ref{gl11ex1} and the definition, from section \ref{wrapped}, of homs in the wrapped Fukaya category. 
The result is the twisted complex $(E_{{\cal U}_1}, \delta)$, 
with differential:
\beq\label{}
\delta= \begin{pmatrix} 0 & \adjincludegraphics[valign=c, height=0.3cm, raise=0.1\baselineskip,
    set vsize={0.3cm}{0cm}]{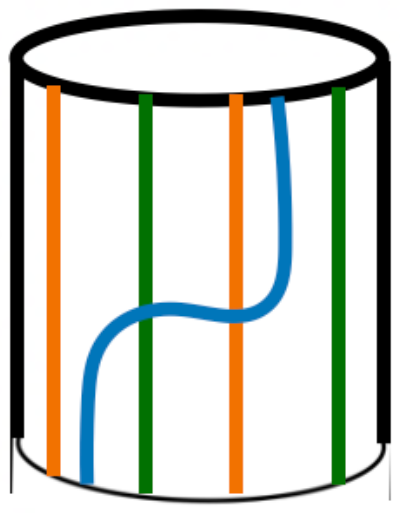}\\
 \adjincludegraphics[scale=0.1]{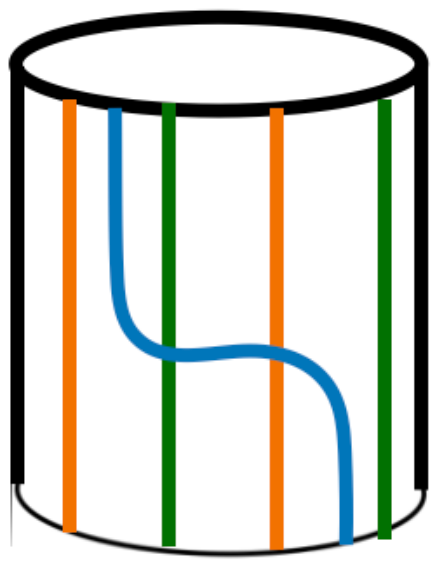}  &0 
\end{pmatrix}
\eeq
The differential squares to zero thanks to the $A$-algebra relations:
\begin{figure}[H]
\begin{center}
  \includegraphics[scale=0.35]{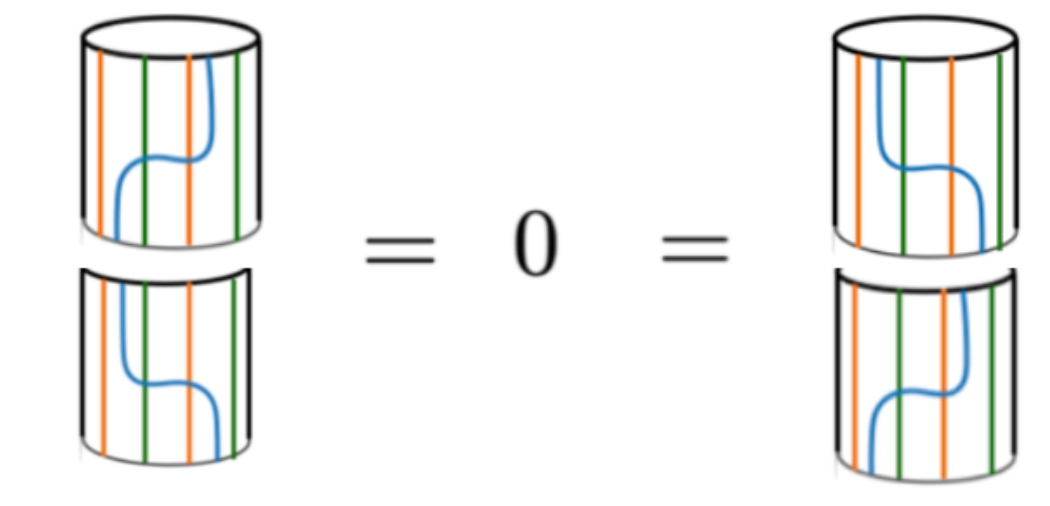}
\end{center}
\vspace*{-20pt}
\end{figure}
{\noindent } 
The complex is often written as 
\begin{figure}[H]
\begin{center}
  \includegraphics[scale=0.33]{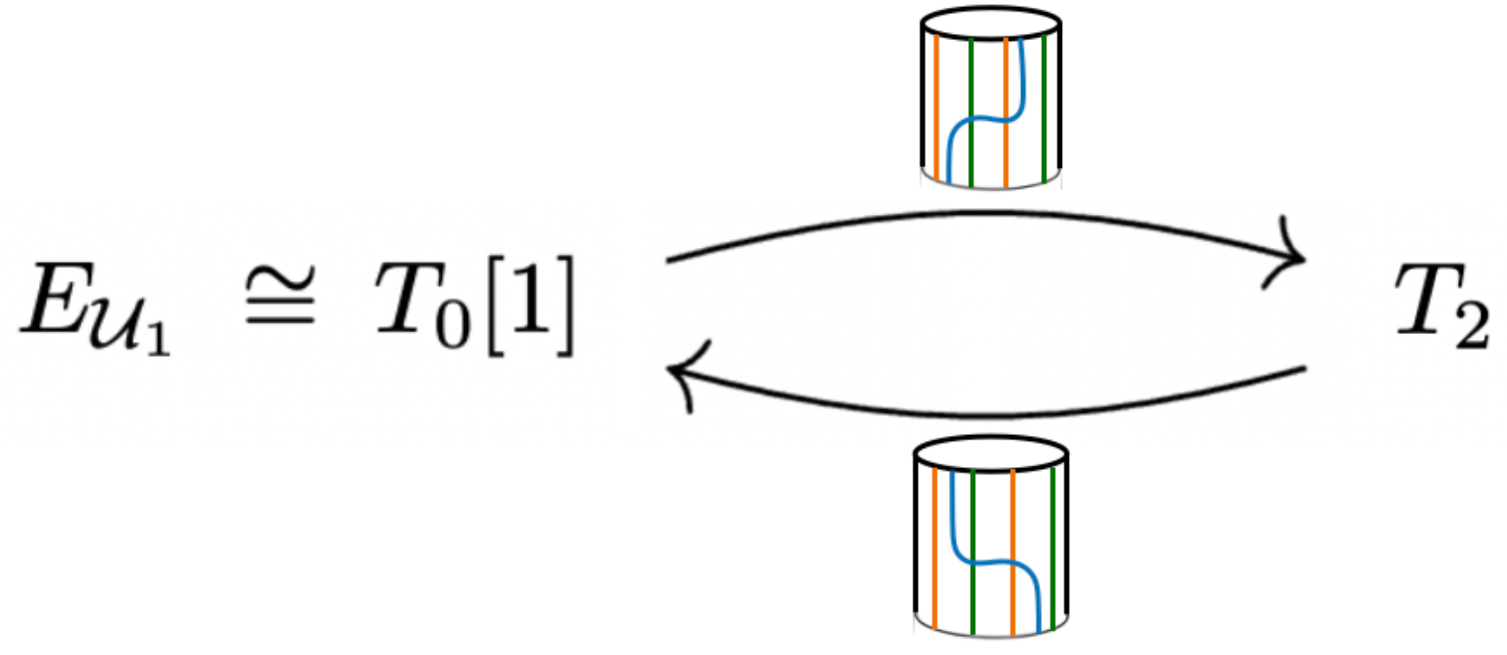}
\end{center}
\vspace*{-20pt}
\end{figure}
{\noindent}For clarity, for twisted complexes, we will write explicitly not only the equivariant degrees of all the terms in the complex, but also the homological ones. (An alternative is to write explicitly the degrees of all the maps. For ordinary complexes, degrees of all the maps are zero, so degrees of direct summand branes are fixed by knowing a degree of any one of its components. In a twisted complex, this is not the case.)  When a degree is not written, it vanishes. For example, the equivariant degrees of both $T$-branes in \eqref{gl1cup1} vanish.

The inequivalent choices of spin structure on a brane, viewed as a smooth Lagrangian on $Y$ should, as in section \ref{sec:signs1}, correspond to gauge inequivalent choices of signs in the differential. In this simple example, gauge transformations of the two $T$-branes in the complex can change signs of both maps in the differential simultaneously, leaving a ${\mathbb Z}_2$ choice for the sign of the product.  In writing $\delta$, we picked one. 

The two choices of spin structure on the $E_{{\cal U}_1}$ brane of $S^1$ topology are related by changing holonomy of the flat $U(1)$ connection on the brane by $e^{i \pi}$. More generally, changing the holonomy of the complexified flat $U(1)$ connection, corresponds to a deformation of $\delta$ that scales the product of the two maps in the differential, as gauge transformations scale the two maps oppositely. The example at hand is in fact a variant of the following canonical example: replace $Y$ by a cylinder without any punctures, and the $E_{{\cal U}_1}$ brane by a brane wrapping the $S^1$. That brane has a similar two-term resolution of the in terms of the real line Lagrangian which generates the category. The differential has an analogous deformation by scaling the product of terms in it,  which corresponds to sliding the brane along the cylinder together with varying the holonomy of the $U(1)$ flat connection on the brane. Note that some definitions of twisted complexes \cite{Seidel} ask for the differential to be upper or lower triangular. In our setting this is not what one wants since it would eliminate this simple brane, which clearly exists as an object of $\MDy.$

From the complex resolving the $E_{{\cal U}_1}$ brane, we can compute all its graded Homs to any of the $I$-branes with no further work. For example, it follows immediately that $Hom^{*,*}_{\MDY}(E_{{\cal U}_1}, I_1)$ vanishes, as applying $hom_Y^*(-, I_1\{\vec J\})$ to the complex sends all its terms to zero. This is because as no $T_1$ branes, which are dual to the $I_1$ brane, enter $E(T)$ in \eqref{gl1cup1}. This reproduces the fact the Floer complex $hom^{*,*}_Y(E_{{\cal U}_1}, I_{1})$ vanishes, as there the $E_{{\cal U}_1}$ and the $I_1$ branes do not intersect. The homology 
$$Hom^{*,*}_{\MDY}(E_{{\cal U}_1}, I_{{\cal U}_1})=0,
$$ 
with $I_{{\cal U}_1}= I_1\{1/2\}$, is the reduced homology of a pair of unlinked unknots.
By removing two of the four punctures, we would find that the unreduced homology of a single unknot also vanishes.  We will explain the degree shift between the cup brane and the basic $I_1$ brane in section \ref{Mgl11}.

\subsubsection{}
\begin{figure}[H]
	\centering
	\includegraphics[scale=0.3]{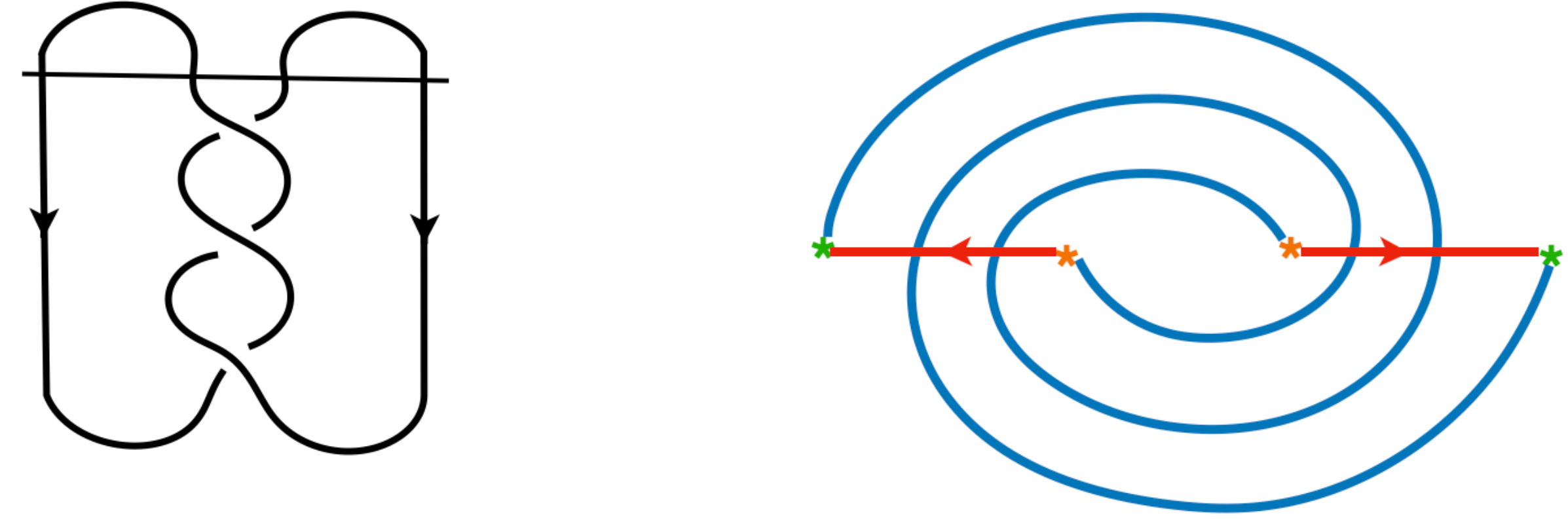}
	\caption{Left handed trefoil, and the corresponding matchings.}
	\label{trefoil}
\end{figure}

\begin{figure}[H]
	\centering
	\includegraphics[scale=0.7]{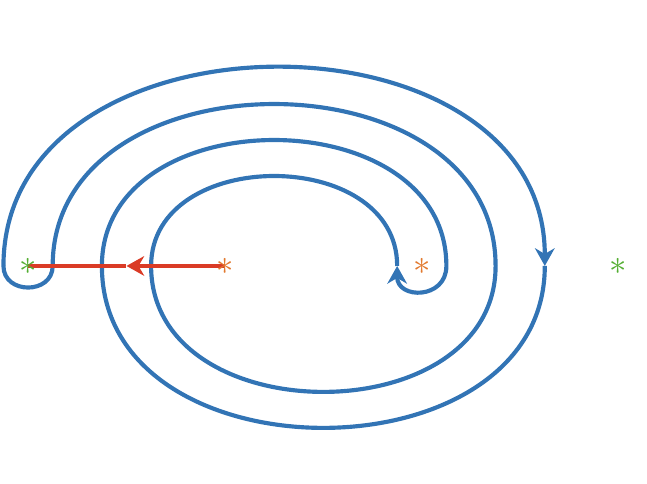}
	\caption{The branes for the reduced trefoil.}
	\label{redtrefoil}
 \vskip -0.5cm
\end{figure}

\begin{figure}[H]
	\centering
	\includegraphics[scale=0.7]{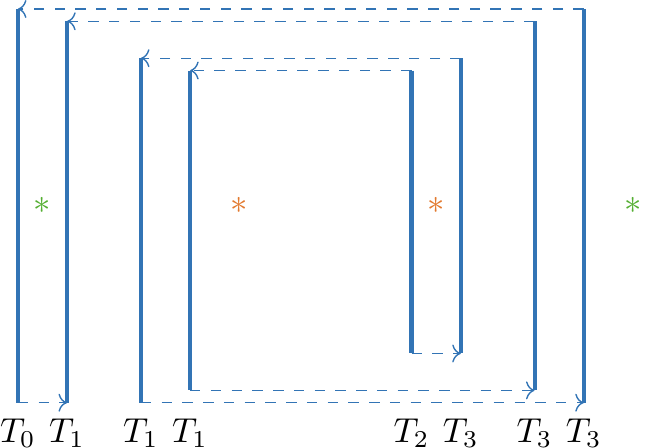}
	\caption{The resolution of the cap brane of the trefoil.}
	\label{trefoilResolution}
\end{figure}

For another simple example, consider the left-handed trefoil. The knot has a representation as a plat closure of a braid of four strands, resulting is the pair of matchings shown in Fig.~\ref{trefoil}. To get the matchings for reduced trefoil, we remove a pair of blue and red segments, per section \ref{sreduced}.  
By replacing the one remaining blue segment by an oval we get a pair of branes ${\mathscr B}E_{\cal U}$ and $I_{\cal U} = I_1\{{1\over 2}\}$ on $Y$, which is the Riemann surface ${\cal A}$ with four punctures in Fig.~\ref{redtrefoil}. We have rotated the figures clockwise by $90^\circ$ and suppressed the cylinder.
\noindent{}We can resolve the ${\mathscr B}E_{\cal U}$  brane into thimbles, as shown in Fig.~\ref{trefoilResolution}.  
Starting with any one thimble, we can work our way around the cap brane, using Fig.~\ref{trefoilResolution} to write down the thimbles and the maps connecting them, which gives:
\vskip -0.2cm
\begin{equation*}
	\includegraphics[scale=0.8]{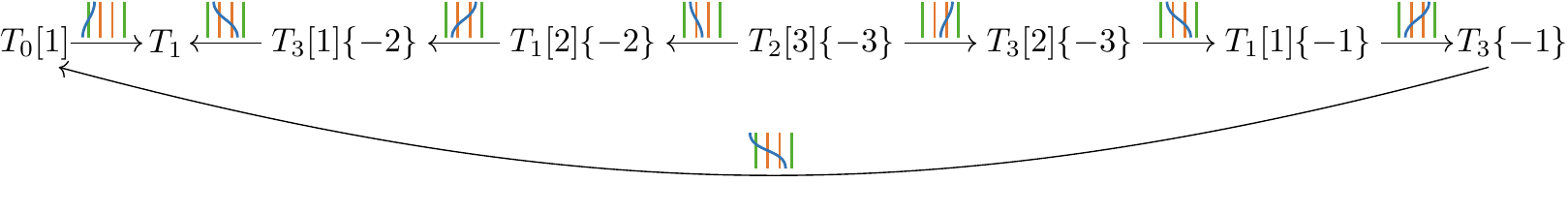}
\end{equation*}
\vskip -0.5cm
\noindent{}By construction, all relevant aspects of the Lagrangian ${\mathscr B}E_{\cal U}$ are encoded in complex. The relative degrees of all the $T$-branes in the complex follow from degrees of the maps between them, shown in Fig.~\ref{gl11}, and general facts from section \ref{coneone}.
\noindent{}
For example, all but one map in the complex have Maslov degree zero and so in the notation of \eqref{deltae}, correspond to $d_{k, \ell}$ which sends a thimble in degree $k$ to a thimble in degree $\ell=k-1$. The exception is the map
$\,\vcenter{\hbox{\includegraphics{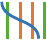}}}\,$
which has Maslov degree $2$ and therefore sends a thimble in Maslov degree $k$ to a thimble in Maslov degree $\ell=k+1$. 

While the relative degrees of the $T$-branes in the complex depend only on the underlying Lagrangian, the absolute degree depends on the braid group element whose action ${\mathscr B}$ represents. We will explain how to find absolute degrees of the branes in section \ref{Mgl11}. For this, it suffices to determine the absolute degree of any one $T$-brane in the complex. In this example, the absolute degree of the $T_1$ brane is $[1]$, unchanged from the $E_{\cal U}$ brane complex, as the functor ${\mathscr B}$ acts trivially on it.

There is an additional, non-geometric degree shift which accompanies the action of ${\mathscr B}$ on $\MDy$, as explained in section \ref{Mgl11}. Since it is the same for all the branes in ${\MDy}$, and in that sense non-geometric, we will not include it when we write the complex resolving the ${\mathscr B}E_{\cal U}$ brane.  It is important, however, to get the link homology whose Euler characteristic is the Alexander polynomial in standard normalization, and we will need to include it in the very end. In this case, there is overall degree shift of $\Delta M=0$, $\Delta J_0 ={\frac{3}{2}} $, coming from the fact ${\mathscr B}$ represents the element $\sigma_2^{-3}$, in the standard braid group notation, which braids a pair of even punctures.

The result is the direct sum brane 
$${\mathscr B}E_{\cal U}(T) = \bigoplus_k {\mathscr B}E_k(T)[k]$$ as in equation \eqref{aproBgl}, which has 
$${\mathscr B}E_3 = T_2\{-3\},\;
{\mathscr B}E_2 = \begin{pmatrix}{}T_3\{-3\} \\  T_1\{-2\}\end{pmatrix},\;
{\mathscr B}E_1 =  \begin{pmatrix}{}T_1\{-1\}\\ T_0 \\ T_3\{-2\}\end{pmatrix}, \;
{\mathscr B}E_0 = \begin{pmatrix}{}T_3\{-1\} \\  T_1\end{pmatrix}.
$$
Then, we can write
\begin{equation}
	\delta= \vcenter{\hbox{\includegraphics{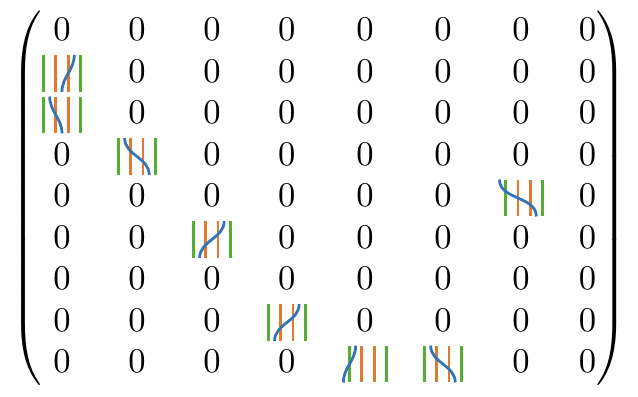}}}.
\end{equation}
The $d=1$ algebra is an ordinary associative algebra whose differential $\partial$ vanishes identically. Indeed, the differential squares to zero: $\delta\cdot\delta =0$.

The differential $\delta$ would still square to zero if we rescaled the maps in the complex by arbitrary ${\mathbb C}^{\times}$-valued parameters. Complexified gauge transformations of the direct sum brane reduce this to one parameter's worth, and correspond to the choice of complexified flat connection on the ${\mathscr B}E_{\cal U}$ brane of $S^1$ topology. Varying the spin structure is equivalent to tensoring with an additional ${\mathbb Z}_2$ flat connection.

The link homology $Hom^{*,*}_{\MDy}({\mathscr B}E_{\cal U}, I_{\cal U})$ is, up to overall degree shifts, the cohomology of the complex of vector spaces obtained by applying the $hom^{*,*}_Y(-, I_1)$ functor to the complex resolving the ${\mathscr B}E_{\cal U}$ brane. Non-vanishing contributions come only from the terms in the complex involving the $T_1$ brane,  since it is the dual to the $I_1$ brane. The complex that results is
\beq\label{ccomplex}
\textstyle {\mathbb C} [0] \xrightarrow{0}{\mathbb C}\{-1\} [1]\xrightarrow{0} {\mathbb C}\{-2\}[2]\rightarrow{}0,
\eeq
with the differential $\delta$ that gets mapped to zero.  
(The complex is really extended by $0$'s on both sides, as is the brane we obtained it from.) 

\begin{figure}[H]
	\centering
 \vskip -0.5cm
	\includegraphics[scale=0.7]{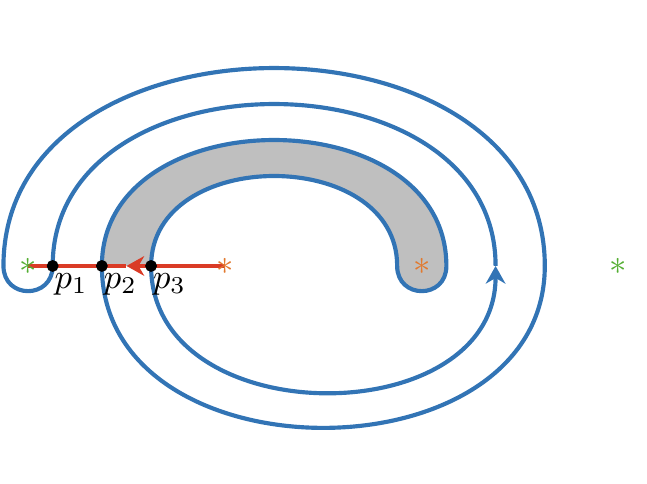}
	\caption{A disk interpolating from the intersection point $p_2$ to $p_3$ when the latter are viewed as elements of $Hom^{*,*}_{\MDy}(\mathscr{B}E_{\cal U}, I_{\cal U})$. }
	\label{trefoilDisks}
\end{figure}

The three copies of ${\mathbb C}$ in the complex \eqref{ccomplex} correspond, respectively, to the three intersection points which we labeled $p_1$, $p_2$, and $p_3$ in Fig.~\ref{redtrefoil}, which generate the Floer complex. As in section \ref{FR}, one establishes this by isotoping the $\mathscr{B}E_{\cal U}$ brane to the connected sum of $T$-branes. One can verify the identification by considering disks interpolating between the three intersection points. For example, the disk in Fig.~\ref{trefoilDisks} interpolates from intersection point $p_2$ to $p_3$, and has $J_0(A)=-1$ and $M(A)=1$ using \eqref{Jgrade} and \eqref{indcompare}, reproducing their relative degrees from \eqref{ccomplex}.

\begin{comment}
We will explain in section \ref{Mgl11} how to fix the absolute degrees of the $T$-branes in the complex. The absolute grading of ${\mathscr B}E_{\cal U}$ brane is data that the defines the brane object of $\MDy$ which is not encoded in the underlying Lagrangian. It is determined uniquely once one understands the ${\mathscr B}E_{\cal U}$ brane as obtained from the basic $E_{\cal U}$ brane in section \ref{simplest} by action of a specific braid group element that ${\mathscr B}$ represents.
\end{comment}

The absolutely graded homology differs from \eqref{ccomplex} by an overall grading shift. It comes from two places, as explained in section \ref{Mgl11}. The first is the overall degree shift of branes that accompanies the action of ${\mathscr B}$, mentioned above, which in this case adds $+\frac{3}{2}$ to $J^0$, and nothing to $M$. The second is the fact the cap brane is related to the $I_1$ brane by a canonical grading shift: $I_1 = I_{\cal U}\{-\frac{1}{2}\}$. The net result is a shift of equivariant grading by $\{1\}$. The result is the reduced ${\mathfrak{gl}_{1|1}}$ homology of the trefoil:
\beq\label{innorma}
Hom_{\MDy}^{*,*}({\mathscr B}E_{\cal U}, I_{\cal U}) ={\mathbb C} \{1\}\oplus {\mathbb C}[1]\oplus {\mathbb C} \{-1\}[2].
\eeq
Our $J^0$ grading is, as we will explain in section \ref{gl11 skein}, related to the usual one in (1.1) by $J^0\rightarrow -J^0$. The graded Euler characteristic recovers the Alexander polynomial ${\fq}-1+{\fq}^{-1}$.

\subsubsection{}
Computing the reduced $U_{\fq}({{\mathfrak{gl}}_{1|1}})$ link homology of any two-bridge link is as simple as the current trefoil example, because the target $Y$ is always one-dimensional. Stretching the ${\mathscr B} E_{\cal U}$ brane along the cylinder, breaking it up into $T$-branes and recording which maps one needs to turn on to glue the brane back together, results in  the full differential ${\delta}$ from the outset, as there is no diagonal to deform it. It is also manifest that the brane differential $\delta$ computes the Floer differential $\delta_F$ which in this case is easy to compute directly. The Floer differential either trivial, or receives contributions from one dimensional disks connecting pairs of intersection points which can cancel out in pairs by isotoping the branes.

In what follows, we will explain how to compute the reduced $U_{\fq}({{\mathfrak{gl}}_{1|1}})$ link homology for arbitrary links $K$, and any corresponding $d$. We will do so without relying on any special link presentations; the resulting theory is completely general.

\subsubsection{}\label{secexg}
To begin with, we will start with a very simple example which, even though not relevant for knot theory, is a good starting point to illustrate how to find projective resolutions and how they compute the Floer differential without counting solutions to instanton equations.

We can recover the brane $L_1\times L_2$ below 
\begin{figure}[H]
	\centering
	\includegraphics[scale=0.7]{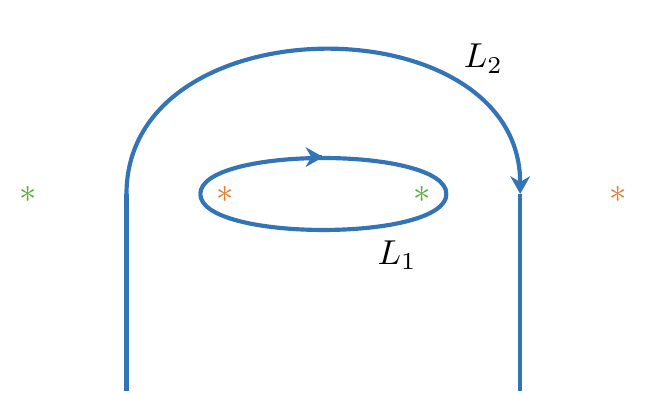}
\vskip -0.5cm
\end{figure}
\noindent{}by starting with a direct sum of $T$-branes, and taking connected sums.
\begin{figure}[H]
	\centering
	\includegraphics[scale=0.7]{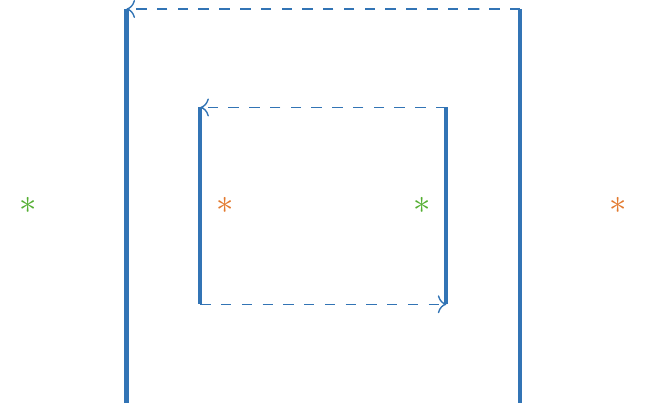}
	\caption{The resolution of the $L_1\times L_2$ brane in terms of $T$-branes.}
	\label{d2exampleResolution}
\end{figure}
To start, pick one of the two factor branes, say the $L_1$ brane, and write down its resolution. Keep track of the morphisms themselves, but ignore any degree shifts until all the factor branes have been incorporated. The resolution of the $L_1$ brane is 
\begin{equation}
	L_1 \qquad \cong \qquad \vcenter{\hbox{\includegraphics{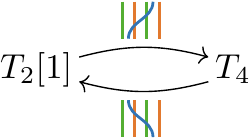}}}
\end{equation}
Then, take the product of $L_1$ with each of the two $T$-brane that $L_2$ branes breaks into, namely $T_2$ and $T_4$. The product with $T_2$ gives the complex:
\begin{equation}\label{fc}
	L_1 \times T_2 \qquad \cong \qquad \vcenter{\hbox{\includegraphics{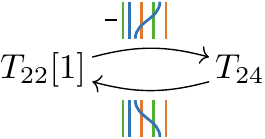}}}.
\end{equation}
The specific morphisms follow from Fig.~\ref{d2exampleResolution}. In writing down the maps, we keep track of the positions of $T$-branes relative to each other and to the punctures.

Similarly, the product of $L_1$ with $T_4$ gives:
\begin{equation}\label{sc}
	L_1 \times T_4 \qquad \cong \qquad \vcenter{\hbox{\includegraphics{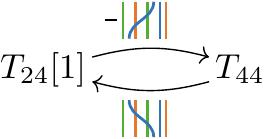}}}.
\end{equation}
The complexes encodes all aspects of the brane relevant to the A-model. For example, because the $T_2$ brane is to the left of the $L_1$ brane, and hence of both of its $T$-branes, the complex resolving $T_2\times L_1 $ does not contain blue strands that cross. 

To get $L_1\times L_2$, we take a connected sum of $L_1 \times T_2$ and $L_1 \times T_4$, using the specific maps that follow from Fig.~\ref{d2exampleResolution}. The maps are identity morphisms of the summands of $L_1$, times the morphism that makes up $L_2$. The corresponding string diagrams now involve blue strings that cross, since the open strings connecting the two $T$-branes in $L_2$ cross the open strings corresponding to identity of $L_1$. This gives:
\begin{equation}
\vcenter{\hbox{\includegraphics{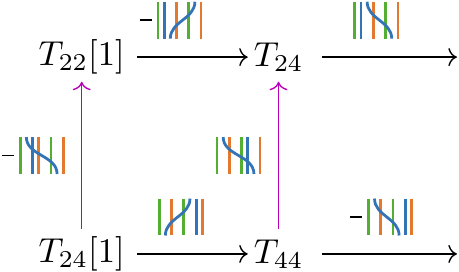}}}
  \label{example1complex}
\end{equation}

To avoid cluttering, we now present the complexes as periodic $d$-dimensional grids. The top and the bottom row of \eqref{example1complex} are rewriting of the complexes in \eqref{fc} and in \eqref{sc}. In this specific example, the complex ``lives" on $S^1\times {\mathbb R}$, where the periodic direction is the horizontal one. Correspondingly, rightmost arrows implicitly point to the beginning of the complex on the left. The new maps we have turned on run vertically, and glue $L_1\times T_2$ and $L_1\times T_4$ to get $L_1\times L_2$. All the relative Maslov and equivariant degrees of the $T$-branes in the complex are determined from the maps in the complex. For example, the morphism that takes the $T_{44}$ brane to lower left $T_{24}$ has Maslov degree $2$, and equivariant degree zero. Consequently if we were to place $T_{44}$ brane in Maslov degree $k$, the lower left $T_{24}$ brane would be in degree $k +2-1=k+1$ and the same equivariant degree. We chose the overall cohomological degree to be $k=0$; a different choice gives a resolution of a brane obtained from $L_1\times L_2$ by a degree shift.

The resulting complex 
has the direct sum brane
$$L(T) = \bigoplus_k L_k(T)[k] =\begin{pmatrix}{}T_{22}  \\  T_{24}\end{pmatrix}[1] \oplus  \begin{pmatrix}{}T_{24}  \\  T_{44} \end{pmatrix}
$$
and differential
\begin{equation}
	\delta_0 = \vcenter{\hbox{\includegraphics{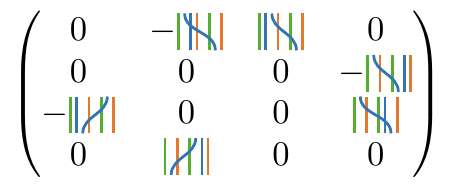}}}
\end{equation}
The differential $\delta_0$ squares to zero in the algebra $A_{\hbar =0}$ and gives a projective resolution of the brane on $Y_0 = Y\backslash \Delta.$  The brane has one complex parameter worth of deformations, corresponding to turning on a generic flat connection on the $L_1$ brane, of which we picked one. 

The signs in $\delta_0$ are assigned according to the discussion in appendix \ref{ssigns}. (On the first pass, it is instructive to ignore the signs and work with ${\mathbb Z}_2$ coefficients instead.) Specializing the signs in the appendix to the current $d=2$ case, we start with assigning the $+1$ sign to all the arrows in the first row as well as to all the arrows in the left-most column in (\ref{example1complex}). We then assign the same $+1$ sign to all columns, but alternate the overall signs in subsequent rows. Finally, we twist the sign of map going from a $T$-brane of odd Maslov degree by an additional $-1$. In checking that $\delta_0\cdot \delta_0=0$, one has to recall that the composition of strand-algebra elements, viewed as the Homs in a category of twisted complexes of $A$-modules, come with extra signs. These signs, explained in appendix \ref{A}, amount to replacing the Floer theory $m_2$, with its sign-twisted version $m_2^p$.

The differential $\delta$ that squares to zero in the full $A=A_{\hbar}$ algebra satisfies $\partial \delta+\delta\cdot \delta=0$, or more precisely eqn.~\eqref{square}. As explained in section \ref{TAg}, for $\hbar\neq 0$ the algebra product stays the same, but the algebra differential $\partial$ deforms. We have that $\partial \delta_0 + \delta_0\cdot \delta_0  = \partial \delta_0$, where
\begin{equation}
	\partial \delta_0 = \vcenter{\hbox{\includegraphics{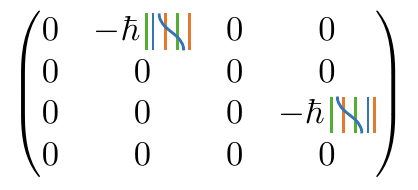}}},
\end{equation}
or, written as a complex:
\begin{equation} \label{example1Q}
    \vcenter{\hbox{\includegraphics{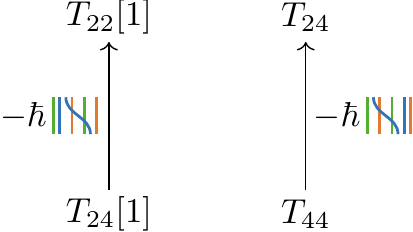}}}
\end{equation}
Since $\partial \delta_0$ does not vanish, the actual differential $\delta$ must be a non-trivial deformation of $\delta_0$. In fact, we will show momentarily that the exact differential $\delta$, one which satisfies \eqref{square} in the algebra  $A_{\hbar}$, is a first order deformation of $\delta_0$.  

On equivariant and cohomological degree grounds, there are only two new maps that one could turn in $\delta$. Solving for $\partial \delta +\delta \cdot \delta =0$ fixes their coefficients. One coefficient gets set to zero and the other to $-\hbar$, so that the exact brane differential $\delta$ equals
\begin{equation}
	\delta = \delta_0 + \hbar \delta_1 = \vcenter{\hbox{\includegraphics{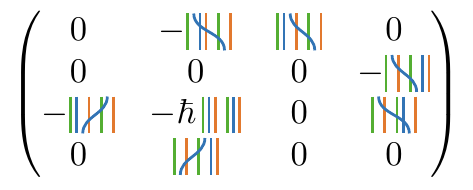}}}
\end{equation}
Written as a complex, the $\hbar$ deformation corresponds adding the orange arrow below:
\begin{equation}\label{hbar}
	\vcenter{\hbox{\includegraphics{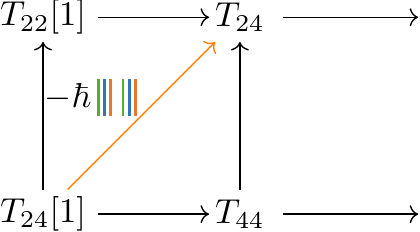}}}
\end{equation}
Per discussion in section \ref{dunique}, in looking for the full differential $\delta$, we excluded deformations of $\delta_0$ by terms already in it. They are either pure gauge, so they do not change the brane, or they change the complexified flat connection.  In deforming the ${\mathscr B}E_{\cal U}$ brane from $Y_0$ to $Y$, we want to keep the flat connection fixed.

The term in the differential \eqref{hbar} proportional to $\hbar$ encodes a non-trivial disk instanton on $Y$. To compute the space 
$Hom^{*,*}_{\MDy}(L, I_{24})$ of morphisms from the brane $L$ to
 $I_{24}=I_2\times I_4$, start by applying $hom^*_Y(-, I_{24}\{\vec{J}\})$ to the complex resolving $L$. 
The result is a complex of vector spaces one for each $\{\vec{J}\}$, to which only the terms with $T_{24}\{\vec{J}\}$ contribute.  The $k$-th cohomology of this complex is $Hom_{\MDy}(L, I_{24}[k]\{\vec J\})$. Recall the functor is contravariant, so reverses the direction of maps relative to the complex we started with.
The only non-zero contribution is a two-term complex in degree ${\vec J}=0$
\begin{eqnarray}\label{die}
\delta_F: \qquad    \mathbb{C} \xrightarrow{\;\;-\hbar\;\;} \mathbb{C}[1]
\end{eqnarray}
with vanishing cohomology. This complex coincides with the Floer complex. 
\begin{figure}[H]
	\centering
	\includegraphics[scale=0.7]{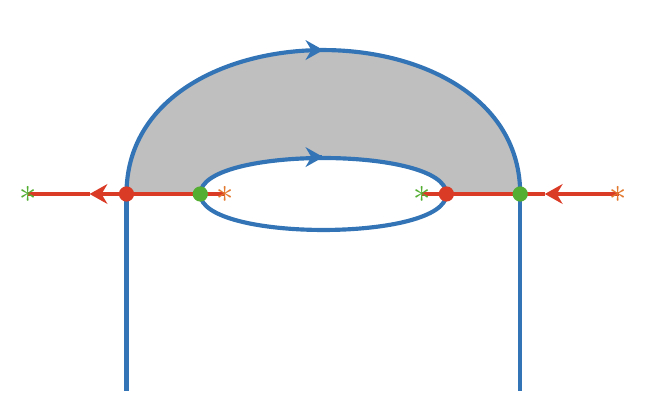}
	\caption{A domain which supports a holomorphic disk that generates the Floer differential acting on $hom_Y^{*,*}(L, I_{24})$. The red intersection point corresponds to the ${\mathbb{C}}$ factor in \eqref{die} and the green one to $\mathbb{C}[1]$.}
	\label{example1disk}
\end{figure}

The two terms in the complex \eqref{die} are the two intersection points of the $I_{24}$-brane with $L$. 
The differential in the complex is proportional to $\hbar$ and so counts maps to $Y$ that intersect the diagonal once. It accounts for the disk  which projects to the shaded domain $A$ shown in Fig.~\ref{example1disk}. Every such ``empty rectangle" has $M(A)=1$ and $J_0(A)=0$ and is known \cite{R} to contribute $\pm 1$ to the Floer differential, by an explicit calculation. Here have re-derived this, together with the sign, from the resolution of the brane $L\cong (L(T), \delta)$.

\subsubsection{}\label{2dexd}
Here is a simple example relevant to knot theory. 

\begin{figure}[H]
	\centering	\includegraphics[scale=0.3]{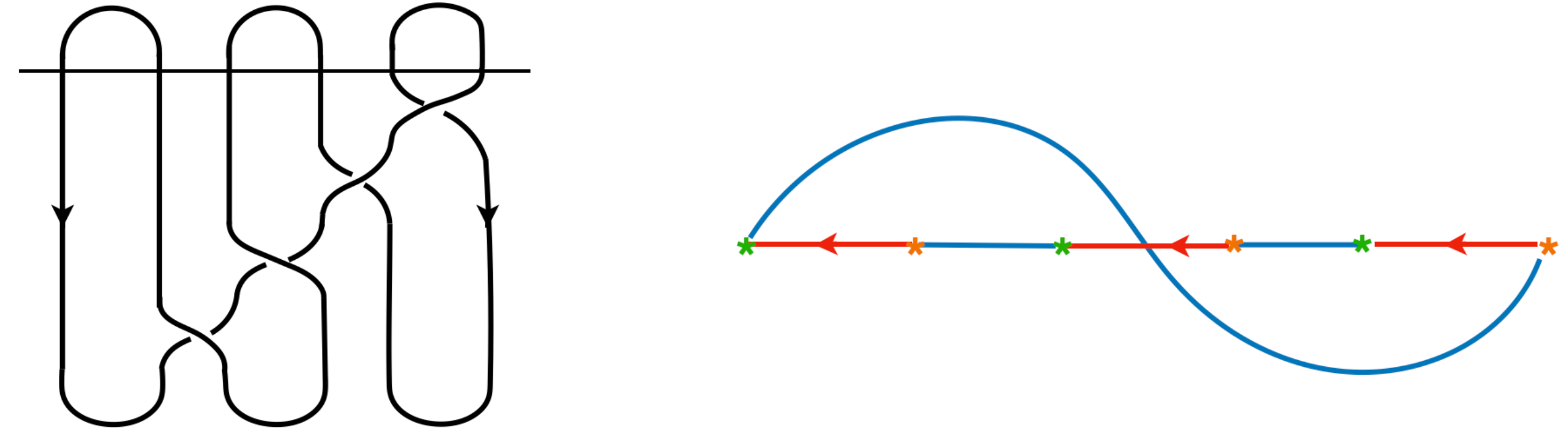}
	\caption{A $d=3$ presentation of the unknot as a plat closure of a braid, and the corresponding matchings.}
	\label{d2example}
\end{figure}
\begin{figure}[H]
	\centering
	\includegraphics[scale=0.7]{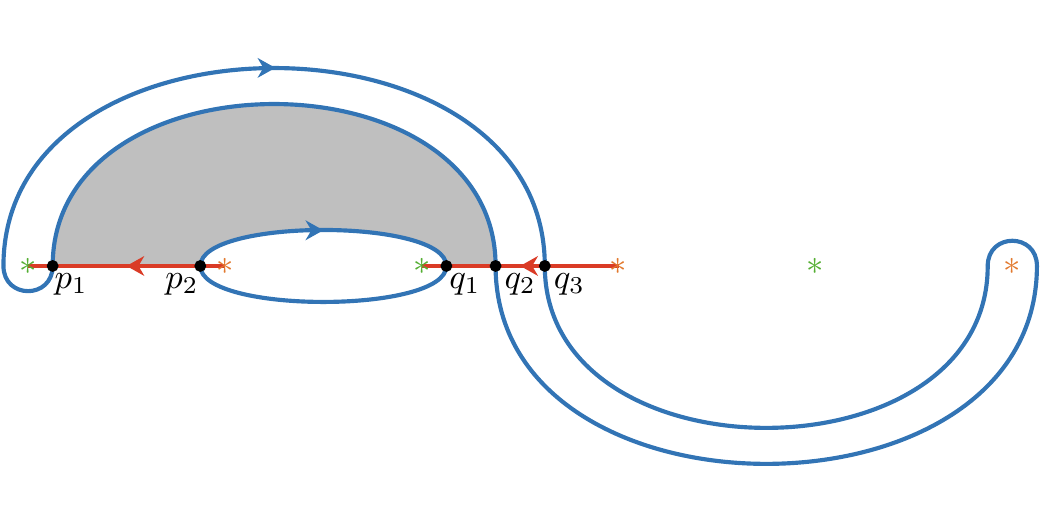}
	\caption{Branes for the unknot, and a disk which generates the Floer differential.}
	\label{unknotDisk}
\end{figure}

Reduced homology of unknot presented as in Fig.~\ref{d2example} comes from the brane configuration in Fig.~\ref{unknotDisk}, which is obtained from by removing a pair of red/blue matchings, and replacing the remaining blue matchings with ovals.

The cap and the cup branes have $3$ intersection points: $p_1 q_1$, $p_2 q_2$, and $p_2 q_3$. The disk shown in Fig.~\ref{unknotDisk} is an empty rectangle, which therefore has $M=1$ and $J_0=0$, and interpolates from $p_2 q_2$ to $p_1 q_1$. As we explained in our toy example above, such empty rectangles are known to contribute $\pm 1$ to the differential in Heegaard-Floer theory. So we expect to find that
\begin{equation}\label{expT}
	\delta_F \, p_2 q_2 = \pm \hbar\, p_1 q_1
\end{equation}
and homology that is one dimensional, as expected for the unknot.

This kind of counting of disks ``by hand" is possible only for very special examples because in general, maps of disks to symmetric product can be arbitrarily complicated. We will now show how the algorithm from section \ref{Algorithm}, which is completely general, recovers this disk and moreover fixes the sign of the disk contribution in \eqref{expT}.

\begin{figure}[H]
	\centering
	\includegraphics[scale=0.7]{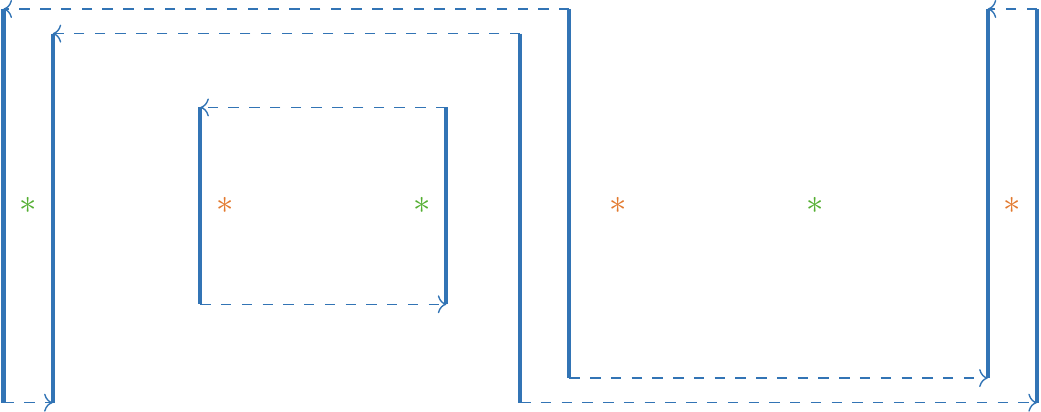}
	\caption{The cap brane for the unknot, broken up into thimbles.}
	\label{unknotResolution}
\end{figure}

Pick one of the two factor branes in ${\mathscr B} E_{\cal U} = {\mathscr B} E_1 \times E_2$, and find its resolution. For example, the resolution of ${\mathscr B} E_1$ is
\begin{equation*}
	\vcenter{\hbox{\includegraphics{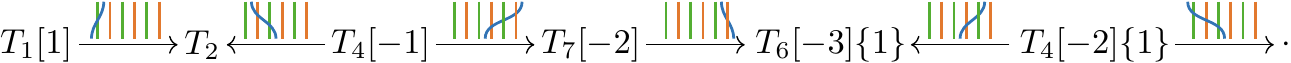}}}
\end{equation*}
Since the $E_2$ brane breaks into $T_2$ and $T_4$, to find the resolution of  $ {\mathscr B} E_1 \times E_2$, begin by reading off the complex for the products 
${ \mathscr B} E_1 \times T_2$ and  $ {\mathscr B} E_1 \times T_4$ 
using Fig.~\ref{unknotResolution}. As before, it is important to keep track of the placement of the $T_2$ and $T_4$ branes relative to the branes in the complex for $ {\mathscr B} E_1 $. Just as picture of the brane encodes the actual Lagrangian, so does the resolution read off from it. Taking these products will change some of the cohomological degrees of $ {\mathscr B} E_1$ because some of the maps end up containing crossings, each of which changes the Maslov degree by $-1$.

Taking these shifts into account, we find that the resolution of ${\mathscr B} E_1 \times T_2$ is
\begin{equation*}
    	\vcenter{\hbox{\includegraphics[scale=0.95]{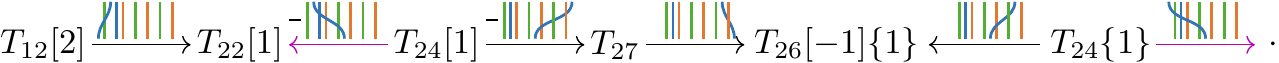}}}
\end{equation*}
and the resolution of ${\mathscr B} E_1 \times T_4$ is
\begin{equation*}
	\vcenter{\hbox{\includegraphics[scale=0.95]{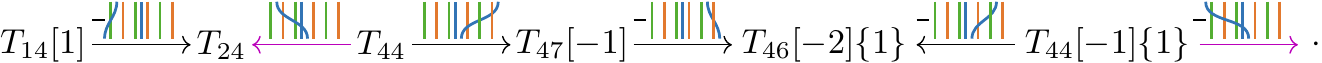}}}
\end{equation*}

Next, we take the direct sum of ${\mathscr B} E_1 \times T_2 $ and ${\mathscr B} E_1 \times T_4$ and turn on the geometric maps between $T_2$ and $T_4$ to combine ${\mathscr B} E_1 \times T_2$ and ${\mathscr B} E_1 \times T_4$ into ${\mathscr B} E_1 \times E_2$. Unfolding the complex into a toric grid, the result is
\begin{equation}
\vcenter{\hbox{\includegraphics[scale=0.8]{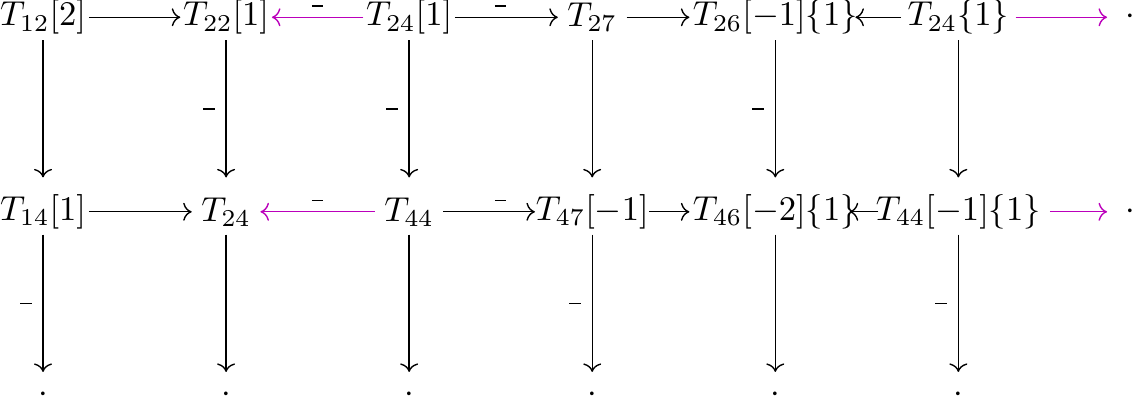}}}
\end{equation}
For the sake of readability, we have suppressed the maps, but they are fixed uniquely by the graded thimbles they map between. Additionally, we colored purple the maps containing strands that cross.

The signs are assigned as described in section \ref{sec:signs1} and appendix \ref{ssigns}. First, we start by fixing signs of the first row and all columns to $+1$. Next, we alternate signs of the horizontal maps according to the row. Finally, we multiply all maps starting at a site of odd Maslov degree by an additional factor of $-1$.

One can easily check that $\delta_0$, the collection of maps we found so far, defines a differential in the algebra $A_0$, which simply requires $\delta_0^2=0$ (taking all the signs from the twisted $A_{\infty}$ products into the account). Turning on $\hbar\neq 0$, $
\delta_0$ still squares to zero, but it no longer defines a differential, as $\partial \delta_0$ does not vanish in $A_{\hbar}$. It consists of:
\begin{equation*}
        \hspace{-0.4cm}
	\vcenter{\hbox{\includegraphics[scale=0.85]{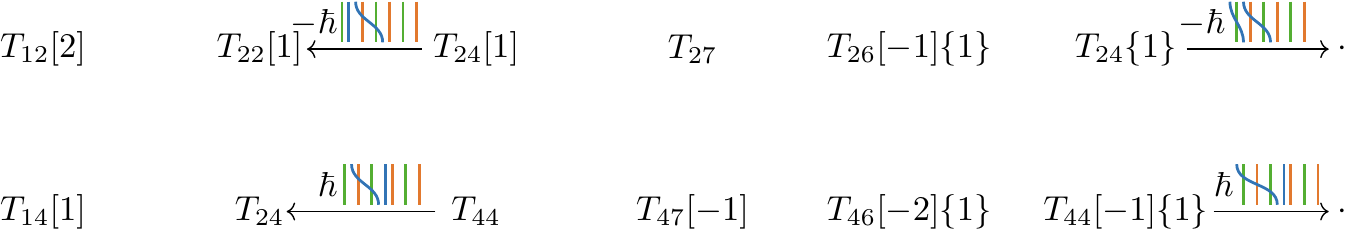}}}
\end{equation*}
The differential $\delta$ that satisfies $\partial \delta + \delta\cdot\delta=0$ in the algebra $A_{\hbar}$ is a deformation of $\delta_0$. In principle, on degree grounds, there are 16 new maps which may be turned on in $\delta$ shown below:
\begin{equation*}
        \hspace{-0.4cm}
	\vcenter{\hbox{\includegraphics[scale=0.85]{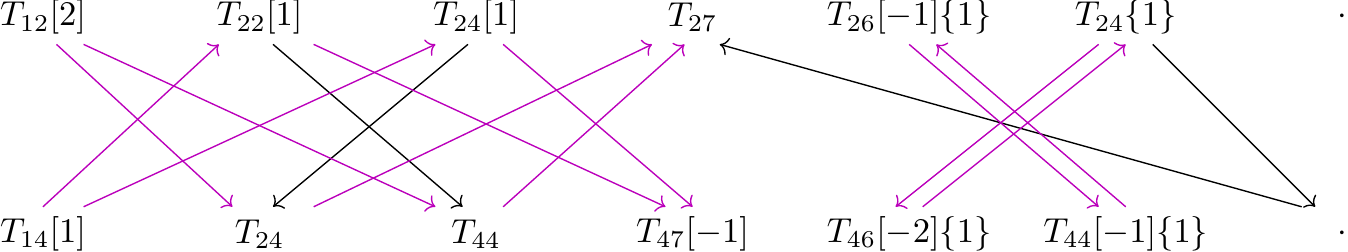}}}
\end{equation*}
%The 4 black maps do not contain crossing strands; the 12 maps colored purple do. 
The ansatz for $\delta_1$, the first order deformation of $
\delta_0$ thus contains 16 coefficients of these maps. 
Solving the first-order equation $\partial\delta_0 + \delta_0 \cdot \hbar\delta_1 + \hbar\delta_1 \cdot \delta_0 = 0$ reduces the 16-dimensional space to a 12-dimensional one.  One can find the exact solution, without introducing higher order terms in $\delta$, by asking that $\hbar\partial\delta_1 + \hbar \delta_1 \cdot \hbar\delta_1 = 0$. This uniquely determines the remaining 12 coefficients in $\delta_1$, leading to $\hbar \delta_1$ composed of only two maps:
\begin{equation*}
        \hspace{-0.4cm} 
	\vcenter{\hbox{\includegraphics[scale=0.85]{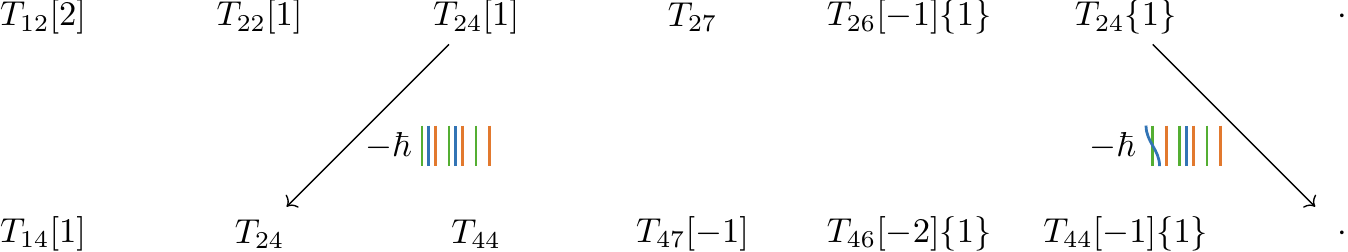}}}
\end{equation*}
One of these maps, the one from $T_{24}[1]$ to $T_{24}$, corresponds to the disk in Fig.~\ref{unknotDisk}. 
The complex resolving the ${\mathscr B}E_1\times E_2$ brane is thus the direct sum brane:
$$T_{12}[2] \oplus \begin{pmatrix}T_{14}\\T_{22}\\T_{24}
    \end{pmatrix}\![1]\oplus \begin{pmatrix}T_{24}\\T_{24}\{1\}\\T_{27}\\T_{44}
    \end{pmatrix}\!\oplus \begin{pmatrix}T_{26}\{1\}\\T_{44}\{1\}\\T_{47}
    \end{pmatrix}\![-1]\oplus T_{46}[-2]
    $$
with differential $\delta=\delta_0+\hbar \delta_1$, which we just computed.
 
To calculate homology,  we take $hom_Y^{*,*}(-, I_{24})$ of the complex, to find
$$0 \longrightarrow 0  \longrightarrow  \begin{pmatrix}  {\mathbb C} \\  {\mathbb C}\{1\} 
\end{pmatrix}  \xrightarrow{ \begin{pmatrix}-\hbar & 0
\end{pmatrix}} {\mathbb C}[1]\longrightarrow 0
$$
The three terms above are the three intersection points of the ${\mathscr B}E_1\times E_2$ brane with the $I_{24}$ brane in Fig.~\ref{unknotDisk}.

Taking cohomology, the the final result is
\begin{equation}
 Hom_{\MDy}({\mathscr B}E_{\cal U}, I_{\cal U}) = \mathbb{C},
\end{equation}
which is what we expect for the unknot. We used that the cap brane differs from the $I_{24}$ brane by $I_{\cal U}=I_{24}\{1\} $, and that the constant shifts $\Delta M$, $\Delta J_0$ from section \ref{Mgl11} vanish for the braid in Fig.~\ref{d2example}.

\subsubsection{}
As the final example, consider the $8_{19}$ knot. The link homology is comes from the following branes: 
\begin{figure}[H]
	\centering
	\includegraphics[scale=0.35]{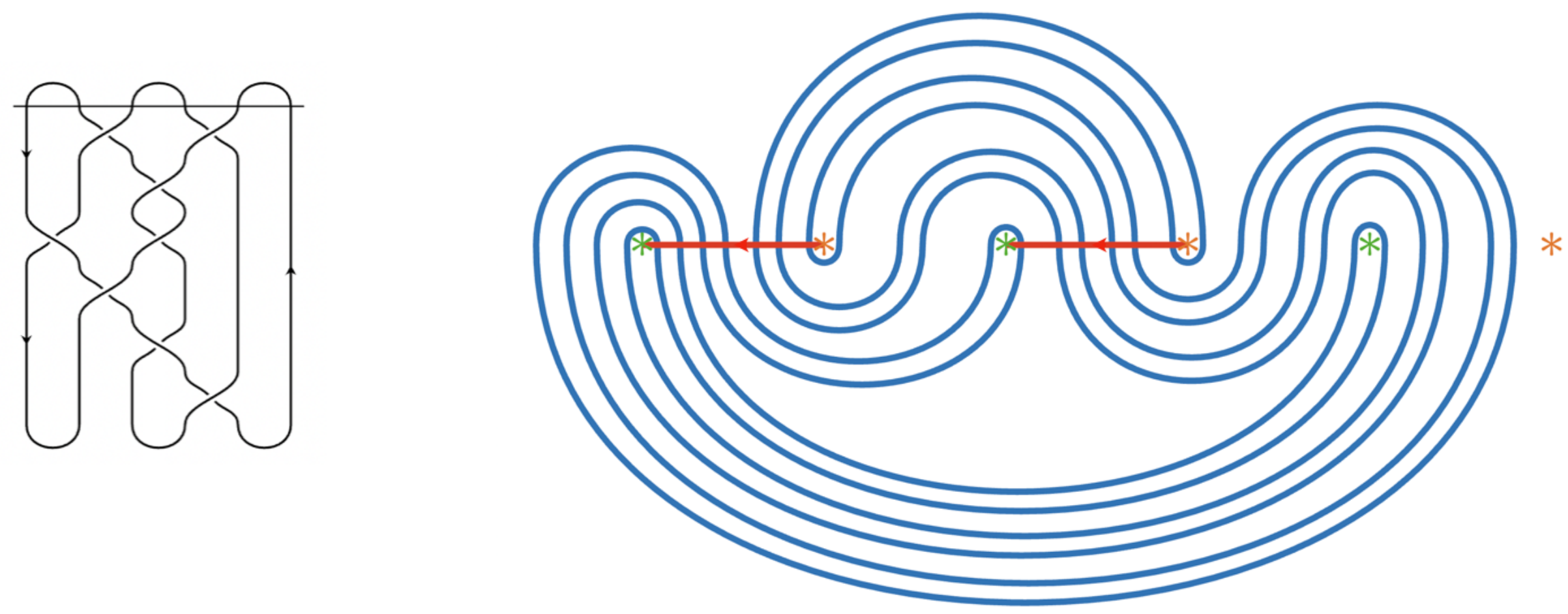}
	\caption{A presentation of $8_{19}$ as a braid closure and the corresponding branes}
	\label{8_19 braid}
\end{figure}

Compared to previous examples, finding the differential $\delta$ requires a second-order correction in $\hbar$ to $\delta_0$. Intersecting with the $I_{24}$ Lagrangian leads to the complex
\begin{equation*}
\begin{matrix}
    \bC\{5\}
\end{matrix}
\xrightarrow{
\begin{pmatrix}
    0 \\ 0 
\end{pmatrix}
}
\begin{matrix}
    \bC \{3\}\\
    \bC \{4\}
\end{matrix}
\xrightarrow{
\begin{pmatrix}
    -\hbar & 0 \\
    0 & 0 \\
    0 & 0 \\
    0 & 0 \\
\end{pmatrix}
}
\begin{matrix}
    \bC\{3\} \\
    \bC\{2\} \\
    \bC\{2\}\\
    \bC\{2\}
\end{matrix}
\xrightarrow{
\begin{pmatrix}
    0 & 0 & \hbar & 0 \\
    0 & \hbar & 0 & -\hbar \\
    0 & 0 & 0 & 0 \\
    0 & 0 & 0 & 0 
\end{pmatrix}
}
\begin{matrix}
    \bC\{2\} \\
    \bC\{2\} \\
    \bC\{1\} \\
    \bC\{1\}
\end{matrix}
\xrightarrow{
\begin{pmatrix}
    0 & 0 & 0 & 0  \\
    0 & 0 & -\hbar & 0  \\
    0 & 0 & -\hbar & \hbar  
\end{pmatrix}
}
\end{equation*}
\begin{equation*}
\begin{matrix}
    \bC \\
    \bC\{1\}\ \\
    \bC\{1\} 
\end{matrix}
\xrightarrow{
\begin{pmatrix}
    0 & 0 & 0   \\
    0 & 0 & 0   \\
    -\hbar & 0 & 0   
\end{pmatrix}
}
\begin{matrix}
    \bC \{-1\} \\
    \bC  \\
    \bC 
\end{matrix}
\xrightarrow{
\begin{pmatrix}
    0 & 0 & 0   \\
    \hbar & 0 & 0   
\end{pmatrix}
}
\begin{matrix}
    \bC \{-1\} \\
    \bC \{-1\} 
\end{matrix}
\xrightarrow{
\begin{pmatrix}
    0 & 0 \\
\end{pmatrix}
}
\begin{matrix}
    \bC \{-2\}  
\end{matrix}
\xrightarrow{
\begin{pmatrix}
    -\hbar
\end{pmatrix}
}
\begin{matrix}
    \bC \{-2\}  
\end{matrix}
\end{equation*}
which has homology
\beq\nonumber{}
 \bC [-2]\{5\}\oplus \bC [-1] \{4\}\oplus \bC \{2\}  \oplus \bC [3]\oplus \bC [4] \{-1\}.
\eeq
The degree shifts that come from the braid, which has $e_{++} = 0$ and $e_{--}=-2$, are $\Delta M = 2$ and $\Delta J_0 = -1$, by eqn.~\eqref{gs}. Accounting for $I_{\cal U} = I_{24} \{1\}$, we find that the regraded homology is
\beq
  Hom^{*,*}_{\MDy}({\mathscr B}E_{\cal U}, I_{\cal U})=     \bC \{3\}\oplus \bC [1] \{2\}\oplus \bC[2] \oplus \bC [5]\{-2\} \oplus \bC [6] \{-3\},
\eeq
which agrees with $\widehat{HFK}$ after replacing $J_0$ by $-J_0$, as expected from section \ref{gl11 skein}.

\subsection{Homological $U_{\fq}({\mathfrak{gl}_{1|1}})$ link invariants from $\mathscr{D}_{Y_{\mathfrak{gl}_{1|1}}}$}\label{Mgl11}	

In this section we will prove that $Hom_{\MDy}^{*,*}({\mathscr B}E_{\cal U}, I_{\cal U})$ are homological link invariants, which is Thm.~\ref{tinv} for $\mathfrak{gl}_{1|1}$. Along the way, we will explain how to assign absolute gradings to the branes and their homology groups.

The key ingredient in the proofs is the following relation, which says that the branes on the left and the right hand sides of:

\begin{figure}[H]
    \centering
	\includegraphics[scale=0.32]{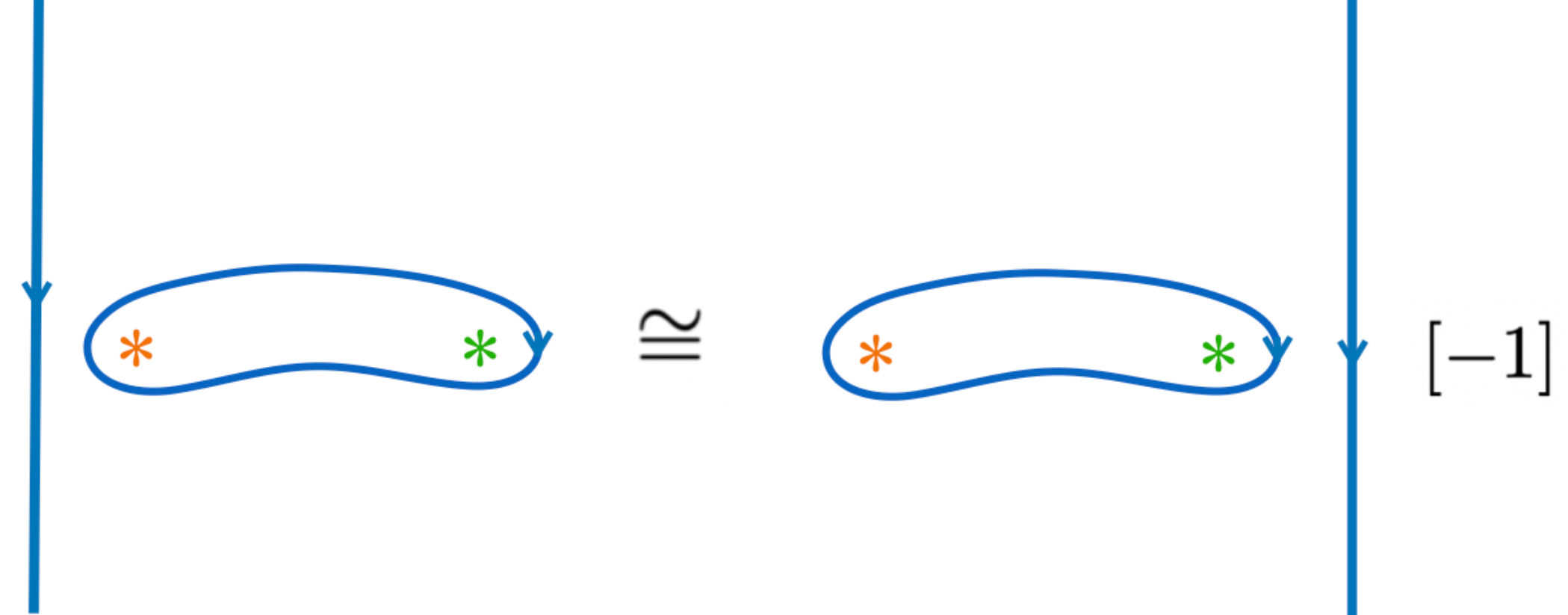}
	\caption{The ``swiping move" relates two equivalent objects of $\MDy$}
	\label{Mgl}
\end{figure}
\noindent{}are equivalent objects of $\MDy$. We prove this in appendix \ref{Markovgl} where we show that the resolutions of the two branes are homotopy equivalent for any $\hbar \neq 0$. 

The fact that the swiping move holds for any $\hbar \neq 0$, but not for $\hbar =0$ is why one must work in the A-model on $Y$, instead of the simpler theory on $Y_0 = Y\backslash \Delta$. 
While the Euler characteristics of the homology theories coming from $Y$ and $Y_0$ are the same,  only the theory on $Y$ will give homological link invariants.

\subsubsection{}
Once we know the swiping move  in Fig.~\ref{Mgl} holds,  all four moves in Figs.~\ref{MI}, \ref{MII} and \ref{stab} follow, up to possible overall Maslov and equivariant degree shifts. They in fact hold with gradings included, as we will explain below.

The moves in Fig.~\ref{MII}, up to possible grading shifts,  hold by isotopy invariance alone. The Markov I type move from Fig.~\ref{MI}, translated to cup branes,
reads 

\begin{figure}[H]
    \centering
\includegraphics[scale=0.35]{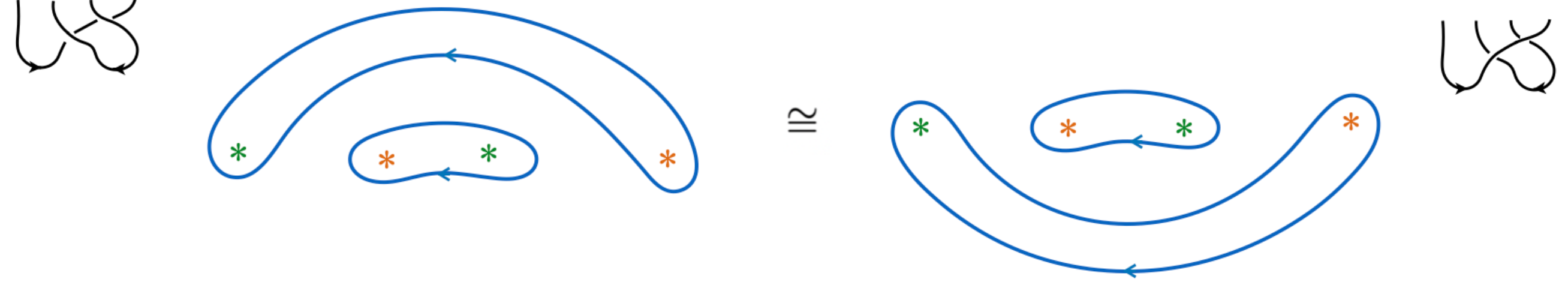}
	\caption{Markov I type move, in terms of branes}
	\label{MIgl11b}
\end{figure}
To show the relation holds, we first show the brane on the left is equivalent, using the swiping move, to the following brane:

\begin{figure}[H]
    \centering
\includegraphics[scale=0.35]{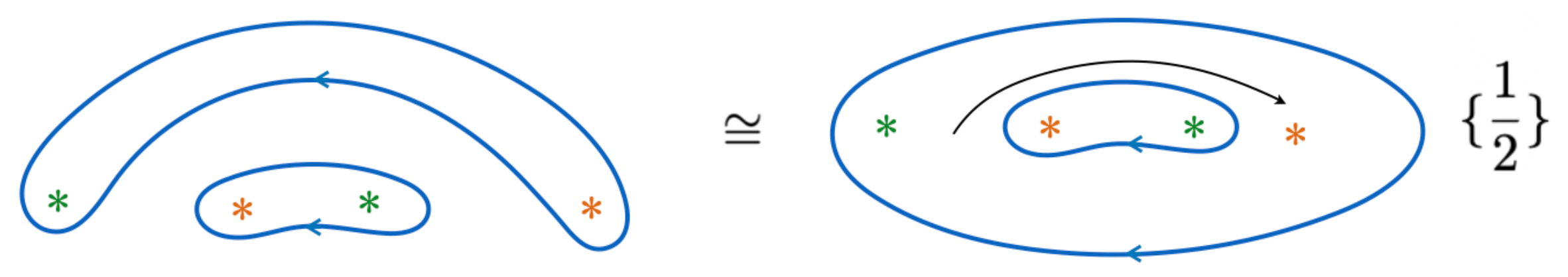}
	\caption{Swiping move simplifies the brane.}\label{secondbl}
\end{figure}
\noindent{}To show this, start with the $E_{\cal U}$ brane which is a product of two basic oval branes. Next, use the swiping move in Fig.~\ref{Mgl} to get an equivalent description of the brane in terms of one where the left brane ``swallows" the right. Finally, braid both branes to get the equivalence in Fig.~\ref{secondbl}. The resulting brane is based on the product Lagrangian pictured, with the overall degree that needs to be determined. The degree will depend on the braid used to obtain the brane.

The brane on the right is similarly equivalent to:

\begin{figure}[H]
    \centering
	\includegraphics[scale=0.35]{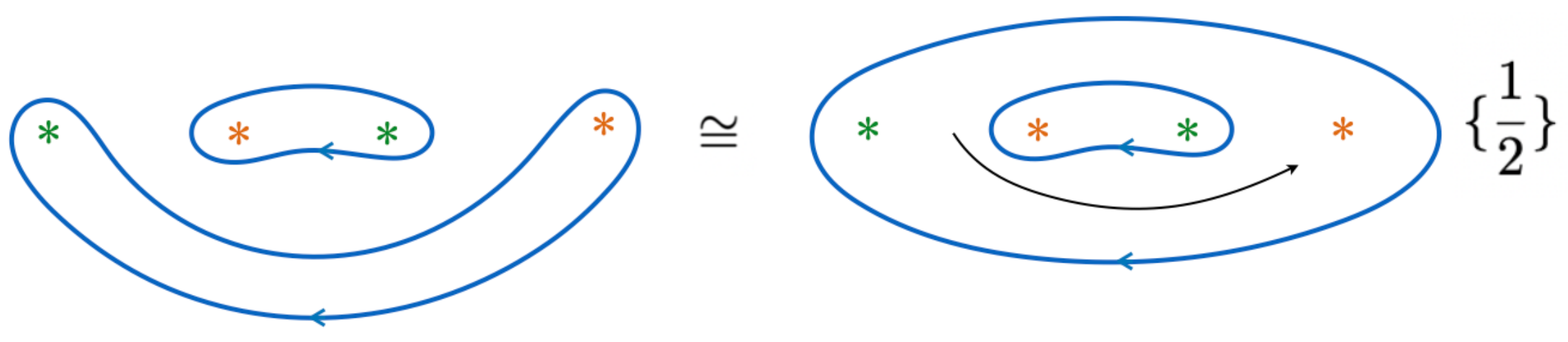}
	\caption{The grading of the resulting brane in principle depends on the path of the green puncture. }\label{secondbr}
\end{figure}
\noindent{}The two branes that result are manifestly equivalent objects of ${\MDy}$, up to possible overall Maslov and equivariant degree shifts which come from the fact we obtained them by braiding in two different ways. We will show below they are equivalent on the nose, with the grading included. There are analogous versions of these relations for different choices of strand orientations, which can be similarly shown to hold exactly.

The stabilization move from Fig.~\ref{stab} reads:
\begin{figure}[H]
    \centering
	\includegraphics[scale=0.28]{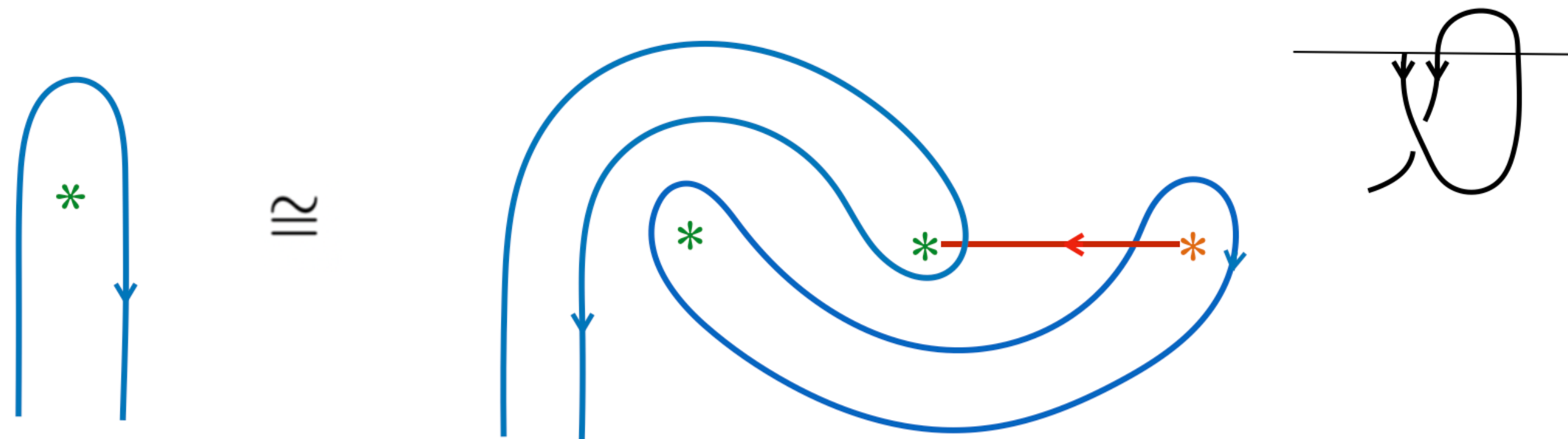}
	\caption{Stabilization move in terms of branes}
	\label{stabgl}
\end{figure}
\noindent{}By first applying the swiping move to the $E_{\cal U}$ brane and then braiding, one gets an equivalent presentation of the brane on the right hand side as:
\begin{figure}[H]
    \centering
	\includegraphics[scale=0.35]{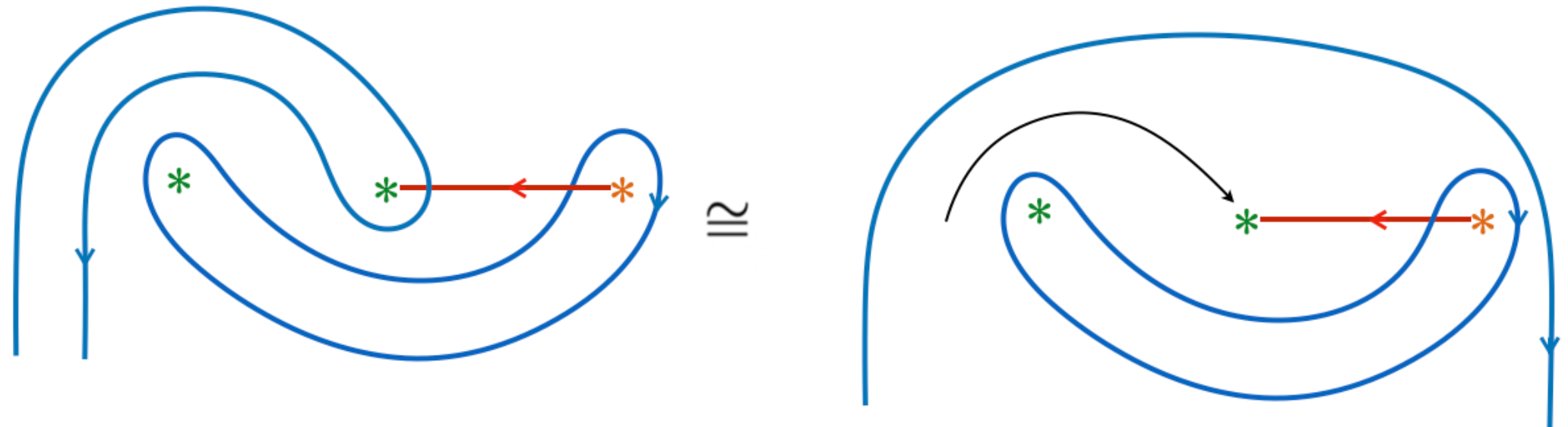}
	\label{}
\end{figure}
\noindent{}Since the $I$-brane intersects the only the interior component of the $E$-brane, and only once, the stabilization move in Fig.~\ref{stabgl} follows up to a grading shift.

\subsubsection{}
The moves in Figs.~\ref{MI}, \ref{MII} and \ref{stab} in fact hold exactly, with Maslov and equivariant gradings  of the branes included.  To explain this, we need to explain how absolute gradings of the branes are assigned.

Absolute equivariant gradings, as well as cohomological gradings mod $2$, are visible already in K-theory, so they are the ``easy part" of the theory, that was not central to us so far.  Now, we will fill the gap.
\subsubsection{}\label{keepingtrack}

The grading of the ${\mathscr B}E_{\cal U}$ brane is determined uniquely from the grading of the $E_{\cal U}$ brane we started with and the braid group element whose action on $\MDy$ the functor ${\mathscr B}$ represents. We will take the $E_{\cal U}$ brane to be the product of $d$ one dimensional $E$-branes from section \ref{secexg}. The degrees of the $T$-branes in the resolution of the brane range from $[0]$ to $[d]$, where $d$ is the dimension of $Y$; the equivariant degrees all vanish.  

Because the relative degrees of the $T$-branes in the complex resolving the ${\mathscr B}E_{\cal U}$ branes are fixed from geometry of the underlying Lagrangian alone, knowing the absolute degree of the ${\mathscr B}E_{\cal U}$-brane is the same as knowing absolute degree of any one $T$-brane in its resolution. 

The absolute grading of the ${\mathscr B}E_{\cal U}$ brane receives two contributions.
The first contribution is the obvious one, due to the geometric braid group action on $\MDy$, whose objects are graded branes:
Pick any $T_{\cal C}$-brane in the direct sum brane resolving the $E_{\cal U}$ brane. Consider its connected sum  with a ``spectator" $T$-brane on which ${\mathscr B}$ acts trivially. Since braiding leaves the spectator brane fixed, the relative gradings of the connected sum $T$-brane determine the absolute gradings of all branes in ${\mathscr B}T_{{\cal C}}$. As an example, the braid in Fig.~\ref{secondbl} gives a brane with no geometric degree shift at all. The braid in Fig.~\ref{secondbr} results in a geometric degree shift of $\{1\}$.

\subsubsection{}
There is a second contribution to the grading of the  ${\mathscr B}E_{\cal U}$ brane, which originates from terms in $\Omega$ and $W$, in 
\eqref{gy} and \eqref{dressed}, which are constant on $Y$, but depend on the positions of the punctures in a non-single-valued way. The constant terms in $\Omega$ and $W$ have a conformal field theory origin, recalling the relation to $\Lfgh$ chiral algebra from section \ref{sCFT}. They become important if we want the quantum group to have the canonical action on $K$-theory and its categorification in ${\MDy}$.

As a result of these terms, the geometric braid group action on ${\MDy}$ is accompanied with an action of a ``constant" degree shift functor. The action of this functor is constant in that it depends only on the braid group element that ${\mathscr B}$ represents, and is the same for all objects in $\MDy$. 

Recall that every braid group element is a word in the generators $\sigma_i^{\pm 1}$, where $\sigma_i$ exchanges $i$-th and $i+1$-st puncture counterclockwise. The generators satisfy $\sigma_i\sigma_{i+1}\sigma_i =\sigma_{i+1}\sigma_{i}\sigma_{i+1}$, and $\sigma_i\sigma_j=\sigma_j\sigma_i$, for $i>j$. Let $e_{++}$ and $e_{--}$ be the sums of exponents of braid group generators that braid a pair of even and odd punctures, respectively. Then terms in \eqref{gy} and \eqref{dressed} result in shift functors that act on every object of $L\in \MDy$ by a shift 
\beq\label{const}L\rightarrow L[\Delta M]\{\Delta J_0\},
\eeq
in cohomological and equivariant degrees, given by
\beq\label{gs}
\begin{aligned} 
\Delta M &= -e_{--} \\
\Delta J_0 &= - \frac{e_{++}}{2} + \frac{e_{--}}{2}.
\end{aligned}
\eeq

For example, the  Markov I move from Fig.~\ref{MIgl11b} involves a pair of braids that have $e_{++}=-1$ and $e_{++}=1$; both have $e_{--}=0$.
The resulting constant shifts of $\{\frac{1}{2}\}$ and $\{-\frac{1}{2}\}$ combine with the geometric degree shifts of zero and $\{1\}$, respectively, to result in both branes having degree $\{\frac{1}{2}\}$. Consequently, the Markov I move holds exactly, with gradings included.
\subsubsection{}\label{absgl11}
The absolute grading on the homology groups $Hom^{*,*}_{\MDy}({\mathscr B}E_{\cal U}, I_{\cal U})$ depends also on the absolute grading of the $I_{\cal U}$ branes.

The grading is determined by asking that the stabilization move in Fig.~\ref{stabgl} holds, as well as its variants obtained by changing the orientations of the strands. The result is that the cap brane $I_{\cal U}$ is not simply the product of basic intervals oriented right to left in degree zero, as is the $I_{\cal C}$ brane, a brane canonically dual to one of the $T$-branes, as in \eqref{dual}.
Rather, it differs by degree shifts which depend on orientation of the caps:
\begin{figure}[H]
    \centering
	\includegraphics[scale=0.38]{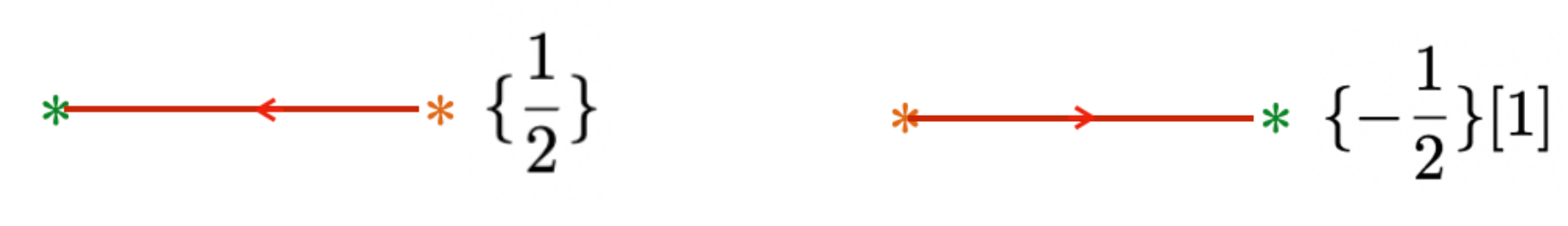}
	\caption{The cap brane $I_{\cal U}$ is related to the   basic corresponding $I$-brane by degree shift, which is either $\{\frac{1}{2}\}$ or $\{-\frac{1}{2}\}[1],$ depending on the orientation. }
	\label{Ibraneo}
\end{figure}
The fact the stabilization move holds comes from three degree shifts canceling out, two shifts coming from the $E$-brane and one from the $I$-brane. There is an overall shift of ${\mathscr B}E_{\cal U}$ brane degree  $[1]\{-\frac{1}{2}\}$ coming from a pair of odd punctures braided an element whose exponent is $-1$.
The geometric degree of the intersection point of the ${\mathscr B}E_{\cal U}$ brane with the basic $I$-brane has degree $[-1]\{1\}$. The two shifts combined give the intersection point of the ${\mathscr B}E_{\cal U}$ brane with a basic $I$-brane net degree $\{\frac{1}{2}\}$. Due to the degree shift between the $I_{\cal U}$ brane and the basic $I$-brane from Fig.~\ref{Ibraneo}, this is equivalent to an intersection point of the ${\mathscr B}E_{\cal U}$ brane with an $I_{\cal U}$ brane of Maslov and equivariant degree zero. It is easy to check that, varying orientations of the strands, we get net zero degree shifts in all other cases as well. 
\subsubsection{}
The remaining two relations in Fig.~\ref{MII} follow similarly. 
The first of the relations Fig.~\ref{MII} is an easy exercise, which amounts to showing the geometric degree shift of the $E_{\cal U}$ brane gets canceled by the constant overall degree shift from \eqref{const}. The second relation in Fig.~\ref{MII}, corresponding to freedom to add a twist to a cup, holds manifestly for $E_{\cal U}$-branes, as it translates to a freedom to reorder the odd and even puncture within it without changing the brane. It also holds for the $I_{\cal U}$ branes, but only after including the shifts from Fig.~\ref{Ibraneo}. These categorify the statement that the change of framing acts trivially on $U_{\fq}(\mathfrak{ gl}_{1|1})$ group invariants. 

\subsubsection{}\label{t1gl11}

To complete the proof of Thm.~\ref{tinv}  for ${\mathfrak{gl}_{1|1}}$, it remains to show the remaining Markov type moves. These are moves obtained from those in Figs.~\ref{MI} and \ref{stab} by exchanging tops and bottoms, also hold. 
 As sketched in section \ref{thm1s}, the proof proceeds in two steps. First, we observe that $Hom_{{\MDy}}^{*,*}({\mathscr B}E_{\cal U}, E_{{\cal U}})$ is invariant under the two moves since the $E_{{\cal U}}$ branes are. Next, we compute how  $Hom_{{\MDy}}^{*,*}({\mathscr B}E_{\cal U}, E_{{\cal U}})$ relates to $Hom_{{\MDy}}^{*,*}({\mathscr B}E_{\cal U}, I_{{\cal U}})$. 
 
Per definition, $Hom_{{\MDy}}^{*,*}({\mathscr B}E_{\cal U}, E_{{\cal U}})$ is the cohomology of a double complex obtained from $hom_Y^{*,*}({\mathscr B}E_{\cal U}, E_{{\cal U}})$ by replacing the ${\mathscr B}E_{\cal U}$ brane its projective resolution, and the $E_{{\cal U}}$ brane by its {\it injective} resolution. The injective resolution is a resolution in terms of $I$-branes, instead of the $T$-branes. Each one-dimensional $E_i$-brane in 
$$E_{\cal U} = E_{1} \times \ldots\times E_d$$
has a resolution 
$E_i \cong (E_i(I), \delta_i)$ in terms of a pair of corresponding $I_i$ branes, each with a relative shift in degree by $[1]\{-1\}$.  Taking the  product of the resulting one-dimensional complexes, we get the injective resolution of the $E_{\cal U}$ brane as $E_{\cal U} = (E(I), \delta_E)$ where the $E(I)$ is a direct sum $I_{\cal U}$ branes in different degrees:
\beq\label{EIr} E(I) =  I_{\cal U} \;\Bigl(\{{1/ 2}\}\oplus [1]\{{-1/ 2}\}\Bigr)^{\bigotimes d} \equiv I_{\cal U}\otimes W.
\eeq
We introduced a $2^d$ dimensional graded vector space $W$ to implement the degree shifts for us. 
Analogously to section \ref{dfd}, the only terms in the corresponding brane differential $\delta_{E}$ that survive to the double complex involve identity maps between $I_{\cal U}$-branes. These do not exist on degree grounds. Correspondingly, it follows
 \beq\label{EI}
 Hom_{{\MDy}}^{*,*}({\mathscr B}E_{\cal U}, E_{{\cal U}}) =
 Hom_{{\MDy}}^{*,*}({\mathscr B}E_{\cal U}, I_{{\cal U}})\otimes  W.
 \eeq
Since $W$ is fixed and independent of the braid ${\mathscr B}$, $ Hom_{{\MDy}}^{*,*}({\mathscr B}E_{\cal U}, I_{{\cal U}})$ must transform in the same way under the outstanding Markov-type moves, as $ Hom_{{\MDy}}^{*,*}({\mathscr B}E_{\cal U}, E_{{\cal U}})$ does. Since the latter is invariant, so is  $ Hom_{{\MDy}}^{*,*}({\mathscr B}E_{\cal U}, I_{{\cal U}})$.
This completes the proof of Thm.~\ref{tinv} for ${\mathfrak{gl}_{1|1}}$.

\vskip 1cm
\subsection{ ${U_{\fq}({\mathfrak{gl}_{1|1}})}$ skein from branes}\label{gl11 skein}

We will now prove Thm.~\ref{tJ}, which says the Euler characteristic of $Hom^{*,*}_{{\MDy}}({\mathscr B}E_{\cal U}, I_{\cal U})$ satisfies the skein of the Alexander polynomial. Since homology groups are independent of the choice of the link projection by Thm.~\ref{tinv}, it suffices to show that Euler characteristic of homological invariants obtained by the very special link presentation in Fig.~\ref{Skeinb} satisfy the skein relation. 

\begin{figure}[H]
    \centering
	\includegraphics[scale=0.38]{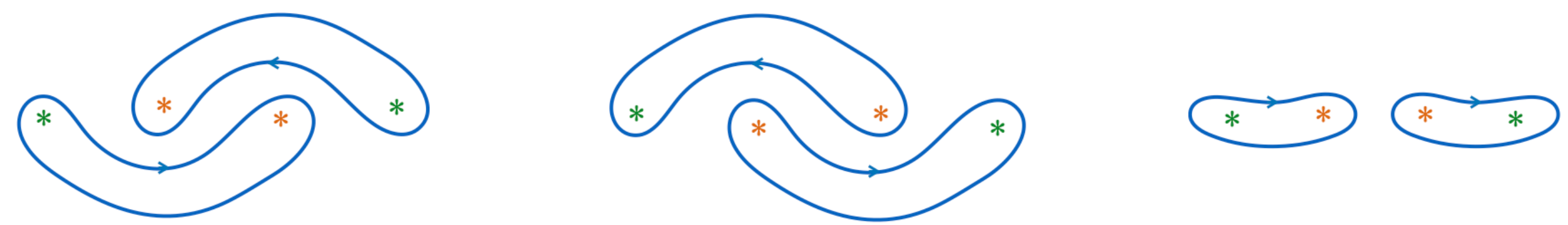}
	\caption{The three branes that the skein relates.}
	\label{skeinbranes}
\end{figure}

\begin{figure}[H]
    \centering
	\includegraphics[scale=0.37]{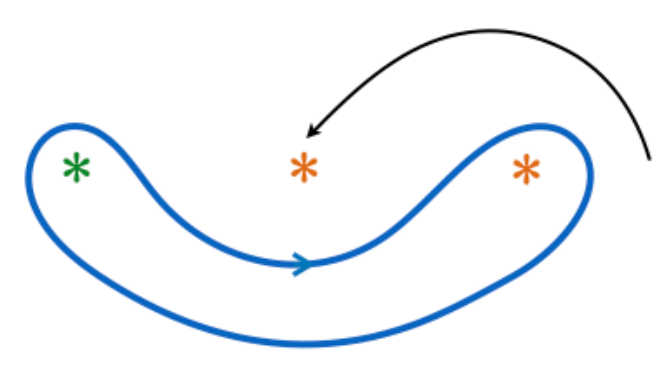}
 \caption{The reduced brane corresponding to $\includegraphics[scale=0.25]{positive}$ crossing}\label{gl11pos}
\end{figure}

\begin{figure}[H]
    \centering
	\includegraphics[scale=0.35]{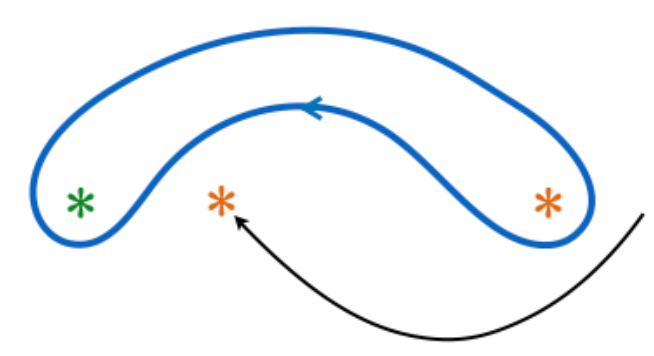}
  \caption{The reduced brane corresponding to $\includegraphics[scale=0.25]{negative}$ crossing}\label{gl11neg}
\end{figure}

\begin{figure}[H]
    \centering
	\includegraphics[scale=0.4]{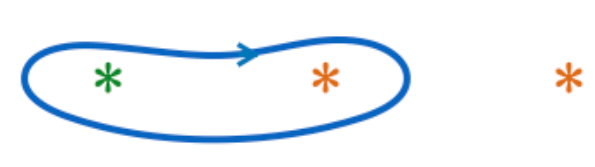}
  \caption{The reduced brane corresponding to $\includegraphics[scale=0.3]{nocrossing}$}\label{gl11no}
\end{figure}

The Euler characters of $Hom^{*,*}_{{\MDy}}({\mathscr B}E_{\cal U}, I_{\cal U})$ will satisfy the skein relation if  K-theory classes of the three branes which represent the braided cups at the bottoms of Fig.~\ref{Skeinb}, satisfy the skein relation. The branes whose K-theory classes we need to consider are given in Fig.~\ref{skeinbranes}.
Using the swiping move, for each of three branes in Fig.~\ref{skeinbranes} can be simplified by letting the rightmost brane  ``swallow" the left. 
\noindent{}Having done so, it suffices to show that K-theory classes of the three branes in Figs.~\ref{gl11pos}, \ref{gl11neg} and \ref{gl11no} satisfy the skein relation.

The first of the three branes, given in Fig.~\ref{gl11pos}, is obtained from the simple $E_{\cal U}$ brane by braiding two positive punctures by $\sigma_2$. Its resolution is given by
\beq \vcenter{\hbox{\includegraphics{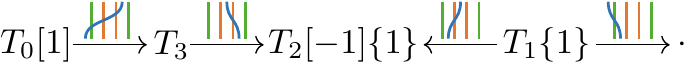}}}
\eeq
\noindent{}
The overall degree of the brane, before any non-geometric shifts, follows from the fact that braiding acted trivially on the $T_0$ brane, hence its degree could not have shifted. With $e_{++}=1$, $e_{--}=0$, there is an additional degree shift of the brane by $\{-1/2\}$. 
The final K-theory class of the brane is ${\fq}^{-1/2}({\fq}[T_1]-{\fq}[T_2]+[T_3]-[T_0])$. 
The second brane, in Fig.~\ref{gl11neg}, is obtained from the simple $E_{\cal U}$ brane by braiding a pair of positive punctures by  $\sigma_2^{-1}$. The resolution of the brane 
\beq
\vcenter{\hbox{\includegraphics{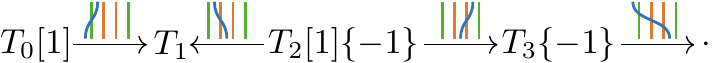}}}
\eeq
gets accompanied by an additional degree shift of $\{1/2\}$, since  $e_{++}=-1$, $e_{--}=0$,  so final K-theory class of the brane is ${\fq}^{1/2}([T_1]-{\fq}^{-1}[T_2]+{\fq}^{-1}[T_3]-[T_0])$. 
The last brane, in Fig.~\ref{gl11no}, is the simple $E_{\cal U}$ brane from section \ref{simplest}, whose K-theory class is $[T_2]-[T_0]$. 
K-theory classes of the three branes in Figs.~\ref{gl11pos}, \ref{gl11neg} and \ref{gl11no} satisfy one relation:
\begin{equation}\nonumber{}
    \centering
	\includegraphics[scale=0.42]{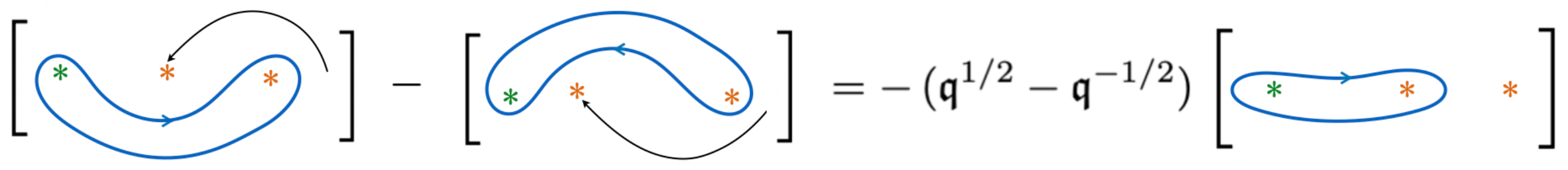}
\end{equation}
\noindent{}This is the  skein relation satisfied by the ${U_{\fq}({\mathfrak{gl}_{1|1}})}$ $R$-matrices from e.g.~\cite{oh}. It  coincides with the $n=0$ skein relation in eqn.~(1.1) after identifying 
\beq\label{glred}{\bf{q}}={\mathfrak{q}}^{-1}.
\eeq
The fact that our ${\fq}$, and ${\bf{q}}$ in terms of which the Alexander polynomial and its skein relation are written, are related in this way is responsible for the flip of $J_0$ grading between our homology theory and the conventions of $\widehat{HFK}$ homology theory which we have seen in all our examples.

\section{ The ${\mathfrak{su}_{2}}$  theory}\label{ssu2}

In this section, we describe the generalization of Heegaard-Floer theory, from previous section, to $^L{\fg} = {\mathfrak{su}_{2}}$, categorifying the Jones polynomial. 

\subsection{The A-model on $Y_{\mathfrak{su}_{2}}$}
We take the Riemann surface ${\cal A}$, with $n=2d$ punctures colored by the two-dimensional defining representation $V$ of ${\mathfrak{su}_{2}}$. The representation is self-conjugate, so $V$ and $V^*$ coincide, and we have only one kind of a puncture, located at $y=a_{i}$, where $i$ runs from $1$ to $2d$. The target 
$Y=Y_{{\mathfrak{su}_{2}}},$
\beq\label{su2t}
Y=Sym^d({\cal A})
\eeq
is again the symmetric product of $d$ points on ${\cal A}$, but equipped with a different top holomorphic form $\Omega_{\mathfrak{su}_2}$ and a different potential $W_{\mathfrak{su}_2}$ than in the ${\mathfrak{gl}_{1|1}}$ case. 
\subsubsection{}
The Maslov grading comes from the top holomorphic form $\Omega =\Omega_{\mathfrak{su}_2} $, 
\beq\label{OSU2}
\Omega =  \bigwedge_{\alpha =1}^{d} \frac{dy_{\alpha}}{y_{\alpha}}, \eeq
which coincides with the holomorphic volume form on the Cartesian product of $d$ copies of ${{\mathbb C}^{\times}}$, with coordinate $y\neq 0, \infty$.
The theory admits a ${\mathbb Z}$-valued Maslov grading since  $\Omega^{\otimes 2}$ is a globally well defined section of $K_Y^{\otimes 2}$, and hence
\beq\label{2c}
2c_1(K_Y) =0.
\eeq
The form $\Omega$ differs from the standard top holomorphic form on $Sym^d({\mathbb C}^{\times})$ by having a first order pole along the diagonal $\Delta$ in $Y$. If $\sigma_{1}, \ldots \sigma_d$ are elementary symmetric functions of $y_1, \ldots, y_d$, the latter equals $\bigwedge_{\alpha =1}^{d} {dy_{\alpha} \over y_{\alpha}} \prod_{\alpha< \beta}(y_{\alpha}-y_{\beta}) = (\bigwedge_{\alpha =1}^{d} d\sigma_{\alpha})/\sigma_d$. Unlike $\Omega$, this canonical top holomorphic form is invariant under the action of the permutation group $S_d$ on $y_1, \ldots, y_d$. 

The fact that $\Omega_{\mathfrak{su}_{2}}$ differs from $\Omega_{\mathfrak{gl}_{1|1}}$ in \eqref{Ogl11} means that the curves one counts in the two theories are not the same, even though they both map to $Y = Sym^{d}({\cal A})$.

\subsubsection{}
The equivariant gradings come from a potential $W= W_{\mathfrak{su}_2}$,
\beq\label{supsu2}
W =   \lambda_{0} W^{0} + \lambda_{1} W^{1}, 
\eeq
which is a multi-valued holomorphic function on $Y$.
We take $\lambda_0$ to be related to ${\fq}$ as in \eqref{qdef},
and $\lambda_1$ is associated to the holonomy around the $S^1$, as in the ${\mathfrak{gl}_{1|1}}$ case. The functions $W^0$ and $W^1$ are given by:
$$
W^{0} = \ln f, \qquad \qquad W^1 =\ln  \prod_{\alpha =1}^{d}y_{\alpha},
$$
for
\beq\label{dressedsu2}
f (y)={ \prod_{ i =1}^{2d} \prod_{\alpha=1}^d(1- a_i/y_{\alpha}) \over \prod_{1\leq \beta \neq \alpha\leq d} (1-y_{\beta}/y_{\alpha})  } {\prod_{1\leq i<j\leq 2d} (1- a_i/a_j)^{1/2}}.
\eeq
For comparison to the general simple Lie algebra $^L{\fg}$ case, if $\mu$ is the highest weight of the fundamental representation coloring the punctures and $^Le$ the positive simple root, the exponents in the numerator are $1={\langle ^Le, \mu \rangle}$ and ${1\over 2}={\langle \mu, \mu  \rangle}$, the exponent in the denominator is $1={\langle ^Le, ^L \! e \,\rangle}/2$.

The factors of \eqref{dressedsu2} that depend only on the positions of punctures on ${\cal A}$ play no role in any local considerations. They become relevant when we consider the action of braiding on $\MDy$, as explained in section \ref{Msu2}. 
\subsubsection{}
A proposal that a similar A-model should categorify the Jones polynomial was made by Gaiotto and Witten  \cite{GW} and pursued in \cite{GMW, GM}. It was meant to be a finite dimensional effective theory for Witten's five dimensional gauge theory \cite{WF}. The theory in \cite{WF} can also be interpreted as an A-model, except coming from a Landau-Ginsburg model with an infinite dimensional target space \cite{GMW}. 
\subsection{The category of A-branes on $Y_{\mathfrak{su}_{2}}$}

The categories of A-branes on  $Y_{\mathfrak{su}_2}$ and on  $Y_{\mathfrak{gl}_{1|1}}$  closely parallel each other.  In particular, for both theories, objects of $\MDy$ are product Lagrangians of the form \eqref{Lo}, which are graded by Maslov and equivariant gradings, where the grading is a choice of a lift of the phase of $\Omega$ and of $e^{-W}$ to real valued functions on $L$. One novelty of the present case is that immersed Lagrangians play a key role.

\subsubsection{}
As in the ${\mathfrak{gl}_{1|1}}$ theory, the branes are allowed to end at punctures of ${\cal A}$, due to the fact potential has logarithmic poles there. An example is a cap brane $I_{\cal U}$ on $Y$, which is a product of $d$ non-intersecting intervals, 
$$I_{\cal U} = I_1 \times \ldots \times I_d,$$
where $I_j$ begins at $y=a_{2j-1}$ and ends at $y=a_{2j}$.

The cup brane $E_{\cal U}$ on $Y$, 
$$E_{\cal U} = E_1 \times \ldots \times E_d,$$
is a product obtained by replacing each $I_j$ brane by an immersed, figure eight Lagrangian,  encircling the corresponding punctures.
The brane has well-defined Maslov grading, since the phase of $dy_j$ comes back to itself around a figure eight. It also has a well defined equivariant grading, as it winds around the two punctures in opposite directions.

\subsubsection{}\label{Fdef}
For a holomorphic map to contribute to A-model amplitudes, as discussed in section \ref{FTD}, its image in $Y$ should have finite area, computed using any symplectic form compatible with $\Omega$, and moreover $W$ should pull back to a regular function on the disk. As always, the latter condition guarantees that the equivariant degree of the map is zero, but in the ${\mathfrak{su}_2}$ theory,  it does not preclude maps from passing through punctures on ${\cal A}$. 

As shown in \cite{A2}, for $W$ to pull back to a regular function on the disk ${\rm D}$ under the map $y:{\rm D}\rightarrow Y$, every order one branch point of the projection $S \rightarrow {\rm D}$ must map to a puncture on ${\cal A}$ with multiplicity $1$, and every node of the projection to $\rm D$ must map to a puncture with multiplicity $2$. 
Here, $S$ is a Riemann surface in ${\rm D}\times {\cal A}$, which the map $y:{\rm D}\rightarrow Y$ corresponds to in the cylindrical formulation \cite{L} of Floer theory, as in section \ref{cylindrical}.

\subsubsection{}
As in Heegaard-Floer theory, there is a simpler cousin of $Y$, the  manifold
\beq\label{su2yo}
Y_O = Y\backslash D_O,
\eeq
obtained from $Y$ by removing the divisor 
\beq\label{su2Do}
D_O = \Delta+ D_a.
\eeq
Along the diagonal $\Delta$, at least two out of $d$ points coincide on ${\cal A}$; along $D_a$, at least one of them coincides with a finite puncture.
The theory on  $Y_O$ is  simple to solve \cite{A2}; in particular, removing $\Delta$ eliminates all but the maps without branching in $S\rightarrow {\rm D}$, as the branching index coincides the intersection of the image with the diagonal $\Delta$.

The theory on $Y_{\mathfrak{su}_2}$ is obtained from the theory on $Y_O$ by carefully understanding the effect of filling in $D_O$. Direct A-model analysis appears in \cite{ADZ}. Equivalently, we can use homological mirror symmetry as a guide, as \cite{ADZ} proves it.
\begin{comment}\subsubsection{}
After deleting $D_O$,
the theories based on ${\mathfrak{su}_2}$ and ${\mathfrak{gl}_{1|1}}$ are not only similar but become equivalent, up to a regrading. The key difference, though not the only one, between the two theories is how divisor $\Delta$ gets filled in on 
$Y_{\mathfrak{su}_2}$ and on $Y_{\mathfrak{gl}_{1|1}}$. In particular, for us on $Y_{\mathfrak{gl}_{1|1}}$, the divisor of punctures never gets filled in.
\end{comment}
 
\subsubsection{}\label{localsu2}
In the theory on $Y=Y_{\mathfrak{su}_{2}}$, the Maslov index of a map $y: {\rm D} \rightarrow Y$, where ${\rm D}$ is a disk with two marked points on the boundary, is given by \cite{A2}
\beq\label{indcompare2}
{ ind}(y) =  2 e(A),
\eeq
where $e(A)$ is the Euler measure defined in \eqref{Euler}. The index of the $\mathfrak{gl}_{1|1}$ theory in \eqref{indcompare} has two additional contributions, corresponding to the two additional contributions to the top holomorphic form of that theory in \eqref{OG11} relative to \eqref{OSU2}. (The indices of $\ell+1$ pointed disks for other values of $\ell$ are related similarly as well.)

The $c^0$-equivariant degree (corresponding to the ${\fq}$ grading) is given by
\beq\label{Jgradeb}
J_0(y) =   i(A) -2n(A). 
\eeq
Above, $n(A)$ is the multiplicity of $A$ at all the finite $a$-punctures, and $i(A)$ is the intersection of the map with the diagonal, defined in \eqref{iD}.
The factor $i(A)$ in the equivariant degree originates from the factor of $\Delta(y)=\prod_{\alpha<\beta}(y_{\alpha} - y_{\beta})$ in the potential $W^0$, through \eqref{dressedsu2}. 
The $c^1$ equivariant degree is the same as in \eqref{J1}.

\subsubsection{}\label{bigelow}
The fact that the Euler characteristic of $Hom_{\MDy}^{*,*}({\mathscr B}E_{\cal U}, I_{\cal U})$, which is the graded count of intersection points of the ${\mathscr B}E_{\cal U}$ and $I_{\cal U}$ branes in~\eqref{EG}, reproduces the Jones polynomial of the link
$${\chi}({\mathscr B} E_{\cal U}, I_{\cal U}) = \sum_{{\cal P}} (-1)^{M({\cal P})} {\fq}^{J( {\cal P})} =  J_K({\fq}).$$
is a theorem of Bigelow \cite{bigelow}, building on the work of Lawrence
\cite{Lawrence1, Lawrence2, Lawrence3}. In section \ref{su2 skein}, we will give a different, direct proof of this.
Our theory explains the homological  origin of Bigelow's peculiar construction and generalizes
it to other $U_{\mathfrak{q}}(^{L}{\mathfrak{g}})$ link invariants. A relation of Bigelow's representation of the Jones polynomial to symplectic Khovanov homology of \cite{SS} was first observed in \cite{M}. 

\subsection{Equivariant mirror symmetry}
Mirror symmetry relating $Y$ to its mirror $X$ fits into a larger framework of mirror symmetries.
The theory on $Y$, and its mirror $X$, are defined by the rectangle
\begin{center}
\begin{tikzcd}[sep= large]
{\mathcal X}\ar[r,<->,"\textup{mirror}"]
\arrow[<->]{rd}[description]{\textup{equiv. mirror}}&{\mathcal Y}
\ar[d, <->,""]
\\
X\ar[r,<->,"\textup{mirror}"]\ar[u,<->,""] & Y
\end{tikzcd}
\end{center}
\label{core2}
\vskip 0.15cm
Here, ${\cal Y}$ is a holomorphic symplectic manifold which fibers over $Y$ with holomorphic Lagrangian $({\mathbb C}^{\times})^d$ fibers. Its mirror ${\cal X}$ is also holomorphic symplectic, and $X$ embeds into it as a  holomorphic Lagrangian called the ``core". We will call $Y$ the equivariant mirror of ${\cal X}$.  The ${\mathfrak{gl}}_{1|1}$ theory is exceptional in this regard, in that there is no ``upstairs". 

The category of branes on ${\cal Y}$ is $\MDY$, the derived Fukaya-Seidel category of ${\cal Y}$ with potential ${\cal W}$, described in \cite{A2, ADZ}. ${\MDX}$ is the derived category of ${\mathbb T}$-equivariant coherent sheaves on ${\cal X}$ from \cite{A1}. 

The equivariant mirror symmetry framework, which provides two descriptions of the theory, one upstairs and one downstairs, is key for proper understanding of many features of the theory, starting with the asymmetry between cup and cap branes. It is also crucial for proper understanding of how the divisor $D_O$ gets filled in.

\subsubsection{}
In our current example, the upstairs ${\cal Y} ={\cal Y}_{{\mathfrak{su}_2}}$ is \cite{A2, ADZ} an open subset of the Hilbert scheme of $d$ points on (a deformation of) the multiplicative $A_{2d-1}$ surface singularity, 
\beq\label{up}
{\cal C}: \qquad \qquad u^+ u^- = \prod_{i=1}^{2d}(1-y/a_{i}),
\eeq
which is a ${\mathbb C}^{\times}$ fibration over ${\cal A}$. The fibration degenerates at $2d$ points where the punctures are. 

A Hilbert scheme of points on the surface is a blow-up of the locus in the symmetric product of $d$ copies of the surface ${\cal C}$ in \eqref{up} where a pair of points comes together. Explicitly, take $(y_\alpha, u^+_\alpha, u^-_\alpha)$ for $\alpha$ running from $1$ to $d$, modulo permutations, to be coordinates on the symmetric product of $d$ copies of the surface in \eqref{up}. The Hilbert scheme is a resolution of it, by remembering the direction ${u_\alpha^+ - u_\beta^+ \over y_\alpha -y_{\beta}}\in {\mathbb P}^1$, in which pairs of points come together, for any two $\alpha\neq \beta$ (and similarly with $+$ and $-$ exchanged, modulo obvious relations). 

To get ${\cal Y}$, one starts with the ``horizontal Hilbert scheme" of $d$ points on the surface ${\cal C}$ in \eqref{up}, which is an open subset of the Hilbert scheme, where one deletes the locus where a pair of points comes together on ${\cal A}={\mathbb C}^{\times}_{y}$, without coming together on the fiber. This corresponds to asking that ${u_\alpha^- - u_\beta^- \over y_\alpha -y_{\beta}}\neq \infty,$ and similarly with $+$ and $-$ exchanged. This space is simply the moduli space of $d$ $SU(2)/{\mathbb Z}_2$ monopoles on ${\mathbb R}^2\times S^1$, in the presence of $2d$ singular ones, or a Coulomb branch of a four-dimensional ${\cal N}=2$ $U(d)$ gauge theory with $2d$ hypermultiplets.
\subsubsection{}
The upstairs ${\cal Y}$ is the horizontal Hilbert scheme of $d$ points on the surface ${\cal C}$, equipped with the potential
\beq\label{supUP}
{\cal W} = \lambda_0 {\cal W}_0+ \lambda_1 {\cal W}_1 + {\cal W}_2
\eeq
where 
$${\cal W}_{0} =\sum_{\alpha =1}^d \log u^+_{\alpha} , \qquad {\cal W}_1 =\sum_{\alpha =1}^d \log y_{\alpha}, \qquad {\cal W}_2 = \sum_{\alpha =1}^d {u^{-}_{\alpha}\over \prod_{\beta\neq \alpha}(1-y_{\beta}/y_{\alpha})}.
$$
The ${\cal W}_0$ and ${\cal W}_1$ terms are mirror to equivariant action on ${\cal X}.$ The ${\cal W}_2$ term is a holomorphic function on the horizontal Hilbert scheme.
For ${\cal W}_0$ to be defined on ${\cal Y}$, we must further remove from it the points where $u_{\alpha}^+=0$. So, ${\cal Y}$ is simply the horizontal Hilbert scheme of $d$ points on 
${\mathbb C}^{\times}_{u^+} \times {\mathbb C}^{\times}_{y},$ or
$$ {\cal Y} = {\rm Hilb}^d({\mathbb C}_{u^+}^{\times} \times {\mathbb C}_{y}^{\times} \longrightarrow {\mathbb C}_{y}^{\times} ).
$$
\subsubsection{}\label{sYF}
The upstairs ${\cal Y}$ fibers over the downstairs $Y$, with generic fiber
$$
Y_F = ({\mathbb C}_u^{\times})^d.
$$
It is helpful to introduce new local coordinates $u_1, \ldots, u_{d} \in {\mathbb C}^{\times}$, related to the old global ones by
\beq\label{udef}
u_{\alpha} = u_{\alpha}^+\,{ \prod_{\beta\neq \alpha}(1-y_{\beta}/y_{\alpha})\over\prod_{i=1}^{2d}(1-y_{\alpha}/a_{i}) }.
\eeq
In terms of $u$'s and $y$'s, the potential ${\cal W}$ on ${\cal Y}$ splits as
\beq\label{supUPb}
{\cal W} = W + W_F.
\eeq
The first term is the downstairs potential $W$ on $Y$, from \eqref{supsu2}. The remainder 
is the potential on the fiber $Y_F$ over $Y$:

\beq\label{WF}
W_F=\sum_{\alpha =1}^d(\lambda_0 \log u_{\alpha} +u_{\alpha}^{-1}).
\eeq
The category ${\mathscr D}_{Y_F}$ of A-branes on $Y_F$ with potential $W_F$ is familiar. It is mirror to the category of B-branes on $X_F$, which is a product of $d$ copies of ${\mathbb C}_{z}$, each with a ${\mathbb C}_{\lambda_0}^{\times}$ action on it.

This has an important consequence. It means that the upstairs category of A-branes ${\MDY}$, has a description purely from the downstairs perspective, as the Fukaya category of $Y$ ``with coefficients in $\mathscr{D}_{Y_F}$".  We will make explicit below what this means.

\subsubsection{}\label{UX}
The mirror of ${\cal Y}$ with potential ${\cal W}$ is 
$${\cal X} = {\rm Hilb}^d({\cal C}^{\vee} \longrightarrow {\mathbb C}_{z}),$$ 
together with an equivariant ${\rm T}$-action. Here, ${\cal C}^{\vee}$ is the resolution of the ordinary $A_{2d-1}$ surface singularity 
$${\cal C}^{\vee}: \qquad\qquad x_+\, x_{-}  = z^{2d}.$$
The equivariant ${\rm T}$-action on ${\cal X}$ is a product ${\mathbb C}_{\lambda_0}^{\times}\times {\mathbb C}_{\lambda_1}^{\times}$, where the first factor comes from scaling ${z}$  with weight ${\fq} = e^{2\pi i \lambda_0},$ and the second scales $x_+$ and $x_{-}$ oppositely, preserving the holomorphic symplectic form on ${\cal C}^{\vee}$ and hence on ${\cal X}$. These two actions mirror the ${\cal W}_0$ and ${\cal W}_1$ terms in the potential, respectively.

The core $X$, the mirror of $Y$, is the fiber of ${\cal X}$ over the point in the base $Hilb^d({\mathbb C}_{z})$,  where $d$ points come together at the origin of the $z$ plane.
\subsubsection{}
The upstairs ${\cal X}$ is the moduli space of $d$ $SU(2)/{\mathbb Z}_2$ monopoles on ${\mathbb R}^3$, in the presence of $2d$ singular ones, or a Coulomb branch of a three-dimensional ${\cal N}=2$ $U(d)$ gauge theory with $2d$ hypermultiplets.
It is based on ${\cal C}^{\vee}$, which is the ordinary  $A_{2d-1}$ surface singularity and not the multiplicative one, due to the ${\cal W}_2$ term in the potential on ${\cal Y}$. This is an example of the fact that filling in a divisor in ${\cal X}$ is mirror to adding a term in the potential on ${\cal Y}$. This interplay between removing/filling in the divisors on one side, and changing the superpotential is first described in \cite{HoriICM}. It is further developed in \cite{AurouxD} and many other works.

\subsubsection{}
The upstairs and the downstairs categories ${\MDX}$ and $\MDy$ are related by a pair of adjoint functors
\begin{center}
\vskip -0.5cm
\begin{tikzcd}[row sep=3em,column sep=3em]
{\mathscr{D}_{\mathcal X}} \arrow[dr,shift right, swap, "h^{*}"]
\\
&{\mathscr{D}_{Y}}\arrow[lu,shift right,swap,"h_{*}"]
\end{tikzcd}
\end{center}
which come from composing the downstairs mirror
symmetry with a pair of functors
$f_{*}: \mathscr{D}_{X}\rightarrow \mathscr{D}_{\mathcal X}$, which interprets a sheaf ``downstairs" on $X$ as a sheaf ``upstairs"
on ${\mathcal X}$, and $f^{*}: \mathscr{D}_{\mathcal X}\rightarrow \mathscr{D}_{X}$ which corresponds to tensoring with
the structure sheaf $\otimes {\mathcal O}_{X}$, and restricting.

Alternatively, $h^{*}$ and $h_{*}$  come by composing the upstairs mirror
symmetry with a pair of functors $k^*$ and $k_*$
\begin{center}
\begin{tikzcd}[row sep=3em,column sep=3em]
{\mathscr{D}_{\mathcal Y}} \arrow[d,shift right, swap, "k^{*}"]
\\
{\mathscr{D}_{Y}}\arrow[u,shift right,swap,"k_{*}"]
\end{tikzcd}
\end{center}
which come from a Lagrangian correspondence on ${\cal Y}\times Y_-$, where $Y_-$ is $Y$ with sign of symplectic form reversed \cite{ATh}. The functors are intersections with the Lagrangian ${\cal K}$ on ${\cal Y}\times Y_-$. The Lagrangian ${\cal K}$ is a product of $Y$ with a real $d-$dimensional fiber over it, constructed in \cite{ADZ}.

\subsubsection{}
Upstairs, a cup and a cap are the same object. From perspective of ${\cal X}$, the brane ${\cal U} \in \MDX$ is the structure sheaf ${\cal U} = {\cal O}_U$ of vanishing cycle. Continuing with the ${\mathfrak{su}_2}$ example, the vanishing cycle is a product of $d$ non-intersecting ${\mathbb P}^1$'s in ${\cal C}^{{\vee}}$, $U=({\mathbb P}^1)^d$.  
In ${\MDX}$ the homological link invariants are simply
$$
Hom_{\MDX}^{*,*}({\mathscr B} {\cal U}, {\cal U}),
$$
where ${\mathscr B}$ is the action of the braiding functor. Equivariant mirror symmetry relates ${\MDX}$ and $\MDy$ and dictates what the cup and cap branes are in $\MDy$.

The ${\cal U} \in {\MDX}$ brane comes from the cap branes $I_{\cal U}$ on $Y$, as the image of the $h_*:{\MDy} \rightarrow \MDX$ functor,
$$ h_*: \;\;I_{\cal U}\in \MDy \;\longrightarrow\;  {\cal U} = h_* I_{\cal U} \in \MDX.$$
The functor $h^*$ that sends the upstairs ${\cal U}$-brane from ${\cal X}$ back down to $Y$ gives the  $E_{\cal U}$ branes, as well as their images under the braiding functor:
$$ h^*: \;\;{\mathscr B} {\cal U}\in \MDX \;\longrightarrow \; {\mathscr B}E_{\cal U}  = h^* {\mathscr B} {\cal U} \in \MDy.$$ 
An explicit computation of this may be found in \cite{A2}.
The adjointness of the functors $h^*$ and $h_*$ implies that the Homs between the branes upstairs are equal to the Homs between the branes downstairs
$$
Hom_{\MDy}^{*,*}({\mathscr B} E_{\cal U}, I_{\cal U}) = Hom_{\MDX}^{*,*}({\mathscr B} {\cal U}, {\cal U}).
$$
The proof that the right hand side gives homological link invariants (for any simply-laced Lie algebra $^L{\fg}$) is in \cite{A1}, and homological mirror symmetry proven in \cite{ADZ} implies the same is true of the left hand side. We will give a second proof that the left hand side is a link invariant later in this section.

\subsubsection{}\label{LS}
In general, the functor $h^*$, or equivalently $k^*: \MDY \rightarrow \MDy$, equips the branes on $Y$ that come from ${\cal Y}$ as the images of this functor with an extra structure of a local system. The familiar case corresponds equipping Lagrangian branes with a unitary bundle of rank one or higher. Here, we find a generalization of this \cite{ADZ}. 

The functor $h^*$ equips the branes on $\MDy$ that come from $\MDY$ with a local system of, possibly trivial, modules for a graded algebra ${{\mathcal B}}$. 
Examples of branes that come from upstairs as images of $h^*$ are $E$-branes and $T$-branes. By contrast, the $I$-branes in $\MDy$ do not arise in this way.

In our case, the algebra ${{\mathcal B}}$ turns out to be the quotient of the algebra of polynomials in $d$ variables $z_{1}, \ldots , z_{d}$, by an ideal ${\cal I}$ which sets their
symmetric functions to zero. The $z$'s have equivariant
${\mathfrak{q}}$-degree equal to one. We will next explain where this structure comes from, following \cite{ADZ}.

\subsection{${\mathfrak{su}_2}$ $T$-branes and their algebra}\label{TAA}
The branes generating ${\MDy}$ are products of $d$ real lines on a cylinder, 
\begin{equation}\label{sT}
T_{\mathcal C} = T_{i_{1}} \times \dots \times T_{i_{d}} \in
\mathscr{D}_{Y}.
\end{equation}
colored by the simple root of $\mathfrak{su}_{2}$.
We get one such brane for each inequivalent ordering of $T$-branes, where passing a component across a puncture changes the brane. Unlike in Heegaard-Floer theory, there is only one kind of a puncture, colored by the fundamental representation. The branes with more than one $T_i$ brane between a pair of punctures are associated to critical points of $W$ at infinity; unlike in the ${\mathfrak{gl}_{1|1}}$ case, all these branes are non-zero objects of the category.
The generator of $\mathscr{D}_{Y}$ is the direct sum of $T_{\cal C}$'s,
\begin{equation*}
T=\bigoplus _{\mathcal C} T_{\mathcal C} \in \mathscr{D}_{Y},
\end{equation*}
over all possible orderings.  Moreover, as an example of the structure in section \eqref{LS}, the $T$-branes get equipped with a local system of modules for the algebra ${\cal B}$, equal to ${\cal B}$ itself. To explain where this extra structure comes from, we will in parallel consider the generators of $\MDY$, the upstairs category of A-branes.

\subsubsection{}
The upstairs category ${\MDY}$ also has a single generator:
$${\cal T}=\bigoplus_{\cal C} {\cal T}_{\cal C}$$
which is a direct sum of
$${\cal T}_{\cal C}=  {\cal T}_{i_{1}} \times \dots \times {\cal T}_{i_{d}} \in
\MDY.$$
Each upstairs ${\cal T}_{\cal C}$ brane fibers over the downstairs ${ T}_{\cal C} \in {\MDY}$ brane with fibers which are a product of $d$ real line Lagrangians, one for each copy of ${\mathbb C}^{\times}$. 

The real line Lagrangians in the fiber $Y_F = ({\mathbb C}_u^{\times})^d$ with potential $W_F$ generate the fiber Fukaya category ${\mathscr D}_{Y_F}$.  The fiber category is in fact a product of $d$ one dimensional categories, one for each copy of ${\mathbb C}_u^{\times}$. 
\begin{figure}[H]
\begin{center}
  \includegraphics[scale=0.37]{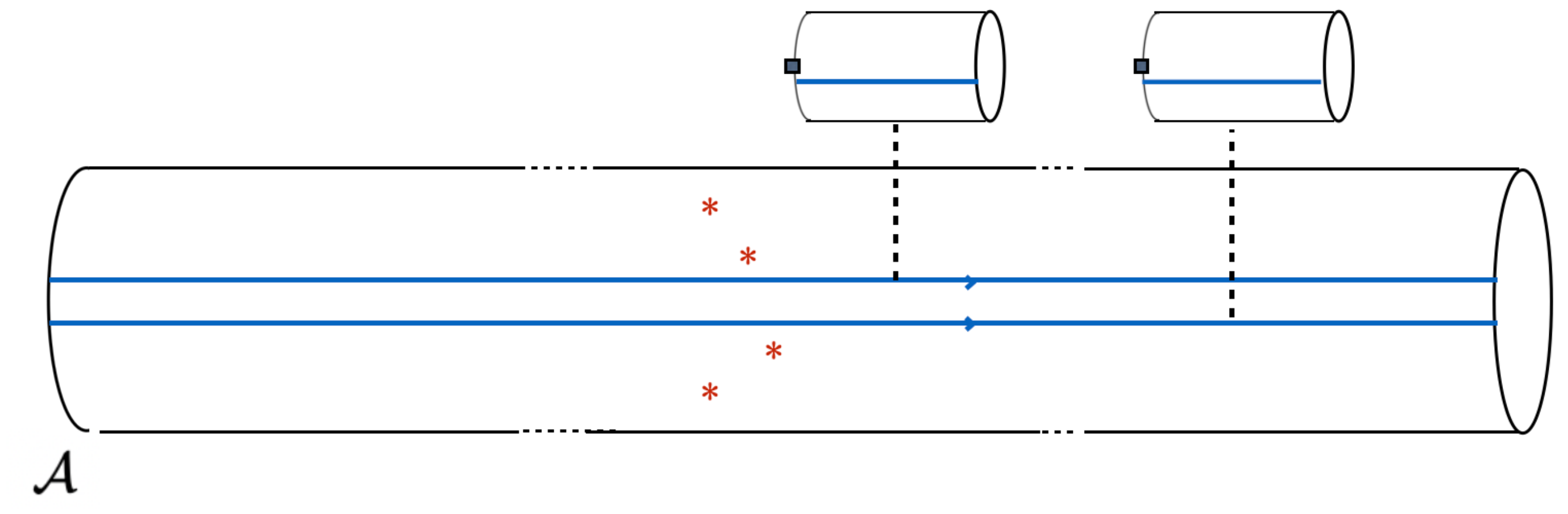}
 \caption{An upstairs ${\cal T}_{\cal C}$ brane pairs every point on the downstairs $T_{\cal C}$ brane in $Y$ (which is a $d$-tuple of points on ${\cal A}$) with a product of $d$ real-line Lagrangians, in the fiber $Y_F$.}
 \label{su2up}
\end{center}
\end{figure}

\subsubsection{}
The endomorphism algebra of the upstairs generators 
\beq\label{AA}
{\mathscr A} = Hom_{\MDY}^{*}({\cal T},{\cal T})
\eeq
is generated by intersection points of ${\cal T}_{\cal C}$ branes in ${\cal Y}$, defined by the wrapped Fukaya category. The kind of wrapping one needs, in the base and in the fiber, is determined by the potential $W$ on $Y$ and $W_F$ on $Y_F$, which add up to the potential ${\cal W}$ on the total space ${\cal Y}$, per \eqref{supUPb}.

Since the upstairs branes fiber over the downstairs, their intersection points are obtained by choosing an intersection point 
of the downstairs generators, 
\beq\label{dP}
{\cal P} \in T_{\cal C}^{\zeta} \cap T_{{\cal C}'},
\eeq
and having chosen ${\cal P}$, picking an intersection point of the branes in the fiber over that point, as in Fig.~\ref{End2up}.

\subsubsection{}\label{sWdots}
The A-model on the fiber $Y_F$ with potential $W_F$ from \eqref{WF}, is very well known. The corresponding category of A-branes ${\mathscr D}_{Y_F}$ is generated by a single brane $T_F$, which is the product of $d$ real line Lagrangians, each on a copy of ${\mathbb C}^{\times}_u$.

As we recalled in section \ref{sYF}, mirror to $Y_F$ with potential $W_F$ is $X_F={\mathbb C}^d$, with the equivariant action that scales each copy of ${\mathbb C}$ by ${\fq}$. Mirror symmetry relates the generator $T_F$ of ${\mathscr D}_{Y_F}$ to the generator of ${\mathscr D}_{X_F}$, which we will also call $T_F$, which is the trivial line bundle on $X_F$.

\subsubsection{}
The endomorphism algebra of the real-line Lagrangian $T_F \in {\mathscr D}_{Y_F}$, defined by the wrapped Fukaya category, is shown in \cite{Auroux} to be
${{\mathcal B}}_{\infty} ={\mathbb C}[z_1, \ldots , z_d] $
the algebra of polynomials in $d$ variables, $z_{1}, \ldots , z_{d}$, with a single grading coming from the equivariant action $z_{\alpha} \rightarrow {\fq} z_{\alpha}$. The endomorphism algebra of the line bundle $T_F$ generating ${\mathscr D}_{X_F}$ is the algebra of homolorphic functions on $X_F$, with grading coming from the ${\mathbb C}^{\times}_{\fq}$ action, in other words the ${\cal B}_{\infty}$ algebra. The fact the two algebras
\beq\label{MFiber} Hom_{{\mathscr D}_{X_F}}^*(T_F, T_F) = {\cal B}_{\infty}\cong Hom_{{\mathscr D}_{Y_F}}^*(T_F, T_F),
\eeq
are the same leads to homological mirror symmetry:
\beq\label{MirFiber}
    {\mathscr D}_{X_F} \cong {\mathscr D}_{{\cal B}_{\infty}} \cong {\mathscr D}_{Y_F},
\eeq
  Any brane in ${\mathscr D}_{X_F}$ or in ${\mathscr D}_{Y_F}$,  gets mapped, by the Yoneda functor $Hom^*(T_F, -)$  as in \eqref{Yoneda}, to a complex of modules of ${\mathcal B}_{\infty}$. Here $T_F$ is viewed as the brane generating the corresponding category.
This identifies as mirror the branes that give rise to the same object of ${\mathscr D}_{{\cal B}_{\infty}}$, the derived category of ${\cal B}_{\infty}$ modules. In particular, the Yoneda functors map the $T_F$ branes to the module for ${\mathcal B}_{\infty}$ equal to itself.

\subsubsection{}\label{key}
Let ${\cal L}$ be any Lagrangian on ${\cal Y}$. Since ${\cal Y}$ fibers over $Y$ with fibers $Y_F$, ${\cal L}$ is a fibration over a Lagrangian $L$ in the base. Let $\Lambda_{{\cal P}}$ be the Lagrangian which is the fiber of ${\cal L}$ over a point ${\cal P} \in L$. The Yoneda functor maps it 
to a module $M_{\cal P}$ for ${\mathcal B}_{\infty}$, 
$$Hom^*_{{\mathscr D}_{Y_F}}(T_F, \Lambda_{\cal P}) = M_{\cal P}.$$

In this way, we identify every Lagrangian ${\cal L} \in \MDY$ upstairs on ${\cal Y}$ with a Lagrangian $L \in \MDy$ downstairs, equipped with a local system $M$ of modules for ${\mathcal B}_{\infty}$,
\beq\label{coeffb}
{\cal L} \; \; \cong \;\; (L, M).
\eeq
The local system is a ${\mathcal B}$-module $M_{\cal P}$ over each point ${\cal P}\in L$ on the Lagrangian $L\in Y$, together with a flat connection that identifies the fibers over different points. 
\begin{figure}[H]
\begin{center}
  \includegraphics[scale=0.44]{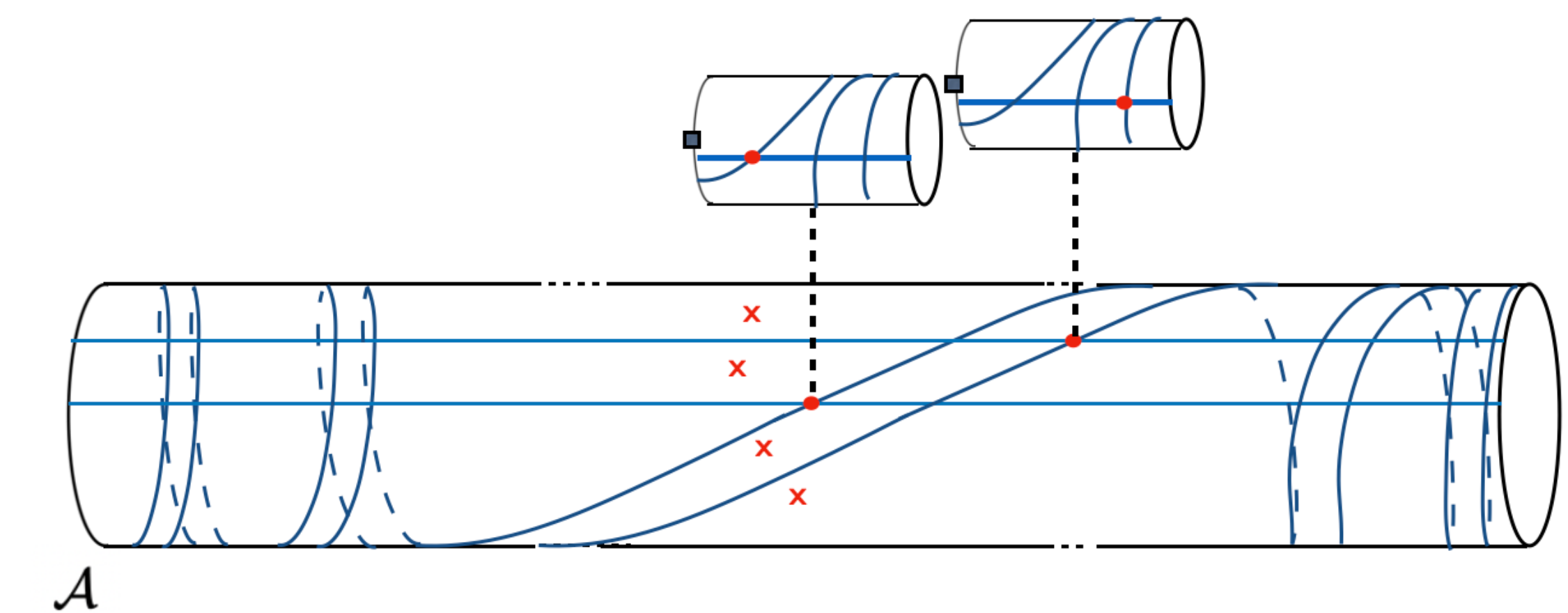}
 \caption{An intersection point of the downstairs $T_{\cal C}$-branes, together with the choice of the intersection in the fiber over it.}
 \label{End2up}
\end{center}
\end{figure}
\noindent{}The Floer complex of ${\cal L}$ and ${\cal L}' \in \MDY$ gets a description as a complex generated by intersection points  ${\mathcal P} \in  L \cap L'$  each tensored by the vector space $hom_{{\mathscr D}_{{\mathcal B}_{\infty}}}(M_{\cal P}, M'_{\cal P})$ 
\beq\label{break}
hom_{\cal Y}^{*,*}({\cal L}, {\cal L}') = hom^{*,*}((L, M), (L', M')) = \bigoplus_{{\cal P} \in {L} \cap {L}'} hom_{{\mathcal B}_{\infty}}(M_{\cal P}, M'_{\cal P}).
\eeq
of homomorphisms of the corresponding ${\mathcal B}_{\infty}$-modules, restricted to ${\mathcal P}$.

\subsubsection{}
As a special case, every upstairs ${\cal T}_{\cal C}$ brane is the downstairs $T_{\cal C}$ brane, equipped with a local system with $M={\mathcal B}_{\infty}$, 
\beq\label{coeffbT}
{\cal T}_{\cal C} \; \; \cong \;\; ({T}_{\cal C}, {\mathcal B}_{\infty}).
\eeq
the latter viewed as a module over itself, using ${\mathcal B}_{\infty}=hom_{{\mathcal B}_{\infty}}({\mathcal B}_{\infty},{\mathcal B}_{\infty}) $.

Furthermore, the Floer complex corresponding to the upstairs ${\cal T}_{\cal C}$ and ${\cal T}_{{\cal C}'}$ branes is generated by intersection points  ${\mathcal P} \in T_{\cal C}^{\zeta} \cap T_{{\cal C}'}$ of the downstairs $T$-branes, tensored with the vector space ${\mathcal B}_{\infty}$, 
\beq\label{upTT}Hom_{\MDY}^*({\cal T}_{\cal C}, {\cal T}_{{\cal C}'}) = \bigoplus_{{\mathcal P} \in T_{\cal C}^{\zeta} \cap T_{{\cal C}'}} \; {\mathcal B}_{\infty} {\cal P}
\eeq
 We equated the Floer complex $hom_{\cal Y}^{*,*}({\cal T}_{\cal C}, {\cal T}_{{\cal C}'})$ with $Hom_{\MDY}^*({\cal T}_{\cal C}, {\cal T}_{{\cal C}'})$, as the space is concentrated in cohomological degree zero with necessarily trivial differential, as we will see momentarily, so the complex and its cohomology coincide.

\subsubsection{}
This translates to a graphical representation of the algebra ${\mathscr A}$.  
By the same argument as for the $\mathfrak{gl}_{1|1}$ theory, the intersection points ${\cal P}$ of the downstairs $T_{\cal C}$-branes in \eqref{dP} translate to diagrams on the cylinder. The choice of an intersection point in the fiber over ${\cal P}$, which corresponds to an element of ${\mathcal B}_{\infty}$ per section \ref{sWdots}, can be represented graphically as well.
To represent an element $z_1^{k_1}\cdots  z_d^{k_d}$ of ${\mathcal B}_{\infty}$, post-compose the first blue string in the diagram with $k_1$ dots, and so on, the last with $k_d$. For example, the intersection point in Fig. \ref{End2up} corresponds to the string diagram:
\begin{figure}[H]
\begin{center}
    \includegraphics[scale=0.39]{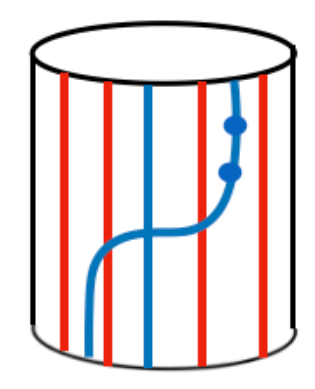}
    \caption{The intersection of the downstairs $T$-branes determines the string diagram; the intersection of the ${\cal T}$-branes in the fiber determines the dots the string diagram gets post-composed with. This string diagram corresponds to the intersection point in Fig.~\ref{End2up}; the intersection in the fiber is $z_1^0z_2^2 \in {\cal B}_{\infty}$.}
    \label{Pup}
\end{center}
\end{figure}
\vskip -1cm

\subsubsection{}
The  relative equivariant gradings of intersection points ${\cal P}$ and ${\cal P}'$ in  $T^{\zeta}_{\mathcal C}\cap T_{{\mathcal C}'} $ can be computed \cite{A2} from a domain $A$ on the Riemann surface that interpolates between them. The gradings are a sum of gradings of string bits, which the string diagram decomposes into. This is as in Heegaard-Floer theory; except the gradings differ.

\begin{figure}[h!]
\begin{center}
  \includegraphics[scale=0.39]{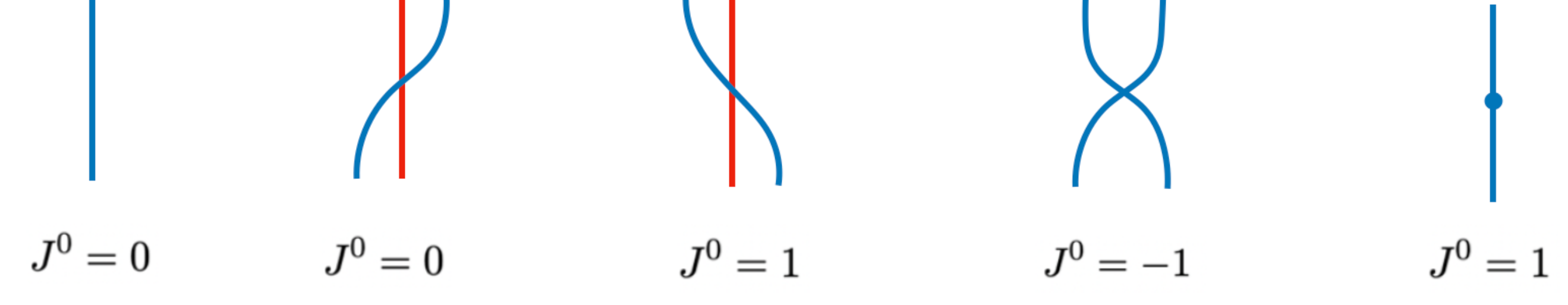}
 \caption{The Maslov degree of every element of $hom(T_{\cal C}, T_{{\cal C}'})$ vanishes. The $J^0$ equivariant degree is determined locally as the sum of the degrees of the string bits in the figure.  The $J^1$ equivariant degree is the winding number of strings in the string diagram, around the cylinder. }\label{su2C}
\end{center}
\end{figure}
\noindent{}The Maslov degrees of all the morphisms of ${\cal T}$-branes vanish \cite{ADZ}. This follows from the fact the holomorphic volume form on ${\cal Y}$ is the Cartesian product of holomorphic forms on $d$ copies of ${\mathbb C}^{\times}\times {\mathbb C}^\times$, with branes that are products of real line Lagrangians in the two copies of ${\mathbb C}^{\times}$.

\subsubsection{}
The algebra ${\mathscr A}$ has a product coming from the product on Floer cohomology,
\beq\label{product2}
m_2: Hom_{{\MDY}}^{*}({\cal T}_{{\cal C}'}, {\cal T}_{{\cal C}''})
\cdot Hom_{{\MDY}}^{*}({\cal T}_{\cal C}, {\cal T}_{{\cal C}'})   \rightarrow Hom^{*}_{{\MDY}}({\cal T}_{\cal C}, {\cal T}_{{\cal C}''}), 
\eeq
defined using the wrapped Fukaya category of ${\cal Y}$ with potential ${\cal W}$. As in the ${\mathfrak{gl}_{1|1}}$ case, the product corresponds to stacking cylinders and rescaling.

Because all elements of ${\mathscr A}$ are in degree zero, only the product in \eqref{product2} needs computing as all other $A_{\infty}$ morphisms vanish. For $m_{\ell}$ products to not vanish the difference between Maslov indices of outgoing intersection point and the $\ell $ incoming ones must be $2-\ell$, and all algebra elements have degree zero.

\subsubsection{}
The associative product on the algebra $\mathscr{A}$ is computed in \cite{ADZ}. The algebra product translates into local relations of strands elements, given in $\#1$ to $\#7$ below. 
\begin{figure}[H]
\begin{center}
    \includegraphics[scale=0.28]{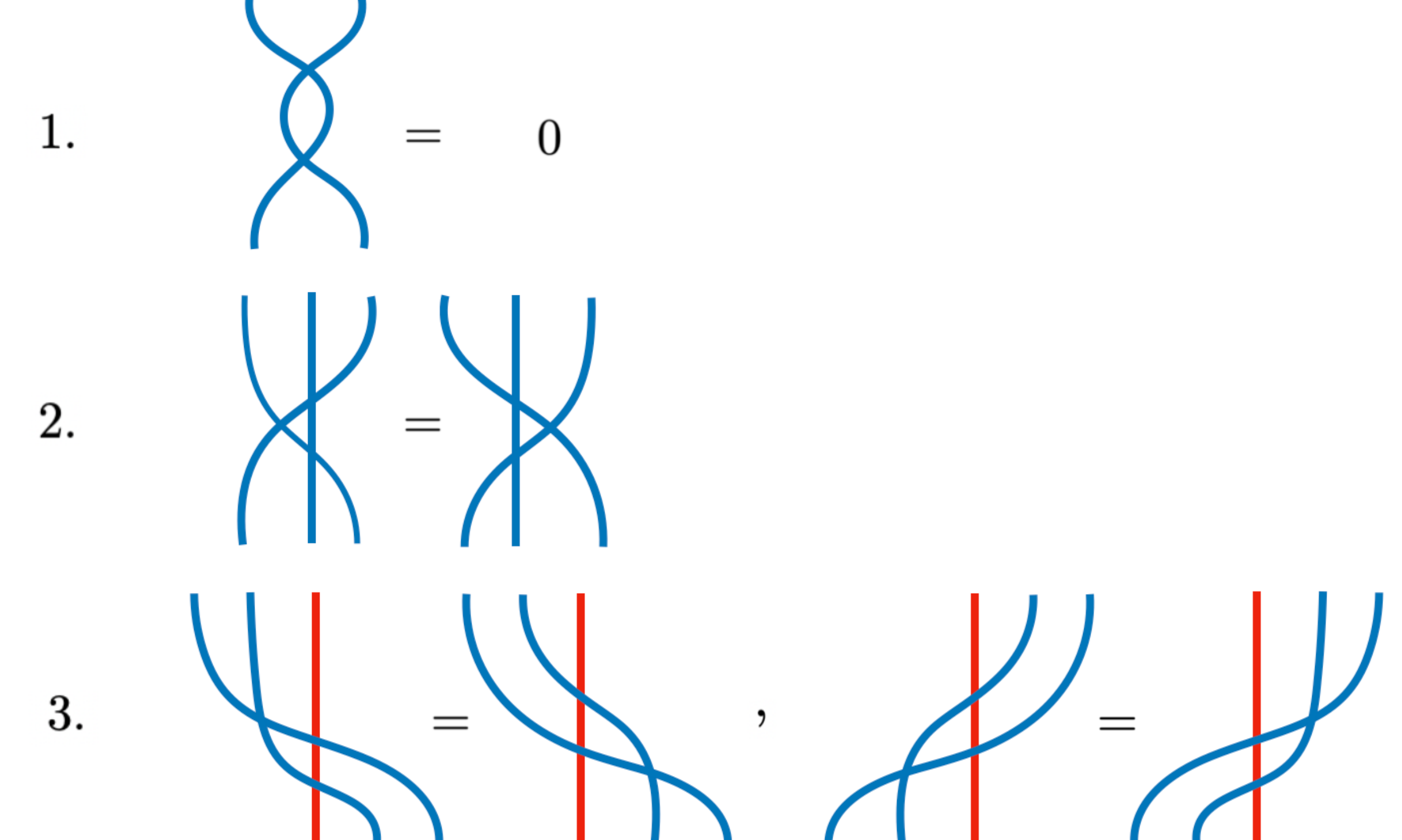}
    \label{su2rel1}
\end{center}
\end{figure}
\vskip -0.5cm
To compute the algebra ${\mathscr A}$, \cite{ADZ} first consider the theory in the complement of the divisor $D_O$, which is the union of the diagonal divisor $\Delta$ and the divisor of all the punctures $D_a=\sum_i D_{a_i}$. More precisely, in the context of ${\cal Y}$, the divisor $D_O$ is really the pullback of the divisor $D_O$ on $Y$ from \eqref{su2Do} by the map that projects ${\cal Y}$ to $Y$. 

Deleting the $D_O$ from ${\cal Y}$ does not affect the generators of the Floer complex, since the ${\cal T}$-branes are disjoint from it. As a result, the algebras as vector spaces do not change, only the relations between their elements may.
In the complement of $D_O$, the theory is extremely simple. The only maps that contribute to the algebra come from counts of one-dimensional triangles, which are easy to evaluate. The result is the algebra 
\beq\label{AO}
    {\mathscr A}_O ={\mathscr A}_{u=0, \hbar=0} 
\eeq
given by relations $\#1-\#7$, with both $u$ and $\hbar$ set to zero. 

Filling in $D_a$ corresponds to deforming the algebra to $u\neq 0$.  
The deformation comes from including disks in ${\cal Y}$ that project to ${{\cal A}}$ as products of one-dimensional triangles passing through the punctures. For the superpotential ${\cal W}$ to remain regular on such a disk, which is the same as saying the disk does not run off to infinity on ${\cal Y}$, the point on the domain that maps to a puncture must also map to $u=\infty$ in the ${\mathbb C}^\times$ fiber over it, in such a way that $u_+$ remains finite by \eqref{udef}. The result was shown in \cite{ADZ} to be the deformation of the product in $\#5$.
\begin{figure}[H]
\begin{center}
    \includegraphics[scale=0.3]{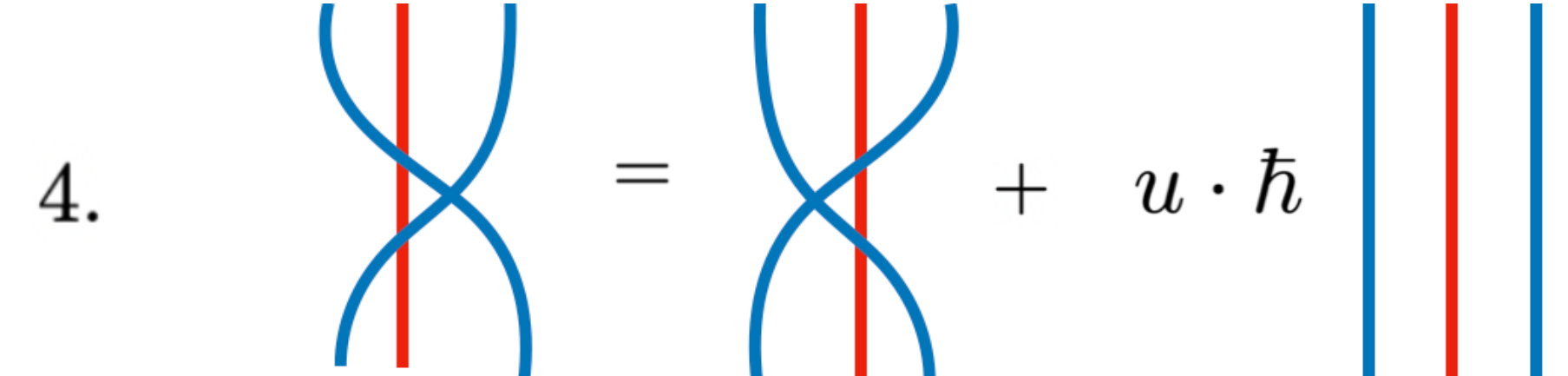}
    \label{su2rel2}
\end{center}
\vskip -0.2cm
\end{figure}
\noindent{}Filling in the diagonal $\Delta$ corresponds to deforming to $\hbar \neq 0$. On equivariant degree grounds, the only possible correction is add a single term to relations $\#4, \#6$ and $\#7$, with coefficient to be computed. The corrections to those relations come from counting disks in the symmetric product of two copies of ${\mathscr A}$. In fact, it suffices to count a single such disk, as knowing the correction to either of the three relations fixes the other two by associativity of the algebra. Finding the $\hbar$ deformation of the algebra thus amounts to a single, but difficult calculation, performed in \cite{ADZ}. 
\begin{figure}[H]
\begin{center}
    \includegraphics[scale=0.32]{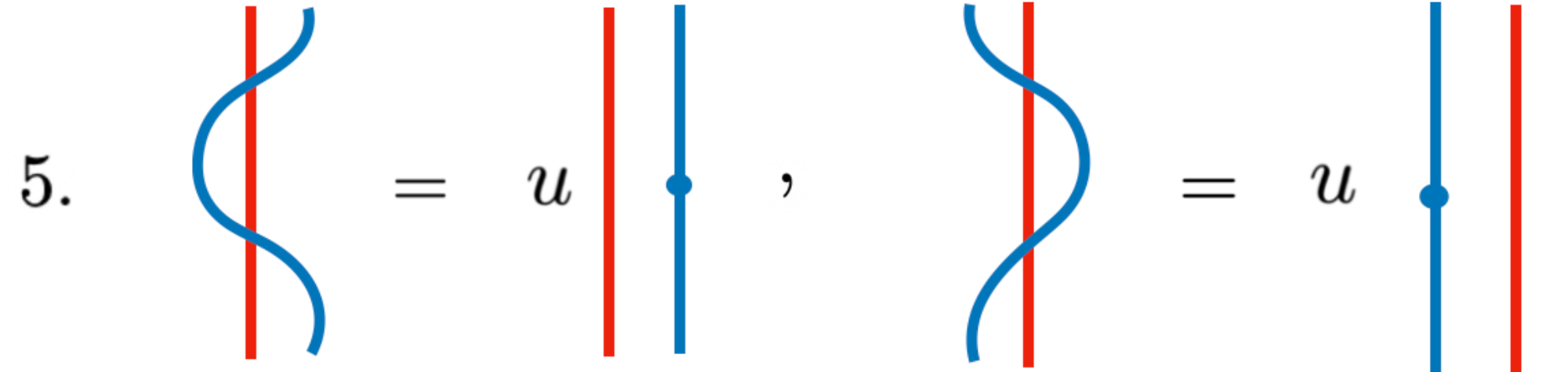}
    \label{su2rel3a}
\end{center}
\vskip -0.2cm
\end{figure}
\noindent{}Both parameters $u$ and $\hbar$ may be set to $1$ by rescaling the algebra generators; if we keep them generic, they are useful as book-keeping parameters.
\begin{figure}[H]
\begin{center}
    \includegraphics[scale=0.32]{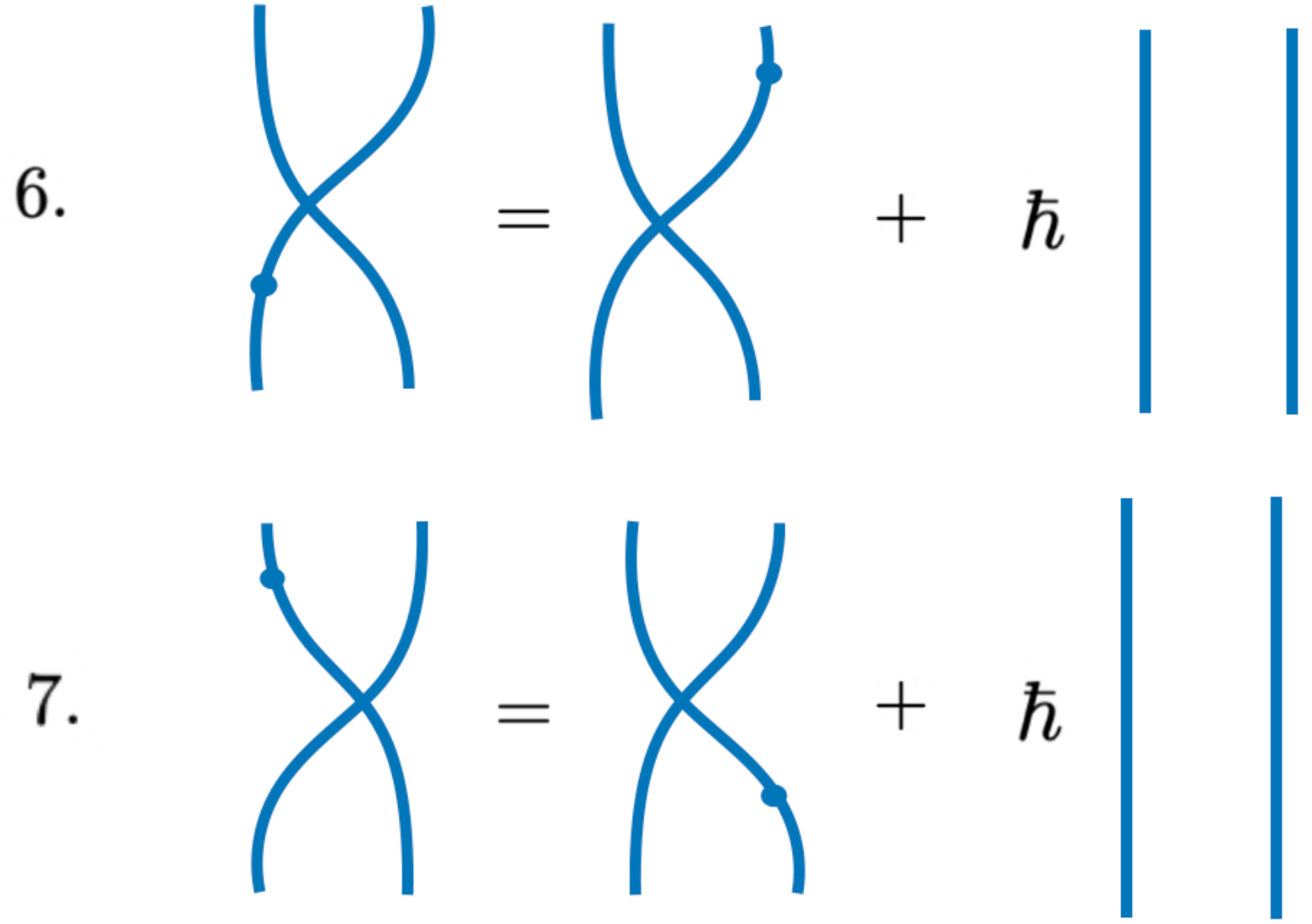}
\end{center}
   \vskip -1cm
\end{figure}
\noindent{}

\subsubsection{}
The algebra relations should be interpreting as stating that the difference between the products of pairs of generators on the left and the right hand sides of relation $\#4$, for example, is not zero, thanks to the contribution of a map ${\rm D}$ to ${\cal Y}$ which intersects the diagonal $\Delta$ and the divisor of the punctures $D_a$ once, and which computes the coefficient of the corresponding identity morphism. 
To show that relations $\#6$ and $\#7$ follow from relations $\#4$ and $\#5$ by associativity of the algebra,
pre- or post-compose relation $\#4$ with an element that, in one of the compositions, creates a crossing of the kind appearing on the left hand side of $\#5$. 

Viewing the ${\cal T}$-brane as the downstairs $T$-brane equipped with the local system ${\cal B}_{\infty}$, the relation $\#5$ says that the product of the two intersection points in $Y$ on the left hand side of the relation  
does not vanish (as it would without the bundle), but rather defines a non-trivial endomorphism of the bundle over the brane.

\subsubsection{}
The relations $\#1, \#2$, $\#6$ and $\#7$ are the basic relations of the ${\mathfrak{su}_2}$ KLR algebra from \cite{KL1, R}. 
The resulting upstairs algebra ${\mathscr A} = Hom_{\MDY}^*({\cal T}, {\cal T})$ is the cylindrical version of the KLRW algebra of \cite{KL1, R}, which \cite{webster} generalized to adding red strands. The KLRW algebras are originally formulated as algebras of strands on a plane, not a cylinder, but the local relations are the same.

This shows that the string diagrams, in terms of which the algebras of \cite{KL1, KL2} and \cite{webster} are phrased, are nothing but Reeb chords associated to $T$-branes that generate the wrapped Fukaya categories of ${\cal Y}$.

\subsubsection{}
The ${\cal T}$-brane, which generates $\MDY$, is related by mirror symmetry to a very special generator of ${\MDX}$, which we will also label ${\cal T}$ and which is known as the tilting vector bundle. The tilting property of the generator ${\cal T}$ is the statement that its endomorphism algebra is vanishes outside of cohomological degree zero, as it does in ${\mathscr A}$. 

The construction of tilting bundle ${\cal T}$ on a holomorphic symplectic ${\cal X}$ is due to Bezrukavnikov and Kaledin \cite{BK1}. It uses methods of characteristic $p$. The ``exotic $t$-structure" which enters the construction of \cite{BK2} corresponds to choosing the 
positions of the punctures on ${\cal A}$ along a fixed $S^1$, as in our setup from section \ref{choice}. The fact that for this choice of $t$-structure, the ${\cal T}$ brane is a good generator of the category is manifest on the mirror ${\cal Y}.$

In \cite{W1, W2}, Webster computed the endomorphism algebra of the Bezrukavnikov-Kaledin tilting generator ${\cal T}$, corresponding to $^L{\fg}$ which is the ordinary simply laced Lie algebra, and showed it equals the cylindrical KLRW algebra ${\mathscr A}$. This, together with the computation of the algebra ${\mathscr A}$ as the endomorphism algebra of ${\cal T}$-branes on ${\cal Y}$ proves \cite{ADZ} the homological mirror symmetry: 
\beq\label{mirrup}
    {\MDX} \cong {\MDA} \cong {\MDY}.
\eeq
The two equivalences separately reflect the fact that ${\cal T} =\bigoplus_{\cal C} {\cal T}_{\cal C}$ generates the category and that ${\mathscr A}$ is its endomorphism algebra. Homological mirror symmetry relating $\MDX$ and $\MDY$ follows from the fact one gets the same algebra from both ${\cal X}$ and ${\cal Y}$.

\subsubsection{}
The functor $k^*$, which takes the branes from upstairs down,
comes from a Lagrangian correspondence. The correspondence acts by intersecting the brane ${\cal L}$ on ${\cal Y}$ with a Lagrangian ${\cal K} \in {\cal Y} \times Y_-$ to get a brane $L$ on $Y$. 

As any A-brane in $Y_F$, the fiber of the brane ${\cal K}$ in $Y_F$ is a module for the algebra ${\cal B}_{\infty}$. The module corresponding to it turns out to not be a trivial module of the algebra ${\cal B}_{\infty}$ \cite{ADZ}. The module is rather a finite rank quotient of ${\cal B}_{\infty}$ 
\beq\label{BIB}
    {\cal B} = {\cal B}_{\infty}/{\cal I},
\eeq
in which the symmetric functions of $z$'s are set to zero. The symmetric functions of $z$'s generate an ideal, which we denoted by ${\cal I}$. The algebra ${\cal B}$ has  rank $d!,$ so is non-trivial for $d>1$.

\subsubsection{}
An immediate implication is that the functor $k^*$ maps the upstairs  ${\cal T}_{\cal C}$  to the 
downstairs ${T}_{\cal C}$ brane but equipped with the extra structure
\beq\label{TTd}
    k^* {\cal T}_{\cal C} \cong ({ T}_{\cal C}, {\cal B}),
\eeq
of a ${\cal B}$-local system of finite rank,
instead of the infinite one in \eqref{coeffbT}. Inherited from the upstairs theory, the local system is not based on ${\cal B}$ simply as a vector space, but rather as a module over itself \cite{ADZ}.

\subsubsection{}
The relation between the upstairs and downstairs generators from \eqref{TTd}
implies that the downstairs algebra $A = Hom^{*}_{\MDy}(T,T)$ has the same elements and the same relations as  ${\mathscr A}=Hom^{*}_{\MDX}({\cal T},{\cal T})$, the latter given by $\#1-\#7$ above, except one quotients by the ideal 
\beq\label{Algebras}
    A = {\mathscr A}/{\cal I},
\eeq
${\cal I}$ since ${\cal B} ={\cal B}_{\infty}/{\cal I}$. 
One can show the ideal ${\cal I}$ is a two sided ideal not only of ${\cal B}_{\infty}$, but also of ${\mathscr A}$, so the quotient makes sense.

\subsubsection{}
Along with the proof of upstairs homological mirror symmetry \eqref{mirrup}, we get the proof of the downstairs one \cite{ADZ},
\beq\label{mirrdwn}
    {\MDx} \cong {\MDa} \cong {\MDy}.
\eeq
by applying the functor $f^*$ and its mirror $k^*$.

From the mirror perspective, the first equivalence in \eqref{Algebras} is a consequence of the fact $X$ embeds into ${\cal X}$, not as a locus where $z$'s are set to zero, but as a locus where one sets their symmetric functions to zero. The action of the symmetric group makes only the latter good variables on ${\cal X}$. The fact that ${\cal I}$ is a two-sided ideal of ${\mathscr A}$ follows \cite{ADZ} from the fact that, since ${\cal T}$ is a vector bundle on ${\cal X}$, the center of its endomorphism algebra is the algebra of holomorphic functions on ${\cal X}$. The two-sided ideal is the subalgebra of it, consisting of those functions that vanish on $X$.

\subsubsection{}
Every Lagrangian ${\cal L}$ of ${\MDY}$ gets mapped by the functor $k^*$ to a Lagrangian $L$ in $\MDy$, but equipped with a possibly non-trivial local system of modules for the algebra ${\cal B}$, as in the case of the $T_{\cal C}$ branes that arise as images of ${\cal T}_{\cal C}$ branes under the functor $k^*$.
The vanishing cycle branes $E_{\cal U}$ and $I_{\cal U}$ come equipped with trivial modules for ${{\mathcal B}}$. For the $I_{\cal U}$ brane, this is obvious, as it does not come from the upstairs. The $E_{\cal U}$ brane does come from the upstairs; however, it comes from a brane ${\cal U} = {\mathscr O}_{U} \in \MDX$ supported on the core $X$. The vanishing cycle $U$, whose structure sheaf is ${\cal U}$, is in $X$. Such a brane is a trivial module for the algebra ${\cal B}_{\infty}$ and hence for ${\cal B}$ as well.

\subsubsection{}\label{sameT}
Since the upstairs and the downstairs generators have direct summands 
\beq\label{same} {\cal T}_{\cal C} \cong (T_{\cal C}, {\cal B}_{\infty}) \in 
\MDY, \qquad\qquad k^*{\cal T}_{\cal C}\cong (T_{\cal C}, {\cal B})\in \MDy,
\eeq
corresponding to same Lagrangian $T_{\cal C}$ on $Y$, just equipped with the local systems ${\cal B}_{\infty}$ and ${\cal B}$ respectively, from now on we will simply denote both by $T_{\cal C}$, leaving implicit the corresponding local system. Most of the time, it is clear from the context which we are referring to; we will spell it out when not.

\subsubsection{}
The structure we found above means that there should be a description of 
the entire upstairs category $\MDY$ as a category of A-branes on $Y$ ``with coefficients" in ${\mathscr D}_{Y_F}$ or, equivalently, in ${\mathscr D}_{{{\mathcal B}}_{\infty}}$. The objects of $
\MDY$ can be described as A-branes on $Y$, equipped with a local system of ${\mathcal B}_{\infty}$-modules, or complexes thereof. Morphisms between its objects are intersection points of the corresponding Lagrangians on $Y$, dressed with homorphisms between the corresponding ${\cal B}_{\infty}$ modules. 

Just as the upstairs category $\MDy$ should have a downstairs description based on a category of A-branes on $Y$ with coefficients in ${\mathscr D}_{{ \cal B}_{\infty}}$, the downstairs category is a category of A-branes on $Y$ with coefficients in ${\mathscr D}_{{ \cal B}}$.

\subsection{Projective resolutions}\label{prsu2}
In this section, we will explain how to find the complex $({\mathscr B}(T), \delta)$ resolving the ${\mathscr B}E_{\cal U}$ brane in $\MDy$ in terms of its $T$-branes. Since all elements of the algebra $A$ have zero cohomological degree, the complexes
resolving the brane are ordinary, rather than twisted, complexes.

A remarkable fact is that we will be able to find the resolution not only of the downstairs brane ${\mathscr B} E_{\cal U}\in \MDy$ but also its upstairs parent ${\mathscr B}{\cal U} \in \MDY$. The two branes are related by 
\beq\label{updwnE}
k^*{\mathscr B}{\cal U} = {\mathscr B} E_{\cal U}.
\eeq 
We will first explain how to find the resolution of the downstairs brane on $Y$, and then how to get the resolution of the upstairs brane on ${\cal Y}$ that it came from as the image of the $k^*$ functor.

\subsubsection{}
The projective resolution of ${\mathscr B} { E}_{\mathcal U} \in \MDy$ in terms of ${T}$-branes is based on the
 direct sum brane
\beq\label{aproBsu2}
    {\mathscr B} { E}(T)=\bigoplus_k  {\mathscr B} { E}_k(T) [k],
\eeq
together with the differential $\delta$. The direct sum brane is obtained by stretching the ${\mathscr B} { E}_{\mathcal U} $ brane straight along the cylinder ${\cal A}$ and allowing it to break at both ends. The differential 
$\delta$, which we turn on to recover the brane we started with, is a cohomological degree $1$ and equivariant degree zero operator 
$$\delta:  {\mathscr B} { E}(T)\rightarrow {\mathscr B} { E}(T)[1]$$
that squares to zero 
\beq\label{orsquare}
    \delta\cdot \delta=0
\eeq
in the algebra $A$. Because all elements of algebra $A$ have cohomological degree zero, the product in \eqref{orsquare} is an ordinary product, and the ordinarily twisted complex becomes an ordinary complex. 
One commonly writes the complex as 
\beq\label{ocomplex}
    {\mathscr B}E_{\cal U} \cong \ldots  \xrightarrow{t_1} {\mathscr B}E_1(T) \xrightarrow{t_0} {\mathscr B}E_0(T),
\eeq
where 
$t_{k} \in Hom^*_{\MDy}( {\mathscr B} { E}_{k+1}(T),  {\mathscr B} { E}_{k}(T)),
$
and $\delta =\sum_{k} t_{k} \in A$. Unlike for a twisted complex, is not necessary to explicitly write the cohomological degrees of all the terms in the complex, as they are fixed by knowing any one.

To find the differential $\delta$, we will proceed in steps, as in Sec. \ref{Algorithm}.

\subsubsection{}
The algorithm from section \ref{Algorithm} starts by finding the resolution of the ${\mathscr B}E_{\cal U}$-brane (or any other product brane) with the divisor $D_O$ in \eqref{su2Do} deleted. The result is a direct sum brane ${\mathscr B}E_{\cal U}(T)$, together with the product differential $\delta_O$ 
\beq\label{rem}
    ({\mathscr B}E_{\cal U}(T),{\delta_O}),
\eeq
both of which can be read off from geometry of the brane. As we explained in section \ref{Algorithm}, the direct sum brane and the product differential form a rectangular toric grid in dimension $d$.

\subsubsection{}\label{downduh}
The differential $\delta$ of the brane in $\MDy$ is found iteratively,
as in section \ref{Algorithm}, by deforming $\delta_O = \delta_{0,0}$ to a differential $\delta = \delta_{u,\hbar}$ that squares to zero in the full algebra $A$. The deformation corresponds to filling the divisor $D_O=\Delta+D_a$ back in.

Filling $D_a$ back in corresponds to deforming the differential $\delta_O=\delta_{0,0}$ to $\delta_{u,0}$ which squares to zero in the algebra ${ A}_{u,0}$, in which $\hbar$, but not $u$ is set to zero:
\beq\label{nbu}
\delta_{u,0} = \sum_{\ell=0} u^\ell\delta_{\ell,0}   = \delta_{0,0}+ u \delta_{1,0}+ \ldots,
\eeq
analogous to the expansion in \eqref{nf}.  
The terms in $\delta_{\ell,0}$ are corrections involving $\ell$ dots, since at $\hbar=0$ the theory has a symmetry that scales the dots and $u$ oppositely. There are at most finitely many terms in \eqref{nbu} that may be non-vanishing, on degree grounds, since $\delta_{u,0}$ must define a differential on the same direct sum brane we started with.
For the deformation problem to make sense, we need to regard the $T$-branes as equipped with a ${\cal B}$-local system. This simply means that their endomorphism algebra is ${ A}_{u,0}$, which however enriches each geometric intersection point ${\cal P}\in T^{\zeta} \cap T$ with a copy of ${\cal B}.$

Next, we fill the diagonal $\Delta$ back in. This corresponds to deforming the differential $\delta_{u,0}$ to $\delta_{u,\hbar}$ 
\beq\label{nbuh}
\delta_{u, \hbar} = \sum_{k=0}  \hbar^k \delta_{u,k} = \delta_{u,0}+ \hbar \delta_{u,1}+ \ldots,
\eeq
which squares to to zero in the algebra full algebra ${A} ={A}_{u,\hbar}$, to get the resolution of the ${\mathscr B}E_{\cal U}$ brane in $\MDy$.

\subsubsection{}\label{updownr}
Along with the resolution of the downstairs brane $ {\mathscr B} E_{\cal U}\in \MDy$, we can also obtain the resolution of its parent, the upstairs brane ${\mathscr B}{\cal U}\in \MDY$.

This is a consequence of two facts. The first is that the same downstairs $T$-branes, just equipped with a different the local system, as in eqn.~\eqref{same}, describe both the upstairs and the downstairs generators. 

The second is that our starting point, the direct sum brane ${\mathscr B}E(T)$ and the product differential $\delta_O$ in \eqref{rem} describes not only a brane downstairs in ${\mathscr D}_{{ Y}_O}$, but also a brane upstairs, in ${\mathscr D}_{{\cal Y}_O}$.
 The differential $\delta_O$, based on the intersection points of $T_{\cal C}$ branes on the Riemann surface ${\cal A}$ and containing no contributions from algebra elements that come from either ${\cal B}$ or from ${\cal B}_{\infty}$, is easily proven to square to zero both in the algebra $A_O$ and in ${\mathscr A}_O$.  

\vskip 1cm
\subsubsection{}
The resolution of the upstairs brane ${\mathscr B}{\cal U}\in \MDY$ is the same the direct sum brane ${\mathscr B}E(T)$ as for the downstairs one, with differential $ \delta =\delta_{u, \hbar}$ that squares to zero in the upstairs algebra ${\mathscr A} = {\mathscr A}_{u,\hbar}$, and reduces to $\delta_O$ uppon setting $u$ and $\hbar$ to zero.

Finding $\delta$ is a problem entirely analogous to finding the differential that squares to zero in the downstairs algebra $A={A}_{u,\hbar}$ from section \ref{downduh}, just the algebra itself is different. 
For the $u\neq 0$ deformation problem to make sense, the $T$-branes on ${\cal A}$ as need to be equipped with a copy of the infinite rank algebra ${\cal B}_{\infty}$, which makes them be equivalent to the upstairs ${\cal T}$-branes, as in \eqref{coeffbT}. 
\subsubsection{}\label{samed}

Since the algebra $A$ is a quotient of the upstairs algebra,  $A={\mathscr A}/{\cal I}$, the differential $\delta: {\mathscr B}E(T) \rightarrow {\mathscr B}E(T)[1] $ that squares to zero in the upstairs algebra ${\mathscr A}$ also always squares to zero in the downstairs algebra $A$, by simply setting any terms in ${\cal I}$ to zero. The converse is of course not true.

The brane $({\mathscr B}E(T), \delta)$ with differential $\delta$ that squares to zero in the upstairs algebra ${\mathscr A}$ thus describes {\it both} the resolution of a brane in $\MDY$ and, after dividing by ${\cal I}$, downstairs in ${\MDy}$. 

\subsubsection{}\label{updowndelta}
In the remainder of this section, when we write the differential $\delta=\delta_{u, \hbar}$ it will always be the differential of the upstairs complex, since it has more information than the downstairs one, and we can find it just as easily. This is particularly pertinent because, taking the hom with the $I_{\cal U}$-branes, the downstairs or the upstairs ones, 
all the dots in $\delta$  get set to zero anyhow. This follows because the $I_{\cal U}$ branes are simples of the corresponding upstairs and downstairs algebras \cite{A2}.

\subsubsection{}\label{sec:signs2}
To get a homology theory over ${\mathbb{Z}}$, we need to assign signs to maps in $\delta_O$, which then determine the signs in $\delta$. 

Up to gauge equivalence, there is a unique choice of signs in the product differential $\delta_O$ that deforms to a theory with $u\neq 0$.  This is natural, as sign choices in the complex should reflect the choices of spin structure on the brane that it describes, and both the upstairs and the downstairs ${\mathscr B}E_{\cal U}$ branes admit a unique spin structure once the divisor $D_a$ of the punctures is filled in. The spin structure is the one inherited from the unique spin structure on the product of $d$ disks in $Y$ which the brane bounds. 

Like in the ${\mathfrak{gl}_{1|1}}$ theory whose signs are discussed in \ref{sec:signs1}, the differential $\delta_O$ is a differential on a product complex whose factors resolve the one-dimensional branes, with the signs derived from signs from the one-dimensional factor branes. Signs of the individual maps in one-dimensional complexes resolving the factor branes may be changed by gauge transformations on the $T$-branes they run between, so they are not meaningful. For the figure eight $E$-brane, only the product of all the signs in the complex is gauge invariant. The choice of signs that does deform to $u\neq 0$ can be shown to be
$(-1)^{1+ n_j/2}$
where $n_j$ is the number, always even, of $T$-branes in the $j$-th one-dimensional complex. 

In a product complex, one gains an additional standard sign needed to ensure the product differential squares to zero if its factors do. This sign is determined recursively. Assume we are given the signs in a product differential on a $k-1$ dimensional grid, which describing a brane which is the product of $k-1$ one-dimensional ${\mathscr B} E_i$-branes.  Taking the product with ${\mathscr B} E_k$, the signs of the maps in the first $k-1$ directions of the grid stay the same, whereas signs in the $k$'th direction get multiplied by an overall alternating sign. For thorough discussion of how  signs are assigned in $
delta_O$, see appendix \ref{ssigns}.

\subsection{Simple examples}
We will now illustrate the ${{\mathfrak{su}}_2}$ algorithm in some simple examples.

\subsubsection{}
\label{sl2cap}
\begin{figure}[H]
\begin{center}
  \includegraphics[scale=0.37]{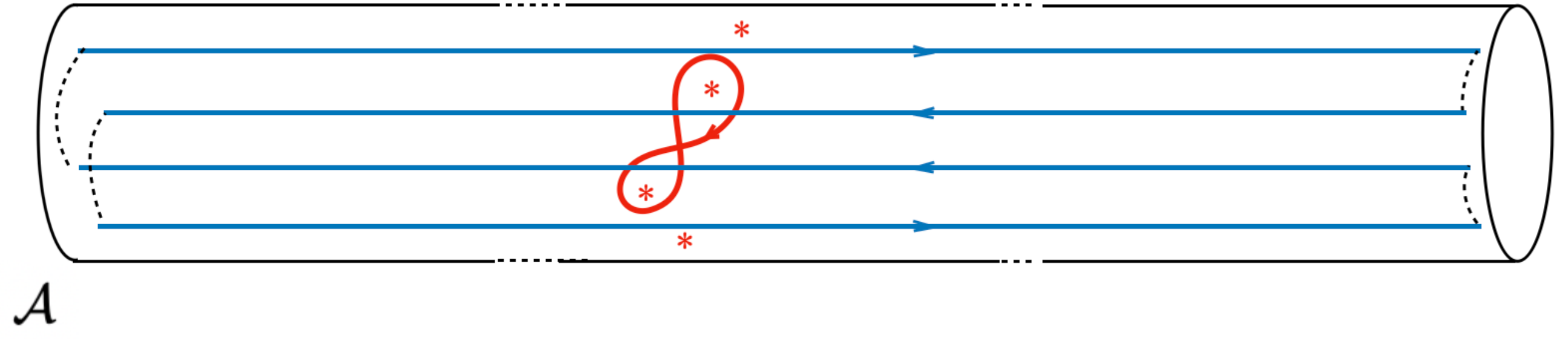}
 \caption{By isotoping the $E_{{\cal U}_2}$ brane, it becomes a connected sum of $T$-branes, which enter the direct sum brane ${E}_{{\cal U}_2}(T)$. }\label{su2ex1}
\end{center}
\end{figure}

Consider $Y$ which is a copy of ${\cal A}$, the Riemann surface in Fig.~\ref{su2ex1}. $Y$ is the equivariant mirror to ${\cal X}$, the resolved $A_3$ surface. There are four $T_j$-branes generating $\MDy$, which are images of the $h^*$ functor of four line bundles which generate $\MDX$ \cite{A2}. They are real line Lagrangians passing between consecutive punctures on the cylinder.

A cap/cup brane on $\MDX$ is a structure sheaf of a ${\mathbb P}^1$.  
We will pick one of them, say ${\cal U}_2 \in \MDX$, and let $E_{{\cal U}_2}\in \MDy$ brane be its equivariant mirror, the image of the $h^*$ functor. (More precisely, only three of four cup branes are structure sheaves of the three vanishing ${\mathbb P}^1$'s in ${\cal X}$ and the fourth one is described in \cite{A2}.)

By isotoping the $E_{{\cal U}_2}$ brane and allowing it to break at $y=0$ and $y=\infty$ we get the direct sum brane:
\beq\label{break2}
{E}_{{\cal U}_2}(T) =T_2\{-1\}[2]  \oplus \begin{pmatrix}T_1 \\ T_3\{-1\} \end{pmatrix}[1] \oplus T_2 ,
\eeq
where the cohomological degree shift comes from the flip of orientation, per Fig.~\ref{su2ex1}. 

\vskip 0.1cm
{\noindent }To recover the $E_{{\cal U}_2}$ brane, take the connected sum of the $T$-branes, over their intersection points at $y=\infty$ and $y=0$.
The intersection points at $y=\infty$ correspond to the following elements of algebra $A$:
\vspace*{-5pt}
\begin{figure}[H]
\begin{center}
  \includegraphics[scale=0.31]{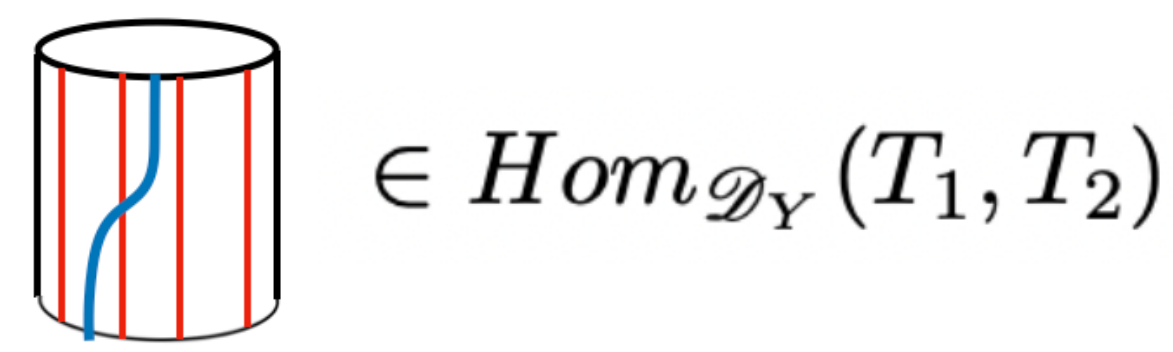}
\end{center}
\vspace*{-5pt}
\end{figure}
\vspace*{-5pt}
 \vskip -0.5cm
\begin{figure}[H]
\begin{center}
  \includegraphics[scale=0.31]{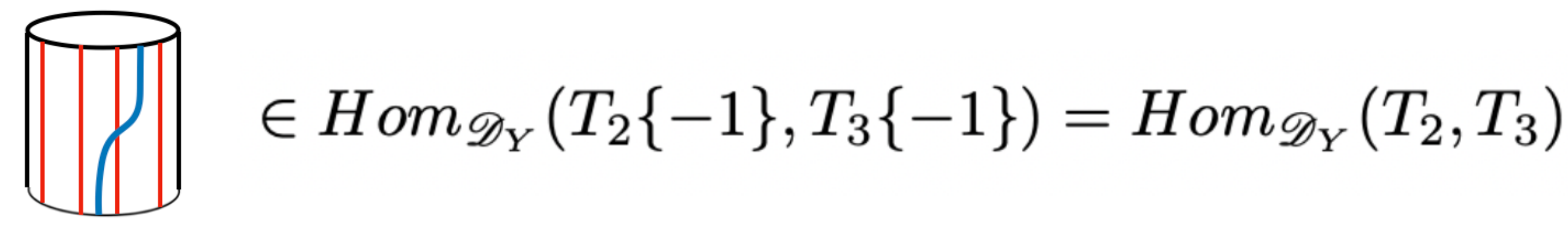}
\end{center}
\vspace*{-13pt}
\end{figure}
{\noindent }The intersection points at  $y=0$,  correspond to the string diagrams going the other way, by the choice of wrapping in Fig.~\ref{su2ex1}:
\vspace*{-5pt}
\begin{figure}[H]
\begin{center}
  \includegraphics[scale=0.31]{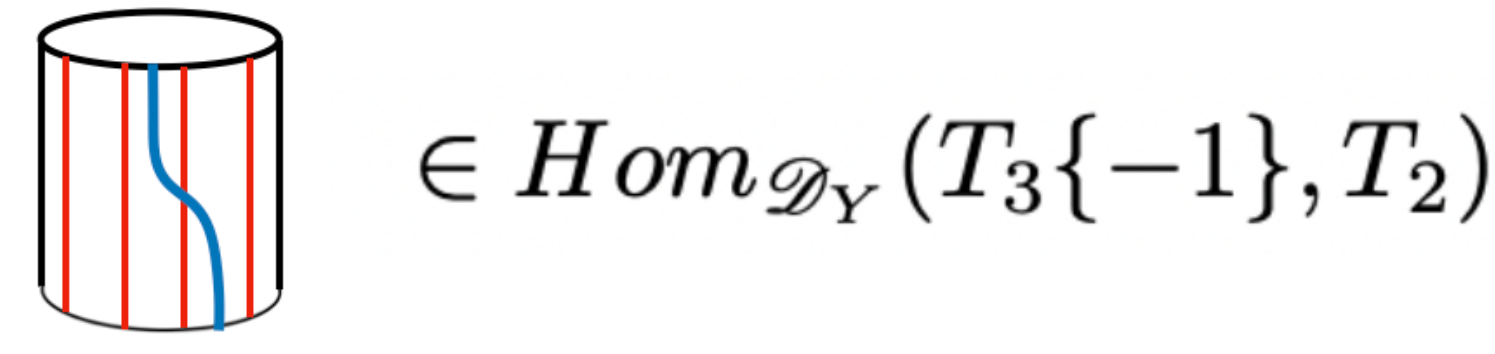}
\end{center}
\vspace*{-5pt}
 \vskip -0.5cm
\end{figure}
\vspace*{-5pt}
\begin{figure}[H]
\begin{center}
  \includegraphics[scale=0.31]{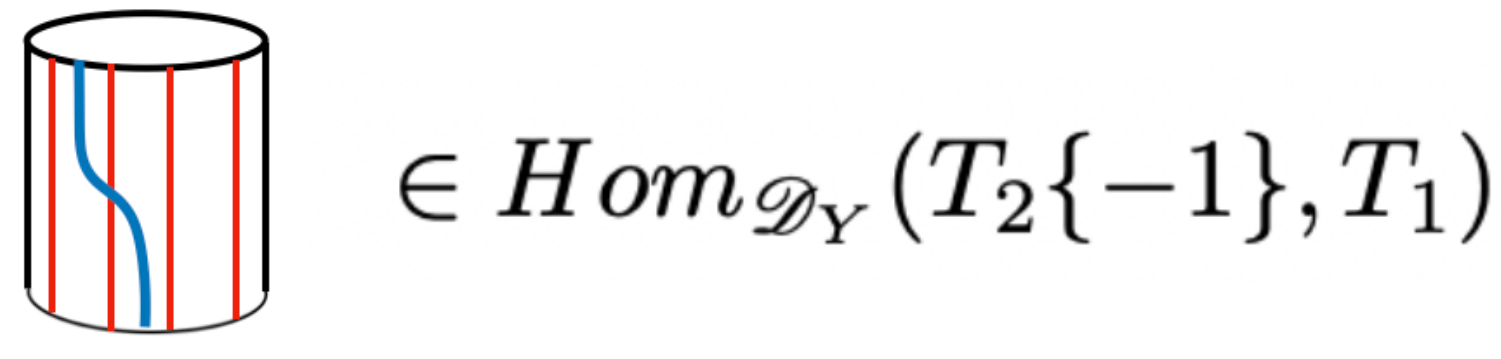}
\end{center}
\vspace*{-5pt}
 \vskip -0.5cm
\end{figure}
{\noindent } The resulting complex
has differential 
\begin{equation}\nonumber
    \delta = \vcenter{\hbox{\includegraphics{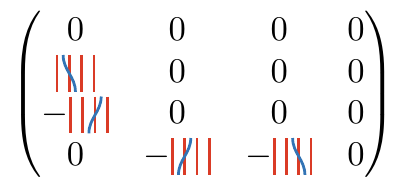}}}
\end{equation}
{\noindent }As before, we suppressed the cylinder to simplify the notation. Thus, the $E_{\cal U}$ brane has a description as the following complex of $T$-branes:
\begin{equation}\label{Esu2}
    \vcenter{\hbox{\includegraphics{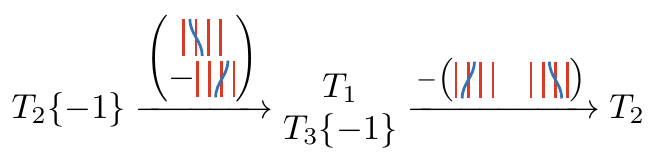}}}
\end{equation}
 Unless stated otherwise, the last term in the complex is placed cohomological degree zero. Since $A$ is an ordinary associative algebra, every map in the complex has degree zero and thus reduces the cohomological degree by $1$. Since cohomological degrees of all terms in the complex are determined from the degree of the first, we do not need to write them. 

In this example, the differential $\delta$ in \eqref{Esu2} squares to zero not only downstairs in $A$, but also upstairs in ${\mathscr A}$. Thus the complex which describes the  downstairs brane $E_{{\cal U}_2}= k^*{\cal U}_2\in \MDy$ also describes its upstairs parent ${\cal U}_2 \in \MDY$. We get one or the other, by simply reinterpreting the downstairs $T$-branes as carrying a trivial local system -- as ${\cal B}$ is trivial in one dimension -- or ${\cal B}_{\infty}$ which is not, as explained in section \ref{dressedsu2}. The complex is the standard Koszul resolution of the structure sheaf of a ${\mathbb P}^1$ in ${\cal X}$.

\subsubsection{}\label{singlecyclesl2}

%\begin{comment}
The case of one dimensional $Y={\cal A}$,  is simple but essential, as it is the basis of the algorithm to solve the theory on $Y=Sym^d({\cal A})$, for $d$ arbitrary.
%\end{comment}

In one dimension, in general, the geometric differential $\delta_O=\delta_0$, which we simply read off from the ${\mathscr B}E_{\cal U}$ Lagrangian, is easily proven to square to zero in the full downstairs algebra $A.$ 
Because there is no diagonal, maps passing through the punctures do not contribute to the A-model on $Y={\cal A}$, and the $u=0$ theory is the full answer. 
The differential $\delta_{u=0}$ thus describes the resolution of the brane in $\MDy$

\begin{figure}[h]
	\centering
\includegraphics[scale=0.65]{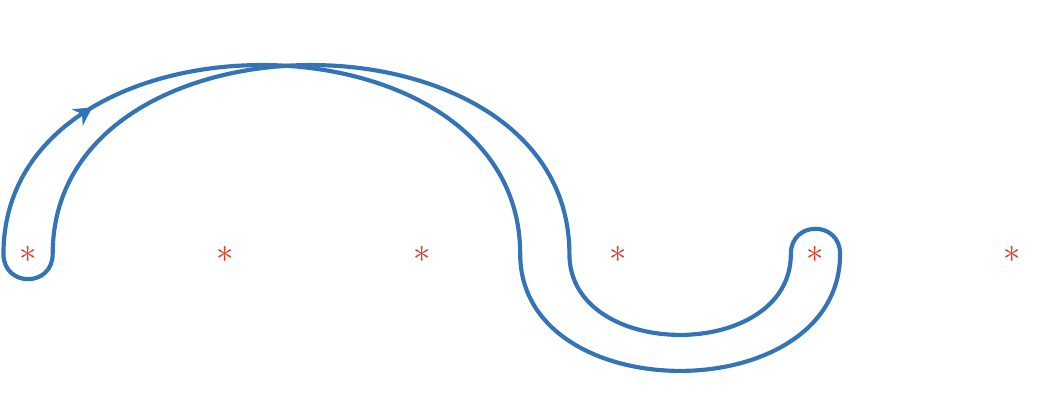}
	\caption{An ${\mathscr B}E_{\cal U}$-brane in the $\mathfrak{su}_2$ theory.}
	\label{unknotsl}
\end{figure}

In general, the differential which squares to zero in ${\mathscr A}={\mathscr A}_{u}$, is a one parameter deformation $\delta_u$, of the geometric differential $\delta_{u=0}$. From $\delta_u$, we get the downstairs differential $\delta_{u=0}$ by either taking the quotient by the ideal ${\cal I}$, which kills the dots, or by setting $u=0.$
The example in Fig.~\ref{unknotsl} illustrates that. The figure has been rotated clockwise by $90$ degrees, and we have also omitted drawing the cylinder. 

\noindent{}The corresponding direct sum brane is 
\beq\label{dex}
{\mathscr B}{E}_{\cal U}(T) = T_4\{ -3\}[2]  \oplus \begin{pmatrix}T_6\{ -3\}\\ T_4\{ -2\} \\ T_1\end{pmatrix}[1]\oplus \begin{pmatrix}T_5\{ -2\}\\ T_2\end{pmatrix}.
\eeq
The geometric differential $\delta_0$ is
\begin{equation}\label{dexb}
    \delta_{u=0} = \vcenter{\hbox{\includegraphics{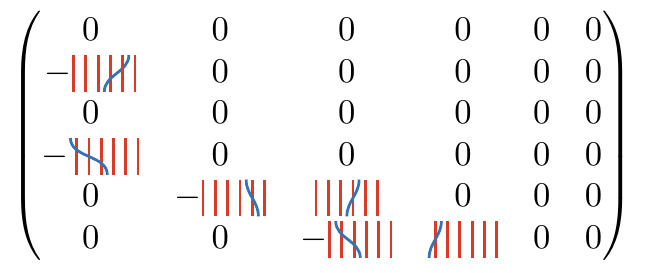}}}
\end{equation}
The differential squares to zero in the downstairs algebra $A$ and describes the resolution of the brane on $\MDy$.

The same direct sum brane from \eqref{dex}, together with the differential $\delta_{u=0}$, describes the resolution of the parent brane upstairs on ${\cal Y}_0 ={\cal Y}/D_a$ with the divisor $D_a$ of punctures removed. The corresponding algebra is ${\mathscr A}_{u=0}$, with the parameter $u$ from relation $\#5$ set to zero.  As explained in section \ref{prsu2}, to get $({\mathscr B}E(T), \delta_0)$ to describe a resolution of the brane on ${\cal Y}_0$,  we merely need to reinterpet the $T$-branes as the upstairs ones. This just changes the local system the branes are implicitly equipped, in this case, from a trivial one to ${\cal B}_{\infty}$, allowing arbitrary numbers of dots in their encomorphism algebra, rather than none. 

The full upstairs differential on ${\cal Y}$, after filling in the divisor $D_a$, is the $u\neq 0$ deformation of $\delta_{u=0}$, as in \eqref{nbu}.  The correction of order $u^{\ell}$ corresponds to introducing maps with $\ell$ dots. In our simple example, on degree grounds, there is only one such map we may introduce: 
\begin{equation}\nonumber
    \delta = \vcenter{\hbox{\includegraphics{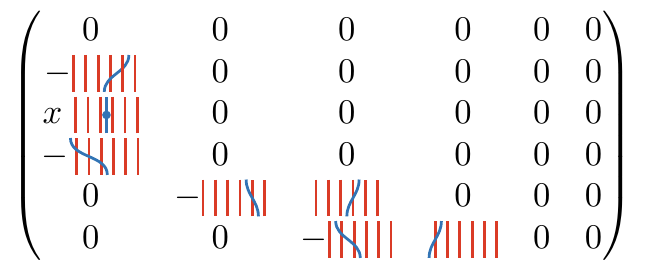}}}
\end{equation}
A short calculation shows that $\delta$ squares to zero in the ${\mathscr A}_{u}$ algebra if we set $x=-u$, as its square equals to
\begin{equation}\nonumber
\delta^2 = (u+x) \vcenter{\hbox{\includegraphics{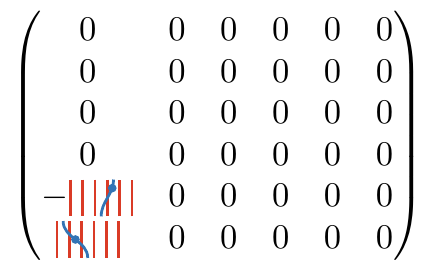}}}
\end{equation}
Thus, the complex resolving the upstairs ${\mathscr B} {\cal U}\in \MDY$ brane is
\beq\label{updot}
    \vcenter{\hbox{\includegraphics{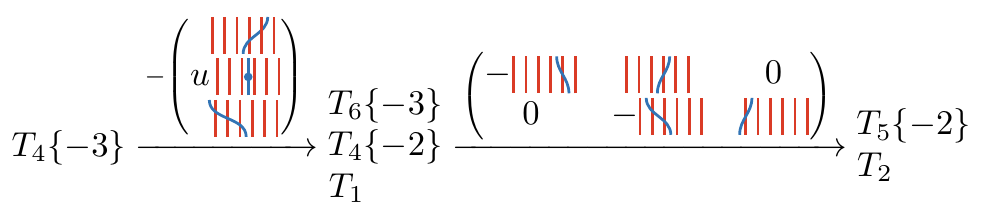}}}
\eeq
The complex resolving the downstairs brane ${\mathscr B} E_{\cal U}\in \MDy$ is obtained from \eqref{updot} as the image of the $k^*:{\MDY} \rightarrow \MDy$ functor. This amounts to thinking of $T$'s as the downstairs generators and setting the dots to zero. Its differential coincides with the $\delta_{u=0}$ in \eqref{dexb}.

\subsubsection{}
\begin{figure}[H]
	\centering
	\includegraphics[scale=0.7]{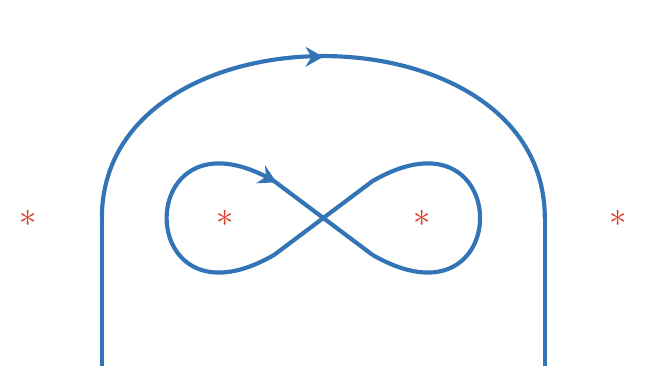}
	\caption{The simplest example of $d=2$ in the $\mathfrak{su}_2$ theory.}
	\label{exampleSL2}
\end{figure}

Projective resolutions encode counts of disk instantons. We will start with a very simple example demonstrating that; later we will turn to examples interesting for knot theory.  Consider a Lagrangian in $d=2$ theory in Fig.~\ref{exampleSL2}.
\noindent{}We found the resolution of the figure-eight brane in section \ref{sl2cap}. The resolution of the outer brane is the complex

\begin{center}
    \includegraphics{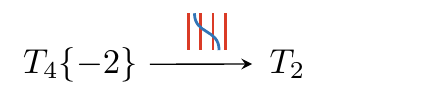}
\end{center}
The product of the two complexes can be represented as a 2-dimensional grid:

\begin{center}
    \includegraphics{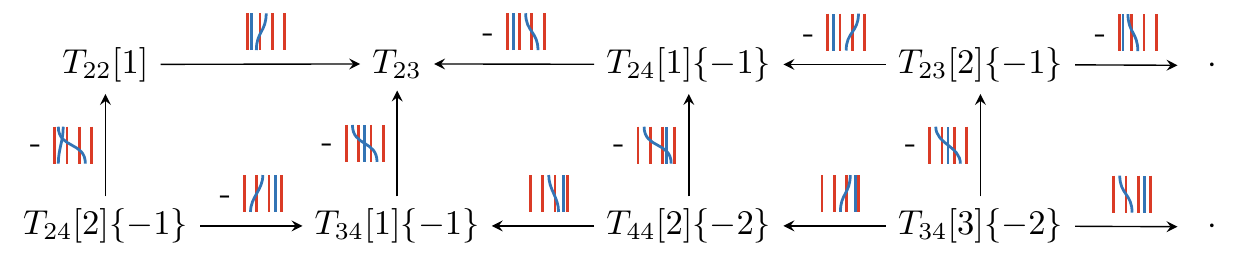}
\end{center}

The maps in the complex encode the relative positions of branes in Fig.~\ref{exampleSL2}, since they represent open strings stretching between them.  A map which is a product of the identity map of the inner brane times a non-trivial morphism from the outer brane is a pair of blue strands, one constant, the other not, which cross. Their crossing reflects the fact that, in this brane configuration, an open string from the differential of the outer brane must cross a thimble of the inner.
Similarly, a map which is a product of a non-trivial morphism from the inner brane complex times the identity map of an outer $T$-brane, is pair of blue strands, one constant and one not, which do not cross.
 
 It is straightforward to read off the Maslov and the equivariant degrees of all the branes in the complex from the degrees of the maps. The signs, which are unique up to gauge transformations, are assigned per section \ref{sec:signs2} and appendix \ref{su2cs}. In particular, the horizontal maps in each line multiply to $(-1)^{4/2+1}$. The sign of the leftmost vertical map gets assigned to all other vertical maps, and the signs in the top row gets inherited by other rows, but multiplied with an overall alternating sign.

Organizing the $T$-branes by their cohomological degrees, we get the direct sum brane 
\beq
 T_{34}\{ -1\}[3]\oplus \begin{pmatrix} T_{44}\{ -1\}\\ T_{23}\\ T_{24}\end{pmatrix} [2]\oplus \begin{pmatrix}T_{24}\\ T_{34}\\T_{22}\end{pmatrix}[1]\oplus T_{23} ,
\eeq
with differential
\begin{equation}
    \delta_{u,0} = \vcenter{\hbox{\includegraphics{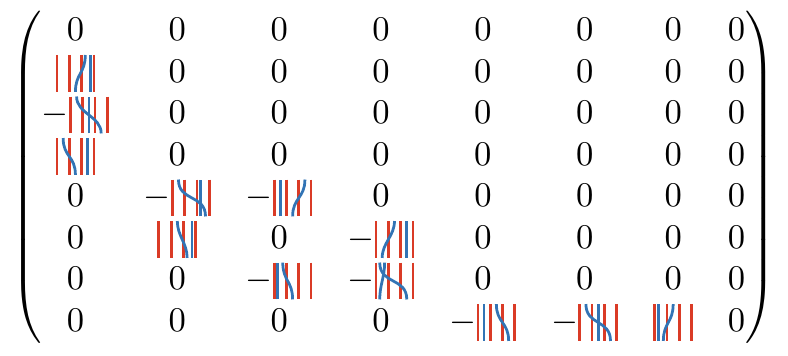}}}.
\end{equation}
In this simple example, the geometric differential happens to also square to zero in ${\mathscr A}_{u,0}$. This follows from the fact that the two one-dimensional complexes we started had geometric differentials that squared to zero for any $u$, as in our first example in this section. 

In general, all $u\neq 0$ corrections in $\delta_{u,0}$ should come from the corrections from one-dimensional complexes, except that not all of them survive to the product on degree grounds. This reflects the fact that, as long as $\hbar=0$, the theory effectively behaves as a product theory.

It is straightforward to check that $\delta_{u,0}$ does not square to zero in the full ${\mathscr A}_{u,\hbar}$ algebra. Thus, we need to look for a deformation of it by terms of order $\hbar$. In fact, the first order deformation $\delta =\delta_{u,0}+\hbar \delta_{u,1}$ turns out to suffice.
The deformation $\delta_{u,1}$ may contain all possible corrections allowed on degree grounds that are not already included in $\delta_{u,0}$, as those correspond to gauge transformations. By uniqueness of $\delta$, discussed in section \ref{dunique}, any two solutions to the deformation problem should be equivalent, so we may try to find the solution in a subspace of the deformation space. In this case, the following simple deformation suffices:
\begin{equation}
    \delta = \vcenter{\hbox{\includegraphics{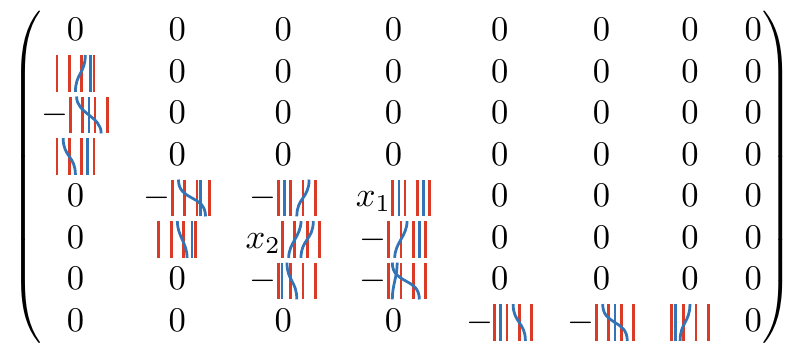}}}
\end{equation}
Computing its square
\begin{equation}
    \delta^2 = \vcenter{\hbox{\includegraphics{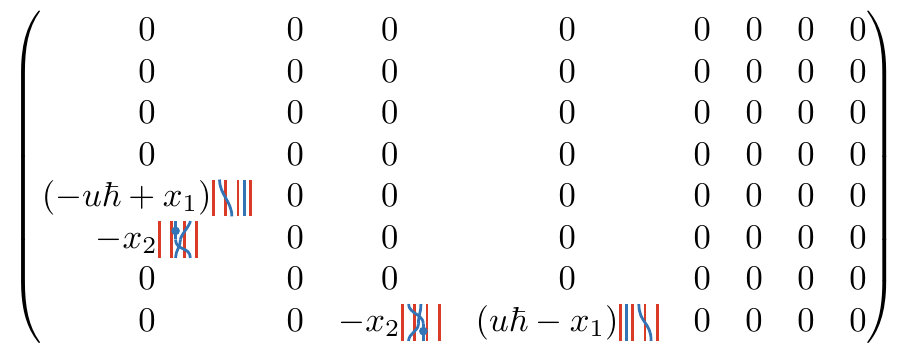}}}
\end{equation}
we can see that setting $x_1=u\hbar$ and $x_2=0$ leads to a differential which, when written as a differential of a standard complex, reads
\beq\label{exsl2c}
\vcenter{\hbox{\includegraphics[scale=1]{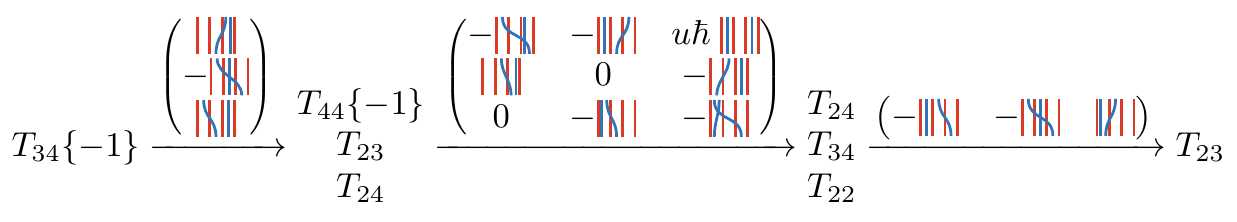}}}
\eeq
\noindent{}The above purely algebraic construction of the brane differential ${\delta}$ accounts for disks contributing to the Floer differential of the brane with any $I$-brane. 

\begin{figure}[H]
	\centering
    \includegraphics[scale=0.73]{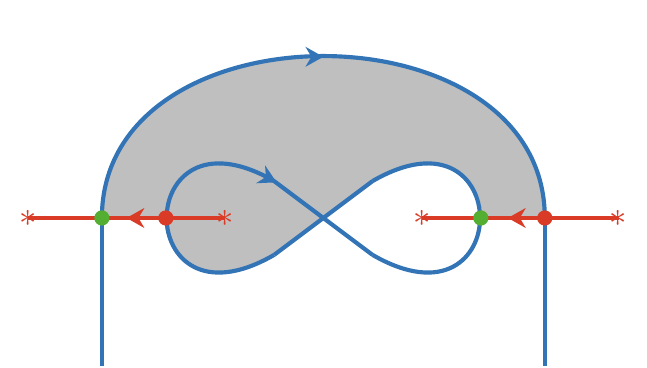}
    \caption{The holomorphic disk counted by the differential in \eqref{ff}.}
	\label{examplesl2disk}
\end{figure}

For example, the Floer complex of the brane configuration in Fig.~\ref{examplesl2disk} should be obtained by applying $hom_{Y}^{*,*}(-, I_{24})$ to the complex in \eqref{exsl2c}. The only terms in the complex that give a non-trivial contribution contain $T_{24}$ branes. The result is the complex 
\beq\label{ff}
{\delta_F}: \qquad \mathbb{C}\xrightarrow{u\hbar} \mathbb{C}
\eeq
The right term is placed in degree $[2]$, the left in degree $[1]$, as the functor is contravariant. Two copies of ${\mathbb C}$ correspond to the two intersection points in Fig.~\ref{examplesl2disk}. The map between them is the Floer differential, together with its sign. The differential is proportional to $u \hbar$ since the map in Fig.~\ref{examplesl2disk} intersects a puncture and the diagonal $\Delta$ once. In particular, the complex has trivial homology whenever $u\hbar\neq 0$. 

One could in principle do the corresponding disk count by hand. The count of holomorphic maps $y: {\rm D} \rightarrow Y$ may be defined using the cylindrical approach to Floer theory \cite{L}. The maps that the theory should count are described in section \ref{Fdef}. We will merely check that the disk is not forbidden on degree grounds. From the shaded domain $A$ which the disk projects to in Fig.~\ref{examplesl2disk}, we can read off its Maslov index and equivariant degree. Using \eqref{indcompare2}, its Maslov index is $ind(A) = 2e(A) =1$ 
since $e(A) = 1- 3\times{1\over 4}+{1\over 4} ={1\over 2}$ per \eqref{Euler}. It has equivariant degree $J^0(A) =0$ per \eqref{Jgradeb} because intersects a puncture once so $n_a(A)=1$, and it also intersects the diagonal once $i(A) = 3\times{1\over 4} +{3\over 4}  -e(A) =1$ per \eqref{iD}.  Thus, the disk has Maslov index $1$ and equivariant degree $0$, as needed for it to contribute to $\delta_F$. The fact that the disk intersects a puncture on ${\cal A}$ and the diagonal $\Delta$ once is in agreement with the fact that its contribution to the differential is at order $u\hbar$.

\subsubsection{}
Here is another example, now more relevant to knot theory. 
\begin{figure}[H]
	\centering 
    \includegraphics[scale=0.45]{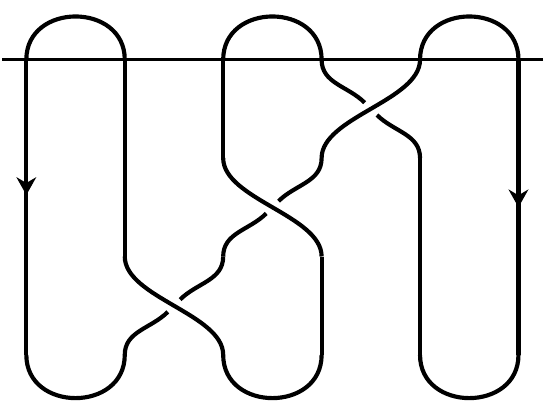} \qquad\qquad
\includegraphics[scale=0.8]{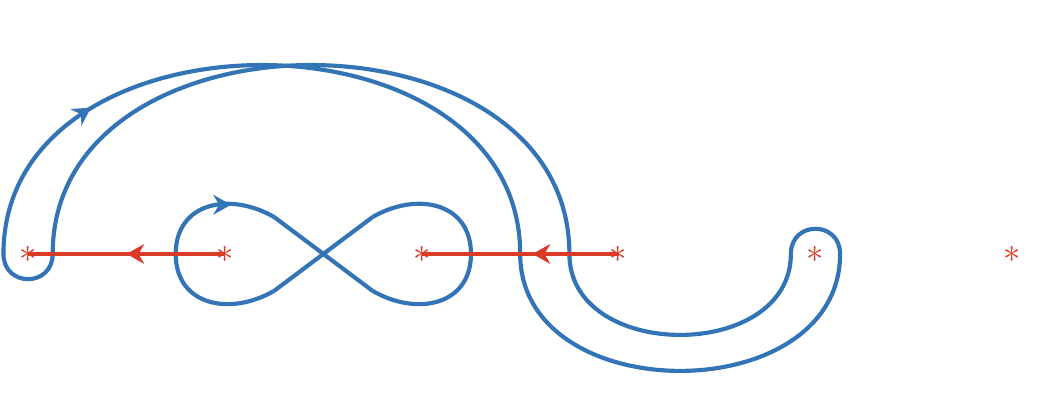}
	\caption{ Branes for a presentation of the reduced unknot.}
	\label{unknotSL2}
\end{figure}
\noindent{}It corresponds to a presentation of the unknot similar to that in section \ref{2dexd}. The brane configuration in Fig.~\ref{unknotSL2} corresponds to the reduced unknot.

Resolutions of the two one-dimensional factor branes were worked out in sections \ref{sl2cap} and \ref{singlecyclesl2}. Taking their product,  since the two one-dimensional branes have the topology of $S^1$, we get a grid on a torus of size $6\times 4$, in \eqref{46}.
We suppressed the maps in the complex, since they are uniquely determined from the two initial one dimensional complexes. We merely indicated the maps in which blue strands cross by purple, instead of black, arrows. They come from maps $T_4\rightarrow T_2$ and $T_4 \rightarrow T_1$ of the longer complex, which intersect the thimbles $T_2,T_3,T_4,T_3$ of the shorter complex.
\beq\label{46}
\vcenter{\hbox{\includegraphics[scale=0.85]{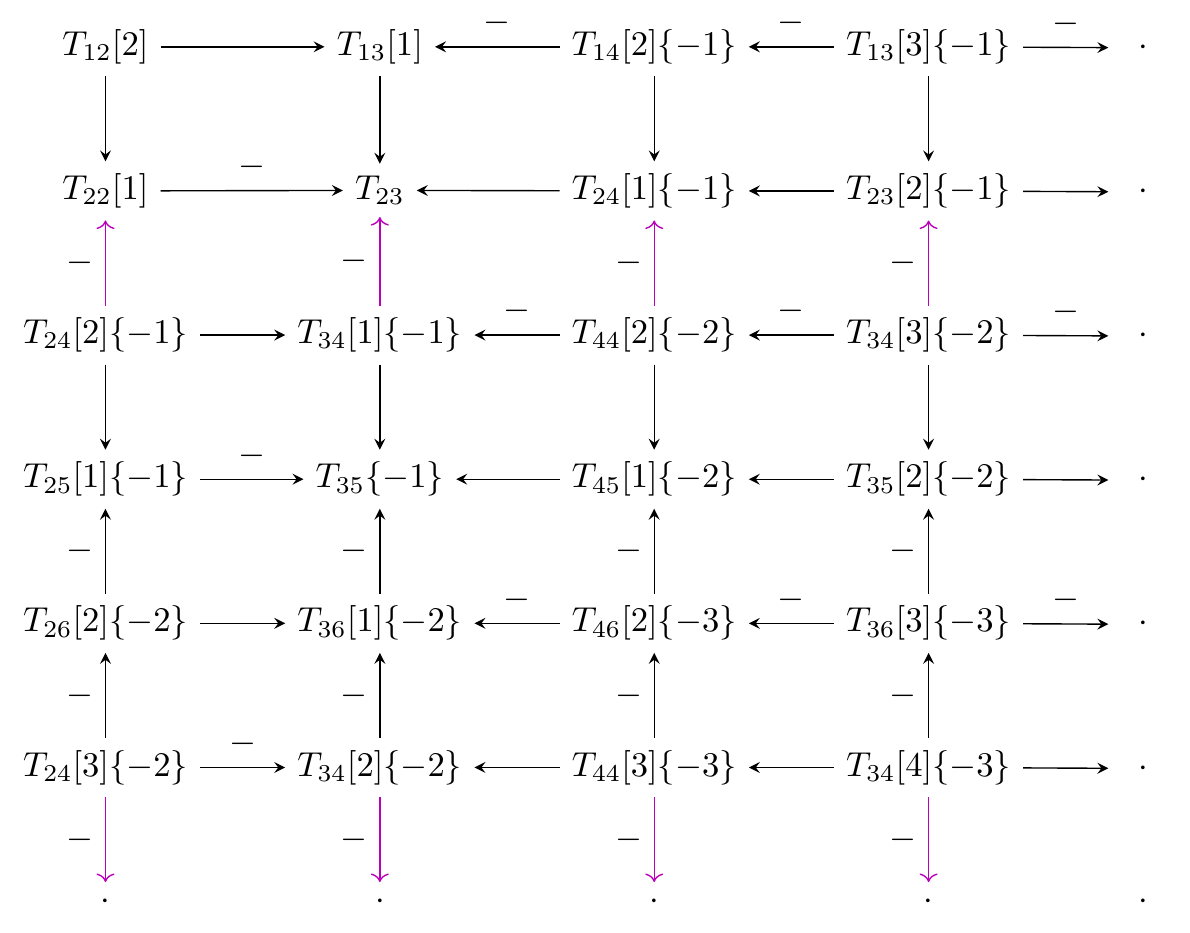}}}
\eeq
 The relative Maslov and equivariant degrees of the thimbles in the complex are the degrees of the maps between them. Per section \ref{su2cs}, we took signs of the vertical maps to be those of the original one-dimensional complex, but then signs of the horizontal maps get multiplied by an overall $\pm$ sign, alternating by row. This is the standard sign on the tensor product of chain complexes.

The differential $\delta_O=\delta_{u=0, \hbar=0}$ of the complex in \eqref{46} is the geometric one; it squares to zero in the algebra ${\mathscr A}_{u=0,\hbar =0}$,  which has $u$ and $\hbar$ both set to zero.

The differential that deforms to $u\neq 0$ is simply the product differential coming from the two one-dimensional $u\neq 0$ complexes. It is easy to check that the product differential squares to zero in the ${\mathscr A}_{u, \hbar=0}$ algebra. It gives the complex in \eqref{udefo}. In a general, only a subset of one-dimensional $u\neq 0$ corrections lift to the product complex because the equivariant degrees of the thimbles in the product complex are not product of their degrees in the one-dimensional complexes. The maps that end up with wrong equivariant degrees get set to zero, as we will see in the later examples. In all cases we checked, this construction leads to a differential that squares to zero in the $\hbar=0$ algebra. 

\beq\label{udefo}
\vcenter{\hbox{\includegraphics[scale=0.85]{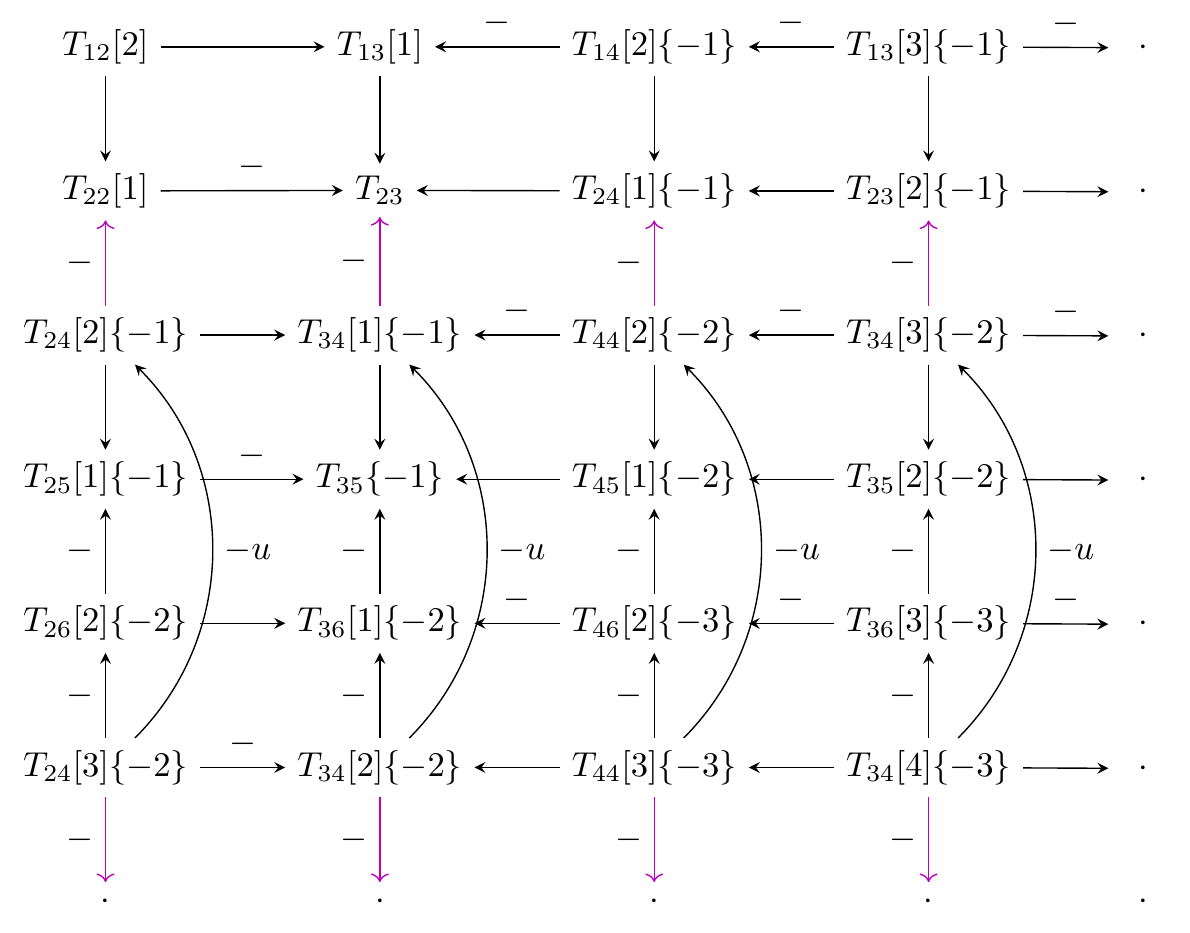}}}
\eeq

Finally, we look for a deformation of the differential that squares to zero in the full ${\mathscr A} ={\mathscr A}_{u, \hbar}$ algebra. To find the first order in $\hbar$ deformation $\delta= \delta_{u,0} +\hbar \delta_{u,1}+\ldots$, we could start by including in $\delta_{u,1}$ all possible maps that have consistent equivariant degree and that are not already present in $\delta_{u,0}$, with some unknown coefficients. We solve for the coefficients by asking that $\delta^2=0$ to order $\hbar$, in the algebra ${\mathscr A}_{u, \hbar}$.

\begin{figure}[H]
	\centering
	\includegraphics[scale=0.7]{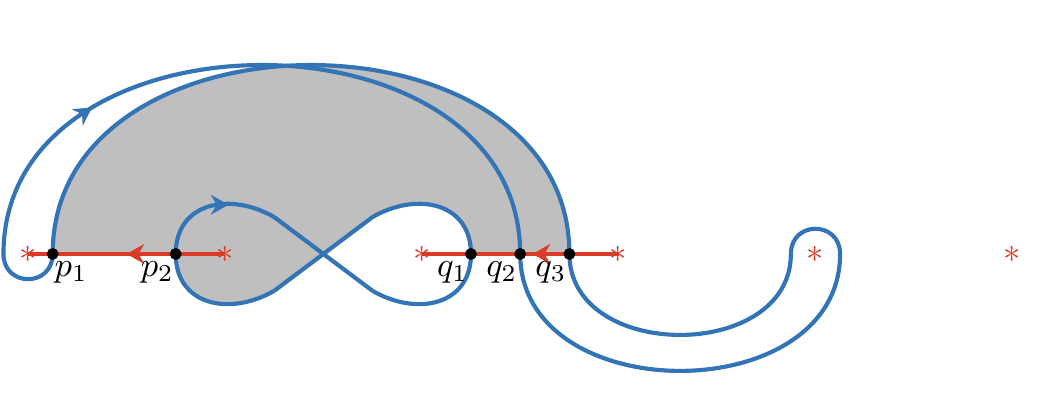}
	\caption{The holomorphic disk counted by the differential $\delta_F$.}
	\label{fig:sl2unknotDisk}
\end{figure}

Since any two solutions to the deformation problem are equivalent, one can start by including in $\delta_{u,1}$ only the simplest terms, without crossings or dots. In the example at hand, there are $19$ such possible entries in $\delta_{u,1}$. Solving the first-order equation, we find a one-parameter family of solutions.  The first-order $\hbar$ deformation we get in this way turns out to be exact: all these solutions turn out to automatically satisfy the higher-order equations as well. One can easily check explicitly that solutions within this family are gauge equivalent, so we pick one. This leads to $4$ non-vanishing coefficients in $\delta_{u,1}$, given in \eqref{4corr}. If we were not not able to find a solution with the initial ansatz, we would relax the assumption about which of the finite number of possible corrections have non-zero coefficients, as we generally do in more complicated examples.

\beq\label{4corr}
\vcenter{\hbox{\includegraphics[scale=0.8]{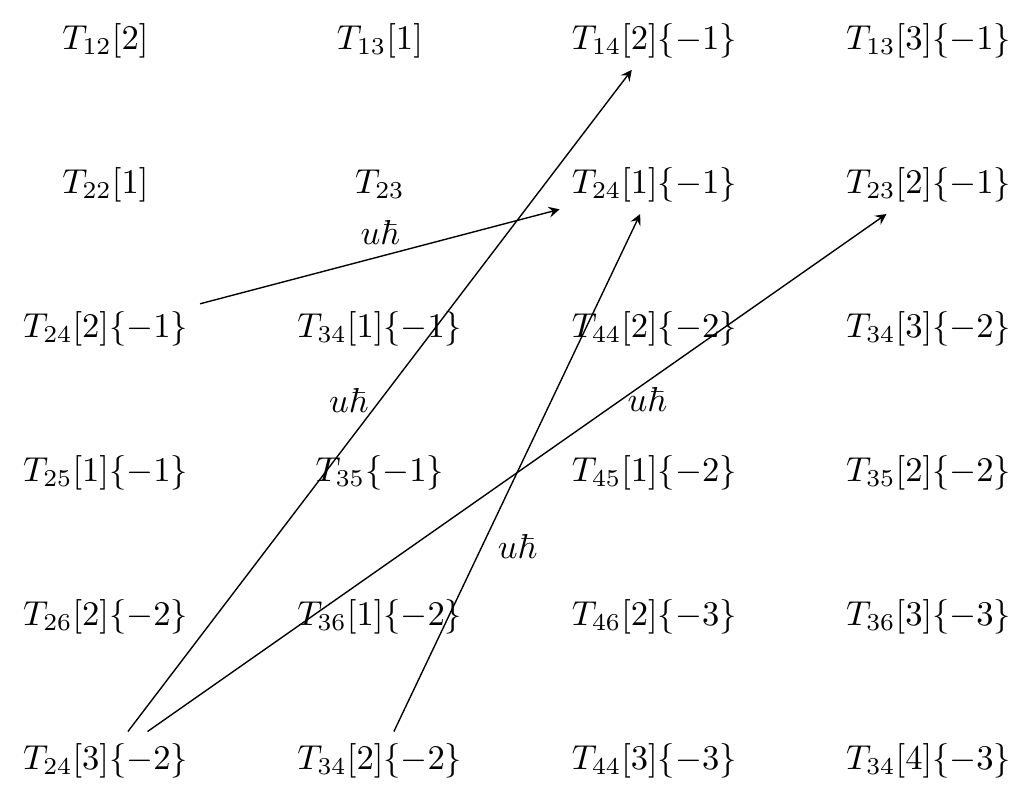}}}
\eeq
The complex, written in standard form, reads:
\beq\label{5.53}
    T_{34}\{ -3\}
    \xlongrightarrow{d_4}
    \begin{matrix}
    T_{36}\{ -3\}\\ T_{44}\{ -3\}\\ T_{34}\{ -2\}\\ T_{24}\{ -2\}\\ T_{13}\{ -1\}
     \end{matrix}
    \xlongrightarrow{d_3}
    \begin{matrix}
    T_{46}\{ -3\}\\ T_{44}\{ -2\}\\ T_{35}\{ -2\}\\ T_{26}\{ -2\}\\ T_{34}\{ -2\}\\T_{14}\{ -1\}\\T_{23}\{ -1\}\\T_{24}\{ -1\}\\T_{12}\;\;\;\;\;\;\;\;\;
         \end{matrix}
    \xlongrightarrow{d_2}
    \begin{matrix}
    T_{45}\{ -2\}\\T_{36}\{ -2\}\\T_{24}\{ -1\}\\ T_{34}\{ -1\}\\T_{25}\{ -1\}\\ T_{13}\;\;\;\;\;\;\;\;\;\\ T_{22}\;\;\;\;\;\;\;\;\;
    \end{matrix}
    \xlongrightarrow{d_1}
    \begin{matrix}
    T_{35}\{ -1\}\\ T_{23} \;\;\;\;\;\;\;\;\;
    \end{matrix}
\eeq
with differential given by
\begin{equation*}
     d_4=   \vcenter{\hbox{\includegraphics{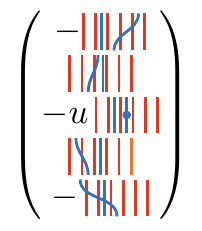}}}, \qquad d_3=\vcenter{\hbox{\includegraphics{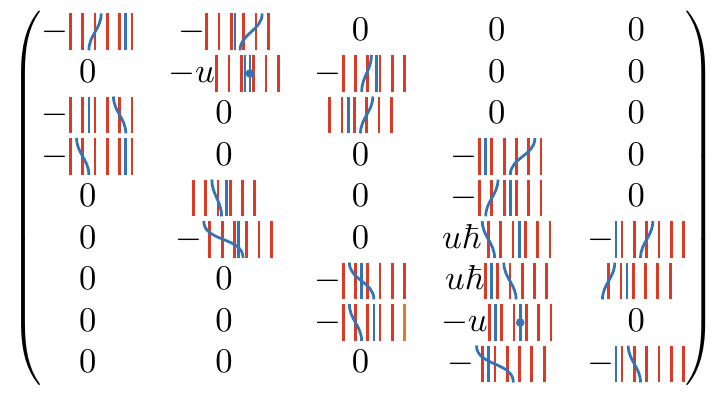}}}
\end{equation*}
\begin{equation*}
 d_2=   \vcenter{\hbox{\includegraphics{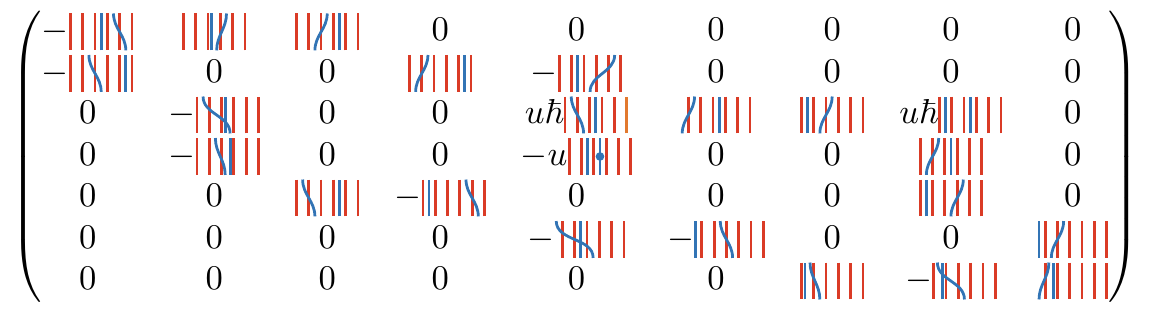}}}
\end{equation*}
\begin{equation*}
  d_1=  \vcenter{\hbox{\includegraphics{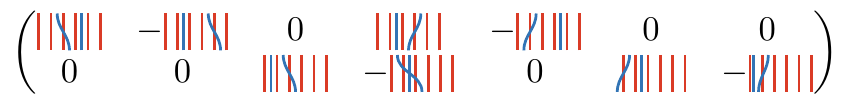}}}, \end{equation*}
Applying the $hom_Y^{*,*}(-,I_{24})$ functor to \eqref{5.53} results in a three-dimensional complex 
\beq
0\rightarrow \mathbb{C}\{-1\}\xrightarrow{u\hbar} \mathbb{C}\{-1\}\xrightarrow{0} \mathbb{C}\{-2\} \rightarrow 0
\eeq
with the rightmost term in homological degree $3$. The three terms in the complex correspond to, respectively, the three intersection points $p_1q_1$, $p_2 q_3$ and $p_2 q_2$ in Fig.~\ref{fig:sl2unknotDisk}. The maps in the complex should compute the action of the Floer differential. 
A disk that could contribute to coefficient of $p_2q_3$ in  $\delta_F p_1q_1$, on degree grounds, projects to the shaded domain in Fig.~\ref{fig:sl2unknotDisk}.  The corresponding disk would have Maslov index $1$ and equivariant degree zero per \eqref{indcompare2} and \eqref{Jgradeb}.  From the complex, we deduce that the count of such disks is $1$. The fact that the map is proportional to $u\hbar$ corresponds to the fact the disk intersects the diagonal and a puncture once, as the one in Fig.~\ref{examplesl2disk}. The homology of the complex is $\mathbb{C}[3]\{-2\}$.

\begin{figure}[H]
    \centering
	\includegraphics[scale=0.45]{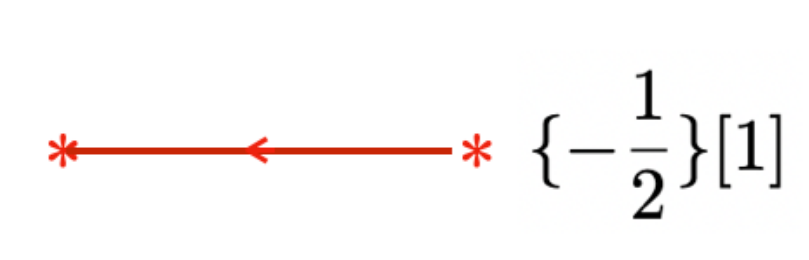}
	\caption{The ${\mathfrak{su}_2}$ cap brane $I_{\cal U}$ equals the corresponding canonical $I$-brane shifted in degree by $\{-{1\over 2}\}[1]$, for each interval it contains. }
	\label{Ibraneosu2}
\end{figure}

To get the absolute grading of the link homology, we need to include the additional, overall grading shifts explained in section \ref{Msu2}. The total shift is the sum of the braid dependent shift in \eqref{su2shift} and the shift associated with grading of the cap branes relative to the canonical $I$-branes, which depends only on the dimension of $Y$. From link diagram in Fig. \ref{unknotSL2} we read off that the writhe of the braid (the difference in the number of positive and negative crossings) is $w=1$, the exponent sum of the braid generators is $e=-1$, so \eqref{su2shift} contributes $\Delta M = -1$ and $\Delta J_0 = 1$ to the shift. Since the dimension is $d=2$, the $I_{\cal U}$ brane is related to the $I_{24}$ brane by the grading shift which is twice that in Fig.~\ref{Ibraneosu2}, so $I_{\cal U} = I_{24}[2]\{-1\}$.  The net shift is $M \to M -3$ and $J^0 \to J^0 + 2$. The result is
$$Hom^{*,*}({\mathscr B}E_{{\cal U}}, I_{{\cal U}}) = {\mathbb C},
$$
which is the reduced homology of the unknot.

\subsubsection{}
\begin{figure}[h]
	\centering
    $\vcenter{\hbox{\includegraphics[scale=0.2]{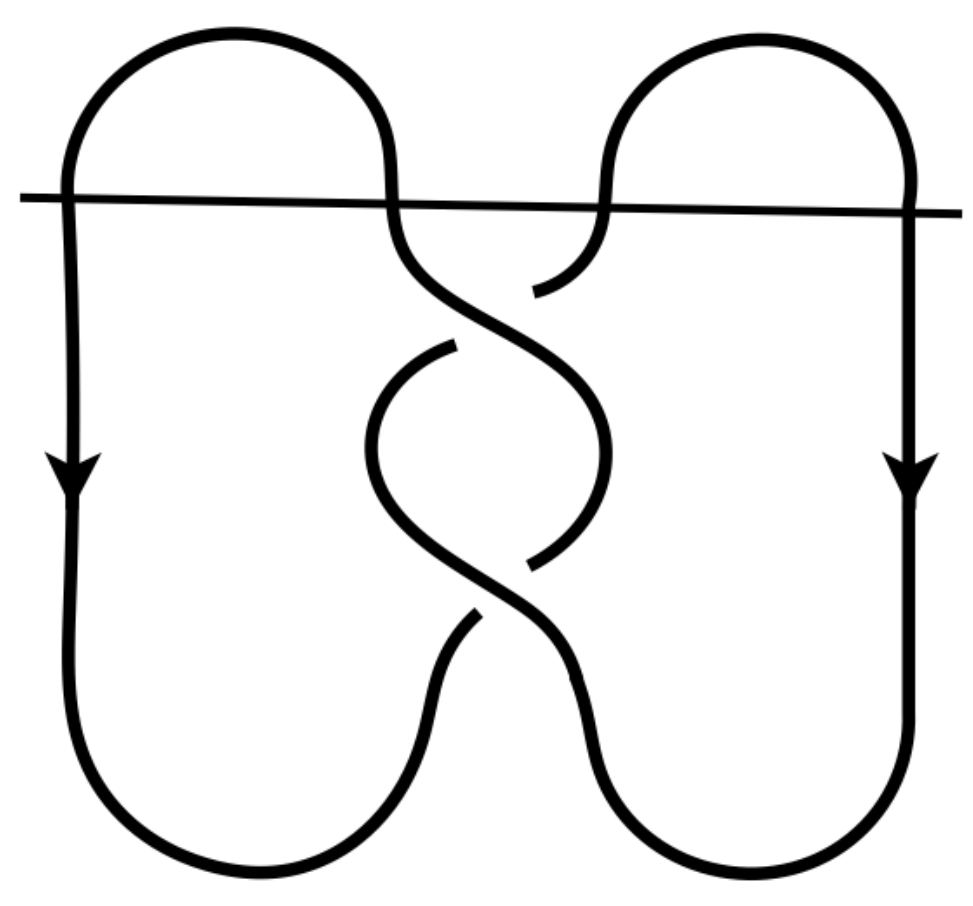}}}$
    \qquad \qquad
$\vcenter{\hbox{\includegraphics[scale=0.8]{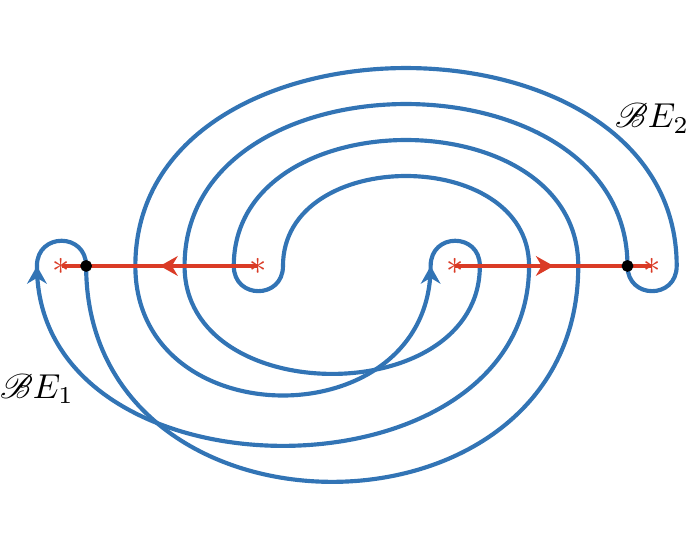}}}$
	\caption{The Hopf link and the corresponding Lagrangians}
	\label{fig:sl2hopflink}
\end{figure}

As our final explicit example, consider the Hopf link as shown in Fig.~\ref{fig:sl2hopflink}.
We first write down the resolutions of the one dimensional Lagrangians ${\mathscr B}E_1$ and ${\mathscr B}E_2$. Including only the geometric terms, the resolution of ${\mathscr B}E_1$ is: 
\begin{equation}\nonumber
\includegraphics[scale=0.95]{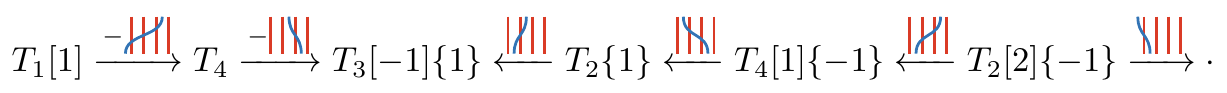}
\end{equation}
For ${\mathscr B}E_2$ we get:
\begin{equation}\nonumber
    \includegraphics[scale=0.95]{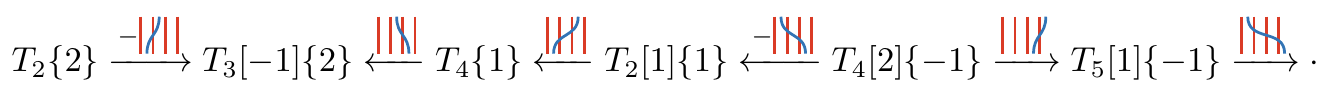}
\end{equation}
The product of these gives is a toric grid whose maps encode the differential $\delta_O=\delta_{0,0}$:
\begin{equation*}
    \includegraphics[scale=0.6]{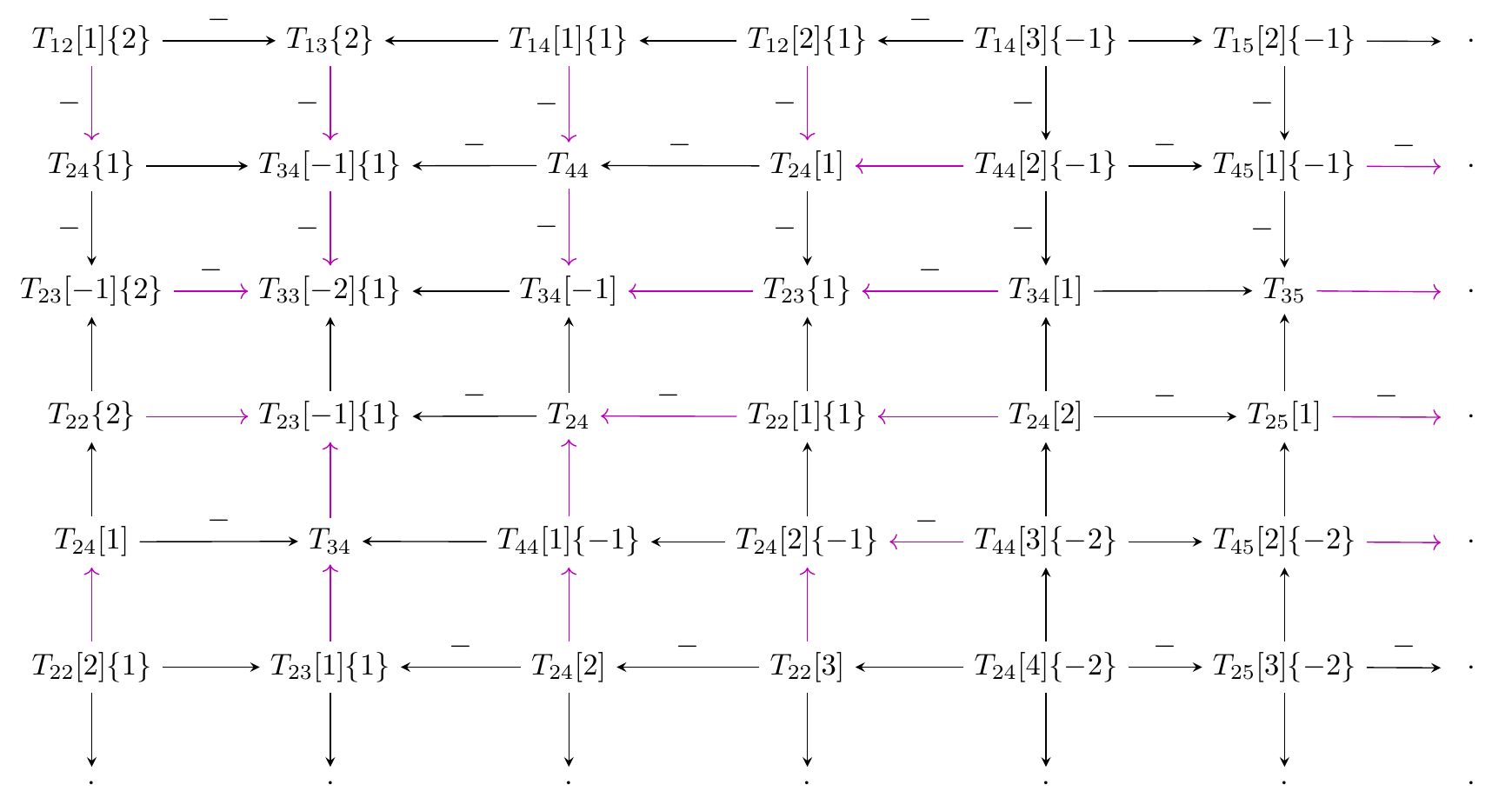}
\end{equation*}
We determined the absolute degrees of the branes in the complex, using the fact that the marked intersection point in Fig.~\ref{fig:sl2hopflink} is fixed under braiding. Therefore, the thimble $T_{24}$ corresponding to this point must have the same degrees before and after braiding. It has $M=4$ and $J^0=-2$. 

The grid can be rearranged into the standard chain complex
\begin{equation}\nonumber
    T_{24}\{-2\}
    \xrightarrow{d_4}
    \begin{matrix}
        T_{44}\{-2\} \\
        T_{25}\{-2\} \\
        T_{14}\{-1\} \\
        T_{22}
    \end{matrix}
    \xrightarrow{d_3}
    \begin{matrix}
        T_{45}\{-2\} \\
        T_{15}\{-1\} \\
        T_{44}\{-1\} \\
        T_{24}\{-1\} \\
        T_{24} \\
        T_{24} \\
        T_{12}\{1\} \\
        T_{22}\{1\}
    \end{matrix}
    \xrightarrow{d_2}
    \begin{matrix}
        T_{45}\{-1\} \\
        T_{44}\{-1\} \\
        T_{24} \\
        T_{34} \\
        T_{25} \\
        T_{24} \\
        T_{14}\{1\} \\
        T_{22}\{1\} \\
        T_{23}\{1\} \\
        T_{12}
    \end{matrix}
    \xrightarrow{d_1}
    \begin{matrix}
        T_{44} \\
        T_{35} \\
        T_{24} \\
        T_{34} \\
        T_{24}\{1\} \\
        T_{23}\{1\} \\
        T_{13} \\
        T_{22}
    \end{matrix}
    \xrightarrow{d_0}
    \begin{matrix}
        T_{34} \\
        T_{34}\{1\} \\
        T_{23}\{1\} \\
        T_{23}
    \end{matrix}
    \xrightarrow{d_{-1}}
    T_{33}\{1\}
\end{equation}
with the rightmost term in homological degree $-2$ and where, for example, the first two differentials are
\begin{align}\nonumber
    &d_4=\vcenter{\hbox{\includegraphics{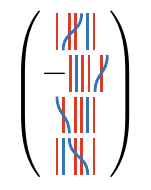}}}
    &d_3^{{\rm geo}}=\vcenter{\hbox{\includegraphics{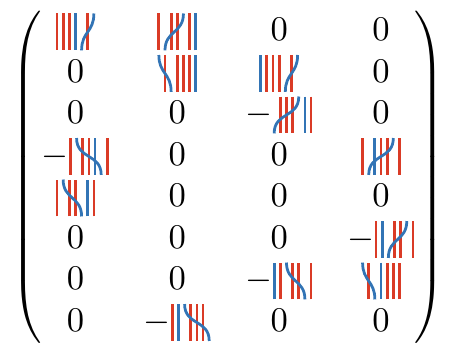}}}
\end{align}

The next step is to find the $u$-deformation of the geometric complex. It is a product of one-dimensional complexes resolving ${\mathscr B}E_1$ and ${\mathscr B}E_2$ for  $u\neq 0$, except that not all of its terms survive to the product complex, as some may have degrees which are inconsistent with the degrees of thimbles they are to map between. The terms that do survive are
\beq\nonumber
    \includegraphics[scale=0.6]{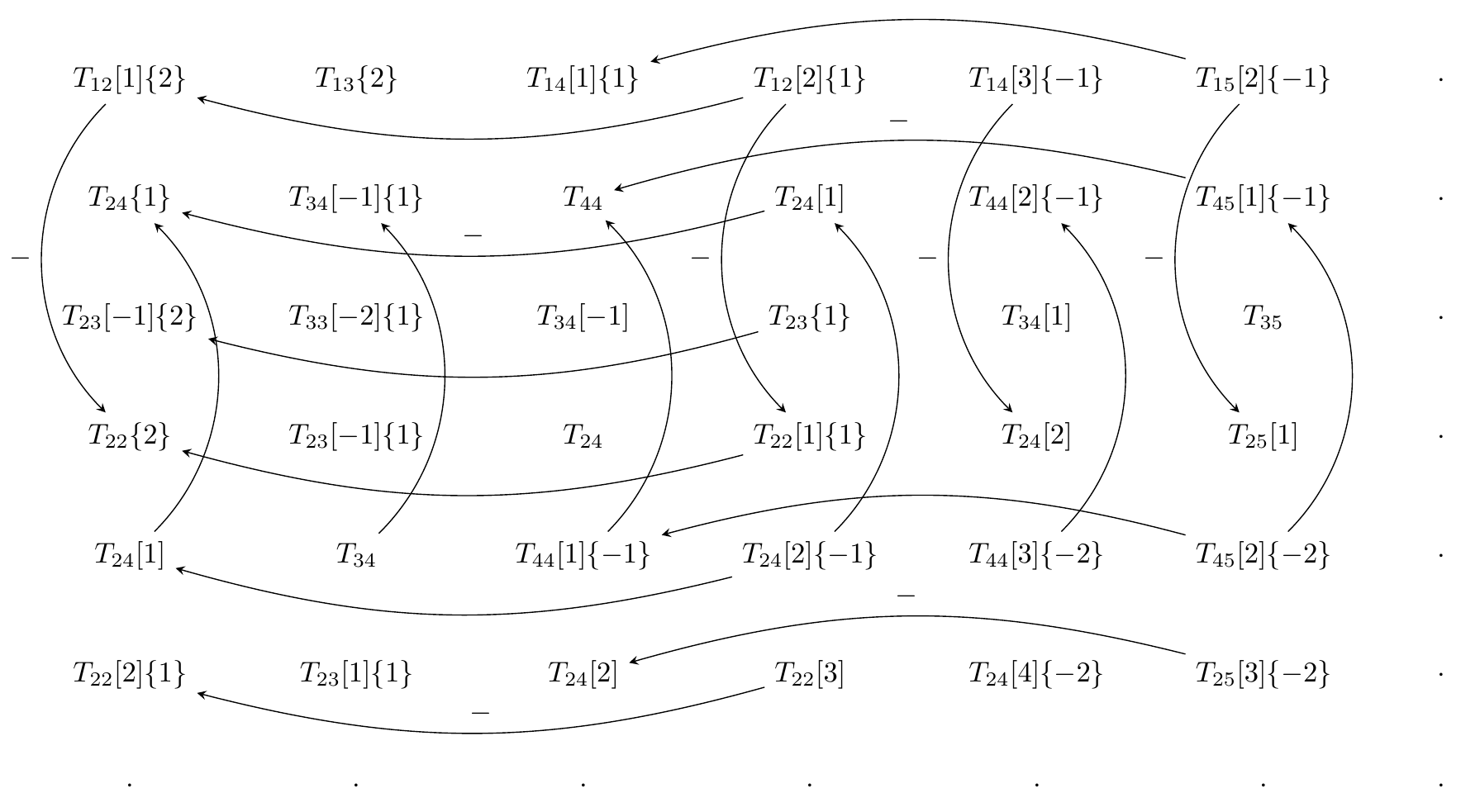}
\eeq
where each map is multiplied by an overall factor $u$.  Continuing with the example of $d_4$ and $d_3$, we find that $d_4$ is unchanged and
\beq\nonumber
d_3^{\hbar=0}=\vcenter{\hbox{\includegraphics{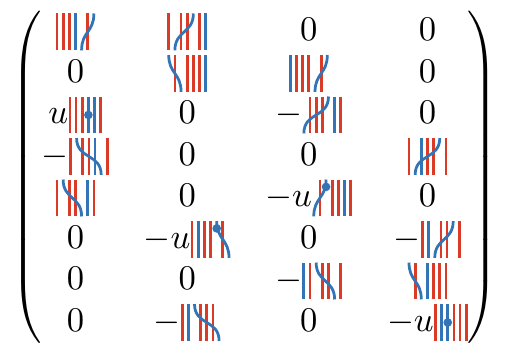}}}
\eeq
The result is a complex that satisfies $\delta_{u,0}^2=0$ at $\hbar=0$.

Finally, we solve for the corrections necessary for $\delta$ to satisfy $\delta^2=0$ with $\hbar \neq 0$. It proves sufficient to turn on only the first order in $\hbar$ deformation, of the form $\delta = \delta_{u,0} + \hbar \delta_{u,1}$, and further restrict the possible terms in $\delta_{u,1}$ to maps without dots so $\delta_{u,1} = \delta_{u=0,1}$. For example, $d_3$ with the possible deformations terms added is given by
\begin{equation}\nonumber
d_3=\vcenter{\hbox{\includegraphics{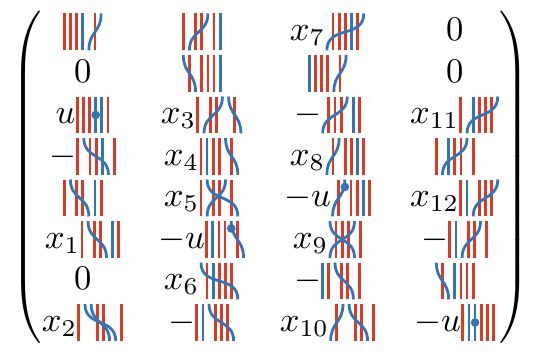}}}
\end{equation}
with coefficients $x_i$, which we want to solve for.
Asking that $d_3 \cdot d_4=0$ leads to a system of equations. For example, the fourth row of $d_3$ gives the equation
\beq\nonumber
\vcenter{\hbox{\includegraphics{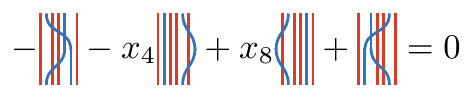}}}
\eeq
which has the solution
\beq\nonumber
	x_4 = -x_8 = - u \hbar.
\eeq
The other coefficients in $d_3$ turn out to vanish, and altogether we find
\beq\nonumber  d_3=\vcenter{\hbox{\includegraphics{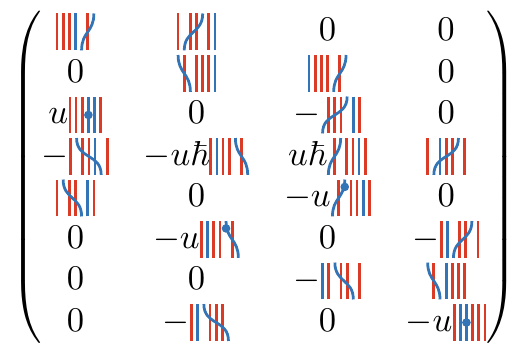}}}.
\eeq
Proceeding further, the $d_3$ deformation we just found will determine the $\delta_{u,1}$ deformation of $d_2$, and so on. Altogether, the maps in $\delta_{u,1}$ are
\begin{equation*}
    \includegraphics[scale=0.6]{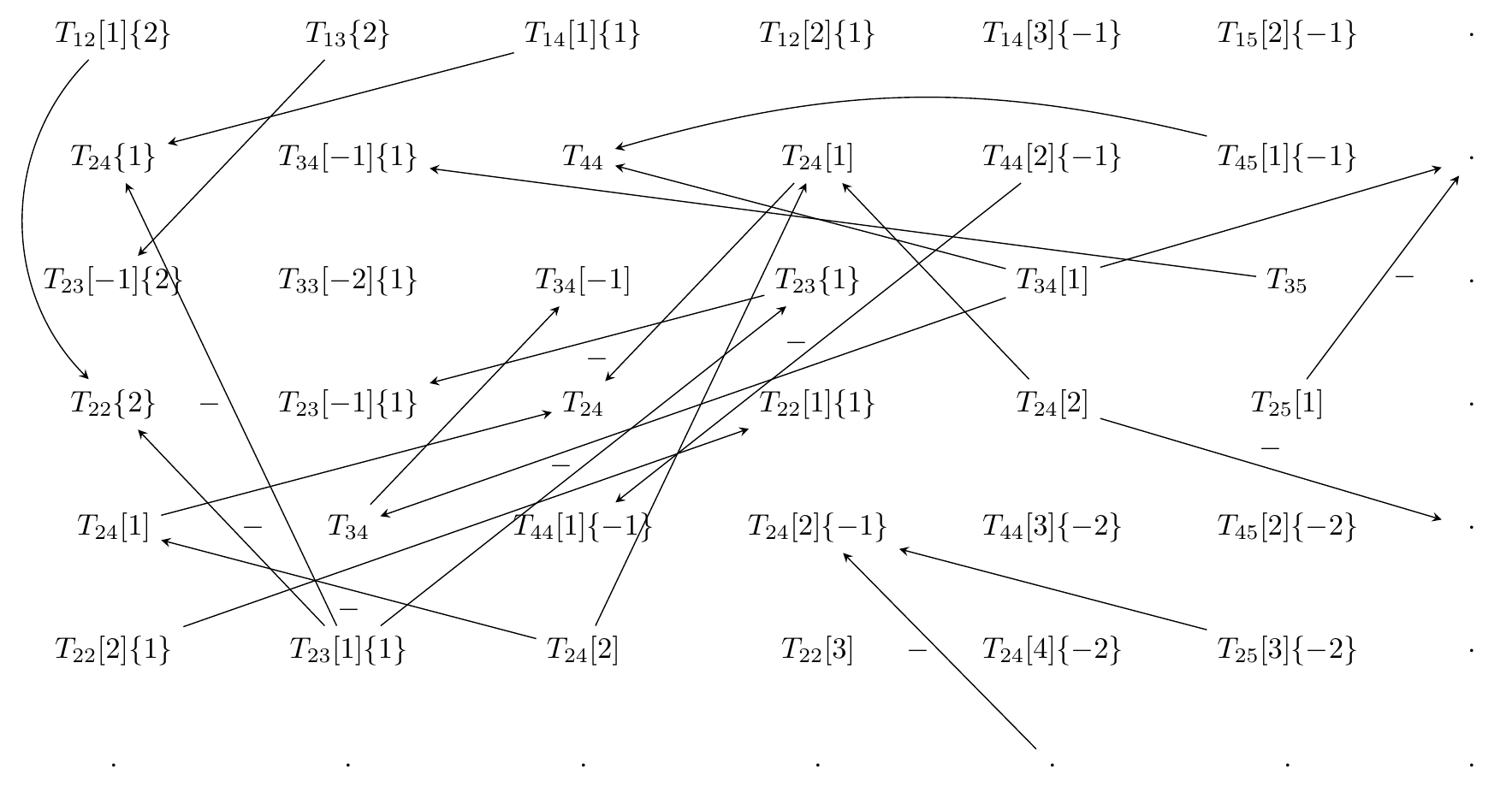}
\end{equation*}

To calculate the ${\mathfrak{su}_2}$ link homology, we apply the $hom^{*}_Y(-, I_{24}\{J\})$ functor to the complex $({\mathscr B}E(T), \delta)$. The functor sends all components the direct sum brane $\mathscr{B}E(T)$ to zero except for the $8$ copies of $T_{24}$, for a suitable choice of $\{J\}$. The result is a complex whose differential comes from identity maps in $\delta$. It is given by:
\beq
\begin{matrix}
    \bC \\
    \bC\{1\}
\end{matrix}
\xrightarrow{  {\scriptsize u\hbar \cdot
\begin{pmatrix}
  1& 0 \\
    1 & 0
\end{pmatrix}}
}
\begin{matrix}
    \bC \\
    \bC
\end{matrix}
\xrightarrow{ {\scriptsize u\hbar \cdot
\begin{pmatrix}
    0 & 0 \\
  1 & -1 \\
 1 & -1
\end{pmatrix}}
}
\begin{matrix}
    \bC\{-1\} \\
    \bC \\
    \bC
\end{matrix}
\to
0
\to
\bC\{-2\}\label{hopfFe}
\eeq
with the rightmost term in homological degree 4.

\begin{figure}[h]
	\centering
    \includegraphics[scale=0.65]{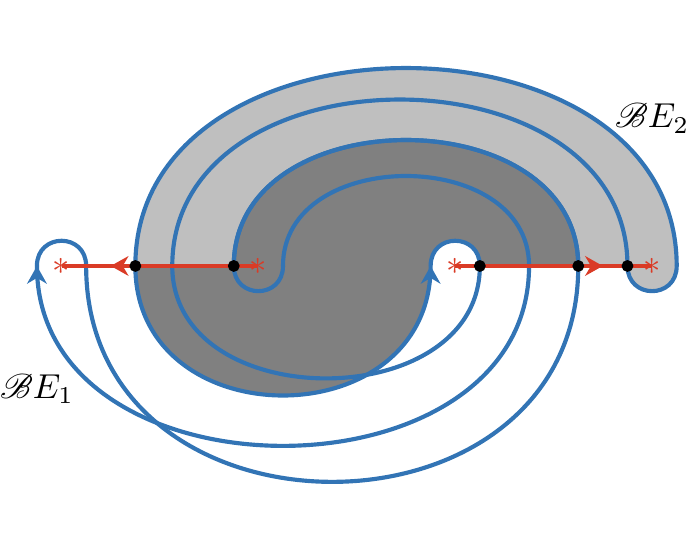} \qquad
    \includegraphics[scale=0.65]{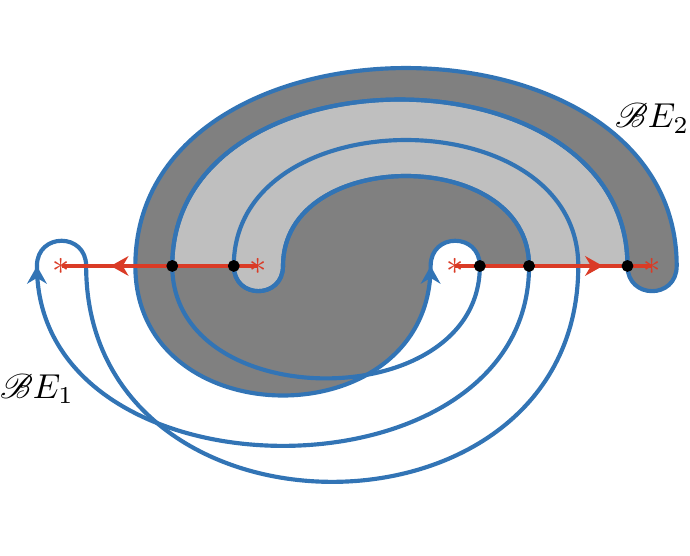}
	\caption{The shaded domains correspond to four terms in the differential $\delta_F$ whose pairwise product cancels in computing $\delta_F^2$.}
	\label{fig:sl2hopflinkDisks}
\end{figure}

The $8$ terms of the complex in \eqref{hopfFe} correspond, per construction of the direct sum brane ${\mathscr B}E(T)$, to the $8$ intersection points of the ${\mathscr B}E_{\cal U}$ brane with the $I_{24}$-brane in Fig.~\ref{fig:sl2hopflinkDisks}. The cohomology of the complex coincides with Floer cohomology, as we explained in section \ref{FR}. It follows that the differential in \eqref{hopfFe} is the Floer differential $\delta_F$, and that its terms count disk instantons. 

The differential $\delta_F$, obtained algebraically, squares to zero by construction. From the geometric perspective, it should square to zero because the cancelling contributions to $\delta_F^2$ correspond to two different ways a Maslov index $2$ map can degenerate into a pair of Maslov index $1$ maps (all with equivariant degree zero). The moduli space of Maslov index $2$ maps is a one real dimensional manifold, an interval, whose two boundary points correspond to the two ways the map can degenerate, and come with opposite orientations. Fig.~\ref{fig:sl2hopflinkDisks} gives an example.

The cohomology of the complex in \eqref{hopfFe} is
\beq
    \bC[4]\{-2\} \oplus\bC[2]\{-1\} \oplus \bC[2] \oplus \bC\{1\}.
\eeq
To get the absolute grading of link homology we need to include the additional, overall grading shifts from section \ref{Msu2}. They depend only on the braid, so can be read off from the link diagram in Fig.~\ref{fig:sl2hopflink}. The net shift is by $M \to M -4 $ and $J^0 \to J^0 + 2$ from $w=2$, $e=2$ and $d=2$,  giving
\beq\label{hfd}
    \bC \oplus\bC[-2]\{1\} \oplus \bC[-2]\{2\} \oplus \bC[-4]\{3\}.
\eeq
This agrees with the known Khovanov homology of the Hopf link after regrading $i=M+2J^0$, $j=2J^0$.

\subsubsection{}\label{trefoilsl2s}
\begin{figure}[h]
	\centering
	\includegraphics[scale=0.85]{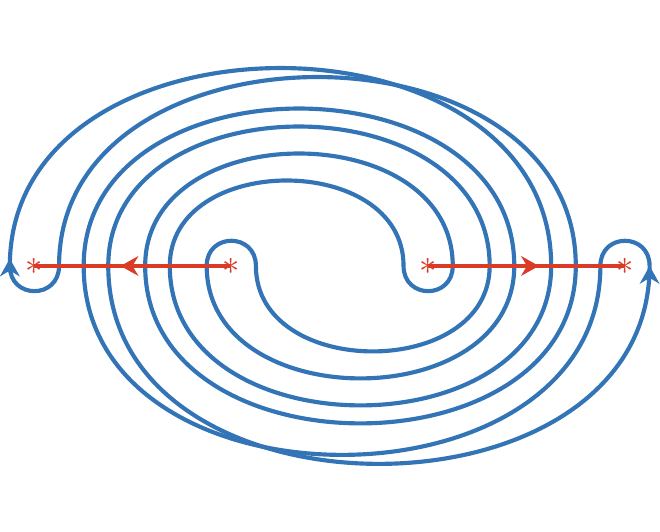}
	\caption{Lagrangians corresponding to left handed trefoil in ${\mathfrak{su}_2}$ trefoil theory.}\label{trefoilsl2}
\end{figure}

The same procedure outlined above can be used for any knot. Let us briefly sketch the analysis for the left-handed trefoil from Fig.~\ref{trefoil}. The Lagrangian corresponding to the unreduced trefoil is depicted in Fig.~\ref{trefoilsl2}.
Each cycle separately can be resolved in terms of a complex consisting of $8$ one-dimensional $T$-branes. The product complex is a toric grid of size $8\times 8$, whose vertices are the $64$ two-dimensional $T$-branes. The geometric differential $\delta_{0,0}$ obtained in this way squares to zero in the ${\mathscr A}_{u=0, \hbar =0}$ algebra.

The differential deforms to $\delta_{u,0}$, which squares to zero in the $u\neq 0$  ${\mathscr A}_{u=0, \hbar =0}$, by including all the $u$-corrections inherited from the product of one-dimensional complexes, provided they are allowed on degree grounds. In this case, each of the two one-dimensional complexes has four $u$-corrections. Out of possible $8\times 4 + 8\times 4 = 64$ $u$-corrections, $12$ end up having wrong equivariant degrees and get set to zero. The deformation is exact at first order in $u$, so $\delta_{u,0} = \delta_{0,0}+ u \delta_{1,0}.$

The differential $\delta=\delta_{u,\hbar}$, which squares to zero in the full algebra ${\mathscr A}_{u, \hbar}$ is the first order in $\hbar$ deformation of $\delta_{u,0}$. 
There are $366$ in total maps that can contribute to $\delta_{0,1}$ in $\delta= \delta_{u,0}+\hbar \delta_{0,1}$, which we include with some arbitrary coefficients $x_i$. The equation $\delta^2=0$ is a linear equation for $x$'s which has a unique solution, exact to all orders in both $u$ and $\hbar$. Note that, compared to the previous examples where none of the  maps in $\delta_{0,1}$ contained any crossings of strands, here they do.

Applying the  $hom^{*,*}_Y(-, I_{24})$ functor, the original $64$-dimensional complex becomes the $18$-dimensional complex:
\begin{center}
\begin{eqnarray}\nonumber
    \mathbb{C} 
    \xrightarrow{} 
    \emptyset 
    \xrightarrow{}
    \begin{matrix}\mathbb{C}\{-2\}\\ \mathbb{C}\{-2\}\\ \mathbb{C}\{-1\} \end{matrix}
    \xrightarrow{{\scriptsize u\hbar \cdot \begin{pmatrix}
    0 & 0 & 0\\
    0 & 0 & 0\\
    1 & 1 & 0\\
    -1 & -1 & 0\\
    \end{pmatrix}}} 
    \begin{matrix} \mathbb{C}\{-3\}\\ \mathbb{C}\{-3\}\\ \mathbb{C}\{-2\}\\ \mathbb{C}\{-2\} \end{matrix} 
    \xrightarrow{{\scriptsize u\hbar \cdot \begin{pmatrix}
   1 & -1 & 0 & 0\\
    -1 & 1 & 0 & 0\\
    -1 & -1 & 0 & 0\\
    0 & 0 & -1 & -1\\
    \end{pmatrix}}}\\
    \begin{matrix} \mathbb{C}\{-3\}\\ \mathbb{C}\{-3\}\\ \mathbb{C}\{-3\}\\ \mathbb{C}\{-2\} \end{matrix} \xrightarrow{{\scriptsize u\hbar \cdot \begin{pmatrix}
    0 & 0 & 0 & 0\\
    0 & 0 & 0 & 0\\
    -1 & -1 & 0 & 0\\
   1 & 1 & 0 & 0\\
    \end{pmatrix}}} 
    \begin{matrix}\mathbb{C}\{-4\}\\ \mathbb{C}\{-4\}\\ \mathbb{C}\{-3\}\\ \mathbb{C}\{-3\} \end{matrix}   \xrightarrow{{\scriptsize u\hbar \cdot  \begin{pmatrix}
    1 & 1 & 0 & 0\\
    0 & 0 & 1 & 1\\
    \end{pmatrix}}} 
    \begin{matrix} \mathbb{C}\{-4\} \\ \mathbb{C}\{-3\} \end{matrix} \nonumber
\end{eqnarray}
\end{center}
with the rightmost term in homological degree 6.

The cohomology of the complex as written is
\begin{equation}\label{trin}
\mathbb{C} \oplus \mathbb{C}[2]\{-1\} \oplus \mathbb{C}[2]\{-2\} \oplus \mathbb{C}[5]\{-4\},
\end{equation}
As a final step, to obtain 
the link homology groups $Hom^{*,*}_{\MDy}({\mathscr B}E_{\cal U}, I_{\cal U})$ we need to include in \eqref{trin} the non-geometric, overall shift of grading from section \ref{Msu2}. 
The net shift is by $M \rightarrow M+1$ and $J^0\rightarrow J^0-\frac{1}{2}$, because the braid has $w=-3$, $e=-3$ and $d=2$. With proper gradings included, the link homology is
\beq\label{overQ}
Hom^{*,*}_{\MDy}({\mathscr B}E_{\cal U}, I_{\cal U})=\textstyle\mathbb{C}[1]\left\{-\frac{1}{2}\right\} \oplus \mathbb{C}[3]\left\{-\frac{3}{2}\right\} \oplus \mathbb{C}[3]\left\{-\frac{5}{2}\right\} \oplus \mathbb{C}[6]\left\{-\frac{9}{2}\right\}.
\eeq
Taking cohomology with coefficients in $\mathbb{Z}$ instead of ${\mathbb C}$, we discover that it has $\mathbb{Z}_2$ torsion:
\beq\label{tZ}
\textstyle\mathbb{Z}[1]\left\{-\frac{1}{2}\right\} \oplus \mathbb{Z}[3]\left\{-\frac{3}{2}\right\} \oplus \mathbb{Z}[3]\left\{-\frac{5}{2}\right\} \oplus \mathbb{Z}_2[5]\left\{-\frac{7}{2}\right\} \oplus \mathbb{Z}[6]\left\{-\frac{9}{2}\right\},
\eeq
which comes from the quotient
\begin{equation*}
\mathbb{Z}_2 = \left \langle \begin{pmatrix}
1 \\ -1 \\ 0 \\ 0
\end{pmatrix},
\begin{pmatrix}
0 \\ 0 \\ 1 \\ 0
\end{pmatrix}\right \rangle /
\left \langle 
\begin{pmatrix}
1 \\ -1 \\ -1 \\ 0
\end{pmatrix},
\begin{pmatrix}
-1 \\ 1 \\ -1 \\ 0
\end{pmatrix}
\right \rangle.
\end{equation*}

The cohomologies in \eqref{overQ} and \eqref{tZ} agree with the Khovanov homology of the left-handed trefoil, when re-written in terms of the $(i,j)$ gradings of \cite{Kh}, which are related to $M$ and $J^0$ by $i=M+2J^0$ and $j=2J^0$. While the cohomologies are the same, the complexes are not. The complex defined by Khovanov in \cite{Kh} (see also \cite{DBN}), is $30$-dimensional in this case, as opposed to ours which is $18$-dimensional.
\subsubsection{}
One can often vastly simplify the ${\mathscr B}E_{\cal U}\in \MDy$ branes by using the swiping move to map the brane to an equivalent but simpler object of ${\MDy}$.  The swiping move plays the key role in proving the theory produces homological link invariants. It is described in section \ref{Msu2} in Fig.~\ref{fig:su2_markov}. Simplicity is measured by the number of $T$-branes needed to describe the brane, or rather, its resolution.  

Having simplified the brane, we get a formulation of $Hom^{*,*}_{\MDy}({\mathscr B}E_{\cal U}, I_{\cal U})$ as the cohomology of a much simpler complex. 
The move should have a version any for Lie algebra; in the ${\mathfrak{gl}_{1|1}}$ case, it was described in section \ref{Mgl11}. 

\begin{figure}[H]
    \centering
	\includegraphics[scale=0.75]{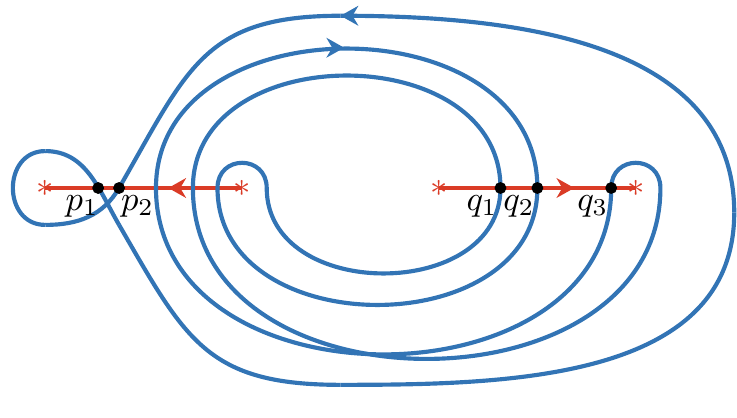}
    \caption{A simplification, using the swiping move, of the branes associated with left-handed trefoil}
	\label{fig:swallowed trefoil}
\end{figure}
\noindent{}For example, to apply it for the branes associated with the left-handed trefoil, start with an unbraided $E_{\cal U}$ brane, which is a product of two simple figure eights. Using the move, one of the two branes can ``swallow" the other, as in the proof of Thm.~1 in section \ref{Msu2}. After that, we braid the brane to get the ${\mathscr B} E_{\cal U}$ brane. For the left handed trefoil, the result is shown in Fig. \ref{fig:swallowed trefoil}. The brane has a resolution consisting of only $32$ thimbles, instead of $64$ we started with:

\begin{equation*}
    T_{23}\{-4\}
    \xrightarrow{d_6}
    \begin{matrix}
        T_{13}\{-3\} \\
        T_{35}\{-5\} \\
        T_{24}\{-4\} \\
        T_{22}\{-3\}
    \end{matrix}
    \xrightarrow{d_5}
    \begin{matrix}
        T_{14}\{-3\} \\
        T_{12}\{-2\} \\
        T_{23}\{-3\} \\
        T_{45}\{-5\} \\
        T_{25}\{-4\} \\
        T_{22}\{-2\} \\
        T_{24}\{-3\}
    \end{matrix}
    \xrightarrow{d_4}
    \begin{matrix}
        T_{12}\{-1\} \\
        T_{14}\{-2\} \\
        T_{24}\{-3\} \\
        T_{22}\{-2\} \\
        T_{25}\{-3\} \\
        T_{45}\{-4\} \\
        T_{22}\{-1\} \\
        T_{25}\{-2\}
    \end{matrix}
    \xrightarrow{d_3}
    \begin{matrix}
        T_{12} \\
        T_{15}\{-1\} \\
        T_{22}\{-1\} \\
        T_{24}\{-2\} \\
        T_{25}\{-2\} \\
        T_{55}\{-3\} \\
        T_{24}\{-1\}
    \end{matrix}
    \xrightarrow{d_2}
    \begin{matrix}
        T_{14} \\
        T_{22} \\
        T_{25}\{-1\} \\
        T_{45}\{-2\}
    \end{matrix}
    \xrightarrow{d_{1}}
    T_{24}
\end{equation*}

To find the differential, we proceed as in the case without swallowing. We start by writing down the differential $\delta_{0,0}$ that squares to zero in $A_{u=0,\hbar=0}$. The price to pay for the simplification of the brane gained by the swallowing move, is that the differential $\delta_{u, \hbar}$ that squares to zero in the full algebra, is a  deformation not of the naive geometric differential $\delta_{0,0}$, but of its gauge transform $\delta_{0,0}\rightarrow e^b \delta_{0,0} e^{-b}$ where $b$ depends on $u\hbar$. Proceeding order by order in $\hbar$ and $u$, we find the final differential $\delta$, as a deformation of $\delta_{0,0}$ by four new terms. 

After applying the $hom^{*,*}_Y(-, I_{24})$ functor, the result is a complex which is just $6$ dimensional, spanned by the intersection points $p_i q_j$ in Fig.~\ref{fig:swallowed trefoil}:
\beq\label{tmani}
    \mathbb{C}
    \xrightarrow{}
    0
    \xrightarrow{}
    \begin{matrix}
        \mathbb{C} \{-2\} \\
        \mathbb{C} \{-1\}
    \end{matrix}
    \xrightarrow{0}
    \mathbb{C}\{-3\}
    \xrightarrow{-2 (u \hbar)^2}
    \mathbb{C}\{-3\}
    \xrightarrow{0}
    \mathbb{C}\{-4\}
\eeq
where the rightmost term is in homological degree 5. After introducing the same degree shifts as in section \ref{trefoilsl2s}, the cohomology of the complex reproduces the homology trefoil homology
\eqref{overQ} over ${\mathbb C}$, and  after shifting to ${\mathbb Z}$ coefficients, the homology with torsion in \eqref{tZ}. The simpler complex in \eqref{tmani} also makes it manifest where the torsion comes from: it is due to a holomorphic disk instanton that connects a pair of intersection points which has multiplicity $-2$. It generates the coefficient of $p_1 q_2$ in $\delta_F p_2 q_1$. The fact the homology is unchanged is a consequence of the fact, proven in appendix \ref{Markovsu2}, that the swiping move leaves the brane invariant as an object of $\MDy$, and hence all its morphisms as well.

\subsubsection{}
Reduced link homology for all two-bridge knots involves a one-dimensional $Y$ and does not require $\hbar$ deformations, so its computation may be done by hand. Our current computer code, implemented in Mathematica and specialized to the two-strand algebra, allows us to compute the reduced homology of 3-bridge links and the full homology (including the $\mathbb{Z}_2$ torsion) of 2-bridge links. Generalizing it to multiple strands and arbitrary links is straightforward, in principle.

We used the code to compute the unreduced homology for all knots up to 6 crossings, and for the first 6 knots with 7 crossings, as well as the reduced homology of several three-bridge knots including $8_{5}$, $8_{19}$ and $9_{42}$. In all the examples we checked, solving the system of linear equations coming from the first order in $\hbar$ and setting all the remaining coefficients to zero sufficed to find the full differential $\delta$.  Furthermore, for two-bridge knots we checked, $\delta$ can be made to square to zero at first-order in $\hbar$ and $u$ which further speeds up the algorithm.

Our homology theory agrees with Khovanov's \cite{Kh}, including torsion, in all cases we checked. More precisely, the theories agree after a regrading: the cohomological index $i$ and the Jones grading $j$ of Khovanov homology are related to ours by $i=M+ 2J^0$ and $j=2J^0$.  The fact our homology theory coincides with Khovanov's to be expected by homological mirror symmetry proven in \cite{ADZ} and abstract theorems of \cite{webster, W1, W2}.

\subsection{Homological $U_{{\fq}}(\mathfrak{su}_2)$ link invariants from ${\mathscr D}_{Y_{\mathfrak{su}_2}}$}\label{Msu2}
In this section we will prove Theorems 1 and 2 for $\mathfrak{su}_2$. The proof is analogous to the $\mathfrak{gl}_{1|1}$ case. 

\begin{figure}[H]
    \centering
	\includegraphics[scale=0.24]{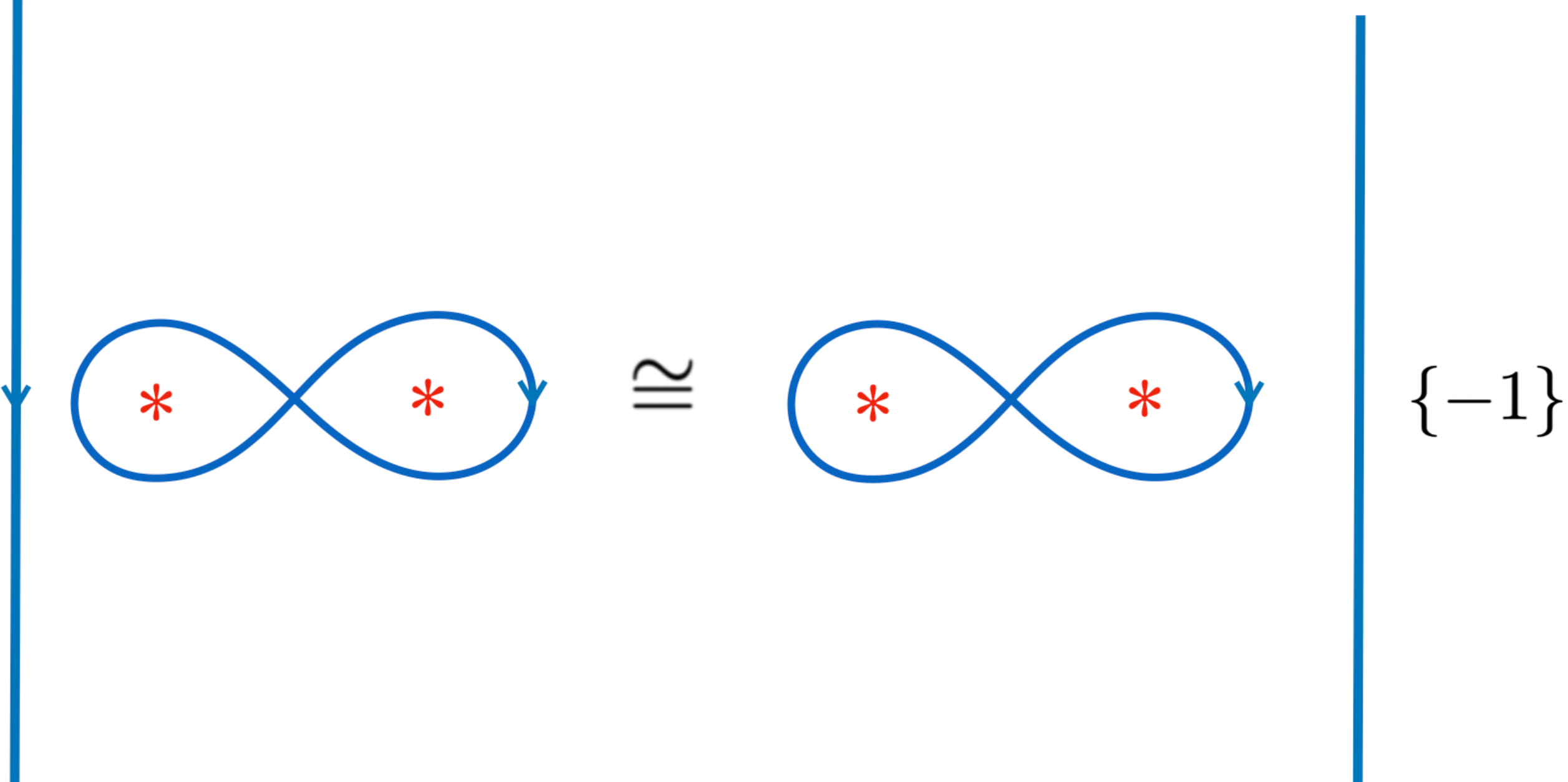}
    \caption{These two brane are equivalent objects of $\mathscr{D}_Y$}	\label{fig:su2_markov}
\end{figure}

The key ingredient is the swiping move, shown for $\mathfrak{su}_2$ in Fig. \ref{fig:su2_markov}. Just as in case of $\mathfrak{gl}_{1|1}$, the move holds for any $u \hbar\neq 0$, so for it to hold and for the theory to give homological link invariants, we need to work on $Y$ without divisors deleted. We prove this move for $\mathfrak{su}_2$ in appendix \ref{Markovsu2}.
\noindent{}

The swiping move holding in $
\MDy$ suffices to ensure that all Markov moves from Figs.~\ref{MI}, \ref{MII} and \ref{stab} hold, up to overall degree shifts. To show that the moves hold exactly, we need to keep track of the gradings of the ${\mathscr B}E_{\cal U}$ brane through the sequence of braid moves. 

The principle for keeping track of degree shifts due to the geometric braid group action on an object of $\MDy$ is described in section \ref{keepingtrack}; it is independent of the choice of a Lie algebra. The other degree shifts depend on the Lie algebra.

\subsubsection{}

As in the ${\mathfrak{gl}_{1|1}}$ case, the geometric action of ${\mathscr B}$ on the $E_{\cal U}$ brane gets combined with a constant degree shift functor. The functor acts on all objects of $\MDy$ in the same way, as in \eqref{const}.  
In the ${\mathfrak{su}_{2}}$ case, only the potential $W$, and not $\Omega$, contains terms that depend on the positions of the punctures. These terms are introduced by \eqref{dressedsu2}. Their contribution to $\Delta J_0$ is equal to $- {1\over 4} e $, where $e$ is the sum of the exponents of braid group generators $\sigma_i^{\pm}$ in the word representing the braid.
 
The one novel ingredient for ${\mathfrak{su}_2}$ case is that framing no longer acts trivially on link invariants in vertical framing. Vertical framing is the framing which the geometry picks, as discussed in section \ref{NormI}. As a consequence, the second relation in Fig.~\ref{stab} is really true only up to a degree shift that depends on the writhe $w$ of the braid. (The framing factor in general depends on the writhe of the link, but in our case, for links represented as plat closures of braids, this is the same as the writhe of the braid. The writhe is number of positive minus $\includegraphics[scale =0.2]{positive}$ minus the number of negative $\includegraphics[scale =0.2]{negative}$ crossings.) The Jones polynomial is an invariant of links in zero framing and satisfies the skein relation with no framing factor. To cancel the framing factor and get the theory whose Euler characteristic is the Jones polynomial, we will introduce an additional, $w$ dependent degree shift which shifts both the Maslov and equivariant degrees. The net constant shift, due to to sources combined, is thus:
\beq
\begin{aligned}\Delta M&= -w,\\
\Delta J_0 &= {3\over 4} w- {1\over 4} e 
\end{aligned}\label{su2shift}
\eeq

\subsubsection{}
The Markov I move from Fig.~\ref{MI} reads
\begin{figure}[H]
    \centering
	\includegraphics[scale=0.35]{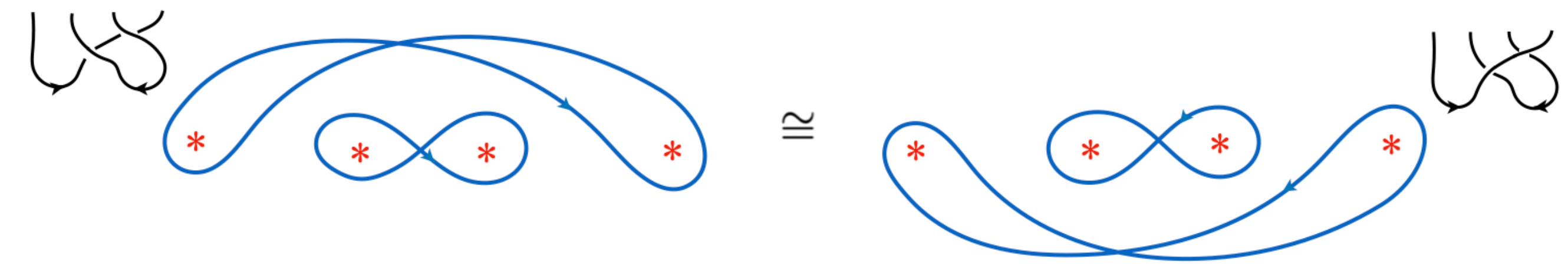}
    \caption{These two brane are equivalent objects of $\mathscr{D}_Y$}	\label{fig:su2_markovr}
\end{figure}
To show that it holds, simplify the two branes, using the swiping move, as follows:
\begin{figure}[H]
    \centering
	\includegraphics[scale=0.35]{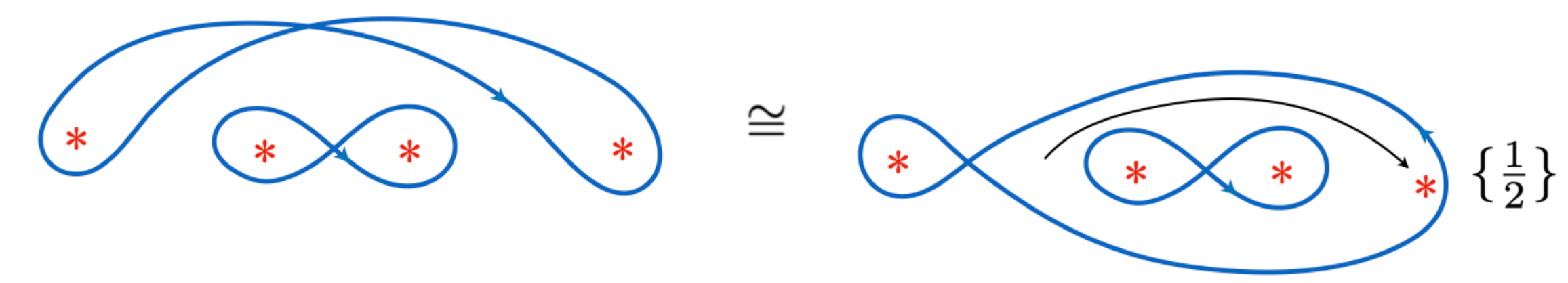}
\vskip -1cm
\end{figure}
and
\begin{figure}[H]
    \centering
	\includegraphics[scale=0.35]{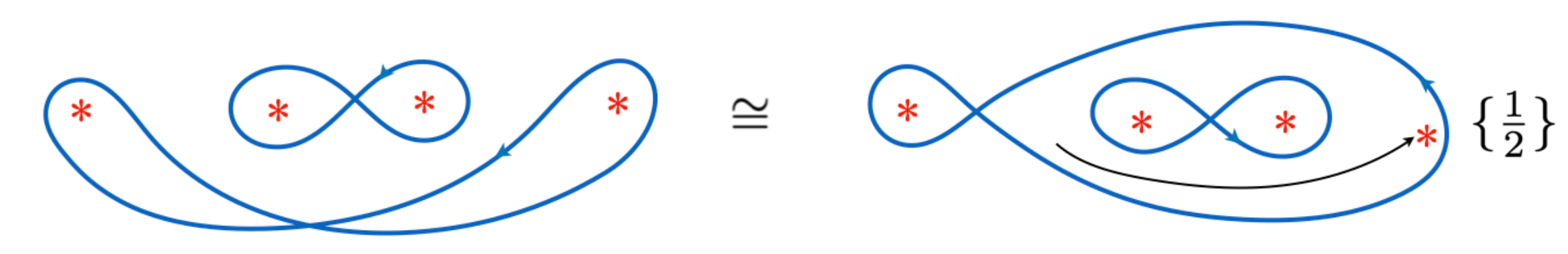}
\end{figure}
To compare the branes, we can ignore the outside figure eight, as it is the same for both. The two branes differ by the path the puncture took around the inner figure eight brane; in particular, they resulting branes are the same modulo shifts of gradings. Using resolutions of the two branes, it is easy to show that the top brane receives no geometric shift due to braiding the puncture, and in particular simply equals the brane on the right hand side, forgetting the path. The bottom brane, by contrast, gets a geometric shift of $\{1\}$. (The particular shifts depend on a  placement of cuts used to define a single-valued branch of ${W_0}$. This is implicit throughout the theory, in particular in the definition of $T$-branes and their algebra.) The final degree of brane has to be supplemented with contributions of the constant degree shift functors in \eqref{const} which are, in the ${\mathfrak{su}_2}$ case, given in \eqref{su2shift}, and which depend only on the braid and not on the brane. The path at the top has $w=0$ and $e=-2$, the path at the bottom has $w=0$ and $e=2$, resulting in shifts of $\{{1\over 2}\}$ and $\{-{1\over 2}\}$. These shifts combine with the geometric ones to ensure the two branes are equivalent objects of $\MDy$. The move in Fig.~\ref{MI} follows.

It is not difficult to show that with degree shifts in 
\eqref{su2shift}, both of the moves in Fig.~\ref{stab} hold as well. For example,  braid group generator $\sigma$ that generates the twist in second move Fig.~\ref{stab} has a geometric action on the $E_{\cal U}$ brane, giving a degree shift of $[-1]\{1\}$. This is canceled by the constant shift in \eqref{su2shift}, since the element has $w=-1$ and $e=1$. Similarly, the constant shift functor of the braid in the first move, which has $w=0$ and $e=4$,  cancels the geometric degree shift of the brane, which equals $\{1\}$. This proves both moves in Fig.~\ref{stab}.

Next, consider the stabilization move in Fig.~\ref{MII}. As in the ${\mathfrak{gl}_{1|1}}$ theory case, asking for the stabilization move to hold as stated  should determine the 
grading of the $I_{\cal U}$ brane, the cap, relative to the basic $I$-branes, which are defined to be dual to $T$-branes via \eqref{dual}. The stabilization move reads:
\begin{figure}[H]
    \centering{
	\includegraphics[scale=0.23]{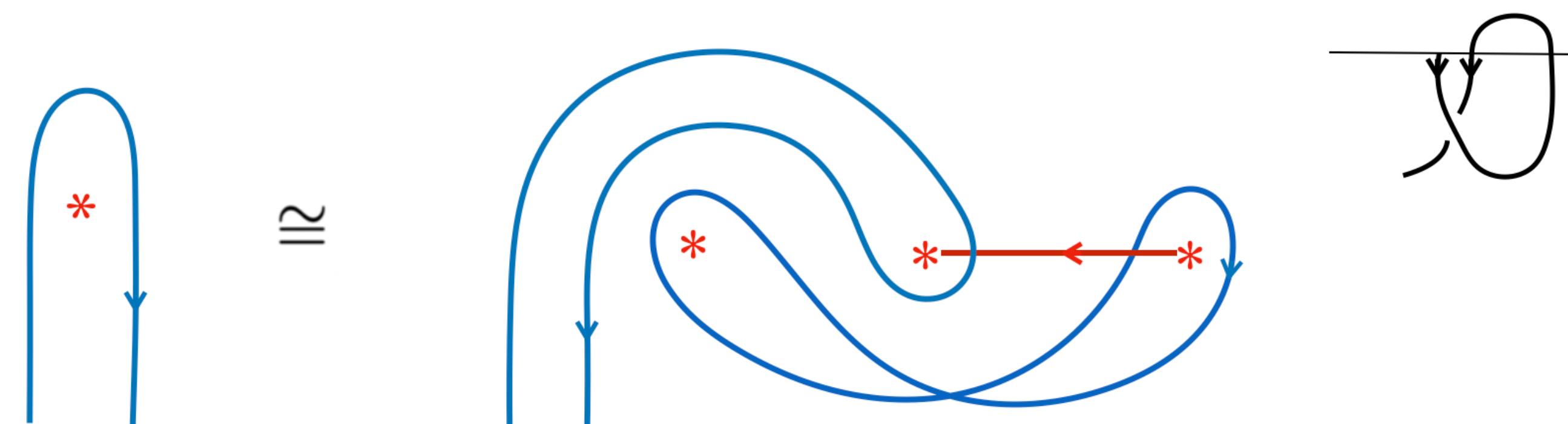}}
\end{figure}
\noindent{}in terms of branes. By starting with the trivial braid, applying the sliding move, and then braiding, the brane on the right is equivalent to, 
\begin{figure}[H]
    \centering
	\includegraphics[scale=0.3]{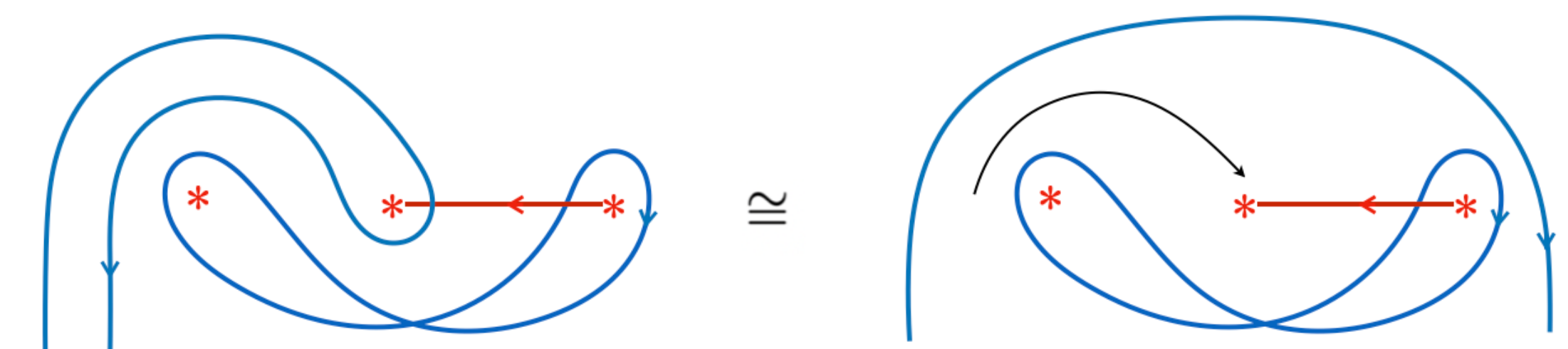}
\end{figure}
The single intersection point of the figure eight brane with the basic $I$-brane has the geometric degree $0$, both equivariant and cohomological. The constant shift functor that accompanies the action of ${\mathscr B}$ is a degree shift of $\{-{1\over 2}\}[1]$, since the braid has $w=-1$ and $e=-1$. Asking for the stabilization move to hold as stated, with no accompanying degree shifts, determines the cap brane $I_{{\cal U}}$ to be equal to the basic $I$-brane, shifted in degree by  $\{-{d\over 2}\}[d]$, where $d$ is the dimension $d$ of $Y$. 

As an aside, the $I_{\cal U}$ brane is invariant under the second move in Fig.~\ref{stab}, just like the corresponding $E_{\cal U}$-brane, and for the same reason: the geometric degree shift, which is the same for both branes, gets canceled by the constant one from \eqref{su2shift}. The overall degree shift of $\{-{d\over 2}\}[d]$, goes along for the ride.

\subsubsection{}\label{t1su2}
To complete the proof of Thm.~\ref{tinv} for ${\mathfrak{su}_2}$ we need to show the remaining Birman moves, obtained from those in Figs.~\ref{MI} and \ref{stab} by exchanging tops and bottoms, also hold. The proof is analogous to that for ${\mathfrak{gl}_{1|1}}$ in section \ref{t1gl11}. 

Since the $E_{{\cal U}}$ branes are invariant under moves in Figs.~\ref{MI} and \ref{stab}, the homology groups $Hom_{{\MDy}}^{*,*}({\mathscr B}E_{\cal U}, E_{{\cal U}})$ are invariant under those moves applied to the second entry in the Hom, as well as the first.  Similarly to  section \ref{t1gl11}, the homology groups $Hom_{{\MDy}}^{*,*}({\mathscr B}E_{\cal U}, E_{{\cal U}})$ are related to $Hom_{{\MDy}}^{*,*}({\mathscr B}E_{\cal U}, I_{{\cal U}})$ by \eqref{EI}, with $W$ which is a fixed graded vector space of rank $2^d$. This comes about because each one dimensional figure eight brane is equivalent to complex $E_i \cong (E_i(I), \delta_i)$  where $E_i(I) = I_i \oplus I_i[1]\{-1\}$. It follows that $Hom_{{\MDy}}^{*,*}({\mathscr B}E_{\cal U}, I_{{\cal U}})$ is invariant under the moves in Figs.~\ref{MI} and \ref{stab} applied to the second entry rather than the first. This proves the invariance under the remaining moves, and hence proves Thm.~\ref{tinv}.

\subsection{$U_{\fq}({\mathfrak{su}_2})$ skein from branes}\label{su2 skein}
Finally, we prove Thm.~\ref{tJ} which says that the Euler characteristics of $Hom^{*,*}_{\MDy}({\mathscr B}E_{\cal U}, I_{\cal U})$ satisfy the ${\mathfrak{su}_2}$ skein relation in eqn.~(1.1), with $n=2$. The strategy of the proof is the same as for ${\mathfrak{gl}_{1|1}}$: Since $Hom^{*,*}_{\MDy}({\mathscr B}E_{\cal U}, I_{\cal U})$ are link invariants, to show their Euler characteristics satisfies the skein relation, it suffices to consider the special link presentations in  Fig.~\ref{Skeinb}. Proving the theorem amounts to showing K-theory classes of the three branes corresponding to the bottom of that figure satisfy the relation. 

\begin{figure}[H]
    \centering
	\includegraphics[scale=0.37]{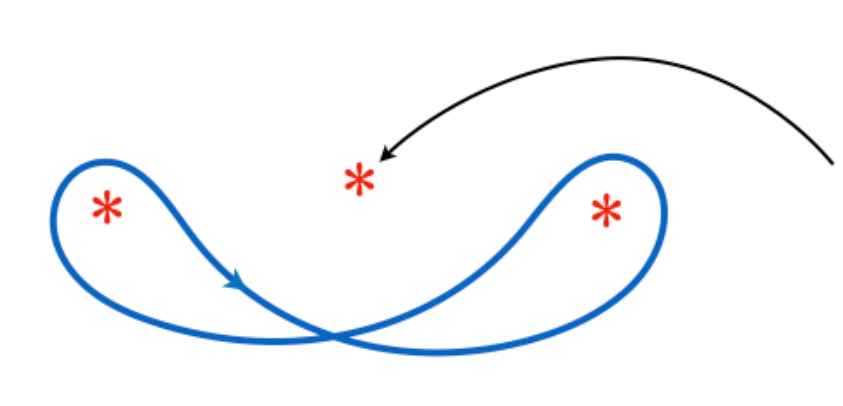}
   \caption{The reduced brane corresponding to positive \includegraphics[scale=0.2]{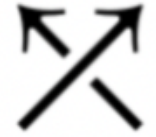} crossing.}\label{su2pos}
\end{figure}

\begin{figure}[H]
    \centering
	\includegraphics[scale=0.37]{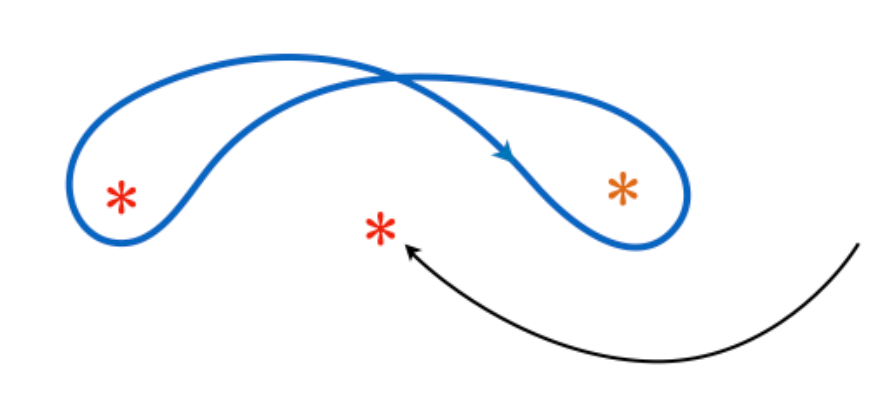}
   \caption{The reduced brane corresponding to the negative \includegraphics[scale=0.2]{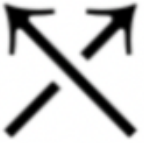}  crossing.}\label{su2neg}
  \vskip -0.5cm
\end{figure}

\begin{figure}[H]
    \centering
	\includegraphics[scale=0.37]{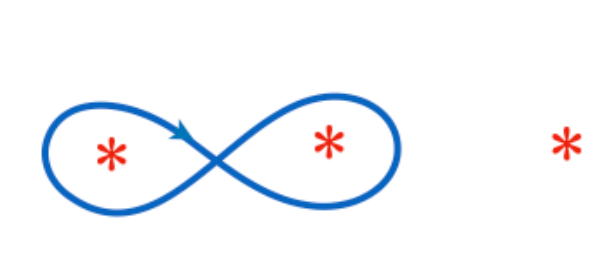}
 \caption{The reduced brane corresponding to \includegraphics[scale=0.25]{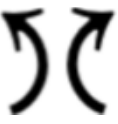} crossing.}\label{su2no}
  \vskip -0.5cm
\end{figure}

The three branes corresponding to the bottom of Fig.~\ref{Skeinb} are:

\begin{figure}[H]
    \centering
	\includegraphics[scale=0.47]{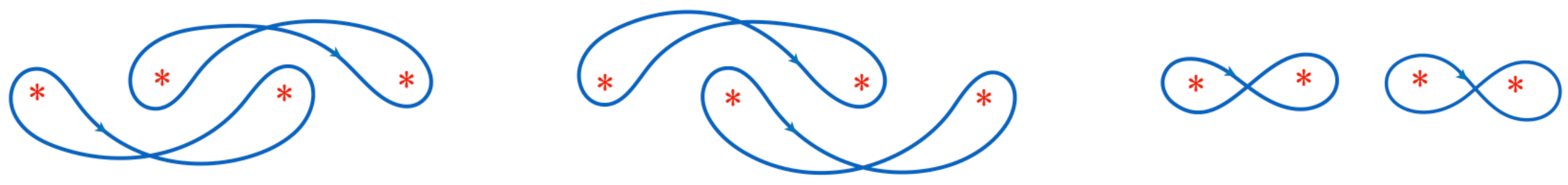}
	\caption{The three branes entering the Skein relation }
	\label{skeinsu2b}
\end{figure}
\noindent{}As in the ${\mathfrak{gl}_{1|1}}$ case, we can simplify the branes using the swiping move. The swiping move lets the right figure eight swallow the left, in each of the three branes, without changing the corresponding object of $\MDy$. Having done so, we learn that to compare K-theory classes of the three branes in Fig.~\ref{skeinsu2b}, it suffices to compare K-theory classes of three simpler branes in Figs.~\ref{su2pos}, \ref{su2neg}, and \ref{su2no}. 

The first of the three branes, the brane in Fig.~\ref{su2pos}, is obtained from the simple figure eight $E_{\cal U}$ brane by braiding the second and third punctures by $\sigma_2$. 
\noindent{}The brane has resolution:
\beq\nonumber{}
    T_2\{-1\} 
    \xrightarrow{\begin{pmatrix}
        \vcenter{\hbox{\includegraphics{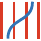}}} \\
        -\vcenter{\hbox{\includegraphics{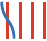}}}
    \end{pmatrix}}
    \begin{matrix}
        T_4\{-1\} \\
        T_1
    \end{matrix}
    \xrightarrow{\begin{pmatrix}
        \vcenter{\hbox{\includegraphics{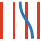}}} &
        \vcenter{\hbox{\includegraphics{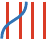}}}
    \end{pmatrix}}
    T_3
\eeq
The overall degree of the brane, before any non-geometric shifts, follows from the fact that braiding acted trivially on the $T_1$ brane, hence its degree remained fixed. The degrees of the other branes in the complex are then fixed by the degrees of the maps in it. K-theory class of the brane is
$(-{\fq}^{1/2})([T_3]-[T_1]-{\fq}^{-1}[T_4]+{\fq}^{-1}[T_2])$, where the prefactor comes from the constant degree shift of the braid that has a single positive crossing, with $e=1$.

The second brane, from Fig.~\ref{su2neg}, has 
resolution:
\beq\nonumber{}
    T_3\{-2\} 
    \xrightarrow{\begin{pmatrix}
        \vcenter{\hbox{\includegraphics{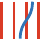}}} \\
        -\vcenter{\hbox{\includegraphics{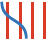}}}
    \end{pmatrix}}
    \begin{matrix}
        T_4\{-2\} \\
        T_1
    \end{matrix}
    \xrightarrow{\begin{pmatrix}
        \vcenter{\hbox{\includegraphics{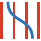}}} &
        \vcenter{\hbox{\includegraphics{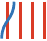}}}
    \end{pmatrix}}
    T_2
\eeq
\noindent{}It is obtained from the $E_{\cal U}$ brane by braiding corresponding to 
$\sigma_2^{-1}$. The braid has has $w=-1$ and $\sigma=-1$,
\noindent{}so K-theory class of the brane equals $(-{\fq}^{-1/2})([T_2]-[T_1]-{\fq}^{-2}[T_4]+{\fq}^{-2}[T_3])$. 

The third brane, from Fig.~\ref{su2no}, is the $E_{\cal U}$ brane itself, with no braiding applied.  
It has resolution:
\beq\nonumber{}
    T_2\{-1\} 
    \xrightarrow{\begin{pmatrix}
        \vcenter{\hbox{\includegraphics{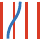}}} \\
        -\vcenter{\hbox{\includegraphics{pdf_figures/su2_b12.pdf}}}
    \end{pmatrix}}
    \begin{matrix}
        T_3\{-1\} \\
        T_1
    \end{matrix}
    \xrightarrow{\begin{pmatrix}
        \vcenter{\hbox{\includegraphics{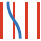}}} &
        \vcenter{\hbox{\includegraphics{pdf_figures/su2_a21.pdf}}}
    \end{pmatrix}}
    T_2
\eeq
and $K$-theory class $(1+{\fq}^{-1})[T_2]-[T_1]-{\fq}^{-1}[T_3]$. 
\noindent{}
K-theory classes of the three branes satisfy one linear relation:
\begin{equation}
    \centering
	\includegraphics[scale=0.47]{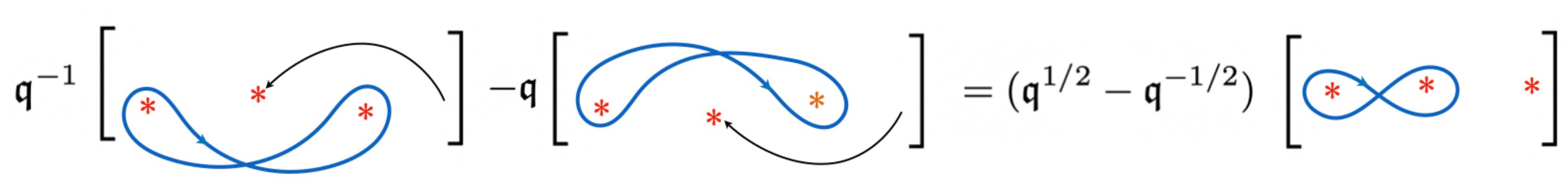}\label{su2fs}
\end{equation}
\noindent{}
The relation \eqref{su2fs} is the skein relation satisfied by the $R$ matrices of $U_{\fq}({\mathfrak{su}_2})$, see e.g.~\cite{oh}.  It coincides with the $n=2$ skein relation in eqn.~(1.1) after setting
\beq\label{sured}{\bf{q}}^{-1/2} = - {\fq}^{1/2}.
\eeq
This proves Thm.~\ref{tJ} for ${\mathfrak{su}_2}$.

\appendix

\section{The swiping move}\label{SM}
The swiping move plays the key role in proving the homology groups are invariants of links. In this appendix we prove that the branes related by the swiping moves are equivalent objects of $\MDy$ by proving their resolutions in terms of $T$-branes are homotopy equivalent. The swiping move is shown in Fig.~\ref{Mgl} for $\mathfrak{gl}_{1|1}$ and in Fig.~\ref{fig:su2_markov} for  $\mathfrak{su}_2$. 

In the derived Fuykaya category a brane $L\in \MDy$ is equivalent to the complex $(L(T), \delta)$ resolving it. Two complexes $(L_1(T), \delta_1)$ and $(L_2(T), \delta_2)$ describe the resolution of the same brane, and are hence equivalent, if they are homotopy equivalent. This means there exist chain maps $f:L_1 \to L_2$ and $f':L_2 \to L_1$ such that $f' \circ f$ and $f \circ f'$ are homotopic to the identity on $(L_1(T), \delta_1)$ and $(L_2(T), \delta_2)$, respectively. The idea is that, if such maps $f$ and $f'$ exist, they are represent the identity maps from $L$ to itself, in terms of the two resolutions of $L$,  see e.g. \cite{Urs, Horie}. Once we specialize to $\mathfrak{su}_2$ and $\mathfrak{gl}_{1|1}$, respectively, we will explain in more detail how the homotopies are defined, in the case of ordinary and twisted complexes. Ordinary complexes are special cases of the twisted ones.

\subsection{The move for $\mathfrak{gl}_{1|1}$}\label{Markovgl}
The swiping move for $\mathfrak{gl}_{1|1}$ is the equivalence of the branes $T_1 \times E_1$ and  $E_1 \times T_3[-1]$ on the left and the right hand sides of Fig.~\ref{Mgl}, repeated here for convenience:
\begin{figure}[H]
    \centering
	\includegraphics[scale=0.3]{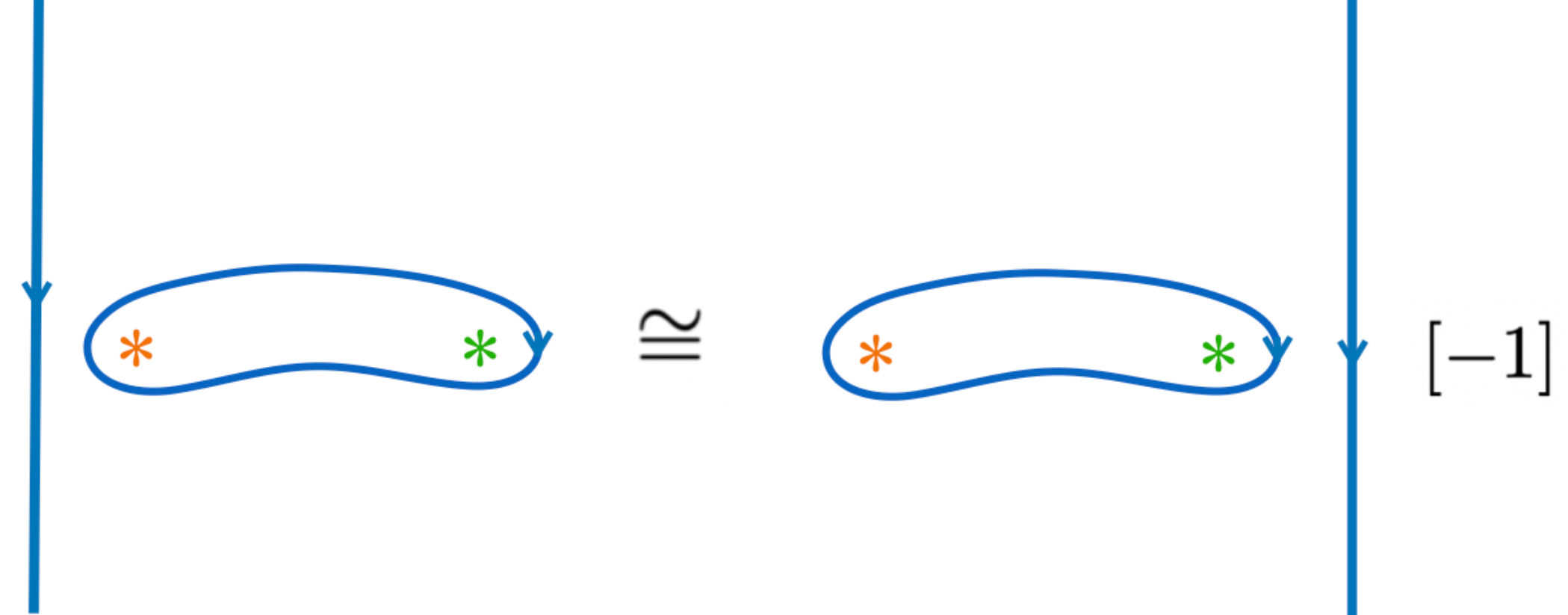}
\end{figure}

\subsubsection{}\label{recall}
Given a pair of branes in $L, \, L'\in \MDy$ and the twisted complexes of $T$-branes resolving them, $L\cong (L(T), \delta)$ and $L'\cong(L'(T), \delta')$, a morphism  
$${\cal P}  \in Hom_{\MDy}(L, L'),$$ 
can be represented as a map of the corresponding complexes, which satisfies \cite{Seidel}:
\begin{equation}\label{tchain}
	0 =m_1^{{\rm Tw}}({\cal P}) = m_1^p({\cal P}) + m_2^p(\delta',{\cal P}) + m_2^p({\cal P},\delta).
\end{equation}
A morphism ${\cal P}$ satisfying \eqref{tchain} plays a role of a chain map for ordinary complexes.
Twisted $A_{\infty}$ products  are sign twists of the ordinary ones, as we recall in appendix \ref{A}. In writing \eqref{tchain}, we take the higher $A_{\infty}$ maps $m_{\ell>2}^p$  to vanish, as they do in our theory, for otherwise they would contribute to the definition of $m_1^{{\rm Tw}}({\cal P})$. 
 
Next, a given pair of morphisms are equivalent in $\MDy$
$${\cal P} \cong {\cal P}' \in Hom_{\MDy}(L, L')$$ 
if they differ by an exact morphism $m^{Tw}_1({\cal Q}) $:
\begin{equation}\label{thom}
	{\cal P} = {\cal P}' + m^{Tw}_1({\cal Q}) 
\end{equation}
where ${\cal Q} \in Hom_{\MDy}(L, L'[-1])$ is analogous to a chain homotopy of ordinary complexes.

\subsubsection{}
We will show the two branes $L_1= E_1 \times T_3$ and $L_2 = T_1 \times E_1[1]$
are equivalent as objects of $\MDy$, by showing that there is a map $f\in Hom_{\MDy}(L_1, L_2)$ with an inverse $f'\in Hom_{\MDy}(L_2, L_1)$
which satisfy
\beq\label{hom}
f'\circ f \cong id_{L_1}, \qquad f\circ f' \cong id_{L_2}.
\eeq
where the equivalence is a homotopy in the sense of \eqref{thom}.

\subsubsection{}
The resolutions of the two branes are:
\beq\label{ETc}
	L_1 \qquad \cong \qquad
	\begin{tikzcd}
		T_{13}[1] \arrow[r, shift left, "\delta'_{01}"] &
		T_{33} \arrow[l, shift left, "\delta'_{10}"]
	\end{tikzcd}
\eeq
where $\delta'_{01} = \vcenter{\hbox{\includegraphics{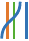}}}$ and  $\delta'_{10} = \vcenter{\hbox{\includegraphics{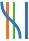}}}$, and:
\beq\label{TEc}
	L_2 \qquad \cong \qquad
	\begin{tikzcd}
		T_{11}[2] \arrow[r, shift left, "\delta_{01}"] &
		T_{13}[1] \arrow[l, shift left, "\delta_{10}"]
	\end{tikzcd}
\eeq
where $\delta_{01} = \vcenter{\hbox{\includegraphics{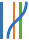}}}$ and $\delta_{10} = \vcenter{\hbox{\includegraphics{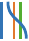}}}$. The signs in the differentials come from the factor of $(-1)^{M(T_\mathcal{C})}$ in taking the product of $T_i$ with $E_1$ combining with the factor of $(-1)^{(n-1)k_0}$ from the individual oval branes. As explained in section \ref{simplest}, we have the freedom to twist the differential by a sign in one of the maps, corresponding to a different choice of spin structure on the brane we started with, or more generally, by a complex parameter which corresponds to choosing a generic holonomy on the brane of $S^1$ topology which the complex resolves. The discussion that follows readily generalizes, as is simple to verify.

One can easily find the morphisms $f$ and $f'$ we need.
The morphism $f\in Hom_{\MDy}(L_1, L_2)$, going from 
the resolution of $L_1$ to the resolution of $L_2$,
 is given by following diagram:
\begin{center}
	\begin{tikzcd}[row sep = large]
		& T_{13}[1] \arrow[r, shift left, "\delta'_{01}"] \arrow[d, "f_{01}"] \arrow[dl, "f_{11}" above left] &
		T_{33} \arrow[l, shift left, "\delta'_{10}"] \arrow[dl, "f_{00}"] \\
		T_{11}[2] \arrow[r, shift left, "\delta_{01}"] &
		T_{13}[1] \arrow[l, shift left, "\delta_{10}"] &
	\end{tikzcd}
\end{center}
where $f_{11} = \vcenter{\hbox{\includegraphics{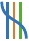}}}$,  $f_{01} = \hbar\; \vcenter{\hbox{\includegraphics{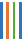}}}$ and $f_{00} = \vcenter{\hbox{\includegraphics{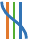}}}.$ The morphism $f'\in Hom_{\MDy}(L_2, L_1)$
is given by:
\begin{center}
	\begin{tikzcd}[row sep = large]
		T_{11}[2] \arrow[r, shift left, "\delta_{01}"] \arrow[dr, "f'_{11}" below left] &
		T_{13}[1] \arrow[l, shift left, "\delta_{10}"] \arrow[d, "f'_{10}"] \arrow[dr, "f'_{00}"] & \\
		& T_{13}[1] \arrow[r, shift left, "\delta'_{01}"] &
		T_{33} \arrow[l, shift left, "\delta'_{10}"] 
	\end{tikzcd}
\end{center}
where $f'_{11} = \vcenter{\hbox{\includegraphics{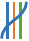}}}$,  $ f'_{10} = \hbar\; \vcenter{\hbox{\includegraphics{pdf_figures/gl11_markov_f_01}}}$ and $f'_{00} = \vcenter{\hbox{\includegraphics{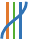}}}.$  One easily checks that
$$m^{Tw}_1(f) =0 =  m^{Tw}_1(f').$$

\subsubsection{}
To show that the compositions $ f  \circ f' $ and $ f' \circ f $ are each homotopic to the corresponding identity, 
let 
\beq\label{keyr} g = f \circ f' - \hbar^2 \cdot id_{L_2} ,  \qquad g' = f' \circ f - \hbar^2 \cdot  id_{L_1}.
\eeq
We will now show that $g$ and $g'$ are exact in the sense of \eqref{thom}. The power of $\hbar$ multiplying the identity elements in \eqref{keyr} is why the homomorphisms in 
\eqref{hom} only hold for $\hbar \neq 0$. For any non-zero $\hbar$, its value is in 
\eqref{keyr} immaterial: and we can set it to $1$ by appropriately rescaling $f, f'$ and $g, g'$. For $\hbar=0$, we cannot.
\subsubsection{}

The maps $g$ and $g'$ can be simplified, as shown:
\begin{center}
	\begin{tikzcd}[row sep = huge, column sep = large]
		T_{11}[2] \arrow[r, shift left, "\delta_{01}"] \arrow[d, "g_{11}" left] \arrow[dr, "g_{01}" xshift=2.5ex, labels=below right,  labels={yshift=-.5ex}] &
		T_{13}[1] \arrow[l, shift left, "\delta_{10}"] \arrow[dl, "g_{10}" xshift=2.5ex, labels=above right,  labels={yshift=.5ex}] \\
		T_{11}[2] \arrow[r, shift left, "\delta_{01}"] &
		T_{13}[1] \arrow[l, shift left, "\delta_{10}"] 
	\end{tikzcd}
\qquad \qquad
	\begin{tikzcd}[row sep = huge, column sep = large]
		T_{13}[1] \arrow[r, shift left, "\delta'_{01}"] \arrow[dr, "g'_{01}" xshift=-2.2ex, labels=above left, labels={yshift=.1ex}] &
		T_{33} \arrow[l, shift left, "\delta'_{10}"] \arrow[d, "g'_{00}"] \arrow[dl, "g'_{10}" xshift=-2.9ex, labels=below left, labels={yshift=-.5ex}] \\
		T_{13}[1] \arrow[r, shift left, "\delta'_{01}"] &
		T_{33} \arrow[l, shift left, "\delta'_{10}"] 
	\end{tikzcd}
\end{center}
with components
\begin{align}
	& g_{11} = -\hbar^2 \; \vcenter{\hbox{\includegraphics{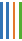}}}
	&& g_{10} = -\hbar \; \vcenter{\hbox{\includegraphics{pdf_figures/gl11_markov_f_11}}}
	&& g_{01} = -\hbar \; \vcenter{\hbox{\includegraphics{pdf_figures/gl11_markov_fp_11}}} \\
	& g'_{10} = -\hbar \; \vcenter{\hbox{\includegraphics{pdf_figures/gl11_markov_f_00}}}
	&& g'_{01} = -\hbar \; \vcenter{\hbox{\includegraphics{pdf_figures/gl11_markov_fp_00}}}
	&& g'_{00} = -\hbar^2 \; \vcenter{\hbox{\includegraphics{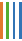}}}
\end{align}
To show that the maps $g$ and $g'$ are exact, we show that there exist $s$ and $s'$ such that
\begin{equation}\label{gg'exact}
g=  m_1^{\rm Tw}(s), \qquad 
g' = m_1^{{\rm Tw}}(s') ,
\end{equation}
Explicitly, the maps $s$ and $s'$ are given by
\begin{center}
	\begin{tikzcd}[row sep = large]
		T_{11}[2] \arrow[r, shift left, "\delta_{01}"] \arrow[d, "s_{11}"] &
		T_{13}[1] \arrow[l, shift left, "\delta_{10}"] \\
		(T_{11}[2] \arrow[r, shift left, "\delta_{01}"] &
		T_{13}[1] \arrow[l, shift left, "\delta_{10}"])[-1]
	\end{tikzcd}
	\qquad \qquad
	\begin{tikzcd}[row sep = large]
		T_{13}[1] \arrow[r, shift left, "\delta'_{01}"] &
		T_{33} \arrow[l, shift left, "\delta'_{10}"] \arrow[d, "s'_{00}"] \\
		(T_{13}[1] \arrow[r, shift left, "\delta'_{01}"] &
		T_{33} \arrow[l, shift left, "\delta'_{10}"])[-1]
	\end{tikzcd}	
\end{center}
where $s_{11} = \hbar \; \vcenter{\hbox{\includegraphics{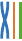}}}$ and  $s'_{00} = \hbar \; \vcenter{\hbox{\includegraphics{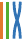}}}$. One can then verify \eqref{gg'exact} by explicit computation.

\vskip 1cm
\subsection{The move for $\mathfrak{su}_2$}\label{Markovsu2}
\begin{figure}[h]
    \centering
	\includegraphics[scale=0.23]{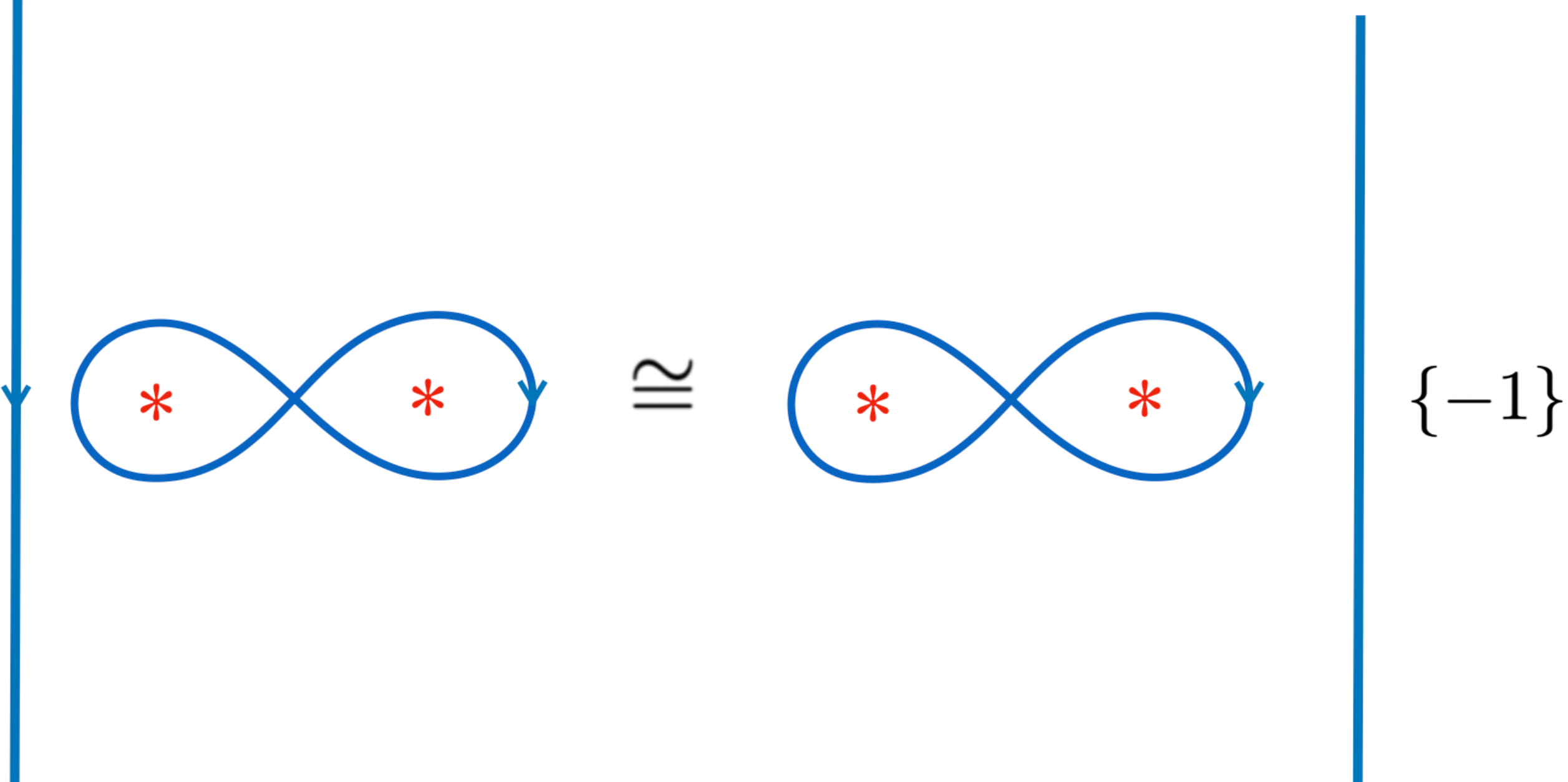}
    \caption{The ``swiping move" for $\mathfrak{su}_{2}$}
	\label{fig:su2_markov 2}
\end{figure}

The swiping move for $\mathfrak{su}_{2}$ is the equivalence of branes
$T_1 \times E_2$ and $E_2 \times T_3\{-1\}$ branes on the left and right hand sides of Fig.~\ref{fig:su2_markov}, repeated in Fig.~\ref{fig:su2_markov 2} for convenience, as objects of $\MDy$.

\subsubsection{}
The resolution of a brane $L\in \MDy$ in terms of $T$-branes is an ordinary complex $L\cong (L(T), \delta)$, as we have seen in section \ref{ssu2}, since the $T$-brane endomorphism algebra is concentrated in degree zero. Specializing the section \ref{recall} to this case, we get the familiar statements, as follows. 

Given a pair of branes in $L, \, L'\in \MDy$ and their projective resolutions $(L(T), \delta)$ and  $(L'(T), \delta')$, a morphism  
${\cal P}  \in hom_{Y}(L, L'),$ 
is represented by chain map of the resolving complexes. This is a map that satisfies
$$\delta'\cdot {\cal P} - {\cal P}\cdot \delta=0,
$$
where $\cdot$ coincides with $m_2$ product. 
Further, a pair of morphisms ${\cal P}, {\cal P}'  \in hom_{Y}(L, L'),$ are equivalent in $\MDy$
if the difference between them 
	$${\cal P} = {\cal P}' + \delta'\cdot {\cal Q} + {\cal Q}\cdot \delta$$ 
is a null homotopic chain map, for some ${\cal Q} \in hom_{Y}(L, L'[-1])$.

\subsubsection{}
The resolutions for the $L_1=T_1 \times E_2$ and $L_2=E_2 \times T_3\{-1\}$ branes on the left and right hand sides of Fig.~\ref{fig:su2_markov}, respectively are
\begin{center}
	$L_1\qquad \cong \qquad$
	\begin{tikzcd}
		T_{12}\{-1\}
		\arrow[r, "d_2"] &
		\begin{matrix}
			T_{13}\{-1\} \\  T_{11}
		\end{matrix}
		\arrow[r, "d_1"] &
		T_{12}
	\end{tikzcd}\\
	$L_2 \qquad \cong \qquad$
	\begin{tikzcd}
		T_{23}\{-2\}
		\arrow[r, "d'_2"] &
		\begin{matrix}
			T_{33}\{-2\} \\ T_{13}\{-1\}
		\end{matrix}
		\arrow[r, "d'_1"] &
		T_{23}\{-1\}
	\end{tikzcd}
\end{center}
where
$d_2 = \vcenter{\hbox{\includegraphics{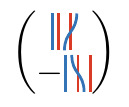}}}$, $d_1 = \vcenter{\hbox{\includegraphics{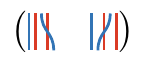}}}$,  and $d'_2 = \vcenter{\hbox{\includegraphics{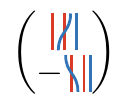}}}$, $d'_1 = \vcenter{\hbox{\includegraphics{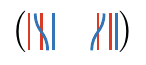}}}.$

To show homotopy equivalence, we must find chains maps $f$ and $f'$, satisfying that $f' \circ f$ is homotopic to the identity on $L_1$ and $f \circ f'$ is homotopic to the identity on $L_2$.

The chain map $f$ is found by asking for the maps in the following diagram to commute:
\begin{center}
	\begin{tikzcd}
		T_{12}\{-1\}
		\arrow[r, "d_2"] \arrow[d, "f_2"] &
		\begin{matrix}
			T_{13}\{-1\} \\  T_{11}
		\end{matrix}
		\arrow[r, "d_1"] \arrow[d, "f_1"] &
		T_{12} 
		\arrow[d, "f_0"] \\
		T_{23}\{-2\}
		\arrow[r, "d'_2"] &
		\begin{matrix}
			T_{33}\{-2\} \\ T_{13}\{-1\}
		\end{matrix}
		\arrow[r, "d'_1"] &
		T_{23}\{-1\}
	\end{tikzcd}
\end{center}
One finds easily that the components of $f$ are $f_2 = \vcenter{\hbox{\includegraphics{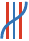}}}$, $f_1 = \vcenter{\hbox{\includegraphics{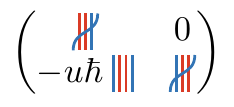}}}$ and $f_0 = \vcenter{\hbox{\includegraphics{pdf_figures/su2_markov_f0.pdf}}}.$ The chain map $f'$ is found from a similar commutative diagram and has components $f'_2 = -\vcenter{\hbox{\includegraphics{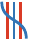}}}$, $f'_1 = \vcenter{\hbox{\includegraphics{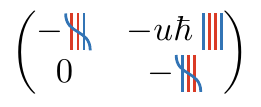}}}$, and $f'_0 = -\vcenter{\hbox{\includegraphics{pdf_figures/su2_markov_f0p.pdf}}}.$

\subsubsection{}
To prove that $f \circ f'$ and $f' \circ f$ are each homotopic to the identify on corresponding brane, let
\beq\label{twogs}
g = f \circ f' - (u \hbar)^2 \cdot id_{L_2}, \qquad
g' = f' \circ f -  (u \hbar)^2\cdot id_{L_1}.
\eeq
We want to find chain homotopies 
$s$ and $s'$
such that 
\beq g = \delta' \circ s + s \circ \delta', \qquad g' = \delta \circ s' + s' \circ \delta,
\eeq
where $\delta$ and $\delta'$ are the differentials in the resolutions of the two branes. 
 The first of these chain homotopies is given by the following diagram:
\begin{center}
	\begin{tikzcd}
		T_{23}\{-2\}
		\arrow[r, "d'_2"] \arrow[d, "g_2"] &
		\begin{matrix}
			T_{33}\{-2\} \\  T_{13}\{-1\}
		\end{matrix}
		\arrow[r, "d'_1"] \arrow[d, "g_1"] \arrow[dl, "s_1"] &
		T_{23}\{-1\} 
		\arrow[d, "g_0"] \arrow[dl, "s_0"] \\
		T_{23}\{-2\}
		\arrow[r, "d'_2"] &
		\begin{matrix}
			T_{33}\{-2\} \\  T_{13}\{-1\}
		\end{matrix}
		\arrow[r, "d'_1"] &
		T_{23}\{-1\}
	\end{tikzcd}
\end{center}
where $g$ is given by 
\begin{align*}
	& g_2 = -u\hbar \; \vcenter{\hbox{\includegraphics{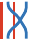}}}
	&& g_1 = \vcenter{\hbox{\includegraphics{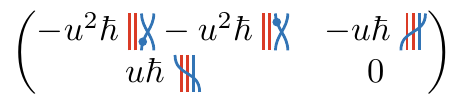}}}
	&& g_0 = -u\hbar \; \vcenter{\hbox{\includegraphics{pdf_figures/su2_markov_g0.pdf}}}.
\end{align*}
The needed chain homotopy $s$ is given by
\beq\nonumber
 s_1 = \vcenter{\hbox{\includegraphics{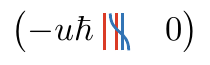}}}
	\qquad s_0 = \vcenter{\hbox{\includegraphics{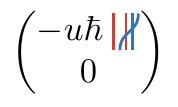}}}.
\eeq
One can then explicitly verify that the first of two equalities in \eqref{twogs} holds. Similarly, we find that $s'$ is given by
$s'_1 = \vcenter{\hbox{\includegraphics{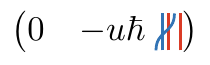}}}$ and $s'_0 = \vcenter{\hbox{\includegraphics{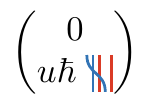}}}.$

Provided $u\hbar \neq 0$, one can always rescale $f$, $f'$, $g$, and $g'$ to set the factor $(u \hbar)^2$ to $1$. This thus completes the proof that the swiping move holds in the ${\mathfrak{su}_2}$ theory. The move does not hold if either of the two components of the divisor $D_O$ is deleted, corresponding to setting either $u$ or $\hbar$ to zero.

\section{Signs in the category of twisted complexes}\label{A}
The most technical aspect of the construction of a consistent assignment of signs, which allows us to obtain link homologies over ${\mathbb Z},$ and not merely over ${\mathbb Z}_2$ coefficients. The appropriate way to do this follows the familiar construction from standard treatments of derived Fukaya categories, for example in \cite{Seidel, Fukaya}. In our setting, there is a slight simplification in the construction, since the algebra $A$ is at most a differential graded algebra, and not a general $A_{\infty}$ algebra. Because the sign assignments are so crucial for us, we will give a self contained summary. We will suppress equivariant gradings, since they play no role for this aspect of the story.

This section is aimed at the ${\mathfrak{gl}_{1|1}}$ theory, whose strands algebra is a differential graded algebra. The ${\mathfrak{su}_2}$
algebra is an ordinary associative algebra, with trivial differential, all of whose elements have Maslov degree $0$ so the discussion of signs simplifies vastly.

\subsection{Differential-graded algebra as an $A_{\infty}$ algebra}

A differential-graded algebra with multiplication $a\cdot b$ and a differential $\partial$ satisfying
\begin{eqnarray}\label{standardr}
\partial^2=0,\qquad \partial(a\cdot b)=\partial(a)\cdot b+(-1)^{M(a)}a\cdot \partial(b), \qquad (a\cdot b)\cdot c = a\cdot (b\cdot c),
\end{eqnarray}
where $M(a)$ is the Maslov degree of $a$. 

It originates from the $A_{\infty}$ algebra underlying the Fukaya category, after we identify 
\begin{eqnarray}\label{algebrasign}
    m_1(a)=(-1)^{M(a)}\partial(a),\qquad\qquad m_2(a,b)=(-1)^{M(b)}a\cdot b,
\end{eqnarray} 
where $m_1$ and $m_2$ are respectively, the Floer differential and product,
assuming the higher $A_{\infty}$ operations $m_{\ell>2} =0$ vanish.

Let us check, as an exercise, that the differential graded algebra relations in \eqref{standardr} coincide with the $A_{\infty}$ relations. 
We obviously have
\begin{eqnarray}
    m_1(m_1(a))=-\partial^2(a)=0,
\end{eqnarray}
and
\begin{eqnarray}\nonumber
m_1(m_2(a_2,a_1))&=&(-1)^{M(a_2)+2M(a_1)}\partial (a_2\cdot a_1)\\ \nonumber
&=&(-1)^{M(a_2)}(\partial (a_2)\cdot a_1+(-1)^{M(a_2)}a_2\cdot \partial (a_1))\\
&=&(-1)^{M(a_1)}m_2(m_1 (a_2), a_1)-m_2(a_2, m_1 (a_1)),
\end{eqnarray}
Finally, the associativity of the algebra is equivalent to
\begin{eqnarray}\nonumber
m_2(m_2(a_3,a_2),a_1)&=&(-1)^{M(a_1)+M(a_2)}(a_3\cdot a_2)\cdot a_1\\ \nonumber
&=&(-1)^{M(a_1)+M(a_2)}a_3\cdot (a_2\cdot a_1)\\
&=&(-1)^{M(a_1)}m_2(a_3,m_2(a_2, a_1)).
\end{eqnarray}
These are indeed the $A_{\infty}$ relations from \cite{Seidel}, restricted to a theory with  $m_{p>2}=0$.

\subsection{Pre-twisted complexes}\label{mps}
A pre-twisted complex $(V, \delta)$ is 
\begin{eqnarray}
(V,\delta)=\left (\bigoplus_i V_i\;[k_i],\,\sum_{ij} d_{i,j}\right )
\end{eqnarray}
 where $V_i$ are objects of an $A_{\infty}$ category, and $\delta:V\rightarrow V[1]$ is a collection of maps 
\begin{eqnarray}
d_{i,j}:  V_{i}[k_{i}] \rightarrow  V_{j}[k_j+1].
\end{eqnarray}
$d_{i,j}$ are morphisms between the objects of the $A_{\infty}$ category.

A morphism between a pair of pre-twisted complexes $(V_{\alpha}, \delta_{\alpha})$ and $(V_{\beta}, \delta_{\beta})$,  of degree $n$,
$A_{\alpha,\beta}: V_\alpha \rightarrow V_\beta[n]$
is simply a collection of maps $A_{\alpha, \beta} = \bigoplus_{i,j} a_{\alpha i,\beta j}$ between the summands of the form
\begin{eqnarray}
a_{\alpha i,\beta j}:  V_{\alpha i}[k_{i}]\rightarrow V_{\beta j}[k_j+n].
\end{eqnarray}
The sign assiggnments on the category of pre-twisted complexes are determined in \cite{Seidel}. (What we call the category of pre-twisted complexes is called the additive enlargment of the $A_{\infty}$ category in \cite{Seidel}.)
First, one modifies the $A_{\infty}$ morphisms  by replacing 
\begin{eqnarray}\label{addsign1}
a_{0,1} \rightarrow (-1)^{(n-1)k_0} a_{0,1},
\end{eqnarray}
whenever $a_{0,1}: V_{0}[k_0] \rightarrow  V_1[k_1+n]$.
 One further modifies the $A_\infty$ operations by introducing an extra sign, sending 
\begin{eqnarray}\label{addsign2}
m_i(a_{i-1,i},\dots,a_{0,1}) \rightarrow (-1)^{k_0}m_i(a_{i-1,i},\dots,a_{0,1}).
\end{eqnarray}
The maps $m_{i}^p(a_{i-1,i},\dots,a_{0,1})$ will be understood as modifications of the standard $A_{\infty}$ maps of the Fukaya category by the two additional signs from \eqref{addsign1} and \eqref{addsign2}.
The sign twisted maps satisfy the standard  $A_\infty$ relations  \cite{Seidel}. 

 \subsection{Twisted objects}
The category of twisted complexes, denoted by ${\rm Tw}A$ in \cite{Seidel}, is itself an $A_{\infty}$ category, whose operations we will denote by $m_\ell^{Tw}.$

Objects in the category are pre-twisted complexes $(V,\delta)$ introduced above, satisfying 
\begin{eqnarray}\label{twdiff}
m_1^{Tw}(\delta) \equiv m^{p}_1(\delta)+m^{p}_2(\delta,\delta)=0.
\end{eqnarray}
A morphism $a$ of degree $n$ from $(V_1,\delta_1)$ to $(V_2,\delta_2)$ in the category of twisted complexes,  an element of $hom_{\rm{Tw}(A)}(V_1, V_2[n])$, is a morphism from $V_1$ to $V_2[n]$ in the category of pre-twisted complexes, on which
the twisted differential of $
{\rm Tw} A$ acts by
\begin{eqnarray}\label{m1t}
m_1^{Tw}( a)=m^{p}_1(a)+m^{p}_2(a,\delta_1)+m^{p}_2(\delta_2,a).
\end{eqnarray}
Further, if $b$ is a map from $(V_0,\delta_0)$ to $(V_1,\delta_1)$ of any degree,
\begin{eqnarray}\label{m2t}
m_2^{Tw}( a,b)=m^{p}_2(a,b),
\end{eqnarray}
because we assumed $A$ is a dga, with only $m_1$ and $m_2$ non-vanishing.
If $A$ were a general $A_{\infty}$ algebra all its higher $A_{\infty}$ operations would also contribute.

The derived category of the $A_{\infty}$ category, or the derived Fukaya category in the text, is the cohomology category $H^0({\rm Tw}A)$, whose objects are the same as in ${\rm Tw}A$, but whose morphisms are cohomology classes of $m_1^{\rm Tw}$ \cite{Seidel}.  
Rather than a complication, signs introduced in this section are necessary for the category of twisted complexes ${\rm Tw}A$ to have the standard sign assignments. With the added signs, the pre-twisted maps   
$A_{\alpha,\beta}:V_\alpha\rightarrow V_\beta[n]$
become honest cohomological degree $n$ maps between objects in the category of twisted complexes ${\rm Tw}A$.
\section{Signs in the differential}\label{ssigns}
In this section, we explain how to assign signs to maps in $\delta$ for a brane in $\MDy$. Having a consistent sign assignment is crucial as we want a homology theory with a ${\mathbb Z}$-valued homological grading. In our case, it suffices to explain how to assign signs to the differential $\delta_O$ of a brane on $Y_O$, which is $Y$ with the divisor $D_O$ deleted.
Having understood how to assign signs in $\delta_O$, the deformation corresponding to filling in $D_O$ determines the signs in the full differential $\delta$ as well.

A special feature of the differential $\delta_O$ is that the corresponding complex is a product of one dimensional complexes, as explained in section \ref{Algorithm}. Assigning signs to a product complex is essentially a standard problem which is easily solved. The only subtlety is that, when the complex is a twisted, rather than an ordinary complex, there are additional signs that enter, sourced by  signs from  section \ref{A}.

\subsection{Signs in one dimension}
We will start our discussion on $Y_{O}$ which is one dimensional.
As explained in section \ref{Algorithm}, a compact brane $L$ of topology of $S^1$ has the resolution $(L(T), \delta_O)$ which may be described in terms of a one-dimensional periodic toric grid. The sites of the grid are the direct summands in $L(T)= \bigoplus_{i} T_i$, and $\delta_O$ is the collection of maps between them:
\beq\label{1ds}\begin{tikzcd}
T_{1} \arrow[swap]{r} & T_{2} \arrow[swap]{r} \arrow[swap]{l} & \dots \arrow[swap]{r} \arrow[swap]{l} &  T_{n}  \arrow[swap]{r} \arrow[swap]{l} & \cdot \arrow[swap]{l}
\end{tikzcd}
\eeq
We do not specify directions of arrows since they are example-dependent and irrelevant in our present discussion. The one dimensional geometry of $Y_O$ ensures that the complex \eqref{1ds} is a linear chain of maps.

A special feature of the theory on $Y_O$ is that the differential $\delta_O$ squares to zero for every choice of signs of the $n$ maps in the grid. Since gauge transformations of the form $T_{i}\rightarrow -T_{i}$ do not change the brane, there is a single choice of sign that is meaningful, given by the product of all signs in the grid. As discussed in section \ref{sec:signs1}, the ${\mathbb Z}_2$ worth of choices of differentials for a brane of $S^1$ topology should correspond to the choice of the spin structure of the Lagrangian brane $L$ which the complex $(L(T), \delta_{O})$ describes.

\subsubsection{}
If the Lie algebra is ${\mathfrak{su}_{2}}$ only one sign choice deforms from $Y_O = Y_{0,0}$ with both the diagonal and the punctures deleted, to $Y_{u, 0}$, where only the diagonal is deleted, or to ${\cal Y}_{u,0}$, the upstairs space. The choice that deforms to $u\neq 0$ is has a product of all the signs given by $(-1)^{1+n/2}$ 
where $n$ is number of $T$-branes in the complex. One shows this by starting with an elementary figure eight brane, for which $n$ is four, and considering the effect of braiding. This is in line with the identification of the ${\mathbb Z}_2$ choice of signs with that of the spin structures on $L$. The deformation $u\neq 0$, corresponds from perspective of $L$, to filling in the disk which the brane bounds and which picks out a unique spin strucutre that extends to the disk.
\subsubsection{}
If the Lie algebra is ${\mathfrak{gl}_{1|1}}$ and $Y$ is one dimensional, the corresponding a deformation does not exist. The ${\mathbb Z}_2$ worth of choices of signs in the full theory reflects the ${\mathbb Z}_2$ worth of choices spin structure on the brane whose topology is ${S}^1$, which is not contractible in $Y$.

\subsection{Signs in ${\mathfrak{su}_2}$ complexes}\label{su2cs}
Since the complex describing the brane $(L(T), \delta_O)$ on $Y_O$ is a product of ordinary complexes when the Lie algebra is ${\mathfrak{su}_2}$, there is a simple, standard construction for assigning signs so that $\delta_O$ squares to zero, which  works as follows.

\subsubsection{}
The best way to describe the construction is recursive.  We  know how to assign signs if the brane is one dimensional. 
Suppose we are given a $d$ dimensional brane $(L_d(T), \delta_d)$. The brane has summands which are labeled by sites of a $d$-dimensional toric grid
$$L_d(T) = \bigoplus_{i_1 ,...,i_d} T_{i_1 ...i_d},$$ 
with a product differential $\delta_d$   which squares to zero, $\delta_d^2=0$ ($\delta_d$ stands for $\delta_O$ for the $d$ dimensional brane). We suppressed the gradings of the $T$-branes. 

We will explain how to assign the signs on the product $L_{d+1} = L_d\times L_1$ of $(L_d(T), \delta_d)$ with another one-dimensional Lagrangian $(L_1(T), \delta_1)$, where 
$$L_1(T) = \bigoplus_{i_{d+1}} T_{i_{d+1}}.$$
The product brane $L_{d+1}(T)$ lives on a $d+1$-dimensional grid. We claim its differential is given by
\beq\label{productsign}\delta_{d+1} = \sum_{i_1 ,...,i_d, i_{d+1}} (-1)^{i_1 +...+i_d} {\mathds{1}}_{i_1 ,...,i_d} \otimes \delta_{i_{{d+1}}} + \sum_{i_{d+1}} \delta_{d} \otimes {\mathds{1}}_{i_{d+1}}
\eeq 
The sign twist in \eqref{productsign} preserves the ${\mathbb Z}_2$ sign choice we made for the $d+1$-th brane, because for a brane of $S^1$ topology, each one dimensional complex has an even number $T$-branes, and of maps connecting them, and we changed the overall sign only. 

To check that the differential squares to zero, we need to show it squares to zero around every square in the grid. Per assumption, it does so for all squares that lie entirely in the first $d$ directions, those associated with $L_d$. Due to the alternating sign we introduced in \eqref{productsign}, it also squares to zero for squares with vertices 
\beq\label{Tv}
T_{i_1,...,i_j,...,i_d} \times T_{i_d},\;\; 
T_{i_1,...,{i_{j}+1},...,i_d} \times T_{i_{d}},\;\; 
T_{i_1 ,...,i_j ,.,i_d} \times T_{i_{d}+1}, \;\;
T_{i_1 ,...,{i_{j}+1},.,i_d} \times T_{i_{d}+1}.
\eeq
which involve the $d+1$-st direction associated to $L_1$.

\subsection{Signs in $\mathfrak{gl}_{1|1}$ complexes}\label{gl11cs}
Twisted complexes introduce additional signs which we reviewed in appendix \ref{A}. For ordinary complexes, the additional sign we introduced in \eqref{productsign} suffices to ensure the differential squares to zero, because products the of elements, going two ways around a square with vertices in \eqref{Tv}, commute. Thinking of an ordinary complex as a special case of a twisted one, the products commute because all elements of differential have Maslov index $0$.

For a twisted complex, differential consists of maps of arbitrary Maslov index. In taking products, one gets additional sources of signs which we summarized in appendix \ref{A}. Due to these signs, there are additional signs that enter in forming the product complex, to get a differential $\delta_O$ that squares to zero.

\subsubsection{}
The first source of signs comes from \eqref{algebrasign} which says that $a\cdot b = (-1)^{M(b)} m_2(a,b)$. For the differential $\delta_O$ of the product complex to square to zero, we need to cancel this sign.

Consider one of the new squares formed in the $d+1$ dimensional grid with vertices in \eqref{Tv}. If Maslov degrees of branes
$T_{i_1,...,{i_{j}+1},...,i_d} \times T_{i_{d}} $ and $T_{i_1 ,...,i_j ,.,i_d} \times T_{i_{d}+1} $, located at a pair of opposite vertices of the square, are both odd or both even, then the sign in \eqref{algebrasign} does not introduce any new signs, and the differential squares to zero as before. If however one is odd, the other even, the product of maps around the square anti-commutes, instead of commuting, and we have a sign that needs cancelling.

To cancel the sign, it suffices to multiply every map in the complex by $(-1)^{M(T)}$ where $M(T)$ is the Maslov index of the thimble $T$ at the grid site where the map lands at. If Maslov degrees of 
$T_{i_1,...,{i_{j}+1},...,i_d} \times T_{i_{d}} $ and $T_{i_1 ,...,i_j ,.,i_d} \times T_{i_{d}+1} $ branes are both odd, or both even, the new sign introduces an even number of sign changes around the square, so nothing changes.  If Maslov index of one brane is even and the other odd, we get an extra minus sign which cancels the sign from \eqref{algebrasign} and which ensures the new differential 
squares to zero. The signs in the differental are no longer of the simple form in \eqref{productsign}, but they are a simple modification of it.

\subsubsection{}
There are two more sources of signs in the differential of the twisted complex, due to the fact the ordinary $A_{\infty}$ products get replaced with the twisted ones from section \ref{mps}.
One can show that both of these additional sign contributes equally to the two paths around the square with vertices \eqref{Tv}. As a result, it does not lead to any further modification of the differential $\delta_O$ from previous subsection.


\newpage
\begin{thebibliography}
%\normalsize\vskip0.5cm
\raggedright

\bibitem{A1} M.~Aganagic, ''Knot Categorification from Mirror Symmetry, {\it Part I: Coherent Sheaves}", arXiv:2004.14518. 

\bibitem{A2} M.~Aganagic, ''Knot Categorification from Mirror Symmetry, {\it Part II: Lagrangians}", arXiv:2105.06039. 

\bibitem{A3} M.~Aganagic, ''Knot Categorification from Mirror Symmetry, {\it Part III: String theory origins}", to appear. 

\bibitem{AICM}
M.~Aganagic,
``Homological Knot Invariants from Mirror Symmetry,''
[arXiv:2207.14104 [math.GT]].

\bibitem{ADZ} M.~Aganagic, I.~Danilenko, Y. Li, P.~Zhou and V.~Shende, ``Hecke algebras from Floer homology in Couloumb branches,'' to appear.

\bibitem{AFO}{M.~Aganagic, E.~Frenkel and A.~Okounkov,
``Quantum $q$-Langlands Correspondence,''
Trans. Moscow Math. Soc. \textbf{79}, 1-83 (2018),
arXiv:1701.03146.}

\bibitem{ATh}{M.~Aganagic, M.~McBreen, V.~Shende and P.~Zhou, to appear.}

\bibitem{ESE}
M.~Aganagic and A.~Okounkov,
``Elliptic stable envelopes,''
J. Am. Math. Soc. \textbf{34} (2021) no.1, 79-133
[arXiv:1604.00423 [math.AG]].

\bibitem{AO}
M.~Aganagic and A.~Okounkov,
``Quasimap counts and Bethe eigenfunctions,''
Moscow Math. J. \textbf{17} (2017) no.4, 565-600
[arXiv:1704.08746 [math-ph]].

\bibitem{AT}
M.~Aganagic and S.~Tamagni,
``Stable envelopes for Verma modules from vortex moduli spaces,'' to appear.

\bibitem{Alexander}
J.~W.~Alexander,  ``Topological invariants of knots and links". Transactions of the American Mathematical Society. 30 (1923): 275–306. 

\bibitem{Douglas} 
  P.~S.~Aspinwall {\it et al.},
  ``Dirichlet branes and mirror symmetry,'' 681 pages,
Clay Mathematics Monographs, 4
Providence, RI: AMS (2009)

\bibitem{Auroux}
D.~Auroux, ``A beginner's introduction to Fukaya categories", 
arXiv:1301.7056.

\bibitem{Auroux1} D.~Auroux, ``Fukaya categories of symmetric products and bordered Heegaard-Floer homology." arXiv: 1001.4323.

\bibitem{Auroux2}D.~Auroux, ``Fukaya categories and bordered Heegaard-Floer homology", arXiv:1003.2962.


\bibitem{AurouxD} D.~Auroux, ``Special Lagrangian fibrations, wall-crossing, and mirror symmetry", arXiv:0902.1595.

\bibitem{DBN} D. Bar-Natan,``On Khovanov’s categorification of the Jones polynomial",
Algebraic and Geometric Topology Volume 2 (2002) 337–370.

\bibitem{bigelow} S.~Bigelow, ``A homological definition of the Jones polynomial",  Geom. \& Top. Monogr. 4, 29-41, 2002. math.GT/0201221.
 
\bibitem{bigelowN} S.~Bigelow, ``A homological definition of the HOMFLY polynomial", Algebr. Geom. Topol. 7(3): 1409-1440 (2007)

\bibitem{BK1} R.~Bezrukavnikov, D.~Kaledin, "Fedosov quantization in positive characteristic,''  2005, arXiv:math/0501247. 


\bibitem{BK2}R.~Bezrukavnikov, ``Noncommutative Counterparts of the Springer Resolution," math/0604445.

\bibitem{Birman}
J.~Birman, ``On the stable equivalence of plat representations of knots and links", Canad. J. Math. 28 (1976), no. 2, 264-290.

\bibitem{caldararu} A.~Caldararu, ``Derived categories of sheaves: a skimming," arXiv:math/0501094.

\bibitem{CK1} S.~Cautis, J.~Kamnitzer, ``Knot homology via derived categories of coherent sheaves I, $sl(2)$ case", 2007, arXiv:math/0701194

\bibitem{CK2}S.~Cautis, J.~Kamnitzer, ``Knot homology via derived categories of coherent sheaves II, ${sl}_{m}$ case", J.\ 2008, Inventiones Mathematicae, 174, 165.  

\bibitem{Cho} C.-H.~Cho, ``Holomorphic disc, spin structures and Floer cohomology of the Clifford torus", arXiv.math/0308224

\bibitem{Cho2}
C.-H.~Cho and Y.-G.~Oh,
``Floer cohomology and disc instantons of Lagrangian torus fibers in Fano toric manifolds,'' Asian J. Math. \textbf{10} (2006), arXiv:math/0308225.



\bibitem{CF}
L.~Crane and I.~Frenkel,
``Four-dimensional topological field theory, Hopf categories, and the canonical bases,''
J. Math. Phys. \textbf{35} (1994), hep-th/9405183.

\bibitem{Danilenko}
  I.~Danilenko, ``Slices of the Affine Grassmannian and Quantum
    Cohomology", PhD thesis, Columbia University. 
    
\bibitem{Drinfeld}
V.~G.~Drinfeld, ``Quasi-Hopf Algebras and Knizhnik-Zamolodchikov Equations". In: Belavin A.A., Klimyk A.U., Zamolodchikov A.B. (eds) Problems of Modern QFT,  Research Reports in Phys. Springer, (1989).

\bibitem{EFK} P.~Etingof, I.~Frenkel, and A.~A.~Kirillov, Jr.,  ``Lectures on representation theory and Knizhnik-Zamolodchikov equations", Math. Surveys and Monographs, vol. 58, Amer. Math. Soc., Providence, RI, 1998, 198 pp.
 
\bibitem{EG}
P.~Etingof, N.~Geer,
``Monodromy of trigonometric KZ equations", arXiv:math/0611003.

\bibitem{FF}B.~Feigin, E.~Frenkel, ``A family of representations of affine Lie algebras", Uspekhi Matem. Nauk, 43:5 (1988), [English translation: Russ. Math. Surv., 43:5 (1988)].

\bibitem{Floer} A.~Floer, ``Morse Theory For Lagrangian Intersections", J.Diff.G, 28 (1988).

\bibitem{qKZ}
I.B.~Frenkel, N.Yu.~Reshetikhin, ``Quantum affine algebras and holonomic difference equations,'' Comm. Math. Phys. {\bf 146} (1992) 1-60.

\bibitem{Fukaya} K.~Fukaya, ``Floer Homology and Mirror Symmetry II",  Advanced Studies in Pure Mathematics 34, 2002
Minimal Surfaces, Geom. Ana. and Symp. Geom.

\bibitem{FOOO}K.~Fukaya, Y.-G.~Oh, H.~Ohta and K.~Ono, Lagrangian intersection Floer theory-anomaly
and obstruction, Kyoto University preprint, 2000.

\bibitem{GW}
D.~Gaiotto and E.~Witten,
``Knot Invariants from Four-Dimensional Gauge Theory,''
Adv. Theor. Math. Phys. \textbf{16} (2012) no.3, arXiv:1106.4789.

\bibitem{GMW}D.~Gaiotto, G.~W.~Moore and E.~Witten,
  ``Algebra of the Infrared: String Field Theoretic Structures in Massive ${\cal N}=(2,2)$ Field Theory In Two Dimensions,''
  ar{X}iv:1506.04087.

\bibitem{GM}
D.~Galakhov and G.~W.~Moore,
``Comments On The Two-Dimensional Landau-Ginzburg Approach To Link Homology,''
arXiv:1607.04222.
\bibitem{GV}R.~Gopakumar and C.~Vafa,
``M theory and topological strings. 1. and 2.,''
hep-th/9809187 and
hep-th/9812127.




\bibitem{GSV}
S.~Gukov, A.~S.~Schwarz and C.~Vafa,
``Khovanov-Rozansky homology and topological strings,''
Lett. Math. Phys. \textbf{74} (2005),
arXiv:hep-th/0412243].


\bibitem{Urs} P.~Hilton, U.~Stammbach, ``A course in homological algebra," Springer-Verlag, New York, 1971, Graduate Texts in Mathematics, Vol. 4.

\bibitem{HIV}K.~Hori, A.~Iqbal and C.~Vafa, ``D-branes and mirror symmetry,'' hep-th/0005247.


\bibitem{HoriICM}
K.~Hori,
``Mirror symmetry and quantum geometry,'' arXiv:hep-th/0207068.

\bibitem{Horie}M.~Herbst, K.~Hori, D.~Page, ``Phases Of N=2 Theories In 1+1 Dimensions With Boundary," arXiv.0803.2045

\bibitem{Yagi}
N.~Ishtiaque, S.~F.~Moosavian, S.~Raghavendran and J.~Yagi,
``Superspin chains from superstring theory,''
SciPost Phys. \textbf{13} (2022) no.4, 083, arXiv:2110.15112.

\bibitem{JonesVFR}V.~F.~R.~ Jones, ``A polynomial invariant for knots via von Neumann
algebras", Bull. Amer. Math. Soc. 12.

\bibitem{Kac}
V.~G.~Kac,
``Lie Superalgebras,''
Adv. Math. \textbf{26} (1977).
%885 citations counted in INSPIRE as of 20 May 2023

\bibitem{Kh}M.~Khovanov, ``A categorification of the Jones polynomial", Duke Math. J. 101 (2000), no. 3.
	
\bibitem{KhICM}M.~Khovanov, ``Link homology and categorification", arXiv:math/0605339.  
 
\bibitem{KL1}M.~Khovanov, A.~D.~Lauda, ``A diagrammatic approach to categorification of quantum groups" I.\ arXiv:0803.4121.

 \bibitem{KL2} M.~Khovanov, A.~D.~Lauda, ``A diagrammatic approach to categorification of quantum groups" II.\ arXiv:0804.2080.



\bibitem{KR1} M.~Khovanov, L.~Rozansky, ``Matrix factorizations and link homology", arXiv.math/0401268.

\bibitem{KR2} M.~Khovanov, L.~Rozansky, ``Matrix factorizations and link homology II", arXiv.math/0505056.




\bibitem{KS}M.~Khovanov, P.~Seidel,  ``Quivers, Floer cohomology, and braid group actions", math/0006056;
J. Amer. Math. Soc. 15 (2002), no. 1.

\bibitem{KZ}V.~G.~Knizhnik and A.~B.~Zamolodchikov, ``Current Algebra and Wess-Zumino Model in Two-Dimensions,''
  Nucl.\ Phys.\ {\bf B247} (1984).
 
\bibitem{Koh}T.~Kohno, ``Monodromy representations of braid groups and Yang Baxter equations". Ann. Inst. Fourier 37, 139 160 (1987)
 
\bibitem{Ko}M.~Kontsevich, ``Homological Algebra of Mirror Symmetry,''  Chatterji, S.D. (eds) Proc. of the ICM, 1994. Birkhäuser, arXiv:alg-geom/9411018.

\bibitem{Lawrence1}R.~J.~Lawrence, ``Homology representations of braid groups", D.Phil. Thesis, University of Oxford (June 1989).

\bibitem{Lawrence2}R.~J.~Lawrence, ``A functorial approach to the one-variable Jones Polynomial", J. Diff. Geom. 37 (1993) 689-710.

\bibitem{Lawrence3}R.~J.~Lawrence, ``The homological approach applied to higher representations," Harvard preprint (1990), available at http://www.ma.huji.ac.il/~ruthel/

\bibitem{L}R.~Lipshitz, ``A cylindrical reformulation of Heegaard Floer homology," Geom. Topol., 10:955–1097, 2006.

\bibitem{LOT}R.~Lipshitz, P.~Ozsvath, D.~Thurston, ``Bordered Heegaard Floer homology: invariance and
pairing," arXiv:0810.0687.

\bibitem{Lit}
A.~Litvinov and L.~Spodyneiko,
``On W algebras commuting with a set of screenings,''
JHEP \textbf{11} (2016), 138, arXiv:1609.06271.

\bibitem{grid} 
  C. Manolescu, P. S. Ozsvath, and S. Sarkar, ``A combinatorial description of knot Floer homology", Ann. of Math. (2) 169 (2009), no. 2.

\bibitem{M} C.~Manolescu, ``Nilpotent slices, Hilbert schemes, and the Jones polynomial," arXiv:math/0411015.

\bibitem{MR} C.~Manolescu, ``An introduction to knot Floer homology." arXiv:1401.7107.

\bibitem{MO}D.~Maulik and A.~Okounkov,
  ``Quantum Groups and Quantum Cohomology,''
  arXiv:1211.1287.

\bibitem{NS}
N.~Nekrasov and S.~Shatashvili, ``Supersymmetric vacua and Bethe ansatz", Nuclear Phys. B Proc. Suppl. 192/193 (2009). 

\bibitem{N}
N.~Nekrasov,
``Superspin chains and supersymmetric gauge theories,''
JHEP \textbf{03} (2019), 102
arXiv:1811.04278.

 \bibitem{OV} 
  H.~Ooguri and C.~Vafa,
  ``Knot invariants and topological strings,''
  Nucl.\ Phys.\ {\bf B577} (2000),
  arXiv:hep-th/9912123.
  
\bibitem{OF} Y.-G. Oh and K. Fukaya, ``Floer homology in symplectic geometry and mirror symmetry", Proceedings for ICM 2006 Madrid.

\bibitem{oh} T.~Ohtsuki, ``Quantum Invariants", World Scientific, 2002.

\bibitem{Or}
A.~Okounkov,
``Lectures on K-theoretic computations in enumerative geometry,'' arXiv:1512.07363.

\bibitem{OkS}
A.~Okounkov and A.~Smirnov,
``Quantum difference equation for Nakajima varieties,''
Invent. Math. \textbf{229} (2022) no.3, 
arXiv:1602.09007.

\bibitem{Or2} A.~Okounkov, ``Enumerative geometry and geometric representation theory". arXiv.1701.00713

\bibitem{Or3}  A.~Okounkov, ``Nonabelian stable envelopes, vertex functions with descendents, and integral solutions of $ q $-difference equations," arXiv:2010.13217.

\bibitem{OS0a}P.~S.~Ozsvath and Z.~Szabo, ``Holomorphic disks and topological invariants for closed three-manifolds", arXiv:math/0101206.

\bibitem{OS0b}P.~S.~Ozsvath and Z.~Szabo, ``Holomorphic disks and three-manifold invariants: properties and applications", arXiv:math/0105202.

\bibitem{OS0c}P.~S.~Ozsvath and Z.~Szabo, ``
Holomorphic disks, link invariants and the multi-variable Alexander polynomial", Algebraic \& Geometric Topology 8 (2008).

\bibitem{OS1}P.~S.~Ozsvath and Z.~Szabo, ``An overview of knot Floer homology," arXiv:1706.07729.

\bibitem{OS}P.~S.~Ozsvath and Z.~Szabo,  ``Algebras with matchings and link Floer homology," arXiv:2004.07309.

\bibitem{Rasmussen}J.~A.~Rasmussen, ``Floer homology and knot complements," PhD thesis, Harvard University, 2003.

\bibitem{RT}N.~Yu.~Reshetikhin and V. ~G. ~Turaev, ``Ribbon graphs and their invariants derived from quantum groups", Comm. Math. Phys. 127(1): 1-26 (1990).

\bibitem{RR}
R.~Rimanyi and L.~Rozansky,
``New Quiver-Like Varieties and Lie Superalgebras,''
Commun. Math. Phys. \textbf{400} (2023) no.1,  arXiv:2105.11499.

\bibitem{R}R.~Rouquier, ``2-Kac-Moody algebras," 
arXiv:0812.5023. 

\bibitem{SV1}V.~Schechtman and A.~Varchenko, ``Quantum groups and
  homology of local systems,'' in Alg. Geom. and Anal.
  Geom., ICM-90 Sat. Conf. Proceedings, Springer, 1991.

\bibitem{SV2}V.~Schechtman and A.~Varchenko, ``Arrangements of hyperplanes and Lie algebra homology,'' in  Inventiones mathematicae 106, 139-194 (1991)

\bibitem{Seidel}P.~Seidel, ``Fukaya categories and Picard-Lefschetz theory", 2007, book, 292 pp., Vol. 10 of Zurich lectures in advanced mathematics.

\bibitem{Seideld} P.~Seidel, ``Fukaya categories and deformations", 	arXiv:math/0206155, To appear in the Proceedings of the Beijing ICM.

 \bibitem{SS} P.~Seidel, I.~Smith, ''A link invariant from the symplectic geometry of nilpotent slices,'' math/0405089.
 
\bibitem{SSo} P.~Seidel, J.~P.~Solomon, ``Symplectic cohomology and q-intersection numbers", arXiv:1005.5156

\bibitem{Thomas}
R.~P. ~Thomas, ``An exercise in mirror symmetry", Proceedings of the International Congress of Mathematicians, Hyderabad, India, 2010. 

\bibitem{webster}B.~Webster,``Knot invariants and higher representation theory'', arXiv:1309.3796.

\bibitem{W1}B.~Webster, ``Coherent sheaves and quantum Coulomb branches I: tilting bundles from integrable systems,'' arXiv:1905.04623.

\bibitem{W2}B.~Webster, ``Coherent sheaves and quantum Coulomb branches II: quiver gauge theories and knot homology," to appear.

\bibitem{WittenM}
 E.~Witten, ``Supersymmetry and Morse Theory", J. Diff. Geom., 17 (1982).

\bibitem{Jones}E.~Witten, ``Quantum Field Theory and the Jones Polynomial,'' Commun.\ Math.\ Phys.\  {\bf 121} (1989) 351.

\bibitem{WA}
E.~Witten,
``Analytic Continuation Of Chern-Simons Theory,''
AMS/IP Stud. Adv. Math. \textbf{50} (2011), 
arXiv:1001.2933.

\bibitem{WF}E.~Witten, ``Fivebranes and Knots,'' arXiv:1101.3216.

\end{thebibliography}
\end{document}